\documentclass[11pt]{article}

\usepackage[margin=1in]{geometry}

\usepackage{graphicx}
\usepackage{hyperref}
\hypersetup{colorlinks=true,linkcolor=black}

\usepackage{amsmath}
\usepackage{amssymb}
\usepackage{amsthm}
\usepackage{ascmac}
\usepackage{wrapfig}
\usepackage{pgf,tikz}
\usepackage{mathrsfs}
\usepackage{mathtools}
\usetikzlibrary{arrows}
\usepackage{comment}
\usepackage{enumerate}
\usepackage{booktabs}
\usepackage{rotating}
\usepackage{appendix}
\usepackage{multirow}
\usepackage{bbm}
\usepackage{framed}
\usepackage{enumitem}
\DeclareSymbolFont{lettersA}{U}{txmia}{m}{it}
\DeclareMathSymbol{\Indi}{\mathord}{lettersA}{'211}
\usepackage{here}
\usepackage{graphicx}
\usepackage{subfigure}
\usepackage[hang,small,bf]{caption}
\usepackage {textcomp}
\usepackage{array,arydshln}
\usepackage{amssymb}

\newtheoremstyle{mystyle}
{}
{}
{\normalfont}
{}
{\bfseries}
{.}
{ }
{}
\theoremstyle{mystyle}

\newtheorem{Thm}{Theorem}[section]
\newtheorem{prop}{Proposition}[section]
\newtheorem{Lem}{Lemma}[section]

\newtheorem*{Rem_n}{Remark}

\renewcommand{\vec}[1]{\boldsymbol{#1}}
\makeatletter
\renewcommand{\theequation}{\arabic{section}.\arabic{equation}}
\renewcommand{\thesection}{\arabic{section}}
\allowdisplaybreaks[4]

\makeatletter
\renewenvironment{proof}[1][\proofname]{\par
  \normalfont \topsep6\p@\@plus6\p@\relax
  \trivlist
  \item\relax
  {\itshape
  #1\@addpunct{.}}\hspace\labelsep\ignorespaces
}{
  \endtrivlist\@endpefalse
}
\makeatother

\usepackage[normalem]{ulem}

\renewcommand{\thefigure}{\thesection.\arabic{figure}}

\usepackage{bm}

\DeclareRobustCommand{\blackbigcirc}{\tikz[baseline=-0.6ex]{\fill (0,0) circle (1.05ex);}}
\makeatletter
\DeclareRobustCommand{\blackbigtriangledown}{\mathbin{\mathchoice
  {\mathbigtriangledown@style{1.0}}
  {\mathbigtriangledown@style{1.0}}
  {\mathbigtriangledown@style{0.7}}
  {\mathbigtriangledown@style{0.5}}
}}
\DeclareRobustCommand{\mathbigtriangledown@style}[1]{%
  \vcenter{\hbox{\tikz[scale=#1]{
    \fill (-0.42em, 0.36em) -- (0.42em, 0.36em) -- (0, -0.36em) -- cycle;
  }}}
}
\makeatother

\DeclareMathOperator*{\argmin}{argmin}

\DeclareMathOperator*{\sol}{solution}

\numberwithin{table}{section}

\usepackage{natbib}
\usepackage{apalike}

\makeatletter
  \@addtoreset{equation}{section}
\makeatother

\title{Semiparametric regression with \\stagewise empirical risk minimization}
\author{Daisuke Matsuno and Kanta Naito \\
\\
Graduate School of Information Sciences, Tohoku University, Japan}
\date{}

\begin{document}

\pagestyle{plain}

\maketitle
\begin{abstract}
Maintaining both interpretability and accuracy remains a fundamental challenge in regression analysis.
This paper addresses these two objectives through a semiparametric approach that combines a parametric crude guess with its nonparametric adjustment.
In particular, the nonparametric adjustment is implemented via a machine learning technique called ``stagewise empirical risk minimization algorithm.''
Once a set (dictionary) of functions (words) is prepared, this algorithm constructs a convex combination of words selected iteratively from the dictionary.
This selection is based on the empirical risk derived from the Bregman divergence, which enables the proposed estimators to be applied in broad settings, including generalized linear models.
This paper examines these proposed semiparametric estimators theoretically and numerically.
In the theoretical study, we derive a non-asymptotic error bound for the estimators constructed from a general data-dependent dictionary.
Among the concrete dictionaries proposed in this paper, some can be utilized to correct variable selection in LASSO-based parametric regression, which is investigated in the numerical study.
Moreover, simulation studies investigate practical performance of the proposed estimators,
while real data applications further confirm their predictive performance.
These studies indicate that the proposed semiparametric estimators can address the above challenge in several settings.
\end{abstract}

\section{Introduction}
\renewcommand{\thefigure}{\arabic{figure}}
\counterwithout{figure}{section}
Regression analysis is one of the most widely used statistical methods and has extensive applications across many fields.
It investigates functional relationships between output (response) and input (explanatory) variables, so that one can model and describe these relationships or can predict new response values, see \citet*{CH2006}, \citet*{K2014}, \citet*{JWDT2021}.

\begin{figure}[tbp]
	\centering
	\subfigure[Data points and estimators.]{\label{subfig:esti}
		\includegraphics[width=.47\columnwidth]{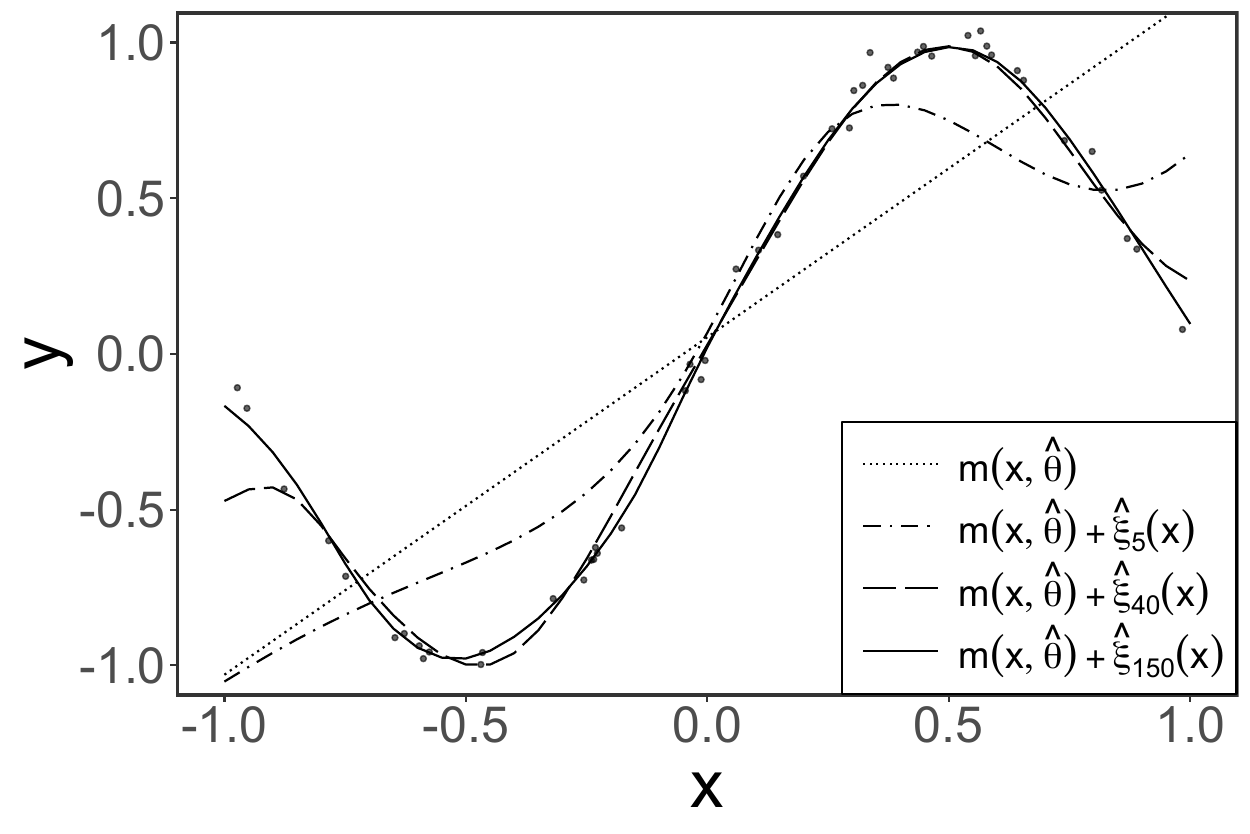}
	}
	\subfigure[Adjustment functions $\widehat{\xi}_I \left( x \right)$.]{\label{subfig:xi}
		\includegraphics[width=.45\columnwidth]{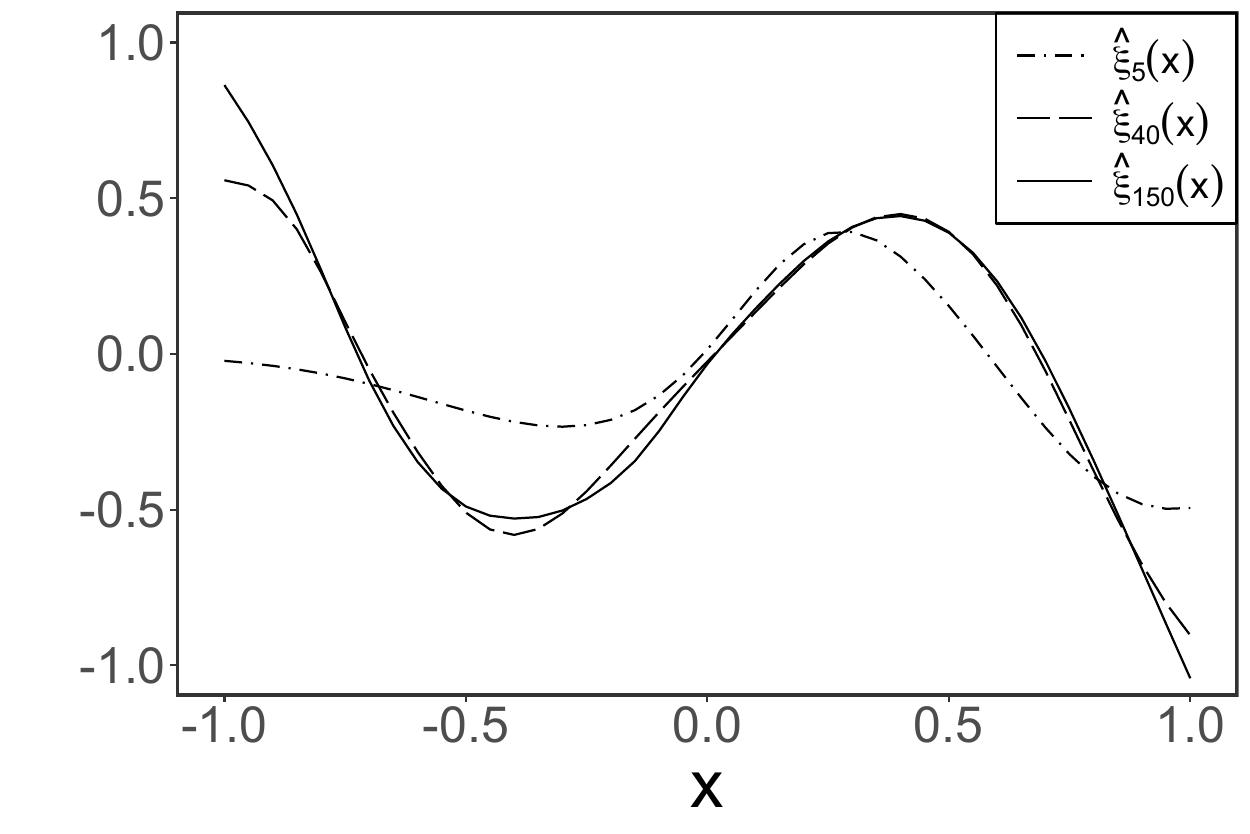}
	}
	\caption{
		An example of regression with our proposed method.
		``Parametric'' regression is conducted by using the least squares regression line $m \left( x, \widehat{\vec{\theta}} \right) = \widehat{\theta}_0 + \widehat{\theta}_1 x$.
		Its ``nonparametric'' adjustment function $\widehat{\xi}_I \left( x \right)$ is implemented via the ``stagewise empirical risk minimization algorithm'' with the number of iterations $I = 5, 40, 150$.
		Then we obtain a proposed semiparametric estimator $m \left( x, \widehat{\vec{\theta}} \right) + \widehat{\xi}_I \left( x \right)$ with additive adjustment.
	} \label{fig:esti_exa}
\end{figure}

In this paper, we propose a semiparametric regression method that seeks both interpretability and flexibility by incorporating a machine learning technique.
To outline the proposed method, Figure \ref{fig:esti_exa} illustrates its application to a simple estimation problem in a one-dimensional normal regression setting.
Our regression estimators are based on a ``parametric'' crude guess and its ``nonparametric'' adjustment:
panel \subref{subfig:esti} displays a scatterplot of the data points and the least squares regression line $m \left( x, \widehat{\vec{\theta}} \right) = \widehat{\theta}_0 + \widehat{\theta}_1 x$ with $\widehat{\vec{\theta}} = \left[ \widehat{\theta}_0 ~ \widehat{\theta}_1 \right]^\top$, an example of a parametric regression estimator,
 represented by a dotted line.
Here the parameters $\vec{\theta} = \left[ \theta_0 ~ \theta_1 \right]^\top$ in $m \left( x, \vec{\theta} \right) = \theta_0 + \theta_1 x$ are estimated to be $\widehat{\theta}_0 \thickapprox 0.05$ and $\widehat{\theta}_1 \thickapprox 1.08 > 0$,
suggesting that the data points globally increase with $x$ in a sense.
By using parametric estimators as in linear models, we can interpret such global behavior of the regression function based only on a model function $m \left( \vec{x}, \vec{\theta} \right)$ and estimated parameters $\widehat{\vec{\theta}}$.
However, as observed in panel \subref{subfig:esti}, the linear model above does not fit the data points well locally.
Then, allowing for model misspecification, we adjust this initial ``parametric'' regression with a ``nonparametric'' approach,
which enables estimators to capture broad classes of regression functions flexibly.
After this combination of parametric and nonparametric approaches, the proposed method is referred to as ``semiparametric.''
In this paper, we implement the adjustment via a machine learning technique called ``stagewise empirical risk minimization algorithm.''
Briefly, from a user-prepared set (dictionary) of functions (words), this algorithm selects one word in each iteration,
and then forms a convex combination of the selected words for estimation or prediction.
Based on this algorithm and using a dictionary introduced in Section \ref{sec:impl}, we now construct a term $\widehat{\xi}_I \left( x \right)$ for additive adjustment with the number of iterations $I = 5, 40, 150$, as presented in panel \subref{subfig:xi} of Figure \ref{fig:esti_exa}.
Then we compute a proposed semiparametric estimator $m \left( x, \widehat{\vec{\theta}} \right) + \widehat{\xi}_I \left( x \right)$ shown in panel \subref{subfig:esti},
demonstrating that the regression line is adjusted and that the proposed estimator seems to fit the data points better as $I$ increases.

Now we describe each of the parametric, nonparametric and semiparametric approaches.
The parametric approach is fundamental in regression analysis.
It models the target regression function as a function determined by a finite-dimensional parameter, as in linear models.
This approach reduces the problem of estimating a ``function'' to that of estimating a ``parameter vector'' contained in the model,
and parametric regression estimators are constructed by plugging an estimator for the parameters (such as the maximum likelihood estimator (MLE) in \citet*{MN1989}, \citet*{DS2018}) into the model function.
Its simple structure facilitates interpretation of the underlying regression function:
for example, whether it increases, decreases or hardly changes globally with some explanatory variables.
Particularly in linear regression, the LASSO estimator is known to estimate several coefficients for explanatory variables to be exactly 0, so that this estimator can be utilized for variable selection, see \citet*{T1996}, \citet*{HTW2015}, \citet{HQT2026}.
Although this parametric approach has the merits of simplicity and interpretability and is widely used in statistics,
this approach has a drawback:
the assumption that the underlying structure can be captured by a finite-dimensional parameter vector is sometimes very restrictive.
It is well known that parametric estimators generally lack consistency under model misspecification.

On the other hand, the nonparametric approach offers flexibility in terms of consistency.
This approach includes several methods for regression analysis, such as kernel smoothing, spline smoothing, gradient boosted regression trees (GBRTs).
In general, by appropriately tuning smoothing parameters, these methods can achieve consistency for broad classes of regression functions,
which makes the nonparametric approach advantageous.
However, the structure of nonparametric estimators is sometimes more difficult to interpret than that of parametric models.
In addition, several nonparametric methods suffer from data sparsity for high-dimensional explanatory variables, a phenomenon referred to as ``curse of dimensionality,'' see, for example, Chapter 3.2.6 of \cite{K2014}.

From the above observations, the parametric and nonparametric approaches each have their own advantages and disadvantages.
This motivates us to consider a fine combination of these two approaches, in order to make regression estimators both easily interpretable and flexible.
Specifically, we consider a semiparametric approach in which an initial crude estimation with the parametric approach is subsequently adjusted through the nonparametric approach.
The underlying idea can be found in several pioneering works.
Focusing on the additive adjustment above, this approach coincides with performing an additional regression on the residuals resulting from the initial regression.
This strategy appeared, for example, in Section 5C of \citet*{T1977}, which fits lines to an original data set and then to the resulting residuals.
Boosting also applies this strategy by iteratively fitting the residuals, see \citet*{F2001} and \citet*{BY2003}.
The semiparametric approach based on this idea has been explored in several studies, including \citet*{N2002}, \citet*{FWF2009}, \citet{Y2018} and references therein.
In particular, \citet*{FWF2009} unified broad classes of semiparametric estimators in kernel smoothing:
they incorporated a quasi-likelihood method including Bernoulli (logistic) and Poisson regression;
they addressed both additive and multiplicative adjustment;
and, through local polynomial smoothing, they proposed semiparametric variants of the Nadaraya--Watson estimator, local linear fitting and so on.
\citet*{FWF2009} also noted on p.4154 that ``Here we focus on local polynomial techniques although the idea can be extended to other nonparametric methods,'' thereby indicating that the nonparametric part of the semiparametric approach can be implemented with some techniques other than local polynomial ones.

As the nonparametric adjustment in the above semiparametric approach, we utilize a stagewise empirical risk minimization algorithm studied by \citet*{K2007}, \citet*[Section 19]{K2009}, \citet*{NE2013}, \citet*{NN2024} in density estimation and by \citet*{YN2019} in regression analysis.
This machine learning technique has several advantages, as pointed out in these studies:
(i) \citet*{YN2019} utilized the empirical risk derived from the Bregman divergence, which allows their proposed estimators to be applied widely in various settings, such as Bernoulli and Poisson regression.
(ii) There are almost no constraints on preparing the dictionary.
(ii-1) For instance, dictionaries can be designed to automate smoothing parameter selection in conventional estimators, through the algorithmic selection of words.
In density estimation, \citet*{NN2024} utilized a dictionary composed of scaled kernel functions with several bandwidths and they conducted bandwidth selection analogous to that in kernel density estimation.
(ii-2) Another utilization of dictionaries is for variable selection.
\citet*{YN2019} constructed a dictionary of univariate  $B$-spline bases corresponding to each explanatory variable and demonstrated variable selection with this dictionary for sparse additive models.

In addition to proposing such a semiparametric method, we aim to investigate its performance both theoretically and numerically.
In the theoretical study, we develop a non-asymptotic error bound for the proposed estimators as in \citet*{K2009}, \citet*{NE2013}, \citet*{NN2024} and \citet*{YN2019}.
At the same time, we examine the asymptotic behavior of the bound, even under general settings where the dictionary and words may depend on the data and sample size.
The numerical study explores accuracy of the proposed estimators with concrete dictionaries, by comparing them with several other regression estimators.
It also addresses behavior of our estimators in terms of variable selection, when utilizing a dictionary composed of univariate words.

This paper is organized as follows.
Section \ref{sec:Bregman} describes the Bregman divergence.
Using this divergence, the proposed semiparametric estimators are formulated in Section \ref{sec:esti}.
Their theoretical results are developed in Section \ref{sec:theo}.
Section \ref{sec:impl} provides some concrete dictionaries for implementing the estimators.
Section \ref{sec:simul} discusses simulations investigating the finite-sample performance of our estimators in several settings.
Applications to real data sets are presented in Section \ref{sec:real}.
Section \ref{sec:concl} offers concluding remarks.
All proofs and calculations for the theoretical results are provided in the \hyperref[sec:appe_proofs]{Appendix}.
\counterwithin{figure}{section}
\renewcommand{\thefigure}{\thesection.\arabic{figure}}

\section{Risk functions based on the Bregman divergence} \label{sec:Bregman}
Let $Y \in \mathbb{R}$ be a response variable and $\vec{X} \in \mathbb{R}^d$ be a vector of explanatory variables.
We are given i.i.d. pairs of observations
$\mathbb{X} = \left\{ \left( Y_1, \vec{X}_1 \right), \ldots, \left( Y_n, \vec{X}_n \right) \right\} \subset \mathbb{R} \times \mathbb{R}^{d}$
drawn from a joint distribution with the density function
		$f \left( y, \vec{x} \right)
=
		p \left( y \middle| \vec{x} \right) q \left( \vec{x} \right)$,
where $p \left( \cdot \middle| \vec{x} \right)$ denotes the conditional density of $Y$ given $\vec{X} = \vec{x}$,
and $q$ is the marginal density of $\vec{X}$ with the support $\mathbb{D} \subset \mathbb{R}^d$.
Our goal is to estimate the regression function $\mu \left( \vec{x} \right) = \mathbb{E} \left[Y \middle| \vec{X} = \vec{x} \right] $ at points $\vec{x} \in \mathbb{D}$.

To formulate the proposed estimator, we introduce a divergence measure between functions. 
For a strictly convex real-valued function $U \colon \mathbb{R} \to \mathbb{R}$,
the functional Bregman divergence between the target $\mu$ and a candidate estimator $\Bar{\mu}$ is defined as
\begin{align*}
		D_U \left( \mu, \Bar{\mu} \right)
&=
		\int_{\mathbb{R}^{d}}
			d_U \left( \mu \left( \vec{x} \right), \Bar{\mu} \left( \vec{x} \right) \right)
			q \left( \vec{x} \right)
		d \vec{x},
\end{align*}
where
\[
		d_U \left(\mu \left( \vec{x} \right), \Bar{\mu} \left( \vec{x} \right) \right)
=
		U \left(\mu \left( \vec{x} \right) \right)
	-
		U \left(\Bar{\mu} \left( \vec{x} \right) \right)
	-
		u \left(\Bar{\mu} \left( \vec{x} \right) \right)
		\left\{ \mu \left( \vec{x} \right) -\Bar{\mu} \left( \vec{x} \right) \right\}
\]
and $u = U'$, see \citet{B1967}, \citet{BSP2011}, \citet{YN2019}, \citet{NP2021}.
It is known that the inequality $d_U \left(\mu \left( \vec{x} \right), \Bar{\mu} \left( \vec{x} \right) \right) \geq 0$ always holds,
and that $d_U \left( \mu \left( \vec{x} \right), \Bar{\mu} \left( \vec{x} \right) \right) = 0$ holds if and only if $\mu \left( \vec{x} \right) = \Bar{\mu} \left( \vec{x} \right)$, see \citet{B1967} and \citet{BSP2011}.
These two properties of $d_U$ are also satisfied by the functional Bregman divergence $D_U$.
Therefore, $D_U$ can be regarded as a measure of discrepancy between functions,
although it is not a metric in the mathematical sense.
As discussed below, different strictly convex functions $U$ yield different forms of $D_U$, such as the squared error and the loss function used in Bernoulli regression.

We next derive an empirical risk function based on $D_U$.
To do this, we use an alternative expression for $d_U$:
\begin{align}
		&d_U \left(\mu \left( \vec{x} \right), \Bar{\mu} \left( \vec{x} \right) \right) \nonumber \\
&=
		d_{U^\ast} \left( u \left( \Bar{\mu} \left( \vec{x} \right) \right), u \left( \mu \left( \vec{x} \right) \right) \right) \nonumber \\
&=
		U^\ast \left( u \left(\Bar{\mu} \left( \vec{x} \right) \right) \right)
	-
		U^\ast \left( u \left(\mu \left( \vec{x} \right) \right) \right)
	-
		\mu \left( \vec{x} \right)
		\left\{
			u \left(\Bar{\mu} \left( \vec{x} \right) \right)
		-
			u \left(\mu \left( \vec{x} \right) \right) \right\} \nonumber \\
&=
		\left\{
			U^\ast \left( u \left(\Bar{\mu} \left( \vec{x} \right) \right) \right)
		-
			\mu \left( \vec{x} \right)
			u \left( \Bar{\mu} \left( \vec{x} \right) \right)
		\right\}
	-
		\left\{
			U^\ast \left( u \left(\mu \left( \vec{x} \right) \right) \right)
		-
			\mu \left( \vec{x} \right)
			u \left(\mu \left( \vec{x} \right) \right)
		\right\}, \label{eq:conj}
\end{align}
where $U^\ast \left( s \right) = \sup_{z \in \mathbb{R}} \left\{ z s - U \left( z \right) \right\}$ is the convex conjugate function of $U$,
with useful properties such as $\left( U^\ast \right)' = u^{- 1}$, see \citet{H-UL1993}.
By using \eqref{eq:conj} and defining
		$\ell \left( y, \vec{x} \middle| \Bar{\mu} \right)
=
		U^\ast \left( u \left( \Bar{\mu} \left( \vec{x} \right) \right) \right)
	-
		y \cdot u \left( \Bar{\mu} \left( \vec{x} \right) \right)$,
$D_U$ can be calculated as
\begin{align*}
		D_U \left( \mu, \Bar{\mu} \right)
&=
		D_{U^\ast} \left( u \left( \Bar{\mu} \right), u \left( \mu \right) \right) \\
&=
		\int_{\mathbb{R}^{d}}
			\left\{
				\ell \left( \mu \left( \vec{x} \right), \vec{x} \middle| \Bar{\mu} \right)
			-
				\ell \left( \mu \left( \vec{x} \right), \vec{x} \middle| \mu \right)
			\right\}
			q \left( \vec{x} \right)
		d \vec{x} \nonumber \\
&=
		\int_{\mathbb{R} \times \mathbb{R}^{d}}
			\left\{
				\ell \left( y, \vec{x} \middle| \Bar{\mu} \right)
			-
				\ell \left( y, \vec{x} \middle| \mu \right)
			\right\}
			f \left( y, \vec{x} \right)
		d y d \vec{x}.
\end{align*}
Then the empirical version of $D_U$ based on the observations $\mathbb{X}$ is given by
\begin{align*}
		\widehat{D}_U \left( \mu, \Bar{\mu} \right)
&=
		\frac{1}{n}
		\sum_{i = 1}^{n}
		\left\{
			\ell \left( Y_i, \vec{X}_i \middle| \Bar{\mu} \right)
		-
			\ell \left( Y_i, \vec{X}_i \middle| \mu \right)
		\right\} \\
&=
		\frac{1}{n}
		\sum_{i = 1}^{n}
		\ell \left( Y_i, \vec{X}_i \middle| \Bar{\mu} \right)
	+
		\left\langle \text{Constant for $\Bar{\mu}$} \right\rangle
\end{align*}
and we obtain the empirical risk function
\begin{align}
		\widehat{L} \left( \Bar{\mu} \right)
=
		\frac{1}{n}
		\sum_{i = 1}^{n}
		\ell \left( Y_i, \vec{X}_i \middle| \Bar{\mu} \right). \label{eq:emp_risk}
\end{align}

This function \eqref{eq:emp_risk} takes different forms depending on the strictly convex function $U$:
if $U^\ast \left( z \right) = z^2 / 2$,
then $u = \mathrm{id}$ and
		$\ell \left( y, \vec{x} \middle| \Bar{\mu} \right)
=
		\left\{ \Bar{\mu} \left( \vec{x} \right) - y \right\}^2 / 2 - y^2 / 2$,
so that \eqref{eq:emp_risk} is equivalent to the squared loss;
the function $U^\ast \left( z \right) = \ln \left[ 1 + e^z \right]$ leads to $u \left( z \right) = \ln \left[ z / \left( 1 - z \right) \right]$
and
		$\ell \left( y, \vec{x} \middle| \Bar{\mu} \left( \vec{x} \right) \right)
=
		- \left\{ y \ln \left[ \Bar{\mu} \left( \vec{x} \right) \right] + \left( 1 - y \right) \ln \left[ 1 - \Bar{\mu} \left( \vec{x} \right) \right] \right\}$,
making \eqref{eq:emp_risk} the Kullback--Leibler divergence between Bernoulli distributions
$Be \left( \mu \left( \vec{x} \right) \right)$ and $Be \left( \Bar{\mu} \left( \vec{x} \right) \right)$;
if $U^\ast \left( z \right) = e^z$,
then $u \left( z \right) = \ln z$
and
		$\ell \left( y, \vec{x} \middle| \Bar{\mu} \right)
=
		\Bar{\mu} - y \ln \left[ \Bar{\mu} \left( \vec{x} \right) \right]$,
so \eqref{eq:emp_risk} is exactly the negative log-likelihood function in Poisson regression, see Section 6.2.2 of \citet{MN1989}.

By utilizing this empirical risk function $\widehat{L} \left( \Bar{\mu} \right)$, an estimator based on the data can be constructed as $\argmin_{\Bar{\mu}} \widehat{D}_U \left( \mu, \Bar{\mu} \right) = \argmin_{\Bar{\mu}} \widehat{L} \left( \Bar{\mu} \right)$, that is, as an optimal function that minimizes the Bregman divergence $\widehat{D}_U$ from the target $\mu$.

\section{Proposed semiparametric estimators} \label{sec:esti}
In this section, we use the empirical risk introduced in the previous section to construct semiparametric estimators through stagewise minimization.
Before detailing the construction procedure, we first describe the functional form of the estimators and their motivation.
We prepare a model function $m \left( \cdot, \vec{\theta} \right) \colon \mathbb{D} \to \mathbb{R}$ determined by a finite-dimensional parameter $\vec{\theta} \in \Theta \subset \mathbb{R}^p, p \geq 1$.
We also prepare functions (called ``words'') $B_i \colon \mathbb{R}^d \to \mathbb{R}, i = 1, \ldots, N, N \in \mathbb{N}$ and
the set (called ``dictionary'') of these functions $\mathcal{D} = \left\{ B_i \middle| i = 1, \ldots, N \right\}$.
Setting a parameter $\gamma \in \left\{ 0, 1 \right\}$, we define an operator
		$T \left( \vec{\theta}, B \right) \left( \vec{x} \right)
=
		m \left( \vec{x}, \vec{\theta} \right)
	+
		m \left( \vec{x}, \vec{\theta} \right)^\gamma \cdot B \left( \vec{x} \right)$
for a function $B$,
and also define a set $T \left( \vec{\theta}, \mathcal{D} \right) = \left\{ T \left( \vec{\theta}, B \right) \middle| B \in \mathcal{D} \right\}$.
For these $m, \mathcal{D}$ and $\gamma$, we consider a class of functions having the form
\begin{align}
		&\Biggl\{
			u^{-1} \left( m \left( \cdot, \vec{\theta} \right) + m \left( \cdot, \vec{\theta} \right)^\gamma \cdot \xi \right)
		\Biggm|
			\xi = \sum_{k = 1}^{N} \lambda_k B_k,
			B_k \in \mathcal{D},
			\sum_{k = 1}^{N} \lambda_k = 1,
			\lambda_k \geq 0,
			k = 1, \ldots, N
		\Biggr\} \nonumber \\
&=
		\Biggl\{
			u^{-1} \left( \sum_{k = 1}^{N}\lambda_k T \left( \vec{\theta}, B_k \right) \right)
		\Biggm|
			B_k \in \mathcal{D},
			\sum_{k = 1}^{N} \lambda_k = 1,
			\lambda_k \geq 0,
			k = 1, \ldots, N
		\Biggr\} \nonumber \\
&=
		u^{-1} \left( \mathrm{co} \left( T \left( \vec{\theta}, \mathcal{D} \right) \right) \right), \label{eq:form_of_esti}
\end{align}
where $u$ is the function $u = U'$ associated with $D_U$ in Section \ref{sec:Bregman},
and $\mathrm{co} \left( T \left( \vec{\theta}, \mathcal{D} \right) \right)$ is the convex hull of the set $T \left( \vec{\theta}, \mathcal{D} \right)$.

The functions \eqref{eq:form_of_esti} are based on a combination of parametric and nonparametric approaches.
In the parametric approach, the target regression function $\mu$ is modeled as a function determined by a finite-dimensional parameter: $\left\{ u^{-1} \left( m \left( \cdot, \vec{\theta} \right) \right) \middle| \vec{\theta} \in \Theta \right\}$.
By plugging in an estimator $\widehat{\vec{\theta}}$ of the parameter $\vec{\theta}$, we obtain the parametric regression estimator $u^{-1} \left( m \left( \cdot, \widehat{\vec{\theta}} \right) \right)$.
Typical examples include generalized linear models (GLMs).
In GLMs, the hyperplane is used as $m \left( \vec{x}, \vec{\theta} \right) = \left[ 1 ~ \vec{x}^\top \right] \vec{\theta}$, $\vec{\theta} \in \mathbb{R}^{d+1}$.
The function $u^{-1}$ serves as a known inverse link function.
Moreover, $\widehat{\vec{\theta}}$ may be taken as, for instance, the MLE (\citet*{MN1989}) or the LASSO estimator (\citet*{HTW2015}), both of which will be formulated below.
This $u^{-1} \left( m \left( \cdot, \widehat{\vec{\theta}} \right) \right)$ provides interpretation of the underlying population regression function $\mu$.
For example, in GLMs, the parameter $\vec{\theta}$ corresponds to the coefficients of hyperplanes, allowing us to get information: whether $\mu$ increases, decreases,
or hardly changes globally with some explanatory variables.
On the other hand, such models with a finite-dimensional parameter are sometimes too restrictive.
It has been understood that parametric estimators generally lack consistency,
and this approach is justified only when the true $\mu$ is exactly within the model $\left\{ u^{-1} \left( m \left( \cdot, \vec{\theta} \right) \right) \middle| \vec{\theta} \in \Theta \right\}$ or at least in its neighborhood.

To retain the interpretability of the parametric approach while improving flexibility, we introduce a nonparametric adjustment to the parametric crude guess.
Unlike the purely parametric model $u^{-1} \left( m \left( \cdot, \vec{\theta} \right) \right)$, the function in \eqref{eq:form_of_esti} includes an adjustment term $\xi$.
Here $\gamma$ in \eqref{eq:form_of_esti} controls the type of adjustment:
$\gamma = 0$ yields additive adjustment $u^{-1} \left( m \left( \cdot, \vec{\theta} \right) + \xi \right)$, while $\gamma = 1$ leads to multiplicative adjustment as $u^{-1} \left( m \left( \cdot, \vec{\theta} \right) \cdot \left( 1 + \xi \right) \right)$, see \citet*{FWF2009}.
Below we construct this $\xi$ as a convex combination of $I$ words $B_1, \ldots, B_I$ selected from the dictionary $\mathcal{D}$:
$\widehat{\xi} = \sum_{i = 1}^{I} \lambda_i B_i$,
in which the same words may be selected repeatedly.
These words are selected by minimizing the empirical risk \eqref{eq:emp_risk} in a stagewise manner.

Now we present the construction of our semiparametric estimators.
Let
\[
		C \left( \Bar{\pi} \middle| g, h \right) \left( \vec{x} \right)
=
		C \left( \Bar{\pi} \middle| g \left( \vec{x} \right), h \left( \vec{x} \right) \right)
=
		\Bar{\pi} \cdot g \left( \vec{x} \right)
	+
		\left( 1 - \Bar{\pi} \right)
	\cdot
		h \left( \vec{x} \right)
\]
for functions $g \colon \mathbb{R}^d \to \mathbb{R}$ and $h \colon \mathbb{R}^d \to \mathbb{R}$ and a coefficient $\Bar{\pi} \in \left[ 0, 1 \right]$.
We introduce the notation
		$\argmin_{B \in \mathcal{D}}^{\varepsilon} g \left( B \right)
=
		\left\{
			B \in \mathcal{D}
		\middle|
			g \left( B \right)
		\leq
				\inf_{B \in \mathcal{D}}
				g \left( B \right)
			+
				\varepsilon
		\right\}$
for a function $g \colon \mathcal{D} \to \mathbb{R}$ and a positive real number $\varepsilon > 0$.
In addition, we set $\xi_0 = 0$, $\pi_k \in \left[ 0, 1 \right]$ ($k \in \mathbb{N}$) and fix a sufficiently small $\varepsilon > 0$.
The semiparametric estimators in this paper are constructed as follows (\citet*{K2009}, \citet*{NE2013}, \citet*{YN2019}, \citet*{NN2024}):
\begin{itemize}
\setlength{\leftskip}{1cm}
\item[$\left\langle Step~1 \right\rangle$] \label{alg:step1}
	Obtain a parametric regression estimator $u^{-1} \left( m \left( \cdot, \Bar{\vec{\theta}} \right) \right)$ with an estimator $\Bar{\vec{\theta}}$ for the parameter $\vec{\theta}$ in the model function.
\item[$\left\langle Step~2 \right\rangle$] \label{alg:step2}
	Prepare a dictionary $\mathcal{D}$.
	Then select $B_1, \ldots, B_I$ and obtain $\xi_I$:
	\begin{align*}
		&B_k \in
			\argmin_{\Bar{B} \in \mathcal{D}}{}^\varepsilon
					\widehat{L} \left( u^{-1} \left(
						T \left( \Bar{\vec{\theta}}, C \left( \pi_k \middle| \Bar{B}, \xi_{k - 1} \right) \right)
					\right) \right), \\
		&\xi_k = C \left( \pi_k \middle| B_k, \xi_{k - 1} \right)
	\end{align*}
	for each $k = 1, \ldots, I$.
\item[$\left\langle Step~3 \right\rangle$] \label{alg:step3}
	Construct the semiparametric regression estimator
		$\widehat{\mu}
	=
		\mu_I \left( \Bar{\vec{\theta}} \right)
	=
		u^{-1} \left(
		T \left( \Bar{\vec{\theta}}, \xi_I \right) \right)$.
\end{itemize}
The above \hyperref[alg:step1]{$\left\langle Step~1 \right\rangle$} is the initial parametric regression.
The estimator $\Bar{\vec{\theta}}$ is, for example, the MLE $\argmin_{\vec{\theta} \in \Theta} \widehat{L} \left( u^{-1} \left( m \left( \cdot, \vec{\theta} \right) \right) \right)$,
or the LASSO estimator
\[
		\argmin_{\vec{\theta} \in \Theta} 
		\left[ \widehat{L} \left( u^{-1} \left( m \left( \cdot, \vec{\theta} \right) \right) \right) + \lambda \left\| \vec{\theta} \right\|_1 \right]
\]
for the $\ell_1$-norm $\left\| \cdot \right\|_1$ and some $\lambda \geq 0$ (and for appropriately transformed data), or others that are not derived from the Bregman divergence $D_U$.
In \hyperref[alg:step2]{$\left\langle Step~2 \right\rangle$}, the adjustment term $\xi$ is introduced through the stagewise minimization of the empirical risk $\widehat{L}$ in \eqref{eq:emp_risk}.
Finally, the proposed estimator $\widehat{\mu}$ has the form
\begin{align*}
		\widehat{\mu}
=
		u^{-1} \left( T \left( \Bar{\vec{\theta}}, \widehat{\xi} \right) \right), \quad
		\widehat{\xi} = \xi_I = \sum_{k = 1}^{I} \lambda_{k, I} B_k, \quad
		\lambda_{k, I} = \pi_k \prod_{j = k + 1}^I \left( 1 - \pi_j \right),
\end{align*}
where $\lambda_{k, I} \geq 0, k = 1, \ldots, I$ and $\sum_{k = 1}^{I} \lambda_{k, I} = 1$ are satisfied.

\section{Theoretical results} \label{sec:theo}
This section provides theoretical results for the proposed estimator $\widehat{\mu}$ in Section \ref{sec:esti}.
We evaluate estimation accuracy of $\widehat{\mu}$ by using a measure called ``divergence risk.''
We derive a non-asymptotic upper bound for this risk, and further investigate this bound asymptotically.
We also compare this non-asymptotic bound for $\widehat{\mu}$ with that for a theoretically ideal estimator.

In what follows, we fix a strictly convex function $U$ and the link function $u = U'$ for $\widehat{\mu}$,
and put the sequence $\pi_k$ forming the coefficients of $\widehat{\xi}$ as
\[
		\pi_k = \frac{\tau}{k - 1 + \tau}
\]
for some $\tau \geq 2$ and for each $k \in \mathbb{N}$.
This $\pi_k$ is similar to those in \citet*{K2009}, \citet*{NE2013}, \citet*{NN2024} and \citet*{YN2019}.
Let $\| \cdot \|_2$ be the $\ell_2$-norm, $\| \cdot \|_\infty$ be the $L_\infty$-norm,
\begin{align*}
		\mathcal{W}
=
		\left\{
			\left[ \lambda_B \right]_{B \in \mathcal{D}} \;\middle|\; \lambda_B \geq 0, \sum_{B \in \mathcal{D}} \lambda_B = 1
		\right\}, \quad N \in \mathbb{N}
\end{align*}
and
\begin{align*}
		\mu \left( \vec{x}, \Lambda, \vec{\theta} \right)
=
		u^{-1} \left(
			T \left( \vec{\theta}, \sum_{B \in \mathcal{D}} \lambda_B B \right) \left( \vec{x} \right)
		\right), \quad
		\vec{x} \in \mathbb{R}^d, \quad
		\Lambda 
= 
		\left[ \lambda_B \right]_{B \in \mathcal{D}} \in \mathcal{W}, \quad
		\vec{\theta} \in \mathbb{R}^p.
\end{align*}
We make the following assumptions:
\begin{enumerate}[label = (A\arabic*), ref=(A\arabic*), leftmargin=*, align = left]
\item
	The density function $q$ for the distribution of the explanatory variables $\vec{X}$ has the compact support $\mathbb{D} \subset \mathbb{R}^d$.
\item
	The response $Y$ satisfies $\mathbb{E} \left[ Y^2 \right] < \infty$.
\item
	The estimator $\widehat{\vec{\theta}}$ for the parameter $\vec{\theta}$ satisfies $\widehat{\vec{\theta}} \stackrel{P}{\to} \vec{\theta}^\ast$ as $n \to \infty$.
\item
	The function $\vec{x} \mapsto m \left( \vec{x}, \vec{\theta} \right)$ is continuous for any $\vec{\theta} \in \Theta$.
	In addition, the function $\vec{\theta} \mapsto m \left( \vec{x}, \vec{\theta} \right)$ is continuously differentiable for any $\vec{x} \in \mathbb{D}$.
\item
	The strictly convex function $U$ is twice continuously differentiable.
\end{enumerate}

We define the divergence risk as the function $\Bar{\mu} \mapsto \mathbb{E} \left[ D_U \left( \mu, \Bar{\mu} \right) \right]$ for the target regression function $\mu$ and an estimator $\Bar{\mu}$, 
where $D_U$ is the Bregman divergence associated with $U$ and the expectation $\mathbb{E}$ is taken with respect to the observations $\mathbb{X}$ constructing $\Bar{\mu}$.
If $\mathbb{E} \left[ D_U \left( \mu, \Bar{\mu} \right) \right]$ is smaller, it means that the estimator $\Bar{\mu}$ has higher accuracy.
In the statements below, words in the dictionary $\mathcal{D}$ are allowed to depend on the observations $\mathbb{X}$.
The divergence risk for $\widehat{\mu}$ is evaluated in Theorem \ref{thm:mu} below, on the basis of Lemmas \ref{lem:ineq1} and \ref{lem:ineq2} establishing the relevant inequalities.
\begin{Lem} \label{lem:ineq1}
For fixed $I \in \mathbb{N}$ and for fixed $\vec{\theta}_1, \vec{\theta}_2 \in \mathbb{R}^p$ which may depend on the observations $\mathbb{X}$, the function $\mu_I \left( \vec{\theta}_1 \right)$ defined in \hyperref[alg:step3]{$\left\langle Step~3 \right\rangle$} of the algorithm in Section \ref{sec:esti} satisfies the following inequality of random variables:
\begin{align*}
		\widehat{L} \left( \mu_I \left( \vec{\theta}_1 \right) \right)
&\leq
		\inf_{\Lambda \in \mathcal{W}}
		\widehat{L} \left( \mu \left( \cdot, \Lambda, \vec{\theta}_2 \right) \right)
	+
		\frac{\tau^2}{I + \tau} A_1 \left( \vec{\theta}_1, \vec{\theta}_2 \right)  \\
	&\hspace{0.5cm}+
		\left\| \vec{\theta}_1 - \vec{\theta}_2 \right\|_2
		A_2 \left( \vec{\theta}_1, \vec{\theta}_2 \right)
	+
		\frac{\tau^2}{I + \tau}
		\left\| \vec{\theta}_1 - \vec{\theta}_2 \right\|_2
		A_3 \left( \vec{\theta}_1, \vec{\theta}_2 \right)
	+
		\varepsilon,
\end{align*}
where
\begin{align*}
		A_1 \left( \vec{\theta}_1, \vec{\theta}_2 \right)
&=
		4
		\sup_{B \in \mathcal{D}} \sup_{\vec{x} \in \mathbb{D}} \left| B \left( \vec{x} \right) \right|^2
		\sup_{\vec{x} \in \mathbb{D}}
			\left|
				m \left( \vec{x}, \vec{\theta}_1 \right)
				m \left( \vec{x}, \vec{\theta}_2 \right)
			\right|^\gamma \\
	&\hspace{1cm}\times
		\sup_{\Lambda \in \mathcal{W}}
		\sup_{\vec{x} \in \mathbb{D}}
		\left( U^\ast \right)''
		\left( u \left( \mu \left( \vec{x}, \Lambda, \vec{\theta}_1 \right) \right) \right), \\
		A_2 \left( \vec{\theta}_1, \vec{\theta}_2 \right)
&=
		\sup_{B \in \mathcal{D}}
		\sup_{\vec{x} \in \mathbb{D}}
		\sup_{\beta \in [0,1]}
			\left\|
				\left.
					\left\{
						1
					+
						\gamma
						m \left( \vec{x}, \vec{\theta} \right)^{\gamma - 1}
						B \left( \vec{x} \right)
					\right\}
					\frac{\partial}{\partial \vec{\theta}}
					m \left( \vec{x}, \vec{\theta} \right)
				\right|_{
					\vec{\theta}
				=
						\vec{\theta}_2
					+ 
						\beta \left( \vec{\theta}_1 - \vec{\theta}_2 \right)}
			\right\|_2 \\
	&\hspace{1cm}\times
		\frac{1}{n}
		\sum_{i = 1}^{n}
		\sup_{B \in \mathcal{D}}
			\left|
				Y_i
			-
				u^{-1} \left( T \left( \vec{\theta}_1, B \right) \left( \vec{X}_i \right) \right)
			\right|, \\
		A_3 \left( \vec{\theta}_1, \vec{\theta}_2 \right)
&=
		2
		\sup_{B \in \mathcal{D}}
		\sup_{\vec{x} \in \mathbb{D}}
			\left| B \left( \vec{x} \right) \right|
		\sup_{\vec{x} \in \mathbb{D}}
			\left|
				m \left( \vec{x}, \vec{\theta}_1 \right)
			\right|^\gamma
		\sup_{\Lambda \in \mathcal{W}}
		\sup_{\vec{x} \in \mathbb{D}}
			\left( U^\ast \right)''
			\left( u \left( \mu \left( \vec{x}, \Lambda, \vec{\theta}_1 \right) \right) \right) \\
	&\hspace{1cm}\times
		\sup_{B \in \mathcal{D}}
		\sup_{\vec{x} \in \mathbb{D}}
		\sup_{\beta \in [0,1]}
			\left\|
				\left.
				\left\{
					1
				+
					\gamma
					m \left( \vec{x}, \vec{\theta} \right)^{\gamma - 1}
					B \left( \vec{x} \right)
				\right\}
				\frac{\partial}{\partial \vec{\theta}}
				m \left( \vec{x}, \vec{\theta} \right)
				\right|_{\vec{\theta}
					=
						\vec{\theta}_2
					+ \beta
					\left( \vec{\theta}_1 - \vec{\theta}_2 \right)}
			\right\|_2.
\end{align*}
\end{Lem}
\begin{Lem} \label{lem:ineq2}
For fixed $\vec{\theta}_1 \left( \mathbb{X} \right), \vec{\theta}_2 \left( \mathbb{X} \right) \in \mathbb{R}^p$ and $\delta \left( \mathbb{X} \right) \geq 0$,
let $\Bar{\Lambda} \left( \mathbb{X} \right) \in \mathcal{W}$
be such that
\begin{align}
		\widehat{L} \left( \mu \left( \cdot, \Bar{\Lambda} \left( \mathbb{X} \right), \vec{\theta}_1 \left( \mathbb{X} \right) \right) \right)
\leq
		\inf_{\Lambda \in \mathcal{W}}
			\widehat{L} \left( \mu \left( \cdot, \Lambda, \vec{\theta}_2 \left( \mathbb{X} \right) \right) \right)
	+
		\delta \left( \mathbb{X} \right). \label{eq:assu_lem2}
\end{align}
Then
\begin{align*}
		&\mathbb{E} \left[ D_U \left( \mu, \mu \left( \cdot, \Bar{\Lambda} \left( \mathbb{X} \right), \vec{\theta}_1 \left( \mathbb{X} \right) \right) \right) \right] \\
&\leq
		\mathbb{E}
		\left[ \inf_{\Lambda \in \mathcal{W}}
		D_U \left( \mu, \mu \left( \cdot, \Lambda, \vec{\theta}_2 \left( \mathbb{X} \right) \right) \right) \right] \nonumber \\
	&\hspace{0.5cm}+
		\mathbb{E}
		\left[ \sup_{\Lambda \in \mathcal{W}}
		\mathcal{E}_n \left( \mu \left( \cdot, \Lambda, \vec{\theta}_2 \left( \mathbb{X} \right) \right) \right) \right]
	+
		\mathbb{E}
		\left[
			\sup_{\Lambda \in \mathcal{W}}
			\left[ - \mathcal{E}_n \left( \mu \left( \cdot, \Lambda, \vec{\theta}_1 \left( \mathbb{X} \right) \right) \right) \right]
		\right]
	+
		\mathbb{E} \left[ \delta \left( \mathbb{X} \right) \right],
\end{align*}
where $\mathcal{E}_n \left( h \right)$ is a centered empirical process
\begin{align*}
		\mathcal{E}_n \left( h \right)
=
		\frac{1}{n}
		\sum_{i = 1}^{n}
		\ell \left( Y_i, \vec{X}_i \middle| h \right)
	-
		\int_{\mathbb{R} \times \mathbb{R}^{d}}
		\ell \left( y, \vec{x} \middle| h \right)
		f \left( y, \vec{x} \right) d y d \vec{x}
\end{align*}
associated with a function $h \colon \mathbb{R}^d \to \mathbb{R}$.
\end{Lem}
We now define a dictionary denoted by $\mathcal{D} \left( \Delta_1, \Delta_2, c \right)$ with some fixed $\Delta_1, \Delta_2, c > 0$, for the theoretical results below.
We first construct its words as follows.
Let the realization of the observation $\mathbb{X}$ be $\mathcal{X} = \left\{ \left( y_1, \vec{x}_1 \right), \ldots, \left( y_n, \vec{x}_n \right) \right\}$.
The words should be allowed to include the estimator $\widehat{\vec{\theta}} \left( \mathcal{X} \right)$ for the parameter in the parametric model, and also to depend on the observations $\mathcal{X}$ themselves.
Further, we introduce an index set $\mathcal{P} \subset \mathbb{R}^\mathfrak{p}$, $\mathfrak{p} \geq 1$, which will determine the words in the dictionary.
For example, the elements of $\mathcal{P}$ may be a specific $i$-th observation $\left( y_i, \vec{x}_i \right)$ or a bandwidth $H$ or their combination $\left( \left( y_i, \vec{x}_i \right), H \right)$.
The general form of the words is a function 
$B \left( \cdot \middle| \vec{\theta}, \mathcal{Z}, \mathfrak{P} \right) \colon \mathbb{R}^d \to \mathbb{R}$,
$\vec{\theta} \in \Theta$,
$\mathcal{Z} = \left\{ \left( v_1, \vec{w}_1 \right), \ldots, \left( v_n, \vec{w}_n \right) \right\}$ with $\left( v_i, \vec{w}_i \right) \in \mathbb{R} \times \mathbb{R}^d$ 
and
$\mathfrak{P} \in \mathcal{P}$.
Then, preparing a finite set $\mathcal{P}^\ast \subset \mathcal{P}$ with $\left| \mathcal{P}^\ast \right| = N$,
we define
		$\mathcal{D}
=
		\left\{
			B \left( \cdot \middle| \widehat{\vec{\theta}} \left( \mathcal{X} \right), \mathcal{X}, \mathfrak{P} \right)
		\middle|
			\mathfrak{P} \in \mathcal{P}^\ast
		\right\}$,
where $N \in \mathbb{N}$ may depend on the sample size $n$.
An example of this dictionary will appear in Section \ref{sec:theo_dict}.
The dictionary utilized in the theorems below is a subset of this $\mathcal{D}$.
The practical dictionary is defined as
\begin{align*}
		\mathcal{D} \left( \Delta_1, \Delta_2, c \right)
=
		&\Biggl\{
				B \left( \cdot \middle| \widehat{\vec{\theta}} \left( \mathcal{X} \right), \mathcal{X}, \mathfrak{P} \right)
			\in
				\mathcal{D}
		\Biggm|
				\sup_{\vec{x} \in \mathbb{D}} 
					\left| T \left( \widehat{\vec{\theta}} \left( \mathcal{X} \right), B \left( \vec{x} \middle| \widehat{\vec{\theta}} \left( \mathcal{X} \right), \mathcal{X}, \mathfrak{P} \right) \right) \right| 
			\leq 
				\Delta_1, \\
		&\hspace{4.5cm}
				\sup_{\vec{x} \in \mathbb{D}}
				\sup_{\vec{\theta} \in \Theta_c}
					\left| B \left( \cdot \middle| \vec{\theta}, \mathcal{X}, \mathfrak{P} \right) \right| \leq \Delta_2
		\Biggr\},
\end{align*}
for some fixed $\Delta_1, \Delta_2, c > 0$, where $\Theta_c = \left\{ \vec{\theta} \in \Theta \middle| \left\| \vec{\theta} - \vec{\theta}^\ast \right\|_2 \leq c \right\}$.

We assume that $\mathcal{D} \left( \Delta_1, \Delta_2, c \right)$ satisfies the following assumption:
\begin{enumerate}[label = (A\arabic*), ref=(A\arabic*), leftmargin=*, align = left, resume]
\item \label{assu:dict}
	There exists a set $\mathcal{B}$ such that $\mathcal{D} \left( \Delta_1, \Delta_2, c \right) \subset \mathcal{B}$ and its covering number $N \left( \epsilon, \mathcal{B}, \| \cdot \|_\infty \right)$ satisfies $\int_0^1 \sqrt{\ln N \left( \epsilon, \mathcal{B}, \| \cdot \|_\infty \right)} d \epsilon < \infty$.
\end{enumerate}
\begin{Rem_n}
The definition of the covering number can be found in Definition 5.1 and Section 5.3.3 of \citet*{W2019}, and Definition 2.1.5 of \citet*{VW1996}.
Examples of sets $\mathcal{B}$ satisfying $\int_0^1 \sqrt{\ln N \left( \epsilon, \mathcal{B}, \| \cdot \|_\infty \right)} d \epsilon < \infty$ can also be found in Section 2.7 of \citet*{VW1996}.
\end{Rem_n}
For the proposed estimator $\widehat{\mu}$ constructed from such a dictionary, we can evaluate the divergence risk $\mathbb{E} \left[ D_U \left( \mu, \widehat{\mu} \right) \right]$  as follows:
\begin{Thm} \label{thm:mu}
For $\varepsilon > 0$ in the algorithm, the stagewise estimator $\widehat{\mu} = \mu_I \left( \widehat{\vec{\theta}} \right)$ satisfies
\begin{align}
		\mathbb{E} \left[ D_U \left( \mu, \widehat{\mu} \right) \right] \nonumber
&\leq
		\mathbb{E}
		\left[
			\inf_{g \in u^{-1} \left( \mathrm{co} \left( T \left( \widehat{\vec{\theta}}, \mathcal{D} \left( \Delta_1, \Delta_2, c \right) \right) \right) \right)}
			D_U \left( \mu, g \right)
		\right] \\
	&\hspace{0.5cm}+
		\sum_{\alpha = 0}^1
		\mathbb{E}
		\left[
			\sup_{g \in u^{-1} \left( \mathrm{co} \left( T \left( \widehat{\vec{\theta}}, \mathcal{D} \left( \Delta_1, \Delta_2, c \right) \right) \right) \right)}
			\left[
				\left( - 1 \right)^\alpha
				\mathcal{E}_n \left( g \right)
			\right]
		\right]
	+
		\frac{\tau^2}{I + \tau}
		C_1 \left( n \right)
	+
		\varepsilon, \label{eq:mu_hat}
\end{align}
where $C_1 \left( n \right)
=
		\mathbb{E}
			\left[ A_1 \left( \widehat{\vec{\theta}}, \widehat{\vec{\theta}} \right) \right]$.
Moreover,
\begin{align*}
		\sum_{\alpha = 0}^1
		\mathbb{E}
		\left[
			\sup_{g \in u^{-1} \left( \mathrm{co} \left( T \left( \widehat{\vec{\theta}}, \mathcal{D} \left( \Delta_1, \Delta_2, c \right) \right) \right) \right)}
			\left[
				\left( - 1 \right)^\alpha
				\mathcal{E}_n \left( g \right)
			\right]
		\right]
=
		O \left( \frac{1}{\sqrt{n}} \right).
\end{align*}
\end{Thm}
We focus on the upper bound \eqref{eq:mu_hat} in Theorem \ref{thm:mu}:
the first term in the bound represents the essential bias in approximating $\mu$ with the function class $u^{-1} \left( \mathrm{co} \left( T \left( \widehat{\vec{\theta}}, \mathcal{D} \left( \Delta_1, \Delta_2, c \right) \right) \right) \right)$, to which the proposed estimator $\widehat{\mu}$ belongs.
The second term approaches 0 as $n \to \infty$, and the third term also converges to 0 as $I \to \infty$.
In short, given that we can take sufficiently small $\varepsilon$, Theorem \ref{thm:mu} states that the divergence risk for $\widehat{\mu}$ can be bounded above only by the essential bias term and asymptotically vanishing terms.

This Theorem \ref{thm:mu} can be better understood with Theorem \ref{thm:mu0} below.
Theorem \ref{thm:mu0} provides an error bound for an ideal estimator $\mu^0$:
\begin{Thm} \label{thm:mu0}
For $\varepsilon > 0$ and for any of the following estimators
\begin{align}
		\mu^0
\in
		\mathop{\mathrm{argmin}\rlap{$^\varepsilon$}}_{g \in u^{-1} \left( \mathrm{co} \left( T \left( \widehat{\vec{\theta}}, \mathcal{D} \right) \right) \right)}
		\widehat{L} \left( g \right), \label{eq:mu0_def}
\end{align}
it holds that
\begin{align}
		&\mathbb{E} \left[ D_U \left( \mu, \mu^0 \right) \right] \nonumber \\
&\leq
		\mathbb{E}
		\left[
			\inf_{g \in u^{-1} \left( \mathrm{co} \left( T \left( \widehat{\vec{\theta}}, \mathcal{D} \right) \right) \right) }
			D_U \left( \mu, g \right)
		\right]
	+
		\sum_{\alpha = 0}^1
		\mathbb{E}
		\left[
			\sup_{g \in u^{-1} \left( \mathrm{co} \left( T \left( \widehat{\vec{\theta}}, \mathcal{D} \right) \right) \right)}
			\left[
				\left( - 1 \right)^\alpha
				\mathcal{E}_n \left( g \right)
			\right]
		\right] 
	+
		\varepsilon. \label{eq:mu0}
\end{align}
\end{Thm}
Comparing the upper bounds in \eqref{eq:mu_hat} and \eqref{eq:mu0}, \eqref{eq:mu_hat} differs from \eqref{eq:mu0} only in that it includes the term $\left\{ \tau^2 / (I + \tau) \right\} \cdot C_1 \left( n \right)$, which converges to 0 as $I \to \infty$.
Therefore, the bound for the practically implemented $\widehat{\mu}$ asymptotically coincides with that for $\mu^0$ in \eqref{eq:mu0_def}, which theoretically minimizes the empirical risk over the convex hull.

\section{Implementation} \label{sec:impl}
In this section, we present practical implementation of the proposed estimator $\widehat{\mu}$ introduced in Section \ref{sec:esti}.
In particular, we propose several dictionaries for constructing the adjustment term $\widehat{\xi}$ in $\widehat{\mu}$.
More specifically, we use three sets containing $d$-variate, univariate and bivariate words, respectively. 
The dictionaries below are based on the estimators previously proposed by \citet{FWF2009} in kernel smoothing.
These dictionaries can be utilized in several settings, including normal, Bernoulli and Poisson regression.

\subsection{Dictionary $\mathcal{D}_0$}
\citet{FWF2009} proposed the following adjustment term for the semiparametric estimator:
\begin{align}
		&\widehat{\xi}_H \left( \vec{x} \right) \nonumber \\
&=
		\sol_{\xi}
		\left[
				\frac{1}{n}
				\sum_{i = 1}^{n}
				K_H \left( \vec{X}_i - \vec{x} \right)
				m \left( \vec{X}_i, \widehat{\vec{\theta}} \right)^\gamma
				\left\{
					Y_i
				-
					u^{-1}
					\left(
							m \left( \vec{X}_i, \widehat{\vec{\theta}} \right)
					+
							m \left( \vec{X}_i, \widehat{\vec{\theta}} \right)^\gamma \xi
					\right)
				\right\}
		=
				0
		\right], \label{eq:FWF_xi}
\end{align}
where $\widehat{\vec{\theta}}$ is an estimator of the parameter $\vec{\theta}$,
$K_H \left( \vec{x} \right) = \left| H \right|^{-1 / 2} K \left( H^{-1 / 2} \vec{x} \right)$,
$K$ is a $d$-dimensional function called ``kernel function,''
and $H$ is a symmetric positive-definite matrix named ``bandwidth.''
\begin{Rem_n}
This $\widehat{\xi}_H$ corresponds to the local constant case $p = 0$ in Sections \ref{sec:Bregman} and \ref{sec:esti} of \citet{FWF2009}.
The estimators in \citet{FWF2009} are constructed via a quasi-log-likelihood function rather than the Bregman divergence.
However, these estimators coincide with those derived from the Bregman divergence, since there is a correspondence between the quasi-log-likelihood function and the Bregman divergence:
for any quasi-log-likelihood function $Q$ satisfying several conditions, there exists a strictly convex function $W$ such that $-Q \left( x, y \right) = d_W \left( y, x \right)$, see Section 3 of \citet{ZJS2009}.
\end{Rem_n}
If $U = x^2 / 2$ as in normal regression, the above solution \eqref{eq:FWF_xi} can be calculated explicitly as
\begin{align}
		\widehat{\xi}_H \left( \vec{x} \right)
=
		\frac{1}{n}
		\sum_{i = 1}^{n}
		\frac{
		K_H \left( \vec{X}_i - \vec{x} \right)
		m \left( \vec{X}_i, \widehat{\vec{\theta}} \right)^\gamma
		\left\{ Y_i - m \left( \vec{X}_i, \widehat{\vec{\theta}} \right) \right\} }
		{(1 / n) \sum_{j = 1}^{n}
		K_H \left( \vec{X}_j - \vec{x} \right)
		m \left( \vec{X}_j, \widehat{\vec{\theta}} \right)^{2 \gamma}}. \label{eq:sol_nor}
\end{align}
On the other hand, if $U \neq x^2 / 2$ as in Bernoulli and Poisson regression, \eqref{eq:FWF_xi} cannot be solved explicitly.
Then we replace the term $u^{-1} \left( m \left( \vec{X}_i, \widehat{\vec{\theta}} \right) + m \left( \vec{X}_i, \widehat{\vec{\theta}} \right)^\gamma \xi \right) $ in \eqref{eq:FWF_xi} with the first-order approximation around $\xi = 0$, that is,
\begin{align*}
	u^{-1}
	\left(
			m \left( \vec{X}_i, \widehat{\vec{\theta}} \right)
	+
			m \left( \vec{X}_i, \widehat{\vec{\theta}} \right)^\gamma \xi
	\right)
\thickapprox
		u^{-1} \left( m \left( \vec{X}_i, \widehat{\vec{\theta}} \right) \right)
	+
		\left( U^\ast \right)'' \left( m \left( \vec{X}_i, \widehat{\vec{\theta}} \right) \right)
		m \left( \vec{X}_i, \widehat{\vec{\theta}} \right)^\gamma \xi,
\end{align*}
and we obtain
\begin{align}
		&\sol_{\xi}
		\biggl[
				\frac{1}{n}
				\sum_{i = 1}^{n}
				K_H \left( \vec{X}_i - \cdot \right)
				m \left( \vec{X}_i, \widehat{\vec{\theta}} \right)^\gamma \nonumber \\
			&\hspace{2cm}\times
				\left[
					Y_i
				-
					\left\{
						u^{-1} \left( m \left( \vec{X}_i, \widehat{\vec{\theta}} \right) \right)
					+
						\left( U^\ast \right)'' \left( m \left( \vec{X}_i, \widehat{\vec{\theta}} \right) \right)
						m \left( \vec{X}_i, \widehat{\vec{\theta}} \right)^\gamma \xi
					\right\}
				\right]
		=
				0
		\biggr] \nonumber \\
&=
		\frac{1}{n}
		\sum_{i = 1}^{n}
		\frac{
			K_H \left( \vec{X}_i - \cdot \right)
			m \left( \vec{X}_i, \widehat{\vec{\theta}} \right)^\gamma
			\left\{ Y_i - u^{-1}\left( m \left( \vec{X}_i, \widehat{\vec{\theta}} \right) \right) \right\}
		}{
			(1 / n) \sum_{j = 1}^{n}
			K_H \left( \vec{X}_j - \cdot \right)
			m \left( \vec{X}_j, \widehat{\vec{\theta}} \right)^{2 \gamma}
			\left( U^\ast \right)'' \left( m \left( \vec{X}_j, \widehat{\vec{\theta}} \right) \right)
		} \label{eq:sol} \\
&=
		\frac{1}{n} \sum_{i = 1}^{n}
		B_0 \left( \vec{x} \middle| \widehat{\vec{\theta}}, i, H \right), \nonumber
\end{align}
where
\begin{align}
		B_0 \left( \vec{x} \middle| \vec{\theta}, i, H \right)
=
		\frac{
			K_H \left( \vec{X}_i - \vec{x} \right)
			m \left( \vec{X}_i, \vec{\theta} \right)^\gamma
			\left\{ Y_i - u^{-1}\left( m \left( \vec{X}_i, \vec{\theta} \right) \right) \right\}
		}{
			(1 / n) \sum_{j = 1}^{n}
			K_H \left( \vec{X}_j - \vec{x} \right)
			m \left( \vec{X}_j, \vec{\theta} \right)^{2 \gamma}
			\left( U^\ast \right)'' \left( m \left( \vec{X}_j, \vec{\theta} \right) \right)
		} \label{eq:B_0}
\end{align}
for $\vec{x} \in \mathbb{R}^d$, $\vec{\theta} \in \mathbb{R}^p$, $i = 1, \ldots, n $ and a positive-definite matrix $H$.
Note that \eqref{eq:sol} is equal to \eqref{eq:sol_nor} when $U = x^2 / 2$.

We adopt $B_0$ above as the words for our $\widehat{\xi}$, and propose a dictionary
\begin{align}
	\mathcal{D}_0
	=
			\left\{
				B_0 \left( \vec{x} \middle| \widehat{\vec{\theta}}, i, H \right)
			\middle|
				i = 1, \ldots, n,
				H \in \mathbb{H}_0
			\right\} 
\end{align}
for a prepared set $\mathbb{H}_0$ of bandwidths.

\subsection{Dictionaries $\mathcal{D}_1$ and $\mathcal{D}_2$}
Here we fix the parametric model function as $m \left( \vec{x}, \vec{\theta} \right) = \left[ 1 ~ \vec{x}^\top \right] \vec{\theta}$, $\vec{\theta} = \left[ \theta_0 ~ \theta_1 ~ \cdots ~ \theta_d \right]^\top \in \mathbb{R}^{d+1}$.
Let $1 \leq i \leq n$ and $1 \leq k < \ell \leq d$.
For the variable $\vec{x} = \left[ x_1 ~ x_2 ~ \ldots ~ x_d \right]^\top$, 
the $i$-th observation of the explanatory variables $\vec{X}_i = \left[ X_{i1} \cdots X_{id} \right]^\top$, $i = 1, \ldots, n$ 
and an estimator $\widehat{\vec{\theta}} = \left[ \widehat{\theta}_0 ~ \widehat{\theta}_1 \cdots \widehat{\theta}_d \right]^\top$ of the parameter $\vec{\theta}$, we denote the three corresponding subvectors as follows:
$\vec{x}_{k \ell} = \left[ x_k ~ x_\ell \right]^\top$,
$\widehat{\vec{\theta}}_{0k \ell} = \left[ \widehat{\theta}_0 ~ \widehat{\theta}_k ~ \widehat{\theta}_\ell \right]^\top$
and
$\vec{X}_{i k \ell} = \left[ X_{ik} ~ X_{i \ell} \right]^\top$.
In addition, we define submodels of $m$ with $\vec{X}$:
$m_k \left( X_{ik}, \widehat{\vec{\theta}}_{0k} \right) = \widehat{\theta}_0 + \widehat{\theta}_k X_{ik}$
and
$m_{k \ell} \left( \vec{X}_{ik \ell}, \widehat{\vec{\theta}}_{0k \ell} \right) = \widehat{\theta}_0 + \widehat{\theta}_k X_{ik} + \widehat{\theta}_\ell X_{i\ell}$.
Using this notation, we define the following univariate and bivariate kernel-based functions
\begin{align}
		B_1 \left( x_k \middle| \widehat{\vec{\theta}}, i, k, h \right)
=
		\frac{
			K_h \left( X_{ik} - x_k \right)
			m_k \left( X_{ik}, \widehat{\vec{\theta}}_{0k} \right)^\gamma
			\left\{
				Y_i
			-
				u^{-1} \left( m_k \left( X_{ik}, \widehat{\vec{\theta}}_{0k} \right) \right)
			\right\}
		}{
			(1 / n) \sum_{j = 1}^{n}
			K_h \left( X_{jk} - x_k \right)
			m_k \left( X_{jk}, \widehat{\vec{\theta}}_{0k} \right)^{2 \gamma}
			\left( U^\ast \right)'' \left( m_k \left( X_{jk}, \widehat{\vec{\theta}}_{0k} \right) \right)
		}, \label{eq:word_B1}
\end{align}
for a positive real number $h$, and
\begin{align*}
		&B_2 \left( \vec{x}_{k \ell} \middle| \widehat{\vec{\theta}}, i, k, \ell, H \right) \\
&=
		\frac{K_H \left( \vec{X}_{i k \ell} - \vec{x}_{k \ell} \right)
			m_{k \ell} \left( \vec{X}_{ik \ell}, \widehat{\vec{\theta}}_{0k \ell} \right)^\gamma
			\left\{
				Y_i
			-
				u^{-1} \left( m_{k \ell} \left( \vec{X}_{ik \ell}, \widehat{\vec{\theta}}_{0k \ell} \right) \right)
			\right\}
		}{
			(1 / n) \sum_{j = 1}^{n}
			K_H \left( \vec{X}_{j k \ell} - \vec{x}_{k \ell} \right)
			m_{k \ell} \left( \vec{X}_{jk \ell}, \widehat{\vec{\theta}}_{0k \ell} \right)^{2 \gamma}
			\left( U^\ast \right)'' \left( m_{k \ell} \left( \vec{X}_{jk \ell}, \widehat{\vec{\theta}}_{0k \ell} \right) \right)}
\end{align*}
for a $2 \times 2$ positive-definite matrix $H$.
These $B_1$ and $B_2$ are univariate and bivariate versions of \eqref{eq:B_0}.
We employ these functions as words in dictionaries.
After preparing sets of bandwidths $\mathbb{H}_1$ and $\mathbb{H}_2$,
we define the following two dictionaries containing univariate or bivariate words:
\begin{align}
		&\mathcal{D}_1
=
		\left\{
			B_1 \left( x_k \middle| \widehat{\vec{\theta}}, i, k, h \right)
		\middle|
			i = 1, \ldots, n,
			k = 1, \ldots, d,
			h \in \mathbb{H}_1
		\right\}, \label{eq:D_1} \\
		&\mathcal{D}_2
=
		\mathcal{D}_1
	\cup
		\left\{
			B_2 \left( \vec{x}_{k \ell} \middle| \widehat{\vec{\theta}}, i, k, \ell, H \right)
		\Biggm|
			i = 1, \ldots, n,
			1 \leq k < \ell \leq d,
			H \in \mathbb{H}_2
		\right\}.
\end{align}
Although trivariate or higher-dimensional words can be defined in a similar way,
we focus on these univariate and bivariate words below.

\subsection{Properties of $\widehat{\mu}$ with dictionaries $\mathcal{D}_0$, $\mathcal{D}_1$ and $\mathcal{D}_2$} \label{sec:theo_dict}
We investigate the theoretical behavior of the proposed estimators $\widehat{\mu}$ constructed from the above dictionaries $\mathcal{D}_0$, $\mathcal{D}_1$ and $\mathcal{D}_2$.
As the kernel function $K$ in these dictionaries, we mainly consider the Gaussian kernel $K \left( \vec{z} \right) = \left( 2 \pi \right)^{- d / 2} \exp \left[ - \left\| \vec{z} \right\|_2^2 / 2 \right]$ here.
Recall the dictionary $\mathcal{D} \left( \Delta_1, \Delta_2, c \right)$ for a dictionary $\mathcal{D}$ and fixed $\Delta_1, \Delta_2, c > 0$ as defined after Lemma \ref{lem:ineq2} in Section \ref{sec:theo}.
Dictionaries $\mathcal{D}_0 \left( \Delta_1, \Delta_2, c \right)$, $\mathcal{D}_1 \left( \Delta_1, \Delta_2, c \right)$ and $\mathcal{D}_2 \left( \Delta_1, \Delta_2, c \right)$ derived from $\mathcal{D}_0$, $\mathcal{D}_1$ and $\mathcal{D}_2$ satisfy the following proposition:
\begin{prop} \label{prop:D012}
For fixed $\Delta_1, \Delta_2, c > 0$, $\mathcal{D}_0 \left( \Delta_1, \Delta_2, c \right)$, $\mathcal{D}_1 \left( \Delta_1, \Delta_2, c \right)$ and $\mathcal{D}_2 \left( \Delta_1, \Delta_2, c \right)$ with the Gaussian kernel $K$ satisfy the assumption \ref{assu:dict}.
\end{prop}
Therefore, the second term of the bound in \eqref{eq:mu_hat} converges asymptotically to 0, when we construct the proposed estimator $\widehat{\mu}$ with $\mathcal{D}_0 \left( \Delta_1, \Delta_2, c \right)$, $\mathcal{D}_1 \left( \Delta_1, \Delta_2, c \right)$ and $\mathcal{D}_2 \left( \Delta_1, \Delta_2, c \right)$.

\section{Simulation} \label{sec:simul}
In this section, we conduct Monte Carlo simulations to investigate the finite-sample performance of the proposed estimator $\widehat{\mu}$ introduced in Section \ref{sec:esti}.
We utilize the dictionaries $\mathcal{D}_0$, $\mathcal{D}_1$ and $\mathcal{D}_2$ defined in Section \ref{sec:impl},
and consider normal, Bernoulli and Poisson regression in both low- and higher-dimensional settings.

Throughout this section, we assess estimation accuracy using the divergence risk $\mathcal{R} \left[ \Bar{\mu} \right] = \mathbb{E} \left[ D_U \left( \mu, \Bar{\mu} \right) \right]$ between the target regression function $\mu$ and an estimator $\Bar{\mu}$, as introduced in Section \ref{sec:theo}.
A smaller value of $\mathcal{R} \left[ \Bar{\mu} \right]$ indicates higher estimation accuracy.
Specifically, we will compute a Monte Carlo approximation $\widehat{\mathcal{R}} \left[ \Bar{\mu} \right]$ of $\mathcal{R} \left[ \Bar{\mu} \right]$ under the following settings:
we assume that the $d$-dimensional explanatory variables $\vec{X}$ follows the uniform distribution $U \left( \left[ -1, 1 \right]^d \right)$.
The conditional distribution of the objective variable $Y| \vec{X} = \vec{x}$ is assumed to be:
the normal distribution $N \left( \mu \left( \vec{x} \right), \left( 0.2 \right)^2 \right)$ in normal regression;
the Bernoulli distribution $Be \left( \mu \left( \vec{x} \right) \right)$ in Bernoulli regression;
and the Poisson distribution $Po \left( \mu \left( \vec{x} \right) \right)$ in Poisson regression,
where $\mu \left( \vec{x} \right)$ is the expectation of each conditional distribution.
We independently generate $n$ observations of $\left( Y, \vec{X} \right)$ and repeat this procedure $R$ times, then writing them as $\left( Y_1^{[r]}, \vec{X}_1^{[r]} \right), \ldots, \left( Y_n^{[r]}, \vec{X}_n^{[r]} \right)$ for $r = 1, \ldots, R$.
We also independently generate $S$ observations of $\vec{X}$: $\vec{X}_1^\ast, \ldots, \vec{X}_S^\ast$.
Calculating the estimator $\Bar{\mu}^{[r]}$ for each repetition $r$, we compute
\begin{align}
		\widehat{\mathcal{R}} \left[ \Bar{\mu} \right] = \frac{1}{R} \sum_{r=1}^{R} D_U^\ast \left( \mu, \Bar{\mu}^{[r]} \right), \quad
		D_U^\ast \left( \mu, \Bar{\mu}^{[r]} \right)
=
		\frac{1}{S}
		\sum_{s=1}^{S}
		d_U \left( \mu \left( \vec{X}_s^\ast\right), \Bar{\mu}^{[r]} \left( \vec{X}_s^\ast\right) \right) \label{eq:R^}
\end{align}
and its standard error
\begin{align}
		\widehat{\mathrm{se}} \left[ \Bar{\mu} \right]
=
		\sqrt{\widehat{\mathrm{Var}} \left[ \widehat{\mathcal{R}} \left[ \Bar{\mu} \right] \right]}
=
		\sqrt{
			\frac{1}{R^2}
			\sum_{r=1}^{R}
			\left\{
				D_U^\ast \left( \mu, \Bar{\mu}^{[r]} \right)
			-
				\widehat{\mathcal{R}} \left[ \Bar{\mu} \right] \right\}^2
			}. \label{eq:se^}
\end{align}
We set $R = S = 1000$.

Throughout this section, the parametric model function is set to $u^{-1} \left( \left[ 1 ~ \vec{x}^\top \right] \vec{\theta} \right)$ for $\vec{\theta} = \left[ \theta_0 ~ \theta_1 ~ \cdots ~ \theta_d \right]^\top \in \mathbb{R}^{d+1}$ and the inverse link function $u^{-1} = U^\ast$.
As the strictly convex function in the Bregman divergence, we use: 
in normal regression, $U \left( z \right) = z^2 / 2$
--- then $u = u^{-1} = \mathrm{id}$ 
and 
$U^\ast \left( z \right) = z^2 / 2$;
in Bernoulli regression, 
$U \left( z \right) = z \ln z + \left( 1 - z \right) \ln \left[ 1 - z \right] - 1$
--- then $u \left( z \right) = \ln \left[ z / \left( 1 - z \right) \right]$,
$u^{-1} \left( z \right) = 1 / \left( 1 + e^{-z} \right)$
and
$U^\ast \left( z \right) = \ln \left[ 1 + e^z \right]$;
in Poisson regression, $U \left( z \right) = z \ln z - z$
--- then $u \left( z \right)= \ln z$, $u^{-1} \left( z \right) = e^z$
and
$U^\ast \left( z \right) = e^z$.
In addition, we write $\vec{x} = \left[ x_1 ~ x_2 ~ \cdots ~ x_d \right]^\top$.

\subsection{Low dimension} \label{subsec:lower_dim}
We compare the estimation accuracy of the proposed estimator $\widehat{\mu}$ with that of other estimators.
The settings and notation for the estimators in this subsection are listed below:
\begin{itemize}
\item Parametric estimator $\widehat{\mu}_{\mathrm{P}}$.
	We use $\widehat{\mu}_{\mathrm{P}} \left( \vec{x} \right) = u^{-1} \left( \left[ 1 ~ \vec{x}^\top \right] \widehat{\vec{\theta}}_{\mathrm{M}} \right)$ with
	\begin{align}
			\widehat{\vec{\theta}}_{\mathrm{M}}
	=
			\argmin_{\vec{\theta} \in \mathbb{R}^{d+1}}
			\widehat{L} \left( u^{-1} \left( \left[ 1 ~ \cdot^\top \right] \vec{\theta} \right) \right), \label{eq:MLE}
	\end{align}
	where $\widehat{L}$ is \eqref{eq:emp_risk} associated with the Bregman divergence $D_U$.
	This estimator $\widehat{\vec{\theta}}_{\mathrm{M}}$ coincides with the MLE in GLMs.
	Under this setting, we compute $\widehat{\mathcal{R}} \left[ \widehat{\mu}_{\mathrm{P}} \right]$ and $\widehat{\mathrm{se}} \left[ \widehat{\mu}_{\mathrm{P}} \right]$ in \eqref{eq:R^} and \eqref{eq:se^} for resulting $\widehat{\mu}_{\mathrm{P}}$.
\item
	Proposed semiparametric estimator $\widehat{\mu}^\gamma_i$.
	The parametric model is the same as that in $\widehat{\mu}_{\mathrm{P}}$.
	We consider both $\gamma = 0$ and 1, corresponding to additive and multiplicative adjustment, respectively.
	As the dictionary, we utilize $\mathcal{D}_i$ for $i = 0, 1 \mathrm{~or~} 2$ in Section \ref{sec:impl} with the Gaussian kernel $K \left( \vec{z} \right) = \left( 2 \pi \right)^{- d / 2} \exp \left( - \left\| \vec{z} \right\|_2^2 / 2 \right)$.
	The bandwidths of these dictionaries are given in each of the subsections below.
	We set the sequence $\pi_k$ forming the coefficients of $\widehat{\xi}$ to $\pi_k = 2 / \left( k - 1 + 2 \right)$, which has already been introduced in Section \ref{sec:theo}.
	For each parameter $\gamma$ and dictionary index $i$, we write the proposed estimators as $\widehat{\mu}^\gamma_i \left( I \right)$ for the number of iterations $I$.
	Hence the adjustment terms in $\widehat{\mu}^\gamma_i \left( I \right)$ are composed of $I$ selected words.
	We compute the smallest divergence risk $\min_{I \in \mathbb{I}} \widehat{\mathcal{R}} \left[ \widehat{\mu}^\gamma_i \left( I \right) \right]$ over the range $\mathbb{I} = \left\{ 1, 2, \ldots, 300 \right\}$.
	We also calculate $\widehat{\mathrm{se}} \left[ \widehat{\mu}^\gamma_i \left( I \right) \right]$ at $\argmin_{I \in \mathbb{I}} \widehat{\mathcal{R}} \left[ \widehat{\mu}^\gamma_i \left( I \right) \right]$.
\item Semiparametric estimator $\widehat{\mu}_{\mathrm{S}}^\gamma$ from \citet{FWF2009}.
	We compare the estimators with another semiparametric estimator from \citet{FWF2009}:
	\[
			\widehat{\mu}_{\mathrm{S}}^\gamma 
	= 
			u^{-1} \left( m \left( \cdot, \widehat{\vec{\theta}}_{\mathrm{M}} \right) + m \left( \cdot, \widehat{\vec{\theta}}_{\mathrm{M}} \right)^\gamma \cdot \widehat{\xi}_H \right),
	\]
	where $m$, $\widehat{\vec{\theta}}_{\mathrm{M}}$ and $u^{-1}$ are the same as those in our $\widehat{\mu}^\gamma_i$, and $\widehat{\xi}_H$ is \eqref{eq:FWF_xi} with the Gaussian kernel.
	We take bandwidths $\mathbb{H}_{\mathrm{S}} = \left\{ \left( 0.01 \cdot s \right)^2 \cdot I_d \middle| s = 1, \ldots, 200\right\}$ and denote the resulting estimator by $\widehat{\mu}_{\mathrm{S}}^\gamma \left( H \right)$ for a bandwidth $H \in \mathbb{H}_{\mathrm{S}}$.
	We compute the divergence risk $\min_{H \in \mathbb{H}_{\mathrm{S}}} \widehat{\mathcal{R}} \left[ \widehat{\mu}_{\mathrm{S}}^\gamma \left( H \right) \right]$
	and its standard error $\widehat{\mathrm{se}} \left[ \widehat{\mu}_{\mathrm{S}}^\gamma \left( H \right) \right]$ at $\argmin_{H \in \mathbb{H}_{\mathrm{S}}} \widehat{\mathcal{R}} \left[ \widehat{\mu}_{\mathrm{S}}^\gamma \left( H \right) \right]$.
\item Nonparametric local linear estimator $\widehat{\mu}_{\mathrm{L}}$.
	We also compare the proposed estimators with the nonparametric local linear estimator in \citet{FHW1995}.
	The link function is set to $u^{-1}$, the same as $\widehat{\mu}^\gamma_i$.
	We use bandwidths $\mathbb{H}_{\mathrm{L}} = \left\{ \left( 0.01 \cdot s \right)^2 \cdot I_d \middle| s = 1, \ldots, 250 \right\}$ and the Gaussian kernel.
	The notation $\widehat{\mu}_{\mathrm{L}} \left( H \right)$ denotes this estimator with a bandwidth $H \in \mathbb{H}_{\mathrm{L}}$.
	We compute $\min_{H \in \mathbb{H}_{\mathrm{L}}} \widehat{\mathcal{R}} \left[ \widehat{\mu}_{\mathrm{L}} \left( H \right) \right]$ and $\widehat{\mathrm{se}} \left[ \widehat{\mu}_{\mathrm{L}} \left( H \right) \right]$ at $\argmin_{H \in \mathbb{H}_{\mathrm{L}}} \widehat{\mathcal{R}} \left[ \widehat{\mu}_{\mathrm{L}} \left( H \right) \right]$.
\item GBRT $\widehat{\mu}_{\mathrm{X}}$ via XGBoost.
	We also compare our estimators with the GBRT implemented in the package \texttt{xgboost} in \texttt{R}, see \citet{C2026} and \citet{CG2016}.
	The common settings for its arguments throughout this Section \ref{sec:simul} are:
	$\texttt{base\_score} = \left( 1 / n \right) \sum_{i = 1}^{n} Y_i$;
	$\texttt{min\_child\_weight} = 0$;
	$\texttt{maxdepth} \in \left\{ 1, 2, \ldots, 6 \right\} = \mathcal{M}_{\mathrm{X}}$;
	and $\texttt{tree\_method} = \text{``exact''}$.
	Moreover, we set the parameter $\texttt{objective} = \text{``reg:squarederror''}$ in normal regression;
	$= \text{``binary:logistic''}$ in Bernoulli regression;
	and $= \text{``count:poisson''}$ in Poisson regression.
	The parameters \texttt{eta} and \texttt{lambda} are:
	in normal regression, $\texttt{eta} = 0.1$ and $\texttt{lambda} = 0$;
	in Bernoulli and Poisson regression, $\texttt{eta} = 0.01$ and $\texttt{lambda} \in \left\{ 0, 1, 10, 100, 1000 \right\} = \Lambda_{\mathrm{X}}$.
	All the other parameters not mentioned here are set to their default values.
	To denote the estimator with $\texttt{maxdepth} = m$ and the number of iterations $\texttt{nrounds} = I$, we use the notation
	$\widehat{\mu}_{\mathrm{X}} \left( m, I \right)$ for normal regression, and
	$\widehat{\mu}_{\mathrm{X}} \left( m, \lambda, I \right)$ for $\texttt{lambda} = \lambda$ in Bernoulli and Poisson regression.
	Its divergence risk
	$\min_{m \in \mathcal{M}_{X}, I \in \mathbb{I}_{\mathrm{X}}} \widehat{\mathcal{R}} \left[ \widehat{\mu}_{\mathrm{X}} \left( m, I \right) \right]$
	or
	$\min_{m \in \mathcal{M}_{X}, \lambda \in \Lambda_{\mathrm{X}}, I \in \mathbb{I}_{\mathrm{X}}} \widehat{\mathcal{R}} \left[ \widehat{\mu}_{\mathrm{X}} \left( m, \lambda, I \right) \right]$
	is calculated, where the range $\mathbb{I}_{\mathrm{X}}$ is as follows:
	in normal regression, $\mathbb{I}_{\mathrm{X}} = \left\{ 1, 2, \ldots, 20000 \right\}$;
	in Bernoulli and Poisson regression, we first set $\mathbb{I}_{\mathrm{X}} = \left\{ 1, 2, \ldots, 5000 \right\}$.
	If the minimizing $I$ is between 4900 and 5000, we additionally search over $\left\{ 5001, 5002, \ldots, 10000 \right\}$ for the minimum value.
	If this minimizing $I$ satisfies $9901 \leq I \leq 10000$, then we again expand the search range to $\left\{ 10001, 10002, \ldots, 15000 \right\}$.
	We also compute
	$\widehat{\mathrm{se}} \left[ \widehat{\mu}_{\mathrm{X}} \left( m, I \right) \right]$
	or
	$\widehat{\mathrm{se}} \left[ \widehat{\mu}_{\mathrm{X}} \left( m, \lambda, I \right) \right]$ at the pair of searched parameters that minimizes the risk.
\end{itemize}

\subsubsection{Normal regression} \label{sec:simu_low_nor}
Here we consider normal regression in $d = 3$ and the following five target regression functions:
\begin{align*}
	&\mu_1 \left( \vec{x} \right) = \exp\left[ x_1 + x_2 + x_3 \right], \\
	&\mu_2 \left( \vec{x} \right) = e^{x_1} + \sin \left( \dfrac{\pi}{2} \left( x_2 + x_3 \right) \right), \\
	&\mu_3 \left( \vec{x} \right) = x_1^2 + e^{x_2} + \sin \left( \pi x_3 \right), \\
	&\mu_4 \left( \vec{x} \right) = 1 + x_1 + 2 x_2 + 3 x_3 + x_1^2 + e^{x_2} + \sin \left( \pi x_3 \right), \\
	&\mu_5 \left( \vec{x} \right) = \left( 1 + x_1 + 2 x_2 + 3 x_3 \right) \cdot \left\{ x_1^2 + e^{x_2} + \sin \left( \pi x_3 \right) \right\}.
\end{align*}
The sample size is $n = 50, 200$.

We utilize the dictionaries $\mathcal{D}_0$, $\mathcal{D}_1$ and $\mathcal{D}_2$ for the proposed estimator $\widehat{\mu}^\gamma_i$, so that $i = 0, 1, 2$.
The sets of bandwidths are given by $\mathbb{H}_0 = \left\{ s^2 \cdot I_d \middle| s^2 = 0.1, 0.2, \ldots, 1.5 \right\}$ for $\mathcal{D}_0$, $\mathbb{H}_1 = \left\{ 0.1, 0.2, \ldots, 1.0 \right\}$ for $\mathcal{D}_1$ and $\mathbb{H}_2 = \left\{ s^2 \cdot I_2 \middle| s^2 = 0.1, 0.2, \ldots, 1.0 \right\}$ for $\mathcal{D}_2$.

The simulation results for these settings are presented in Table \ref{tab:accu_nor}.
\begin{table}[tbp]
\centering
\setlength{\tabcolsep}{2.7pt}
\small
\begin{tabular}{crrrrrrrrrrrr}
\toprule
	$\mu$ & $n$ &
	$\widehat{\mu}^0_0 $ &
	$\widehat{\mu}^0_1 $ &
	$\widehat{\mu}^0_2 $ &
	$\widehat{\mu}^1_0 $ &
	$\widehat{\mu}^1_1 $ &
	$\widehat{\mu}^1_2 $ &
	$\widehat{\mu}_{\mathrm{P}}$&
	$\widehat{\mu}^0_{\mathrm{S}} $ &
	$\widehat{\mu}^1_{\mathrm{S}} $ &
	$\widehat{\mu}_{\mathrm{X}}$&
	$\widehat{\mu}_{\mathrm{L}}$\\
\midrule
\midrule
	$\mu_1$ & $50$ & $\dashuline{84[2]} $ & $517[4] $ & $167[2] $ & $\dotuline{108[1]} $ & $129[2] $ & $127[3] $ & $546[8] $ & $157[3] $ & $132[3] $ & $515[1] $ & $\uline{71[2]} $ \\
	&& 298 & 26 & 297 & 297 & 293 & 247 & & 0.05 & 0.06 & [2, 676] & 0.29 \\
	& $200$ & $\dashuline{16[.4]} $ & $450[.4] $ & $72[.6] $ & $\dotuline{33[.3]} $ & $86[.5] $ & $40[.3] $ & $502[2] $ & $48[.8] $ & $41[.6] $ & $140[.1] $ & $\uline{11[.5]} $ \\
	&& 300 & 59 & 296 & 300 & 297 & 295 & & 0.03 & 0.03 & [3, 20000] & 0.17 \\
\midrule
	$\mu_2$ & $50$ & $\dotuline{24[.3]} $ & $59[.5] $ & $\uline{12[.2]} $ & $34[.3] $ & $62[.2] $ & $46[.9] $ & $68[.4] $ & $32[.3] $ & $37[.3] $ & $73[.2] $ & $\dashuline{22[.2]} $ \\
	&& 298 & 40 & 300 & 297 & 25 & 191 & & 0.05 & 0.06 & [2, 672] & 0.29 \\
	& $200$ & $\dotuline{5[.05]} $ & $50[.07] $ & $\uline{1[.06]} $ & $12[.08] $ & $50[.02] $ & $20[.2] $ & $63[.1] $ & $11[.07] $ & $15[.08] $ & $21[.02] $ & $\dashuline{3[.04]} $ \\
	&& 300 & 119 & 300 & 298 & 95 & 294 & & 0.02 & 0.02 & [2, 20000] & 0.17 \\
\midrule
	$\mu_3$ & $50$ & $82[.6] $ & $\uline{15[.8]} $ & $\dashuline{23[.3]} $ & $90[.6] $ & $55[.3] $ & $54[.9] $ & $176[.2] $ & $97[.5] $ & $101[.6] $ & $\dotuline{26[.7]} $ & $87[.4] $ \\
	&& 297 & 300 & 300 & 300 & 299 & 292 & & 0.23 & 0.24 & [1, 10761] & 0.31 \\
	& $200$ & $25[.1] $ & $\uline{3[.2]} $ & $23[.03] $ & $34[.2] $ & $38[.3] $ & $54[.9] $ & $164[.02] $ & $36[.2] $ & $43[.3] $ & $\dashuline{4[.10]} $ & $\dotuline{15[.08]} $ \\
	&& 300 & 300 & 300 & 300 & 300 & 292 & & 0.02 & 0.02 & [1, 19999] & 0.16 \\
\midrule
	$\mu_4$ & $50$ & $\dotuline{82[.6]} $ & $\uline{7[2]} $ & $\dashuline{19[.1]} $ & $126[1] $ & $113[.3] $ & $126[6] $ & $176[1] $ & $97[.5] $ & $127[1] $ & $122[.7] $ & $87[.4] $ \\
	&& 297 & 300 & 299 & 125 & 258 & 91 & & 0.05 & 0.07 & [1, 15837] & 0.31 \\
	& $200$ & $25[.1] $ & $\uline{2[.2]} $ & $\dashuline{3[.02]} $ & $54[.1] $ & $83[.03] $ & $55[.2] $ & $164[.07] $ & $36[.2] $ & $57[.2] $ & $\dotuline{14[.10]} $ & $15[.08] $ \\
	&& 300 & 300 & 300 & 300 & 300 & 300 & & 0.02 & 0.02 & [1, 20000] & 0.16 \\
\midrule
	$\mu_5$ & $50$ & $644[8] $ & $2812[8] $ & $\uline{378[11]} $ & $746[5] $ & $\dashuline{384[5]} $ & $\dotuline{468[29]} $ & $3236[21] $ & $934[9] $ & $900[10] $ & $1682[7] $ & $706[8] $ \\
	&& 296 & 96 & 300 & 298 & 300 & 288 & & 0.05 & 0.06 & [2, 870] & 0.32 \\
	& $200$ & $172[1] $ & $2340[2] $ & $\uline{86[3]} $ & $256[1] $ & $233[.6] $ & $\dotuline{160[.8]} $ & $2999[2] $ & $313[2] $ & $339[3] $ & $393[1] $ & $\dashuline{143[2]} $ \\
	&& 300 & 299 & 300 & 300 & 300 & 300 & & 0.02 & 0.02 & [3, 1109] & 0.17 \\
\bottomrule
\end{tabular}
\caption{The minima of estimated divergence risks $\left( \times 10^3 \right)$ in low-dimensional normal regression.} \label{tab:accu_nor}
\end{table}
For each $\mu$ and $n$, Table \ref{tab:accu_nor} lists the above-mentioned divergence risks as follows:
in columns 3--5, $\min_{I \in \mathbb{I}} \widehat{\mathcal{R}} \left[ \widehat{\mu}_i^0 \left( I \right) \right]$ for $\gamma = 0$, $i = 0, 1, 2$;
in columns 6--8, $\min_{I \in \mathbb{I}} \widehat{\mathcal{R}} \left[ \widehat{\mu}_i^1 \left( I \right) \right]$ for $\gamma = 1$, $i = 0, 1, 2$;
in column 9, $\widehat{\mathcal{R}} \left[ \widehat{\mu}_{\mathrm{P}} \right]$;
in columns 10 and 11, $\min_{H \in \mathbb{H}_{\mathrm{S}}} \widehat{\mathcal{R}} \left[ \widehat{\mu}_{\mathrm{S}}^\gamma \left( H \right) \right]$ for $\gamma = 0, 1$;
in column 12, $\min_{m \in \mathcal{M}_{\mathrm{X}}, I \in\mathbb{I}_{\mathrm{X}}} \widehat{\mathcal{R}} \left[ \widehat{\mu}_{\mathrm{X}} \left( m, I \right) \right]$;
and in column 13, $\min_{H \in \mathbb{H}_{\mathrm{L}}} \widehat{\mathcal{R}} \left[ \widehat{\mu}_{\mathrm{L}} \left( H \right) \right]$.
Next to these risk values, the corresponding standard errors are given in square brackets.
The parameters minimizing the risks are also reported beneath the corresponding risk values:
in columns 3--5, $\argmin_{I \in \mathbb{I}} \widehat{\mathcal{R}} \left[ \widehat{\mu}_i^0 \left( I \right) \right]$ for $i = 0, 1, 2$;
in columns 6--8, $\argmin_{I \in \mathbb{I}} \widehat{\mathcal{R}} \left[ \widehat{\mu}_i^1 \left( I \right) \right]$ for $i = 0, 1, 2$;
in columns 10 and 11, $\argmin_{H \in \mathbb{H}_{\mathrm{S}}} \widehat{\mathcal{R}} \left[ \widehat{\mu}_{\mathrm{S}}^\gamma \left( H \right) \right]$ for $\gamma = 0, 1$;
in column 12, $\argmin_{m \in \mathcal{M}_{\mathrm{X}}, I \in\mathbb{I}_{\mathrm{X}}} \widehat{\mathcal{R}} \left[ \widehat{\mu}_{\mathrm{X}} \left( m, I \right) \right]$;
and in column 13, $\argmin_{H \in \mathbb{H}_{\mathrm{L}}} \widehat{\mathcal{R}} \left[ \widehat{\mu}_{\mathrm{L}} \left( H \right) \right]$.
In each row, the row-wise minimum, second minimum and third minimum risk values are underlined with solid, dashed and dotted lines, respectively.

Observations from Table \ref{tab:accu_nor} are as follows:
\setlist[enumerate]{label*=\textbf{\arabic*}., listparindent=\parindent}
\begin{enumerate}[label=\textbf{6}.\textbf{\alph*}., leftmargin=*, align = left, series=desc]
\item Comparing $\widehat{\mu}^\gamma_i$ with the other estimators.
	As shown in each row, in almost all cases of $\left( \mu, n \right)$ except $\left( \mu, n \right) = \left( \mu_1, 50 \right), \left( \mu_1, 200 \right)$,
	the risk values underlined with the solid lines are given by $\widehat{\mu}^\gamma_i$ with some adjustment type $\gamma$ and a dictionary $\mathcal{D}_i$.
	\begin{enumerate}
	\item $\widehat{\mu}^\gamma_i$ vs $\widehat{\mu}_{\mathrm{P}}$.
		In columns 3--8 and column 9, the risks for $\widehat{\mu}^\gamma_i$ are smaller than that of $\widehat{\mu}_{\mathrm{P}}$ for any $\gamma$ and $i$ in each $\left( \mu, n \right)$.
		This means that $\widehat{\mu}^\gamma_i$ adjusts $\widehat{\mu}_{\mathrm{P}}$ well in these cases.
	\item \label{desc:alg_vs_S0} $\widehat{\mu}^0_0$ vs $\widehat{\mu}^0_{\mathrm{S}}$.
		Columns 3 and 10 indicate that $\widehat{\mu}^0_0$ yields a smaller risk value than $\widehat{\mu}^0_{\mathrm{S}}$ in each $\left( \mu, n \right)$.
		That is, $\widehat{\mu}_{\mathrm{P}}$ is improved by the adjustment constructed with the algorithm more effectively than by that in $\widehat{\mu}^0_{\mathrm{S}}$.
	\item $\widehat{\mu}^1_0$ vs $\widehat{\mu}^1_{\mathrm{S}}$.
		Except when $\left( \mu, n \right) = \left( \mu_4, 50 \right)$, similar results to those in the description \ref{desc:alg_vs_S0} can be observed between $\widehat{\mu}^1_0$ (column 6) and $\widehat{\mu}^1_{\mathrm{S}}$ (column 11).
	\item $\widehat{\mu}^\gamma_i$ vs $\widehat{\mu}_{\mathrm{X}}$.
		According to columns 3--8 and column 12, several $\widehat{\mu}^\gamma_i$ with some $\gamma$, $i$ have higher accuracy than $\widehat{\mu}_{\mathrm{X}}$.
		On the other hand, in $\left( \mu, n \right) = \left( \mu_3, 50 \right), \left( \mu_3, 200 \right), \left( \mu_4, 200 \right)$, where $\mu$ are additive functions of univariate ones, $\widehat{\mu}_{\mathrm{X}}$ is the second or the third best among all the compared estimators.
	\item $\widehat{\mu}^\gamma_i$ vs $\widehat{\mu}_{\mathrm{L}}$.
		In $\left( \mu, n \right) = \left( \mu_1, 50 \right), \left( \mu_1, 200 \right)$, $\widehat{\mu}_{\mathrm{L}}$ is more accurate than $\widehat{\mu}^\gamma_i$ with any $\gamma$, $i$.
		One possible reason for this is as follows:
		considering the following quantity
		\[
				\mathrm{Ratio} \left( \mu, n \right)
		=
				\frac
				{
					\widehat{\mathcal{R}} \left[ \widehat{\mu}_{\mathrm{P}} \right]
				}{
					\min_{H \in \mathbb{H}_{\mathrm{L}}}
					\widehat{\mathcal{R}} \left[ \widehat{\mu}_{\mathrm{L}} \left( H \right) \right]
				},
		\]
		the resulting $\mathrm{Ratio} \left( \mu_1, 50 \right) \thickapprox 7.7$ for the case $\left( \mu, n \right) = \left( \mu_1, 50 \right)$ is much larger than the other $\mathrm{Ratio} \left( \mu_i, 50 \right)$ for $i = 2, \ldots, 5$.
		This is also seen when $n = 200$, among $\mathrm{Ratio} \left( \mu_1, 200 \right) \thickapprox 45.6$ and the other $\mathrm{Ratio} \left( \mu_i, 200 \right)$, $i = 2, \ldots, 5$.
		Hence, the risks of $\widehat{\mu}_{\mathrm{P}}$ are relatively larger than those of $\widehat{\mu}_{\mathrm{L}}$,
		and significant improvement in such $\widehat{\mu}_{\mathrm{P}}$ cannot be provided by the adjustment in $\widehat{\mu}^\gamma_i$ in these cases.
	\end{enumerate}
\item $\widehat{\mu}^0_i$ vs $\widehat{\mu}^1_i$.
	In each $\left( \mu, n \right)$, the smallest risk among $\widehat{\mu}^\gamma_i$ is achieved by $\widehat{\mu}^0_i$ with $\gamma = 0$ and some $i$.
	That is, in these cases, the additive adjustment is more effective than the multiplicative adjustment.
	In particular, since $\mu_5$ is defined as the product of linear and nonlinear functions, one might intuitively expect that $\mu_5$ would be better estimated by $\widehat{\mu}^1_i$ than $\widehat{\mu}^0_i$,
	but this does not occur in the last rows of Table \ref{tab:accu_nor}.
	\item \label{desc:match} $\widehat{\mu}^0_0$ vs $\widehat{\mu}^0_1$ vs $\widehat{\mu}^0_2$.
	As seen in columns 3--5, $\widehat{\mu}^0_i$ with $\gamma = 0$ and different $i$ yield the best accuracy for different $\mu$ among $\widehat{\mu}^0_0$, $\widehat{\mu}^0_1$ and $\widehat{\mu}^0_2$.
	Intuitively, this can be explained by considering whether the variables in $\mu$ match those in the words constructing the adjustment term in $\widehat{\mu}^0_i$:
	for example, we focus on $\mu_3$ and $\mu_4$.
	Since these targets are additive functions of univariate components,
	the adjustment for $\gamma = 0$ seems to be more appropriately constructed from $\mathcal{D}_1$ of univariate words than $\mathcal{D}_0$ and $\mathcal{D}_2$, which include bivariate or trivariate words.
	In fact, Table \ref{tab:accu_nor} shows that $\widehat{\mu}^0_1$ estimates $\mu_3$ and $\mu_4$ better than $\widehat{\mu}^0_0$ and $\widehat{\mu}^0_2$.
	Next for $\mu_2$, which sums univariate and bivariate functions, it can be said that the adjustment term should be constructed from $\mathcal{D}_2$, which has both univariate and bivariate words.
	Actually, $\widehat{\mu}^0_2$ has the best accuracy among $\widehat{\mu}^0_i$ in Table \ref{tab:accu_nor}.
	An analogous comparison of $\widehat{\mu}^0_0$ to $\widehat{\mu}^0_1$, $\widehat{\mu}^0_2$ can be made when $\mu = \mu_1$,
	indicating that $\widehat{\mu}^0_0$ of trivariate words provides better estimation than $\widehat{\mu}^0_1$ and $\widehat{\mu}^0_2$.
	In this way, there seems to be some appropriate combination of $\gamma$ and a dictionary for estimating $\mu$.
\end{enumerate}

\begin{figure}[htb]
\centering
\includegraphics[width = 0.8\textwidth]{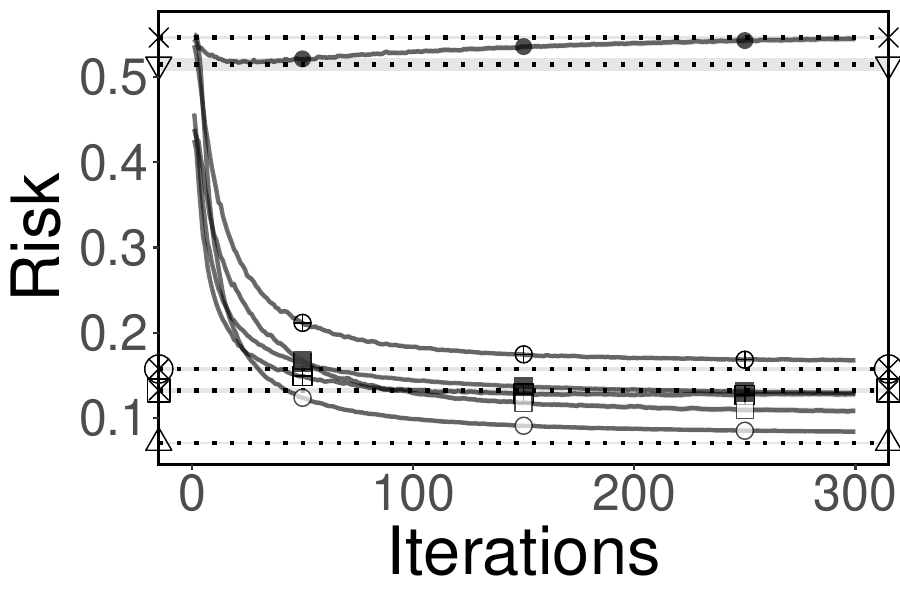}
\caption{
	The curves of estimated divergence risks for each estimator against the number of iterations for the proposed estimators $\widehat{\mu}^\gamma_i$, when $\mu = \mu_1$ and $n = 50$ in normal regression.
	$\bigcirc$: $\widehat{\mu}^0_0$,
	\blackbigcirc : $\widehat{\mu}^0_1$,
	$\oplus$: $\widehat{\mu}^0_2$,
	$\square$: $\widehat{\mu}^1_0$,
	$\blacksquare$: $\widehat{\mu}^1_1$,
	$\boxplus$: $\widehat{\mu}^1_2$,
	$\times$: $\widehat{\mu}_\mathrm{P}$,
	$\otimes$: $\widehat{\mu}^0_{\mathrm{S}}$,
	$\boxtimes$: $\widehat{\mu}^1_{\mathrm{S}}$,
	$\bigtriangledown$: $\widehat{\mu}_{\mathrm{X}}$ and
	$\bigtriangleup$: $\widehat{\mu}_{\mathrm{L}}$.
} \label{fig:nor_d3j}
\end{figure}

\begin{figure}[htb]
\centering
\includegraphics[width = 0.8\textwidth]{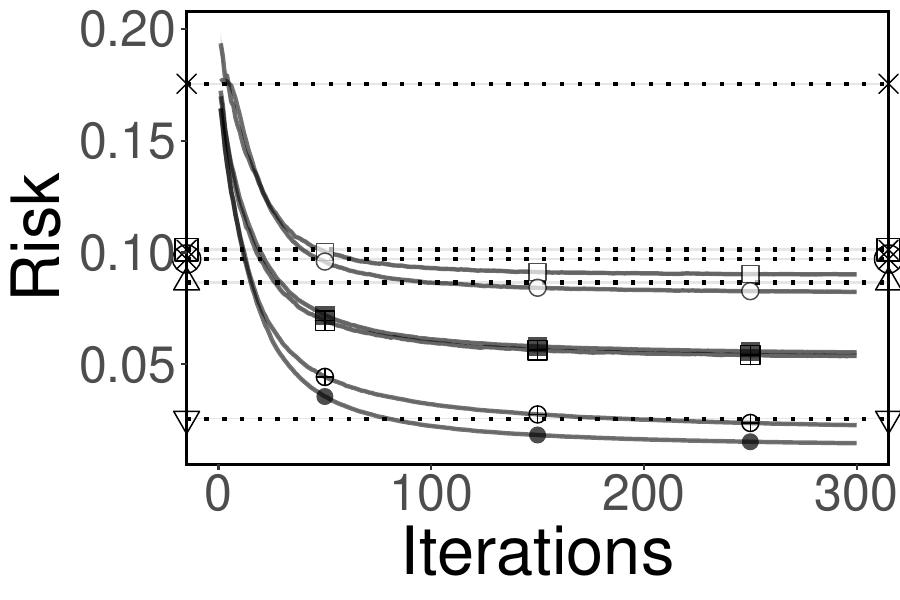}
\caption{
	The curves of estimated divergence risks for each estimator against the number of iterations for the proposed estimators $\widehat{\mu}^\gamma_i$, when $\mu = \mu_3$ and $n = 50$ in normal regression.
	$\bigcirc$: $\widehat{\mu}^0_0$,
	\blackbigcirc : $\widehat{\mu}^0_1$,
	$\oplus$: $\widehat{\mu}^0_2$,
	$\square$: $\widehat{\mu}^1_0$,
	$\blacksquare$: $\widehat{\mu}^1_1$,
	$\boxplus$: $\widehat{\mu}^1_2$,
	$\times$: $\widehat{\mu}_\mathrm{P}$,
	$\otimes$: $\widehat{\mu}^0_{\mathrm{S}}$,
	$\boxtimes$: $\widehat{\mu}^1_{\mathrm{S}}$,
	$\bigtriangledown$: $\widehat{\mu}_{\mathrm{X}}$ and
	$\bigtriangleup$: $\widehat{\mu}_{\mathrm{L}}$.
} \label{fig:nor_d3c}
\end{figure}

In addition to Table \ref{tab:accu_nor}, Figures \ref{fig:nor_d3j} and \ref{fig:nor_d3c} present the curves of the estimated divergence risks for each estimator against the number of iterations $I$ in $\widehat{\mu}^\gamma_i$,
when $\left( \mu, n \right) = \left(  \mu_1, 50 \right)$ and $\left( \mu_3, 50 \right)$.
The risk curve for $\widehat{\mu}^0_0$ (respectively $\widehat{\mu}^0_1$, $\widehat{\mu}^0_2$) is drawn with the points ``$\bigcirc$'' (respectively ``\blackbigcirc,'' ``$\oplus$'').
In contrast, the risk curve for $\widehat{\mu}^1_0$ (respectively $\widehat{\mu}^1_1$, $\widehat{\mu}^1_2$) is depicted with the points ``$\square$'' (respectively ``$\blacksquare$,'' ``$\boxplus $'').
The horizontal dotted lines represent the results for the estimators other than $\widehat{\mu}^\gamma_i$:
at both ends of these lines, the points ``$\times$'' indicate the result for $\widehat{\mu}_\mathrm{P}$;
``$\otimes $'' for $\widehat{\mu}^0_{\mathrm{S}}$;
``$\boxtimes $'' for $\widehat{\mu}^1_{\mathrm{S}}$;
``$\bigtriangledown$'' for $\widehat{\mu}_{\mathrm{X}}$;
and ``$\bigtriangleup $'' for $\widehat{\mu}_{\mathrm{L}}$.
\setlist[enumerate]{label*=\textbf{\arabic*}., listparindent=\parindent}
\begin{enumerate}[label=\textbf{6}.\textbf{\alph*}., leftmargin=*, align = left, resume=desc]
\item Figure \ref{fig:nor_d3j}.
	The risk curves for $\widehat{\mu}^0_0$, $\widehat{\mu}^1_0$, $\widehat{\mu}^1_1$, $\widehat{\mu}^0_2$ and $\widehat{\mu}^1_2$ decrease as $I$ increases.
	On the other hand, the risk of $\widehat{\mu}^0_1$ initially decreases with $I$, but increases from $I \thickapprox 30$,
	eventually taking much the same risk values as $\widehat{\mu}_{\mathrm{P}}$ for sufficiently large $I$.
\item Figure \ref{fig:nor_d3c}.
	The risks of $\widehat{\mu}^\gamma_i$ with each $\gamma$ and $i$ decrease with $I$.
	In particular, the risk of $\widehat{\mu}^0_2$ becomes smaller than that of $\widehat{\mu}_{\mathrm{L}}$ around $I \thickapprox 200$,
	while the risk of $\widehat{\mu}^0_1$ does so before $I \thickapprox 100$, and $\widehat{\mu}^0_1$ always outperforms the other estimators after $I \thickapprox 100$.
	This does not contradict the intuition in the description \ref{desc:match} that, when estimating the additive target $\mu = \mu_3$ of univariate functions, the adjustment for $\gamma = 0$ is more appropriately constructed from $\mathcal{D}_1$ of univariate words than $\mathcal{D}_2$ including bivariate ones.
\item Figure \ref{fig:nor_d3j} vs Figure \ref{fig:nor_d3c}.
	In Figure \ref{fig:nor_d3j}, $\widehat{\mu}^0_1$ improves $\widehat{\mu}_{\mathrm{P}}$ less than the other proposed estimators $\widehat{\mu}^\gamma_i$ for each $I$.
	In contrast, Figure \ref{fig:nor_d3c} shows that $\widehat{\mu}^0_1$ always outperforms the other estimators when $I > 100$.
	As described in \ref{desc:match}, this difference in the behavior of $\widehat{\mu}^0_1$ between $\mu = \mu_1$ and $\mu = \mu_3$ may be explained by considering whether the variables in $\mu$ match those in the words.
\end{enumerate}

In normal regression, it can be claimed that the proposed estimators with dictionaries $\mathcal{D}_0$, $\mathcal{D}_1$ and $\mathcal{D}_2$ have good estimation accuracy in the simulated cases.

\subsubsection{Bernoulli and Poisson regression} \label{sec:simu_low_berpoi}
For Bernoulli and Poisson regression, we consider the following dimension $d$, mean function $\mu$, and sample size $n$:
\begin{itemize}
\item in Bernoulli regression,
	\begin{itemize}
	\item
		$d=3$,
		$\displaystyle \mu \left( \vec{x} \right)
		=
				p_1 \left( \vec{x} \right)
		=
				0.5 + 0.4 \cos \left( \frac{\pi}{3} \left( x_1 + x_2 + x_3 \right) \right)$
		and $n=50, 200$;
	\item
		$d=4$,
		$\displaystyle \mu \left( \vec{x} \right)
		=
				p_2 \left( \vec{x} \right)
		=
				\frac{1}{1 + \exp \left[ - \left\{ x_1^2 + x_2 + e^{x_3} + \sin \left( \pi x_4 \right) \right\} \right]}$
		and
		$n=150$;
	\end{itemize}
\item in Poisson regression,
	\begin{itemize}
	\item
		$d=3$,
		$\displaystyle \mu \left( \vec{x} \right) = \lambda_1 \left( \vec{x} \right) = \exp \left[ 2 \sin \left( \frac{\pi}{12} \left( x_1 + x_2 + x_3 \right) - \frac{\pi}{2} \right) + 4 \right]$
		and $n=50, 200$;
	\item
		$d=4$,
		$\displaystyle \mu \left( \vec{x} \right) = \lambda_2 \left( \vec{x} \right) = \exp \left[ x_1^2 + x_2 + e^{x_3} + \sin \left( \pi x_4 \right) \right]$
		and $n=50$.
	\end{itemize}
\end{itemize}

We utilize the dictionaries $\mathcal{D}_0$ and $\mathcal{D}_1$ for the proposed estimator $\widehat{\mu}^\gamma_i$ (that is, $i = 0, 1$).
We apply several sets of bandwidths defined for $h_{\mathrm{min}} = 0.1, 0.2, \ldots, 0.7$:
\[
		\mathbb{H}_{0, h_{\mathrm{min}}} = \left\{ s^2 \cdot I_d \middle| s^2 = h_{\mathrm{min}}, h_{\mathrm{min}} + 0.1, \ldots, 1.5 \right\}
\]
of $d \times d$ scalar bandwidth matrices for $\mathcal{D}_0$ and $\mathbb{H}_{1, h_{\mathrm{min}}} = \left\{ s^2 \middle| s^2 = h_{\mathrm{min}}, h_{\mathrm{min}} + 0.1, \ldots, 1.0 \right\}$ of positive real numbers for $\mathcal{D}_1$.
For these sets, $h_{\mathrm{min}} < h_{\mathrm{min}}'$ implies $\mathbb{H}_{i, h_{\mathrm{min}}} \supset \mathbb{H}_{i, h_{\mathrm{min}}'}$.
Hence, larger $h_{\mathrm{min}}$ restricts these sets to larger bandwidths.
As will be discussed below, such a restriction on bandwidths is necessary for stable estimation with the proposed estimators.
To emphasize the dependence on $h_{\min}$, we write the proposed estimators as $\widehat{\mu}^\gamma_i \left( h_{\mathrm{min}}, I \right)$ from $\mathcal{D}_i$ with $\mathbb{H}_{i, h_{\mathrm{min}}}$.
For each $h_{\mathrm{min}} = 0.1, 0.2, \ldots, 0.7$, we compute $\min_{I \in \mathbb{I}} \widehat{\mathcal{R}} \left[ \widehat{\mu}^\gamma_i \left( h_{\mathrm{min}}, I \right) \right]$
and its standard error.

The results are summarized in Table \ref{tab:accu_berpoi}, much as in Table \ref{tab:accu_nor}.
\begin{sidewaystable}
\footnotesize
\setlength{\tabcolsep}{1.2pt}
\begin{tabular}{crrrrrrrrrrrrrrrrrr}
\toprule
	$\mu$ & $n$ &
	$\widehat{\mu}^0_0 $ &
	$\widehat{\mu}^0_0 $ &
	$\widehat{\mu}^0_0 $ &
	$\widehat{\mu}^0_1 $ &
	$\widehat{\mu}^0_1 $ &
	$\widehat{\mu}^0_1 $ &
	$\widehat{\mu}^1_0 $ &
	$\widehat{\mu}^1_0 $ &
	$\widehat{\mu}^1_0 $ &
	$\widehat{\mu}^1_1 $ &
	$\widehat{\mu}^1_1 $ &
	$\widehat{\mu}^1_1 $ &
	$\widehat{\mu}_{\mathrm{P}}$&
	$\widehat{\mu}^0_{\mathrm{S}} $ &
	$\widehat{\mu}^1_{\mathrm{S}} $ &
	$\widehat{\mu}_{\mathrm{X}}$&
	$\widehat{\mu}_{\mathrm{L}}$\\
\midrule
\midrule
	$p_1$ & $50$ & $\dotuline{78[2]}$ & $\uline{76[2]}$ & $\dashuline{77[2]}$ & $121[.5]$ & $119[2]$ & $118[2]$ & $104[2]$ & $102[2]$ & $101[2]$ & $122[2]$ & $120[2]$ & $118[.5]$ & $144[2]$ & $111[2]$ & $123[2]$ & $88[2]$ & $98[2]$ \\
	&& [.5, 283] & [.6, 297] & [.7, 296] & [.5, 294] & [.6, 287] & [.7, 296] & [.5, 239] & [.6, 285] & [.7, 276] & [.5, 294] & [.6, 293] & [.7, 294] & & 0.63 & 0.8 & [6, 1, 34] & 1 \\
	& $200$ & $\dotuline{28.1[.3]}$ & $\uline{26.7[.3]}$ & $\dashuline{27.8[.3]}$ & $88[.4]$ & $88[.4]$ & $88[.4]$ & $33[.4]$ & $31[.3]$ & $32[.3]$ & $85[.3]$ & $82[.3]$ & $88[.4]$ & $96[.4]$ & $48[.4]$ & $57[.4]$ & $58[.3]$ & $33[.3]$ \\
	&& [.5, 297] & [.6, 295] & [.7, 271] & [.4, 274] & [.5, 298] & [.6, 299] & [.4, 286] & [.5, 300] & [.6, 297] & [.2, 79] & [.3, 253] & [.5, 298] & & 0.53 & 0.61 & [4, 10, 335] & 0.72 \\
\midrule
	$\lambda_1$ & $50$ & $79[1]$ & $75[1]$ & $72.2[1]$ & $80[.4]$ & $80[1]$ & $80[1]$ & $79[1]$ & $76[1]$ & $71.99[1]$ & $80[1]$ & $80[1]$ & $80[.4]$ & $80[1]$ & $\dotuline{71.94[1]}$ & $72.1[1]$ & $\uline{41[1]}$ & $\dashuline{69[1]}$ \\
	&& [.5, 18] & [.6, 27] & [.7, 35] & [.5, 15] & [.6, 32] & [.7, 37] & [.5, 16] & [.6, 25] & [.7, 31] & [.5, 16] & [.6, 19] & [.7, 50] & & 0.63 & 0.63 & [4, $10^2$, 22] & 1.18 \\
	& $200$ & $24[.4]$ & $22.494[.4]$ & $\dashuline{21.7[.3]}$ & $42[.2]$ & $42[.3]$ & $42[.3]$ & $24[.3]$ & $\dotuline{22.491[.3]}$ & $\uline{21.6[.4]}$ & $42[.4]$ & $42[.3]$ & $42[.2]$ & $44[.3]$ & $27[.3]$ & $27[.3]$ & $31[.3]$ & $23[.4]$ \\
	&& [.5, 34] & [.6, 38] & [.7, 62] & [.5, 61] & [.6, 114] & [.7, 138] & [.5, 31] & [.6, 39] & [.7, 61] & [.5, 60] & [.6, 96] & [.7, 115] & & 0.46 & 0.46 & [4, $10^3$, 175] & 0.82 \\
\midrule
	$p_2$ & $150$ & $45[.5]$ & $44[.5]$ & $44[.5]$ & $\uline{34[.4]}$ & $\dashuline{35[.4]}$ & $\dotuline{36[.4]}$ & $50[.4]$ & $48[.4]$ & $48[.5]$ & $45[.5]$ & $45[.5]$ & $45[.4]$ & $40[.4]$ & $40[.4]$ & $40[.4]$ & $40[.4]$ & $40[.5]$ \\
	&& [.5, 1] & [.6, 1] & [.7, 1] & [.2, 300] & [.3, 300] & [.4, 300] & [.5, 1] & [.6, 1] & [.7, 1] & [.5, 178] & [.6, 173] & [.7, 298] & & 0.83 & 1.41 & [1, $10^2$, 5928] & 2.13 \\
\midrule
	$\lambda_2$ & $50$ & $16[.1]$ & $16[.1]$ & $17[.1]$ & $\uline{5[.1]}$ & $\dashuline{9[.1]}$ & $12[.1]$ & $18[.1]$ & $18[.1]$ & $18[.1]$ & $\dotuline{10[.1]}$ & $11[.1]$ & $13[.1]$ & $19[.1]$ & $17[.1]$ & $17[.1]$ & $11[.1]$ & $17[.1]$ \\
	&& [.2, 93] & [.3, 131] & [.4, 165] & [.1, 300] & [.2, 300] & [.3, 300] & [.5, 227] & [.6, 296] & [.7, 286] & [.1, 64] & [.2, 300] & [.3, 298] & & 0.39 & 0.4 & [1, $10^2$, 15000] & 0.79 \\
\bottomrule
\end{tabular}
\normalsize
\caption{The minima of estimated divergence risks $\left( \times 10^1 ~ \text{for} ~ \lambda_2; \times 10^3 ~ \text{for others} \right)$ in low-dimensional Bernoulli and Poisson regression.}\label{tab:accu_berpoi}
\end{sidewaystable}
We point out the main differences between Table \ref{tab:accu_berpoi} and Table \ref{tab:accu_nor}:
for the proposed estimators, the first, second and third smallest values in
$\left\{
	\min_{I \in \mathbb{I}} \widehat{\mathcal{R}} \left[ \widehat{\mu}^\gamma_i \left( h_{\mathrm{min}}, I \right) \right]
\middle|
	h_{\mathrm{min}} = 0.1, 0.2, \ldots, 0.7
\right\}$
are reported
in columns 3--5 for $\widehat{\mu}^0_0$;
in columns 6--8 for $\widehat{\mu}^0_1$;
in columns 9--11 for $\widehat{\mu}^1_0$;
and in columns 12--14 for $\widehat{\mu}^1_1$.
Below each risk, the corresponding $h_{\mathrm{min}}$ is also given together with $I$ minimizing it:
for example, the entries in rows 2--3 and columns 3--5 indicate that, in $\left( \mu, n \right) = \left( p_1, 50 \right)$,
$\min_{I \in \mathbb{I}} \widehat{\mathcal{R}} \left[ \widehat{\mu}^0_0 \left( 0.5, I \right) \right]$,
$\min_{I \in \mathbb{I}} \widehat{\mathcal{R}} \left[ \widehat{\mu}^0_0 \left( 0.6, I \right) \right]$ and
$\min_{I \in \mathbb{I}} \widehat{\mathcal{R}} \left[ \widehat{\mu}^0_0 \left( 0.7, I \right) \right]$
are smaller than $\min_{I \in \mathbb{I}} \widehat{\mathcal{R}} \left[ \widehat{\mu}^0_0 \left( h_{\mathrm{min}}, I \right) \right]$ with the other $h_{\mathrm{min}}$.
Columns 15--19 report the results of the estimators other than $\widehat{\mu}^\gamma_i$.
For $\widehat{\mu}_{\mathrm{X}}$ in column 18, $\texttt{maxdepth} = m$, $\texttt{lambda} = \lambda$ and $\texttt{nrounds} = I$ minimizing the risk are given as in $\left[ m, \lambda, I \right]$.

Recognitions from Table \ref{tab:accu_berpoi} are as follows:
\setlist[enumerate]{label*=\textbf{\arabic*}., listparindent=\parindent}
\begin{enumerate}[label=\textbf{6}.\textbf{\alph*}., leftmargin=*, align = left, resume=desc]
\item \label{desc:hmin}
	By $\mathcal{D}_{i, h_{\mathrm{min}}}$, we denote the dictionary $\mathcal{D}_i$ with the set of bandwidths $\mathbb{H}_{i, h_{\mathrm{min}}}$.
	In Section \ref{sec:simu_low_nor} on normal regression, we used $\widehat{\mu}^\gamma_i$ constructed only from $\mathcal{D}_{i, 0.1}$ with $h_{\mathrm{min}} = 0.1$, and such $\widehat{\mu}^\gamma_i$ outperforms the other estimators in almost all simulated cases.
	On the other hand, in Bernoulli and Poisson regression, $\widehat{\mu}^\gamma_i$ from $\mathcal{D}_{i, 0.1}$ does not always have the best accuracy among the compared estimators,
	and the restriction for the bandwidth set improves the accuracy of $\widehat{\mu}^\gamma_i$ in some simulated cases:
	for example, focusing on the case $\left( \mu, n \right) = \left( \lambda_1, 200 \right)$ and the estimator $\widehat{\mu}^1_0$, we can see from Table \ref{tab:accu_berpoi} that
	\begin{align}
			\min_{I \in \mathbb{I}}
			\widehat{\mathcal{R}} \left[ \widehat{\mu}^1_0 \left( 0.7, I \right) \right]
	=
			21.6 \pm 0.4
	<
			23 \pm 0.4
	=
			\min_{H \in \mathbb{H}_{\mathrm{L}}}
			\widehat{\mathcal{R}} \left[ \widehat{\mu}_{\mathrm{L}} \left( H \right) \right] \label{eq:evid1}
	\end{align}
	for $\widehat{\mu}^1_0$ with $\mathcal{D}_{i, 0.7}$, while
	\begin{align}
			\min_{I \in \mathbb{I}}
			\widehat{\mathcal{R}} \left[ \widehat{\mu}^1_0 \left( h_{\mathrm{min}}, I \right) \right]
	>
			\min_{H \in \mathbb{H}_{\mathrm{L}}}
			\widehat{\mathcal{R}} \left[ \widehat{\mu}_{\mathrm{L}} \left( H \right) \right] \label{eq:evid2}
	\end{align}
	for $\widehat{\mu}^1_0$ from $\mathcal{D}_{i, h_{\mathrm{min}}}$ with $h_{\mathrm{min}} \leq 0.5$.

	We describe this in further detail.
	First, $h_{\mathrm{min}} > 0.1$ implies $\mathcal{D}_{i, 0.1} \supset \mathcal{D}_{i, h_{\mathrm{min}}}$, so that $\mathcal{D}_{i, 0.1}$ is a thicker dictionary than $\mathcal{D}_{i, h_{\mathrm{min}}}$.
	However, the above results \eqref{eq:evid1} and \eqref{eq:evid2} mean that the estimated ``divergence risk'' does not become smaller,
	even if the ``empirical risk'' in constructing $\widehat{\mu}^\gamma_i$ becomes smaller by using a thicker dictionary.
	More specifically, there is a difference between construction of $\widehat{\mu}^\gamma_i \left( 0.1, I \right)$ and $\widehat{\mu}^\gamma_i \left( h_{\mathrm{min}}, I \right)$,
	in that $\widehat{\mu}^\gamma_i \left( 0.1, I \right)$ may select words $B_\sharp$ such that $B_\sharp \notin \mathcal{D}_{i, h_{\mathrm{min}}}$ but $B_\sharp \in \mathcal{D}_{i, 0.1}$.
	In \hyperref[alg:step2]{$\left\langle Step~2 \right\rangle$} of the algorithm in Section \ref{sec:esti} for $\widehat{\mu}^\gamma_i$,
	the following inequality for the ``empirical risk''
	\begin{align*}
			\inf_{\Bar{B} \in \mathcal{D}_{i, 0.1}}
			\widehat{L} \left( u^{-1} \left(
				T \left( \Bar{\vec{\theta}}, C \left( \pi_k \middle| \Bar{B}, \xi_{k - 1} \right) \right)
			\right) \right)
	\leq
			\inf_{\Bar{B} \in \mathcal{D}_{i, h_{\mathrm{min}}}}
			\widehat{L} \left( u^{-1} \left(
				T \left( \Bar{\vec{\theta}}, C \left( \pi_k \middle| \Bar{B}, \xi_{k - 1} \right) \right)
			\right) \right),
	\end{align*}
	holds for $k = 1, \ldots, I$.
	However, if the above $B_\sharp$ is selected at some iteration $k$ in \hyperref[alg:step2]{$\left\langle Step~2 \right\rangle$},
	the resulting forms of $\widehat{\mu}^\gamma_i \left( 0.1, I \right)$ and $\widehat{\mu}^\gamma_i \left( h_{\mathrm{min}}, I \right)$ differ essentially,
	and the inequality of the ``divergence risk''
	\begin{align*}
			\widehat{\mathcal{R}} \left[ \widehat{\mu}^\gamma_i \left( 0.1, I \right) \right]
	<
			\widehat{\mathcal{R}} \left[ \widehat{\mu}^\gamma_i \left( h_{\mathrm{min}}, I \right) \right]
	\end{align*}
	does not always hold.
	That is, the divergence risk is not always reduced with a thicker dictionary.

	From the above observations, when constructing the proposed estimators in Bernoulli and Poisson regression, it may be preferable to utilize $\mathcal{D}_0$ and $\mathcal{D}_1$ with larger bandwidths than those in normal regression.
\item
	According to columns 3--14 of Table \ref{tab:accu_berpoi}, in almost all cases except $\left( \mu, n \right) = \left( \lambda_1, 50 \right)$, the risks indicated by the solid lines are given by $\widehat{\mu}^\gamma_i$ with some $\gamma$, $i$.
	On the other hand, there are also some $\gamma$, $i$ for which $\widehat{\mu}^\gamma_i$ does not have a smaller risk than other estimators:
	\begin{enumerate}
	\item $\widehat{\mu}^\gamma_i$ vs $\widehat{\mu}_{\mathrm{P}}$.
		The proposed estimators $\widehat{\mu}^0_1$, $\widehat{\mu}^1_1$ in $\left( \mu, n \right) = \left( \lambda_1, 50 \right)$,
		and $\widehat{\mu}^0_0$, $\widehat{\mu}^1_0$, $\widehat{\mu}^1_1$ in $\left( \mu, n \right) = \left( p_2, 150 \right)$ are not more accurate than $\widehat{\mu}_{\mathrm{P}}$.
	\item $\widehat{\mu}^0_0$ vs $\widehat{\mu}^0_{\mathrm{S}}$.
		In $\left( \mu, n \right) = \left( \lambda_1, 50 \right), \left( p_2, 150 \right)$ and $\left( \lambda_2, 50 \right)$, the risks of $\widehat{\mu}^0_0$ are not smaller than those of $\widehat{\mu}^0_{\mathrm{S}}$.
		On the other hand, the risks of $\widehat{\mu}^0_0$ are smaller than those of $\widehat{\mu}^0_{\mathrm{S}}$ in $\left( \mu, n \right) = \left( p_1, 50 \right), \left( p_1, 200 \right)$ and $\left( \lambda_1, 200 \right)$.
	\item $\widehat{\mu}^1_0$ vs $\widehat{\mu}^1_{\mathrm{S}}$.
		In $\left( \mu, n \right) = \left( \lambda_1, 50 \right), \left( p_2, 150 \right)$ and $\left( \lambda_2, 50 \right)$, $\widehat{\mu}^1_0$ does not have smaller risks than $\widehat{\mu}^1_{\mathrm{S}}$.
		On the other hand, $\widehat{\mu}^1_0$ has smaller risks than $\widehat{\mu}^1_{\mathrm{S}}$ in $\left( \mu, n \right) = \left( p_1, 50 \right)$, $\left( p_1, 200 \right)$ and $\left( \lambda_1, 200 \right)$.
	\item $\widehat{\mu}^\gamma_i$ vs $\widehat{\mu}_{\mathrm{X}}$, $\widehat{\mu}_{\mathrm{L}}$.
		In $\left( \mu, n \right) = \left( \lambda_1, 50 \right)$, $\widehat{\mu}_{\mathrm{X}}$ and $\widehat{\mu}_{\mathrm{L}}$ have the best and the second best accuracy among the compared estimators.
		On the other hand, in almost all cases except $\left( \mu, n \right) = \left( \lambda_1, 50 \right)$,
		$\widehat{\mu}^\gamma_i$ with some $\gamma, i$ has higher accuracy than $\widehat{\mu}_{\mathrm{X}}$ and $\widehat{\mu}_{\mathrm{L}}$.
	\end{enumerate}
\end{enumerate}

\begin{figure}[tbp]
\centering
\subfigure[$n = 50$]{\label{subfig:ber_d3j_n50}
	\includegraphics[width=.45\columnwidth]{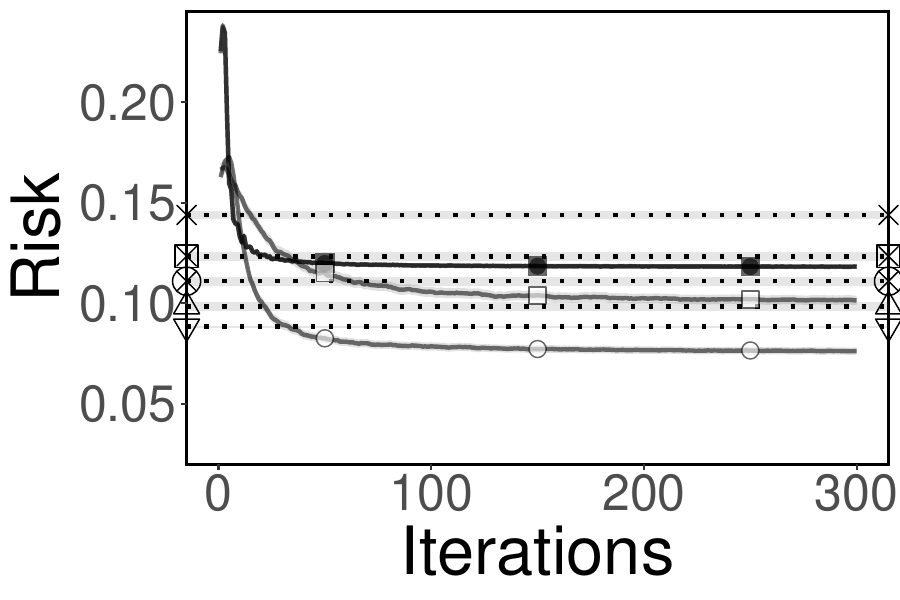}
}
\subfigure[$n = 200$]{\label{subfig:ber_d3j_n200}
	\includegraphics[width=.45\columnwidth]{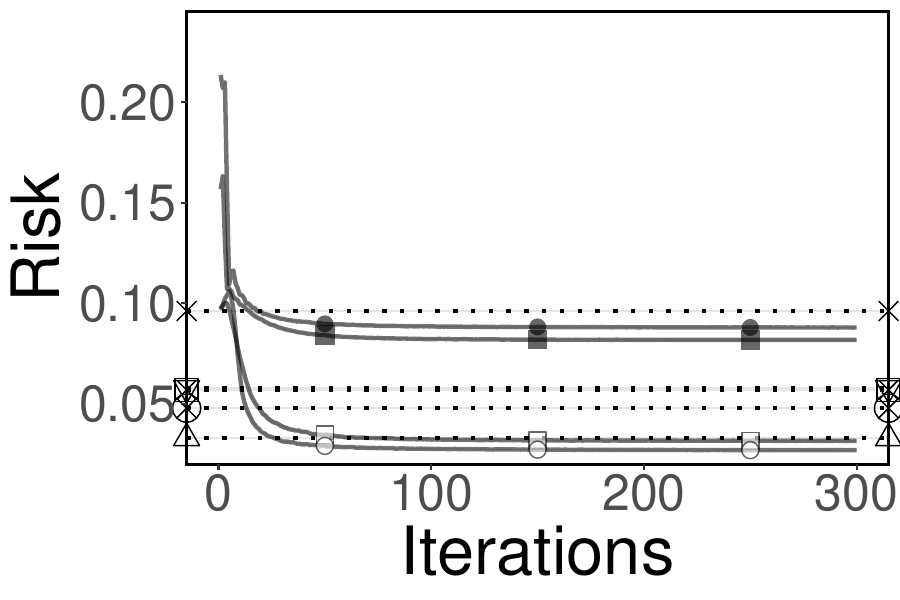}
}
\caption{
	The curves of estimated divergence risks for each estimator against the number of iterations for the proposed estimators $\widehat{\mu}^\gamma_i$, when $\mu = p_1$ in Bernoulli regression. (a) $n = 50$; (b) $n = 200$.
	$\bigcirc$: $\widehat{\mu}^0_0$,
	\blackbigcirc: $\widehat{\mu}^0_1$,
	$\square$: $\widehat{\mu}^1_0$,
	$\blacksquare$: $\widehat{\mu}^1_1$,
	$\times$: $\widehat{\mu}_\mathrm{P}$,
	$\otimes $: $\widehat{\mu}^0_{\mathrm{S}}$,
	$\boxtimes $: $\widehat{\mu}^1_{\mathrm{S}}$,
	$\bigtriangledown$: $\widehat{\mu}_{\mathrm{X}}$ and
	$\bigtriangleup $: $\widehat{\mu}_{\mathrm{L}}$.
} \label{fig:ber_d3j}
\end{figure}

\begin{figure}[tbp]
\centering
\subfigure[$n = 50$]{\label{subfig:poi_d3j_n50}
	\includegraphics[width=.45\columnwidth]{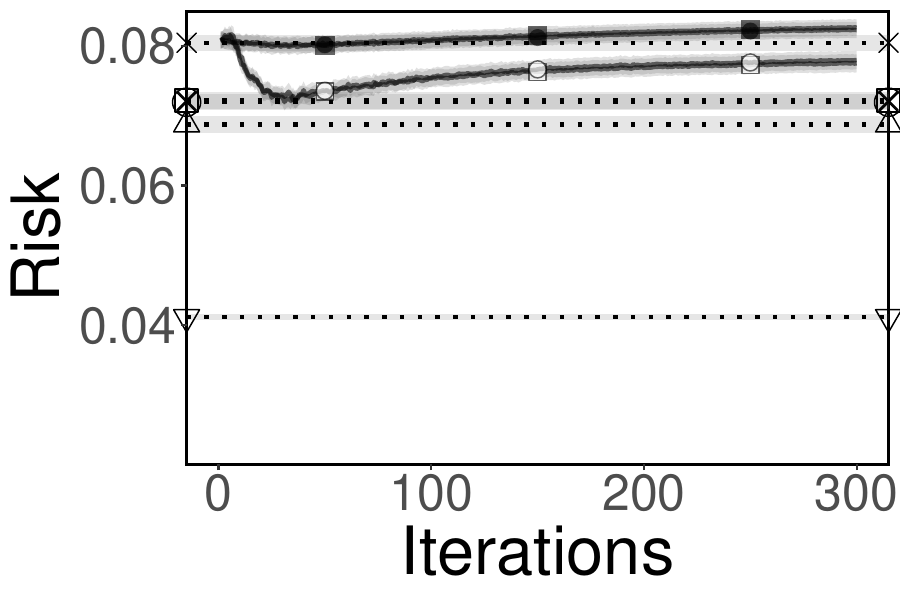}
}
\subfigure[$n = 200$]{\label{subfig:poi_d3j_n200}
	\includegraphics[width=.45\columnwidth]{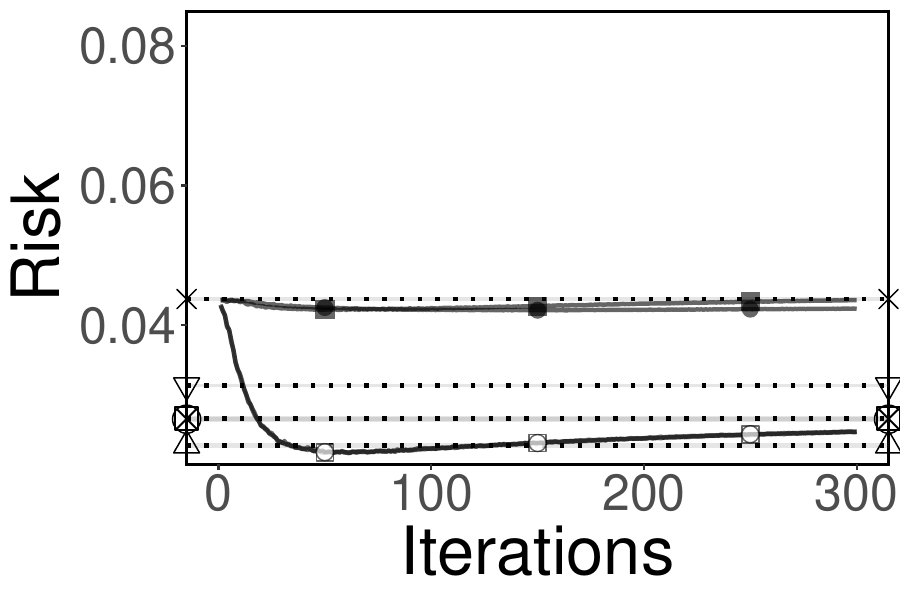}
}
\caption{
	The curves of estimated divergence risks for each estimator against the number of iterations for the proposed estimators $\widehat{\mu}^\gamma_i$, when $\mu = \lambda_1$ in Poisson regression. (a) $n = 50$; (b) $n = 200$.
	$\bigcirc$: $\widehat{\mu}^0_0$,
	\blackbigcirc: $\widehat{\mu}^0_1$,
	$\square$: $\widehat{\mu}^1_0$,
	$\blacksquare$: $\widehat{\mu}^1_1$,
	$\times$: $\widehat{\mu}_\mathrm{P}$,
	$\otimes $: $\widehat{\mu}^0_{\mathrm{S}}$,
	$\boxtimes $: $\widehat{\mu}^1_{\mathrm{S}}$,
	$\bigtriangledown$: $\widehat{\mu}_{\mathrm{X}}$ and
	$\bigtriangleup $: $\widehat{\mu}_{\mathrm{L}}$.
} \label{fig:poi_d3j}
\end{figure}

When $\left( \mu, n \right) = \left( p_1, 50 \right), \left( p_1, 200 \right), \left( \lambda_1, 50 \right)$ and $\left(\lambda_2, 200 \right)$, Figures \ref{fig:ber_d3j} and \ref{fig:poi_d3j} display risk curves for each estimator in exactly the same manner as in Figures \ref{fig:nor_d3j} and \ref{fig:nor_d3c}.
In particular, for the proposed estimators, Figures \ref{fig:ber_d3j} and \ref{fig:poi_d3j} present the results of $\widehat{\mathcal{R}} \left[ \widehat{\mu}^\gamma_i \left( h^\ast_{\mathrm{min}} \left( \gamma, i \right), I \right) \right]$ with
		$h^\ast_{\mathrm{min}} \left( \gamma, i \right)
=
		\argmin \left\{
			\min_{I \in \mathbb{I}} \widehat{\mathcal{R}} \left[ \widehat{\mu}^\gamma_i \left( h_{\mathrm{min}}, I \right) \right]
		\middle|
			h_{\mathrm{min}} = 0.1, 0.2, \ldots, 0.7
		\right\}$:
that is, $\widehat{\mathcal{R}} \left[ \widehat{\mu}^\gamma_i \left( h_{\mathrm{min}}, I \right) \right]$ only with the $h_{\mathrm{min}}$ that yields the best accuracy among $\min_{I \in \mathbb{I}} \widehat{\mathcal{R}} \left[ \widehat{\mu}^\gamma_i \left( h_{\mathrm{min}}, I \right) \right]$ for the simulated $h_{\mathrm{min}}$.
\setlist[enumerate]{label*=\textbf{\arabic*}., listparindent=\parindent}
\begin{enumerate}[label=\textbf{6}.\textbf{\alph*}., leftmargin=*, align = left, resume=desc]
\item Figure \ref{fig:ber_d3j}.
	\begin{enumerate}
	\item
		It can be seen that panel \subref{subfig:ber_d3j_n50} contains only three rather than four risk curves for $\widehat{\mu}^\gamma_i$.
		This is because the curves for $\widehat{\mu}^0_1$ and $\widehat{\mu}^1_1$ overlap for almost all $I$.
	\item
		In panel \subref{subfig:ber_d3j_n50}, in $\left( \mu, n \right) = \left( p_1, 50 \right)$, the risk of each $\widehat{\mu}^\gamma_i$ initially takes larger values than that of $\widehat{\mu}_{\mathrm{P}}$ for sufficiently small $I$,
		and then decreases with $I$.
		In particular, the risk of $\widehat{\mu}^0_0$ is smaller than that of $\widehat{\mu}_\mathrm{L}$ before $I \thickapprox 50$, and then becomes the smallest among the estimators.
		Similar behavior can be seen in panel \subref{subfig:ber_d3j_n200} for $n  = 200$.
	\item
		Comparing the curves of $\widehat{\mu}^\gamma_i$ between $n = 50$ and $n = 200$ for each $\gamma, i$, we can see that the risk in $n = 200$ is smaller than that in $n = 50$ for almost all $I$.
		This means that the estimation accuracy of the proposed estimators improves as the sample size increases.
	\end{enumerate}
\item Figure \ref{fig:poi_d3j}.
	\begin{enumerate}
	\item
		In panel \subref{subfig:poi_d3j_n50}, the curves for $\widehat{\mu}^0_0$ and $\widehat{\mu}^1_0$ overlap for almost all $I$, and similarly the curves for $\widehat{\mu}^0_1$ and $\widehat{\mu}^1_1$ also overlap.
		Hence, the proposed estimators do not differ substantially between $\gamma = 0$ and 1.
		Much the same behavior is observed in panel \subref{subfig:poi_d3j_n200} for $n = 200$.
	\item
		As observed in panel \subref{subfig:poi_d3j_n50} for $\left( \mu, n \right) = \left( \lambda_1, 50 \right)$, $\widehat{\mu}^0_1$ and $\widehat{\mu}^1_1$ do not have smaller risk values than $\widehat{\mu}_\mathrm{P}$.
		On the other hand, both $\widehat{\mu}^0_0$ and $\widehat{\mu}^1_0$ improve $\widehat{\mu}_\mathrm{P}$.
		However, after $\widehat{\mu}^0_0$ and $\widehat{\mu}^1_0$ yield much the same risk values as $\widehat{\mu}^0_{\mathrm{S}}$, $\widehat{\mu}^1_{\mathrm{S}}$ around $I \thickapprox 30$, they increase with $I$.
		In addition, the risks of $\widehat{\mu}^0_0$ and $\widehat{\mu}^1_0$ are higher than those of $\widehat{\mu}_{\mathrm{X}}$ and $\widehat{\mu}_{\mathrm{L}}$ for any $I$.
		\item
		In contrast to the case $n = 50$, panel \subref{subfig:poi_d3j_n200} shows that both $\widehat{\mu}^0_1$ and $\widehat{\mu}^1_1$ slightly improve $\widehat{\mu}_\mathrm{P}$ in $n = 200$.
		Moreover, $\widehat{\mu}^0_0$ and $\widehat{\mu}^1_0$ have the smallest risk around $I \thickapprox 50$ among the estimators.
	\end{enumerate}
\end{enumerate}

In normal regression, $\widehat{\mu}^\gamma_i$ outperforms $\widehat{\mu}_{\mathrm{P}}$ and $\widehat{\mu}^\gamma_{\mathrm{S}}$ for each $\gamma$, $i$.
On the other hand, in Bernoulli and Poisson regression, although $\widehat{\mu}^\gamma_i$ with some $\gamma$, $i$ underperforms $\widehat{\mu}_{\mathrm{P}}$ and $\widehat{\mu}^\gamma_{\mathrm{S}}$,
there are also some $\gamma$, $i$ for which $\widehat{\mu}^\gamma_i$ has the best accuracy among the compared estimators.
Consequently, as in normal regression, there also seems to be some appropriate combination of $\gamma$ and dictionary for estimating $\mu$ in Bernoulli and Poisson regression.

\subsection{Higher dimension} \label{sec:simu_high}
This subsection addresses higher-dimensional settings in which the target $\mu$ is an additive function composed of univariate ones.
We consider both cases in $d=10$ and in a high-dimension low-sample-size setting $n / d = 0.5$.
In detail, we put $\eta \left( \vec{x} \right) = x_1^2 + x_2 + e^{x_3} + \sin \left( \pi x_4 \right)$, which is defined on $\mathbb{R}^d$ but depends only on the first four coordinates $x_1, x_2, x_3, x_4$.
We then define $\mu$ and set $\left( d, n \right)$ as follows:
\begin{itemize}
	\item
		$\mu = \mu_d = \eta$ for $\left( d, n \right) = \left( 10, 50 \right), \left( 100, 50 \right)$ in normal regression;
	\item
		$\mu = p_d = \left( 1 + e^{- \eta} \right)^{-1} $ for $\left( d, n \right) = \left( 10, 150 \right), \left( 300, 150 \right)$ in Bernoulli regression;
	\item
		and $\mu = \lambda_d = e^\eta$ for $\left( d, n \right) = \left( 10, 50 \right), \left( 100, 50 \right)$ in Poisson regression.
\end{itemize}
These target functions can each be expressed as $u^{-1} \left( \eta \right)$ in the three regression settings.

We estimate this $\mu$ with the following estimators:
\begin{itemize}
\item Parametric estimator $\widehat{\mu}_{\mathrm{P}}$.
	In $d = 10$, we use $u^{-1} \left( \left[ 1 ~ \vec{x}^\top \right] \widehat{\vec{\theta}}_{\mathrm{M}} \right)$
	with the MLE $\widehat{\vec{\theta}}_{\mathrm{M}}$ of \eqref{eq:MLE}.
	Here we also utilize the LASSO-based parametric regression estimator:
	$u^{-1} \left( \left[ 1 ~ \vec{x}^\top \right] \widehat{\vec{\theta}}_{\mathrm{L}} \right)$,
	where $\widehat{\vec{\theta}}_{\mathrm{L}}$ denotes the LASSO estimator of the parameter $\vec{\theta}$.
	We write the corresponding parametric estimators as $\widehat{\mu}_{\mathrm{P}} \left( \widehat{\vec{\theta}}_{\mathrm{M}} \right)$ and $\widehat{\mu}_{\mathrm{P}} \left( \widehat{\vec{\theta}}_{\mathrm{L}} \right)$, respectively.
	In the case $n / d = 0.5$, we use only $\widehat{\mu}_{\mathrm{P}} \left( \widehat{\vec{\theta}}_{\mathrm{L}} \right)$.
	We implement $\widehat{\vec{\theta}}_{\mathrm{L}}$ with the package $\texttt{glmnet}$ in $\texttt{R}$,
	where the regularization parameter is selected according to the default cross-validation procedure in the function \texttt{cv.glmnet}, see Section ``Cross-validation'' in \citet{HQT2026}.
	Under this setting, we compute the divergence risks of the resulting parametric estimators.
\item Proposed semiparametric estimator $\widehat{\mu}^\gamma_i$.
	We fix $i = 1$ and $\gamma = 0$.
	That is, the dictionary $\mathcal{D}_1$ is utilized and the adjustment is additive,
	so that $\widehat{\mu}^0_1$ is an additive function composed of univariate words.
	In $d = 10$, we consider $\widehat{\mu}^0_1$ derived from the above two parametric estimators $\widehat{\mu}_{\mathrm{P}} \left( \widehat{\vec{\theta}}_{\mathrm{M}} \right)$ and $\widehat{\mu}_{\mathrm{P}} \left( \widehat{\vec{\theta}}_{\mathrm{L}} \right)$.
	We write them as
	$\widehat{\mu}^0_1 \left( \widehat{\vec{\theta}}_{\mathrm{M}} \right) = u^{-1} \left( m \left( \cdot, \widehat{\vec{\theta}}_{\mathrm{M}} \right) + \widehat{\xi}_{\mathrm{M}} \right)$
	and
	$\widehat{\mu}^0_1 \left( \widehat{\vec{\theta}}_{\mathrm{L}} \right) = u^{-1} \left( m \left( \cdot, \widehat{\vec{\theta}}_{\mathrm{L}} \right) + \widehat{\xi}_{\mathrm{L}} \right)$,
	where $\widehat{\xi}_{\mathrm{M}}$ and $\widehat{\xi}_{\mathrm{L}}$ are the adjustment terms resulting from the corresponding parametric estimators.
	All remaining parameter settings and the calculation of divergence risks are identical to those in Section \ref{subsec:lower_dim}.
\item GBRT $\widehat{\mu}_{\mathrm{X}}$ via XGBoost.
	The parameter settings for $\widehat{\mu}_{\mathrm{X}}$ are exactly the same as in Section \ref{subsec:lower_dim}.
\end{itemize}

Besides the estimation accuracy, we investigate practical behavior of the word selection in constructing the adjustment terms of the $\widehat{\mu}^0_1$.
Since these $\widehat{\xi}_{\mathrm{M}}$ and $\widehat{\xi}_{\mathrm{L}}$ depend on the parametric estimators utilized,
we also examine the behavior of $\widehat{\vec{\theta}}_{\mathrm{M}} $ and $ \widehat{\vec{\theta}}_{\mathrm{L}} $.
Specifically, in addition to the divergence risk $\widehat{\mathcal{R}}$, we compute the quantities $\widehat{\vec{\theta}}^\ast_\mathrm{M}$, $\widehat{\vec{\theta}}^\ast_\mathrm{L}$
and
$\mathcal{C}_s \left( \widehat{\vec{\theta}}_\mathrm{M} \right)$, $\mathcal{C}_s \left( \widehat{\vec{\theta}}_\mathrm{L} \right)$ below, based on the observations $\left( Y_1^{[r]}, \vec{X}_1^{[r]} \right), \ldots, \left( Y_n^{[r]}, \vec{X}_n^{[r]} \right)$ for $r = 1, \ldots, R$ at the beginning of Section \ref{sec:simul}:
\begin{align*}
		\widehat{\vec{\theta}}^\ast_\mathrm{M}
=
		\frac{1}{R} \sum_{r=1}^{R} \widehat{\vec{\theta}}^{[r]}_\mathrm{M}, \quad
		\widehat{\vec{\theta}}^\ast_\mathrm{L}
=
		\frac{1}{R} \sum_{r=1}^{R} \widehat{\vec{\theta}}^{[r]}_\mathrm{L},
\end{align*}
where $\widehat{\vec{\theta}}^{[r]}_\mathrm{M}$ and $\widehat{\vec{\theta}}^{[r]}_\mathrm{L}$ are respectively the MLE and the LASSO estimator in the $r$-th repetition of the simulation.
For $s = 1, 2, \ldots, d$,
\begin{align*}
		&\mathcal{C}_s \left( \widehat{\vec{\theta}}_\mathrm{M} \right)
=
		\frac{1}{R}
		\sum_{r = 1}^{R}
		\left[
			\sum_{k = 1}^{I} \mathbb{I} \left\{ \mathcal{S}^{[r]}_\mathrm{M} \left( k \right) = s \right\}
		 \right],
		&\mathcal{C}_s \left( \widehat{\vec{\theta}}_\mathrm{L} \right)
=
		\frac{1}{R} \sum_{r = 1}^{R}
		\left[
			\sum_{k = 1}^{I} \mathbb{I} \left\{ \mathcal{S}^{[r]}_\mathrm{L} \left( k \right) = s \right\}
		\right],
\end{align*}
where $\mathbb{I}$ is the indicator function, and $\mathcal{S}^{[r]}_\mathrm{M} \left( k \right)$, $\mathcal{S}^{[r]}_\mathrm{L} \left( k \right)$ are defined as follows:
when the number of repetitions for the simulation is $r$,
$\mathcal{S}^{[r]}_\mathrm{M} \left( k \right)$ and $\mathcal{S}^{[r]}_\mathrm{L} \left( k \right)$ denote the coordinate indices corresponding to the univariate word selected in the $k$-th iteration of \hyperref[alg:step2]{$\left\langle Step~2 \right\rangle$} in the algorithm for $\widehat{\xi}_{\mathrm{M}}$ and $\widehat{\xi}_{\mathrm{L}}$, respectively.
Thus, $\mathcal{C}_s$ measures the frequency with which words corresponding to the $s$-th explanatory variable $x_s$ are selected.
We also calculate their standard errors as in \eqref{eq:se^}:
\begin{align*}
		&\widehat{\mathrm{se}} \left[ \widehat{\vec{\theta}}_{\mathrm{M}}^\ast \right]
=
		\sqrt{\frac{1}{R^2} \sum_{r=1}^{R} \left\{ \widehat{\vec{\theta}}_{\mathrm{M}}^{[r]} - \widehat{\vec{\theta}}_{\mathrm{M}}^\ast \right\}^2},\quad
		\widehat{\mathrm{se}} \left[ \widehat{\vec{\theta}}_{\mathrm{L}}^\ast \right]
=
		\sqrt{\frac{1}{R^2} \sum_{r=1}^{R} \left\{ \widehat{\vec{\theta}}_{\mathrm{L}}^{[r]} - \widehat{\vec{\theta}}_{\mathrm{L}}^\ast \right\}^2}, \\
		&\widehat{\mathrm{se}} \left[ \mathcal{C}_s \left( \widehat{\vec{\theta}}_\mathrm{M} \right) \right]
=
		\sqrt{
			\frac{1}{R^2}
			\sum_{r=1}^{R}
			\left\{
				\sum_{k = 1}^{I}
				\mathbb{I} \left\{ \mathcal{S}^{[r]}_\mathrm{M} \left( k \right) = s \right\}
			-
				\mathcal{C}_s \left( \widehat{\vec{\theta}}_\mathrm{M} \right)
			\right\}^2
		}, \\
		&\widehat{\mathrm{se}} \left[ \mathcal{C}_s \left( \widehat{\vec{\theta}}_\mathrm{L} \right) \right]
=
		\sqrt{
			\frac{1}{R^2}
			\sum_{r=1}^{R}
			\left\{
				\sum_{k = 1}^{I} \mathbb{I} \left\{ \mathcal{S}^{[r]}_\mathrm{L} \left( k \right) = s \right\}
			-
				\mathcal{C}_s \left( \widehat{\vec{\theta}}_\mathrm{L} \right)
			 \right\}^2}.
\end{align*}

When comparing the results of $\widehat{\vec{\theta}}_{\mathrm{M}}$ and $\widehat{\vec{\theta}}_{\mathrm{L}}$, we will refer to
\begin{align}
		\vec{\theta}^\ast
=
		\argmin_{\vec{\theta} \in \mathbb{R}^{d+1}}
		D_U \left( \mu, u^{-1} \left( \left[ 1 ~ \cdot^\top \right] \vec{\theta} \right) \right). \label{eq:thata^ast}
\end{align}
Under the above settings, several components of $\vec{\theta}^\ast = \left[ \theta^\ast_0 ~ \theta^\ast_1 ~ \theta^\ast_2 ~ \cdots ~ \theta^\ast_d \right]^\top$ can be obtained explicitly:
\begin{Lem} \label{lem:theta}
If $\mu = u^{-1} \left( \eta \right)$ as defined above and $\vec{X} \thicksim U \left( \left[ -1, 1 \right]^d \right)$,
then the following components of $\vec{\theta}^\ast$ in \eqref{eq:thata^ast} satisfy
\[
		\theta^\ast_1
=
		\theta^\ast_5
=
		\theta^\ast_6
=
		\cdots
=
		\theta^\ast_d =0.
\]
If $u = \mathrm{id}$ additionally, then
\begin{align}
		\vec{\theta}^\ast
=
		\begin{bmatrix}
			1 / 3 + \sinh (1) \\
			0 \\
			1 \\
			3 / e \\
			3 / \pi\\
		\vec{0}_{d-4}
		\end{bmatrix}
\thickapprox
		\begin{bmatrix}
			1.51 \\
			0 \\
			1 \\
			1.10 \\
			0.95\\
		\vec{0}_{d-4}
		\end{bmatrix}. \label{eq:nor_high_theta^ast}
\end{align}
\end{Lem}

\subsubsection{Normal regression} \label{sec:simu_high_nor}
\begin{figure}[htb]
\centering
\includegraphics[width = 0.6\textwidth]{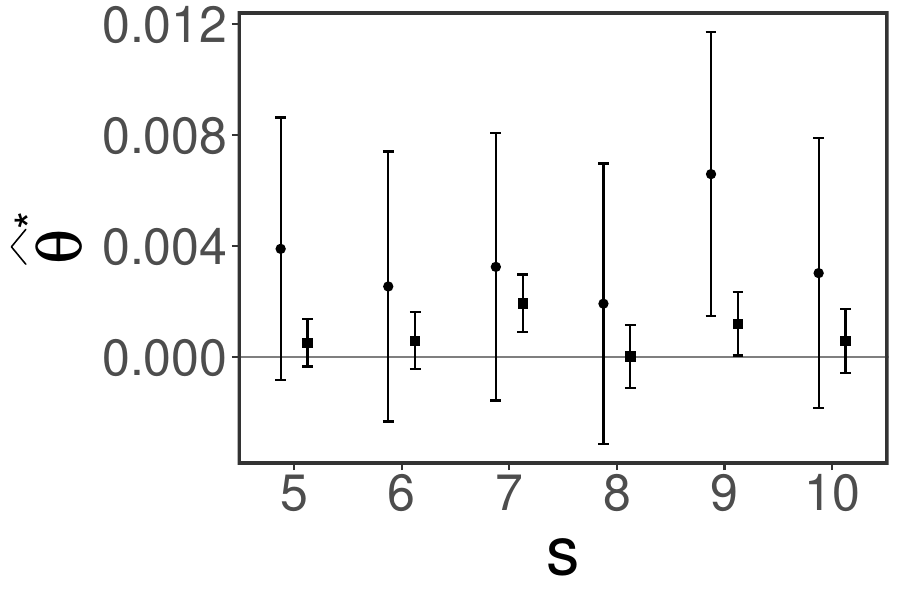}
\caption{
	The results of $\widehat{\theta}^\ast_{\mathrm{M} s}$ (\blackbigcirc) and $\widehat{\theta}^\ast_{\mathrm{L} s}$ ($\blacksquare$) for $s = 5, 6, \ldots, d$ in normal regression with $d=10$.
} \label{fig:nor_d10_thetas}
\end{figure}

The simulation results for normal regression are presented in Tables \ref{tab:accu_nor_high}, \ref{tab:words_nor} and Figure \ref{fig:nor_d10_thetas}.
\begin{table}[tbp]
\centering
\small
\begin{tabular}{crrrrrrr}
\toprule
	$\mu$ &
	$\widehat{\mu}^0_1 \left( \widehat{\vec{\theta}}_{\mathrm{M}} \right)$ &
	$\widehat{\mu}_{\mathrm{P}} \left( \widehat{\vec{\theta}}_{\mathrm{M}} \right)$&
	$\widehat{\mu}^0_1 \left( \widehat{\vec{\theta}}_{\mathrm{L}} \right)$ &
	$\widehat{\mu}_{\mathrm{P}} \left( \widehat{\vec{\theta}}_{\mathrm{L}} \right)$&
	$\widehat{\mu}_{\mathrm{X}}$\\
\midrule
\midrule
	$\mu_{10}$ & $\dashuline{34[.5]}$ & $205[.7]$ & $\uline{22[.4]}$ & $220[1]$ & $\dotuline{83[.7]}$ \\
	& 300 & & 300 & & [1, 20000] \\
\midrule
	$\mu_{100}$ & & & $\uline{124[2]}$ & $\dotuline{259[2]}$ & $\dashuline{203[2]}$ \\
	& & & 69 & & [1, 930] \\
\bottomrule
\end{tabular}
\caption{The minima of estimated divergence risks $\left( \times 10^3 \right)$ in higher-dimensional normal regression.} \label{tab:accu_nor_high}
\end{table}
\begin{table}[tbp]
\centering
\setlength{\tabcolsep}{4pt}
\begin{tabular}{ccrrrrrr}
\hline
	$\mu$ & & $s = 0$ & $s = 1$ & $s = 2$ & $s = 3$ & $s = 4$ & $s = 5, 6, \ldots, d$\\
\midrule
\midrule
	$\mu_{10}$ & $\widehat{\theta}^\ast_{\mathrm{M} s}$ & 1.500[.002] & .007[.005] & 1.001[.004] & 1.110[.004] & .966[.005] & \multirow{2}{*}{Figure \ref{fig:nor_d10_thetas}} \\
	&$\widehat{\theta}^\ast_{\mathrm{L} s}$ & 1.506[.003] & .001[.001] & .729[.005] & .840[.005] & .695[.005] & \\
\cmidrule{2-8}
	&$\mathcal{C}_s \left( \widehat{\vec{\theta}}_{\mathrm{M}} \right)$ & & 66.7[.4] & 6.9[.2] & 31.2[.4] & 152.3[.6] & 7.0[.2] \\
	&$\mathcal{C}_s \left( \widehat{\vec{\theta}}_{\mathrm{L}} \right)$ & & 68.4[.4] & 25.1[.4] & 38.4[.3] & 151.9[.5] & 2.7[.1]\\
\midrule
	$\mu_{100}$ & $\widehat{\theta}^\ast_{\mathrm{L} s} $ & 1.506[.003] & $-.0002$[.0007] & .62[.006] & .721[.006] & .59[.006] & $-.0018$[.0007]\\
\cmidrule{2-8}
	& $\mathcal{C}_s \left( \widehat{\vec{\theta}}_{\mathrm{L}} \right)$ & & 31.3[.4] & 20.2[.3] & 26.6[.3] & 64.8[.5] & 1.7[.1]\\
\hline
\end{tabular}\\
\caption{The results for $\vec{\widehat{\theta}}^\ast_\mathrm{M}$, $\vec{\widehat{\theta}}^\ast_\mathrm{L}$, $\mathcal{C}_s \left( \widehat{\vec{\theta}}_{\mathrm{M}} \right)$ and $\mathcal{C}_s \left( \widehat{\vec{\theta}}_{\mathrm{L}} \right)$ in normal regression.} \label{tab:words_nor}
\end{table}
Table \ref{tab:accu_nor_high} lists the resulting divergence risks in each $d$, corresponding standard errors and the minimizing parameters in the same manner as in Table \ref{tab:accu_nor}:
in column 2 for $\widehat{\mu}^0_1 \left( \widehat{\vec{\theta}}_{\mathrm{M}} \right)$;
in column 3 for $\widehat{\mu}_{\mathrm{P}} \left( \widehat{\vec{\theta}}_{\mathrm{M}} \right)$;
in column 4 for $\widehat{\mu}^0_1 \left( \widehat{\vec{\theta}}_{\mathrm{L}} \right)$;
in column 5 for $\widehat{\mu}_{\mathrm{P}} \left( \widehat{\vec{\theta}}_{\mathrm{L}} \right)$;
and in column 6 for $\widehat{\mu}_{\mathrm{X}}$.
Table \ref{tab:words_nor} summarizes the estimation results of the parameter $\vec{\theta}$ for each $d$ and for each component $s = 1, 2, \ldots, d$: with $\widehat{\vec{\theta}}_{\mathrm{M}}^\ast$ in row 1 and with $\widehat{\vec{\theta}}_{\mathrm{L}}^\ast$ in row 2.
In addition, Table \ref{tab:words_nor} shows $\mathcal{C}_s \left( \widehat{\vec{\theta}}_{\mathrm{M}} \right)$ in row 3 and $\mathcal{C}_s \left( \widehat{\vec{\theta}}_{\mathrm{L}} \right)$ in row 4 for each $s$.
In particular, column 5 for $s = 5, 6, \ldots, d$ in Table \ref{tab:words_nor} displays
$\max_{s = 5, 6, \ldots, d} \left| \theta^\ast_{\mathrm{M}s} \right|$,
$\max_{s = 5, 6, \ldots, d} \left| \theta^\ast_{\mathrm{L}s} \right|$,
$\min_{s = 5, 6, \ldots, d} \mathcal{C}_s \left( \widehat{\vec{\theta}}_{\mathrm{M}} \right)$ and
$\max_{s = 5, 6, \ldots, d} \mathcal{C}_s \left( \widehat{\vec{\theta}}_{\mathrm{L}} \right)$, 
where the `minimum' of $\mathcal{C}_s \left( \widehat{\vec{\theta}}_{\mathrm{M}} \right)$ among $s = 5, 6, \ldots, d$ is displayed unlike the other `maxima', for the comparison below.
Next to these resulting values in Table \ref{tab:words_nor}, their standard errors are given in square brackets.
In addition to $d = 10$, Table \ref{tab:words_nor} summarizes $\widehat{\vec{\theta}}_{\mathrm{L}}^\ast$ and $\mathcal{C}_s \left( \widehat{\vec{\theta}}_{\mathrm{L}} \right)$ in $d = 100$.
Figure \ref{fig:nor_d10_thetas} depicts the results of both values $\left| \widehat{\theta}^\ast_{\mathrm{M} s} \right|$ and $\left| \widehat{\theta}^\ast_{\mathrm{L} s} \right|$ for $s = 5, 6, \ldots, d$ as the points ``\blackbigcirc'' and ``$\blacksquare$,'' respectively.
Figure \ref{fig:nor_d10_thetas} also illustrates the following intervals
$\left[
	\left| \Bar{\theta}_s \right| - \widehat{\mathrm{se}} \left[ \Bar{\theta}_s \right],
	\left| \Bar{\theta}_s \right| + \widehat{\mathrm{se}} \left[ \Bar{\theta}_s \right]
	\right]$
	above and below each $\Bar{\theta}_s = \widehat{\theta}^\ast_{\mathrm{M s}}, \widehat{\theta}^\ast_{\mathrm{L s}}$ for $s = 5, 6, \ldots, d$.

Findings from Table \ref{tab:accu_nor_high} are as follows:
\setlist[enumerate]{label*=\textbf{\arabic*}., listparindent=\parindent}
\begin{enumerate}[label=\textbf{6}.\textbf{\alph*}., leftmargin=*, align = left, resume=desc]
\item $d = 10$:
	\begin{enumerate}
	\item Estimation accuracy.
		As seen in row 1 of Table \ref{tab:accu_nor_high},
		resulting $\widehat{\mu}^0_1 \left( \widehat{\vec{\theta}}_{\mathrm{L}} \right)$ and $\widehat{\mu}^0_1 \left( \widehat{\vec{\theta}}_{\mathrm{M}} \right)$ have respectively the smallest and second smallest risk values among the compared estimators.
		We further detail the behavior of $\widehat{\mu}^0_1 \left( \widehat{\vec{\theta}}_{\mathrm{L}} \right)$ and $\widehat{\mu}^0_1 \left( \widehat{\vec{\theta}}_{\mathrm{M}} \right)$:
		\begin{enumerate}
		\item \label{desc:nor_d10_theta} $\widehat{\theta}^\ast_{\mathrm{M}}$ vs $\widehat{\theta}^\ast_{\mathrm{L}}$.
			In Figure \ref{fig:nor_d10_thetas}, $0 \in
			\left[
				\left| \widehat{\theta}^\ast_{\mathrm{M} 7} \right| - \widehat{\mathrm{se}} \left[ \widehat{\theta}^\ast_{\mathrm{M} 7} \right],
				\left| \widehat{\theta}^\ast_{\mathrm{M} 7} \right| + \widehat{\mathrm{se}} \left[ \widehat{\theta}^\ast_{\mathrm{M} 7} \right]
			\right]$
			holds, whereas
			$0 \notin
			\left[
			\left| \widehat{\theta}^\ast_{\mathrm{L} 7} \right| - \widehat{\mathrm{se}} \left[ \widehat{\theta}^\ast_{\mathrm{L} 7} \right],
			\left| \widehat{\theta}^\ast_{\mathrm{L} 7} \right| + \widehat{\mathrm{se}} \left[ \widehat{\theta}^\ast_{\mathrm{L} 7} \right]
			\right]$.
			However, based on the values indicated by the points and the lengths of the intervals in Figure \ref{fig:nor_d10_thetas}, it is seen that the mean squared error of $\widehat{\theta}^\ast_{\mathrm{L} s}$ is smaller than that of $\widehat{\theta}^\ast_{\mathrm{M} s}$ for each $s = 5, 6, \ldots, 10$.
			This means that $\widehat{\vec{\theta}}^\ast_{\mathrm{L}}$ estimates the coefficients of the variables $x_5, \ldots, x_{10}$ to be closer to 0 than $\widehat{\vec{\theta}}^\ast_{\mathrm{M}}$.
			Hence, compared with $\widehat{\mu}_{\mathrm{P}} \left( \widehat{\vec{\theta}}_{\mathrm{M}} \right)$,
			$\widehat{\mu}_{\mathrm{P}} \left( \widehat{\vec{\theta}}_{\mathrm{L}} \right)$ better captures the underlying fact that $\mu$ does not depend on $x_5, x_6, \ldots, x_{10}$.
		\item $\widehat{\xi}_\mathrm{M}$ vs $\widehat{\xi}_\mathrm{L}$.
			From the entries in rows 3, 4 and column 8 of Table \ref{tab:words_nor}, we can observe that
			$\max_{s = 5, 6, \ldots, 10} \mathcal{C}_s \left( \widehat{\vec{\theta}}_{\mathrm{L}} \right)
					= 2.7 \pm 0.1
					<
						7.0 \pm 0.2
					=
						\min_{s = 5, 6, \ldots, d} \mathcal{C}_s \left( \widehat{\vec{\theta}}_{\mathrm{M}} \right)$.
			Therefore, in the adjustment by $\widehat{\mu}^0_1$, $\widehat{\xi}_\mathrm{L}$ selects univariate words for $x_5, x_6, \ldots, x_{10}$ fewer times than $\widehat{\xi}_\mathrm{M}$.
			One possible factor may be considered from the comparison in the description \ref{desc:nor_d10_theta},
			which indicates that $\widehat{\xi}_\mathrm{L}$ needs to select words for $x_5, x_6,\ldots, x_{10}$ less often than $\widehat{\xi}_\mathrm{M}$.

			On the other hand, as seen from rows 3 and 4 of Table \ref{tab:words_nor}, $\sum_{s = 1}^{4} \mathcal{C}_s \left( \widehat{\vec{\theta}}_{\mathrm{L}} \right) \thickapprox 283.8 > 257.1 \thickapprox \sum_{s = 1}^{4} \mathcal{C}_s \left( \widehat{\vec{\theta}}_{\mathrm{M}} \right)$,
			which means that $\widehat{\xi}_\mathrm{L}$ more often selects words for the variables $x_1, x_2, x_3, x_4$ on which the target $\mu$ essentially depends, compared with $\widehat{\xi}_\mathrm{M}$.
		\end{enumerate}
	\item \label{desc:selection}
		Correcting variable selection in $\widehat{\mu}_{\mathrm{P}} \left( \widehat{\vec{\theta}}_{\mathrm{L}} \right)$.
		As indicated in Table \ref{tab:words_nor}, the resulting estimates with $\widehat{\vec{\theta}}^\ast_{\mathrm{L} s}$ are close to 0 when $s = 1$, and similarly when $s = 5, 6, \ldots, 10$.
		On the other hand, the resulting $\mathcal{C}_s \left( \widehat{\vec{\theta}}_{\mathrm{L}} \right)$ are not similar between $s = 1$ and $s = 5, 6, \ldots, 10$:
		in fact, $\mathcal{C}_1 \left( \widehat{\vec{\theta}}_{\mathrm{L}} \right)$ takes a much larger value than $\max_{s = 5, 6, \ldots, 10} \mathcal{C}_s \left( \widehat{\vec{\theta}}_{\mathrm{L}} \right)$.
		These smaller $\mathcal{C}_s \left( \widehat{\vec{\theta}}_{\mathrm{L}} \right)$ for $s = 5, 6, \ldots, 10$ are thought to be caused by $\widehat{\mu}_{\mathrm{P}} \left( \widehat{\vec{\theta}}_{\mathrm{L}} \right)$,
		which correctly excludes variables $x_5, x_6, \ldots, x_{10}$ from the parametric model as in the description \ref{desc:nor_d10_theta}, suggesting that these variables are not important for variable selection.
		However, the result $\widehat{\vec{\theta}}^\ast_{\mathrm{L}, 1} \thickapprox 0$ in Table \ref{tab:words_nor} means that $\widehat{\mu}_{\mathrm{P}} \left( \widehat{\vec{\theta}}_{\mathrm{L}} \right)$ also discards the variable $x_1$, on which the target $\mu$ actually depends.
		To correct this, it seems that $\mathcal{C}_1 \left( \widehat{\vec{\theta}}_{\mathrm{L}} \right)$ takes a larger value than $\max_{s = 5, 6, \ldots, 10} \mathcal{C}_s \left( \widehat{\vec{\theta}}_{\mathrm{L}} \right)$.
		In this way, it is expected that actual importance of the variables discarded at the stage of $\widehat{\mu}_{\mathrm{P}} \left( \widehat{\vec{\theta}}_{\mathrm{L}} \right)$ can be evaluated, by considering how many words for such variables are included in the adjustment term of $\widehat{\mu}^0_1 \left( \widehat{\vec{\theta}}_{\mathrm{L}} \right)$.
	\end{enumerate}
\item \label{desc:simu_HDLSS} $n / d = 0.5$.
	In Table \ref{tab:accu_nor_high}, the smallest risk is given by $\widehat{\mu}^0_1 \left( \widehat{\vec{\theta}}_{\mathrm{L}} \right)$ among the compared estimators.
	Moreover, as shown in rows 5 and 6 of Table \ref{tab:words_nor}, $\widehat{\vec{\theta}}^\ast_{\mathrm{L} s} \thickapprox 0$ when $s = 1$ and when $s = 5, 6, \ldots, 100$,
	whereas $\max_{s = 5, 6, \ldots, 100} \mathcal{C}_s \left( \widehat{\vec{\theta}}_{\mathrm{L}} \right) < \mathcal{C}_1 \left( \widehat{\vec{\theta}}_{\mathrm{L}} \right)$.
	Therefore, we can draw the same conclusion as in the description \ref{desc:selection} in $d = 10$:
	it is expected that actual importance of the variables discarded at the stage of $\widehat{\mu}_{\mathrm{P}} \left( \widehat{\vec{\theta}}_{\mathrm{L}} \right)$ can be evaluated by considering how many words for such variables are included in the adjustment term of $\widehat{\mu}^0_1 \left( \widehat{\vec{\theta}}_{\mathrm{L}} \right)$.
	In this way, even in the high-dimension low-sample-size setting here,
	$\widehat{\mu}^0_1 \left( \widehat{\vec{\theta}}_{\mathrm{L}} \right)$ estimates $\mu$ better than the other estimators compared, and also corrects variable selection in $\widehat{\mu}_{\mathrm{P}} \left( \widehat{\vec{\theta}}_{\mathrm{L}} \right)$.
\end{enumerate}

\begin{figure}[htb]
\centering
\includegraphics[width = 0.8\textwidth]{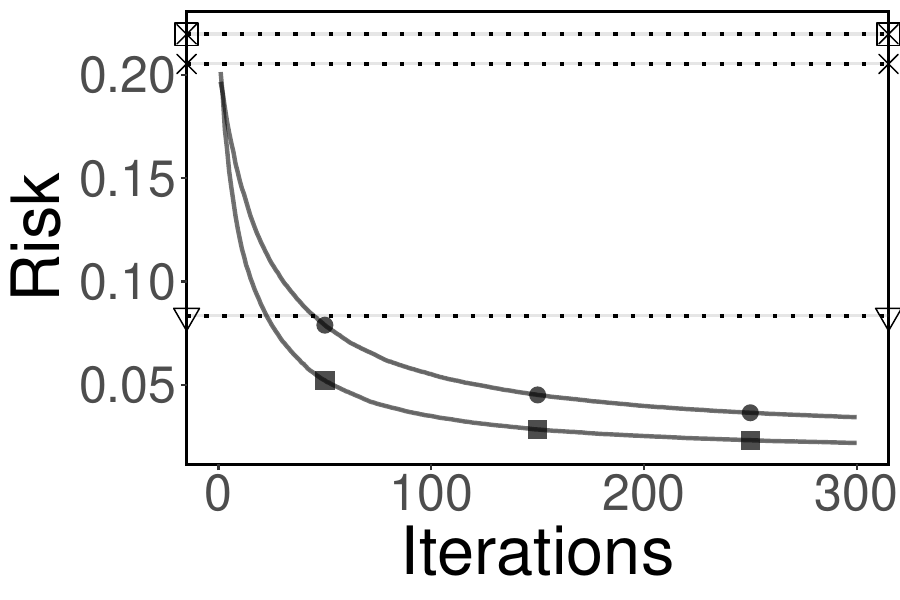}
\caption{
	The curves of estimated divergence risks for each estimator against the number of iterations for the proposed estimators $\widehat{\mu}^\gamma_i$, when $\mu = \mu_{10}$ in normal regression.
	\blackbigcirc : $\widehat{\mu}^0_1 \left( \widehat{\vec{\theta}}_{\mathrm{M}} \right)$,
	$\blacksquare$: $\widehat{\mu}^0_1 \left( \widehat{\vec{\theta}}_{\mathrm{L}} \right)$,
	$\times$: $\widehat{\mu}_{\mathrm{P}} \left( \widehat{\vec{\theta}}_{\mathrm{M}} \right)$,
	$\boxtimes $: $\widehat{\mu}_{\mathrm{P}} \left( \widehat{\vec{\theta}}_{\mathrm{L}} \right)$ and
	$\bigtriangledown$: $\widehat{\mu}_{\mathrm{X}}$.
}\label{fig:nor_d10c}
\end{figure}

Besides Table \ref{tab:accu_nor_high}, Figure \ref{fig:nor_d10c} displays the curves of the estimated divergence risks for each estimator against the number of iterations $I$ for $\widehat{\mu}^0_1$, when $\mu = \mu_{10}$.
The risk curves for $\widehat{\mu}^0_1 \left( \widehat{\vec{\theta}}_{\mathrm{M}} \right)$ and $\widehat{\mu}^0_1 \left( \widehat{\vec{\theta}}_{\mathrm{L}} \right)$ are plotted with the points ``\blackbigcirc'' and ``$\blacksquare$,'' respectively.
On the other hand, the horizontal dotted lines are the results for the estimators other than $\widehat{\mu}^\gamma_i$:
at both ends of these lines, the points ``$\times$'' indicate the result for $\widehat{\mu}_{\mathrm{P}} \left( \widehat{\vec{\theta}}_{\mathrm{M}} \right)$;
``$\boxtimes $'' for $\widehat{\mu}_{\mathrm{P}} \left( \widehat{\vec{\theta}}_{\mathrm{L}} \right)$;
and ``$\bigtriangledown$'' for $\widehat{\mu}_{\mathrm{X}}$.
\setlist[enumerate]{label*=\textbf{\arabic*}., listparindent=\parindent}
\begin{enumerate}[label=\textbf{6}.\textbf{\alph*}., leftmargin=*, align = left, resume=desc]
\item \label{desc:nor_d10} Figure \ref{fig:nor_d10c}.
	\begin{enumerate}
	\item \label{desc:nor_d10_decr}
		Both resulting risk curves for $\widehat{\mu}^0_1 \left( \widehat{\vec{\theta}}_{\mathrm{M}} \right)$ (\blackbigcirc) and $\widehat{\mu}^0_1 \left( \widehat{\vec{\theta}}_{\mathrm{L}} \right)$ ($\blacksquare$) decrease with $I$.
	\item \label{desc:nor_d10_LvsM}
		Focusing on the parametric estimation,
		the estimated divergence risk for $\widehat{\mu}_{\mathrm{P}} \left( \widehat{\vec{\theta}}_{\mathrm{L}} \right)$ ($\boxtimes$) is smaller than that for $\widehat{\mu}_{\mathrm{P}} \left( \widehat{\vec{\theta}}_{\mathrm{M}} \right)$ ($\times$).
		However, after adjusting them, $\widehat{\mu}^0_1 \left( \widehat{\vec{\theta}}_{\mathrm{L}} \right)$ has smaller risk values than $\widehat{\mu}^0_1 \left( \widehat{\vec{\theta}}_{\mathrm{M}} \right)$ for almost all $I$.
		One possible reason for this may be the behavior of $\widehat{\vec{\theta}}_{\mathrm{L}}$ and $\widehat{\vec{\theta}}_{\mathrm{M}}$ as in the description \ref{desc:nor_d10_theta}.
	\end{enumerate}
\end{enumerate}

\subsubsection{Bernoulli and Poisson regression} \label{sec:simu_high_berpoi}
As in Section \ref{sec:simu_low_berpoi}, when constructing $\widehat{\mu}^0_1 \left( \widehat{\vec{\theta}}_{\mathrm{M}} \right)$ and $\widehat{\mu}^0_1 \left( \widehat{\vec{\theta}}_{\mathrm{L}} \right)$,
we apply several sets of bandwidths:
$\mathbb{H}_{1, h_{\mathrm{min}}} = \left\{ s^2 \middle| s^2 = h_{\mathrm{min}}, h_{\mathrm{min}} + 0.1, \ldots, 1.0 \right\}$ for $h_{\mathrm{min}} = 0.1, 0.2, \ldots, 0.7$.

\begin{figure}[htb]
\centering
\includegraphics[width = 0.6\textwidth]{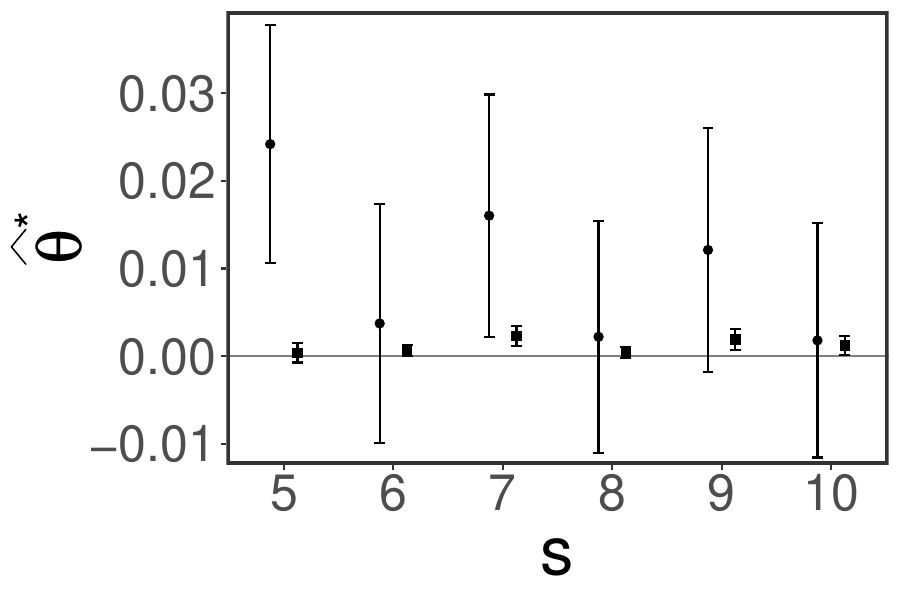}
\caption{
	The results of $\widehat{\theta}^\ast_{\mathrm{M} s}$ (\blackbigcirc) and $\widehat{\theta}^\ast_{\mathrm{L} s}$ ($\blacksquare$) for $s = 5, 6, \ldots, d$ in Bernoulli regression with $d=10$.
} \label{fig:ber_d10_thetas}
\end{figure}

\begin{figure}[htb]
\centering
\includegraphics[width = 0.6\textwidth]{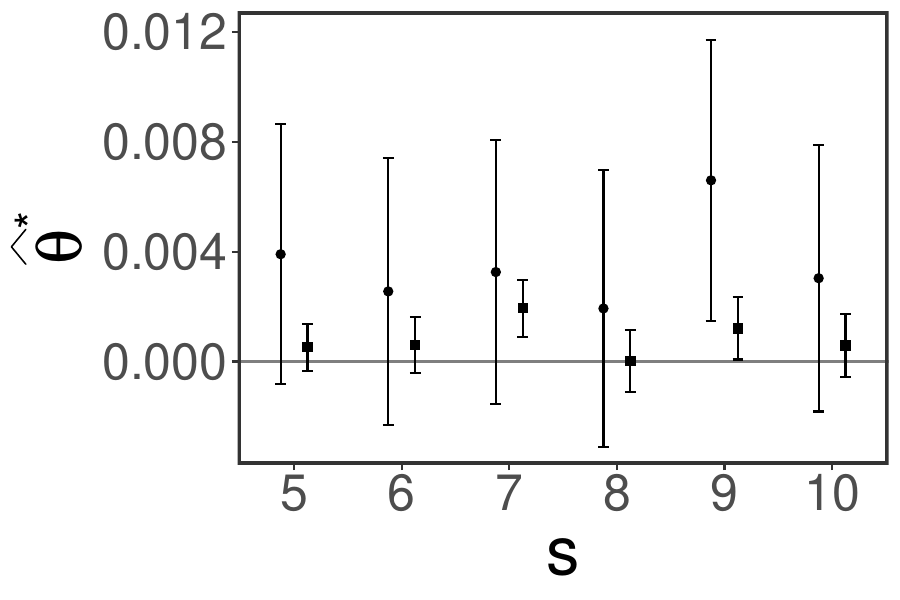}
\caption{
	The results of $\widehat{\theta}^\ast_{\mathrm{M} s}$ (\blackbigcirc) and $\widehat{\theta}^\ast_{\mathrm{L} s}$ ($\blacksquare$) for $s = 5, 6, \ldots, d$ in Poisson regression with $d=10$.
	} \label{fig:poi_d10_thetas}
\end{figure}

\begin{figure}[htb]
\centering
\includegraphics[width = 0.8\textwidth]{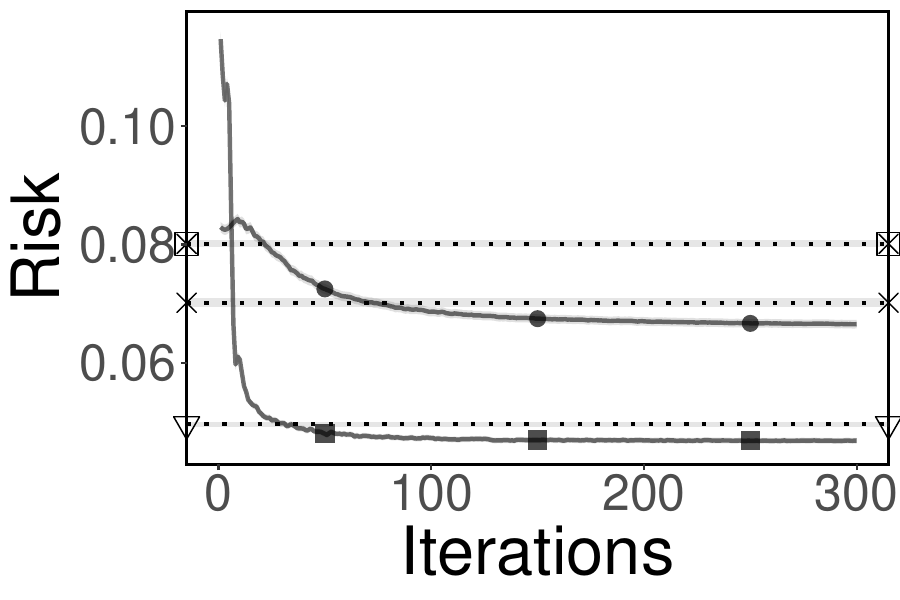}
\caption{
	The curves of estimated divergence risks for each estimator against the number of iterations for the proposed estimators $\widehat{\mu}^\gamma_i$, when $\mu = p_{10}$ in Bernoulli regression.
	\blackbigcirc : $\widehat{\mu}^0_1 \left( \widehat{\vec{\theta}}_{\mathrm{M}} \right)$,
	$\blacksquare$: $\widehat{\mu}^0_1 \left( \widehat{\vec{\theta}}_{\mathrm{L}} \right)$,
	$\times$: $\widehat{\mu}_{\mathrm{P}} \left( \widehat{\vec{\theta}}_{\mathrm{M}} \right)$,
	$\boxtimes $: $\widehat{\mu}_{\mathrm{P}} \left( \widehat{\vec{\theta}}_{\mathrm{L}} \right)$ and
	$\bigtriangledown$: $\widehat{\mu}_{\mathrm{X}}$.
} \label{fig:ber_d10c}
\end{figure}

\begin{figure}[htb]
\centering
\includegraphics[width = 0.8\textwidth]{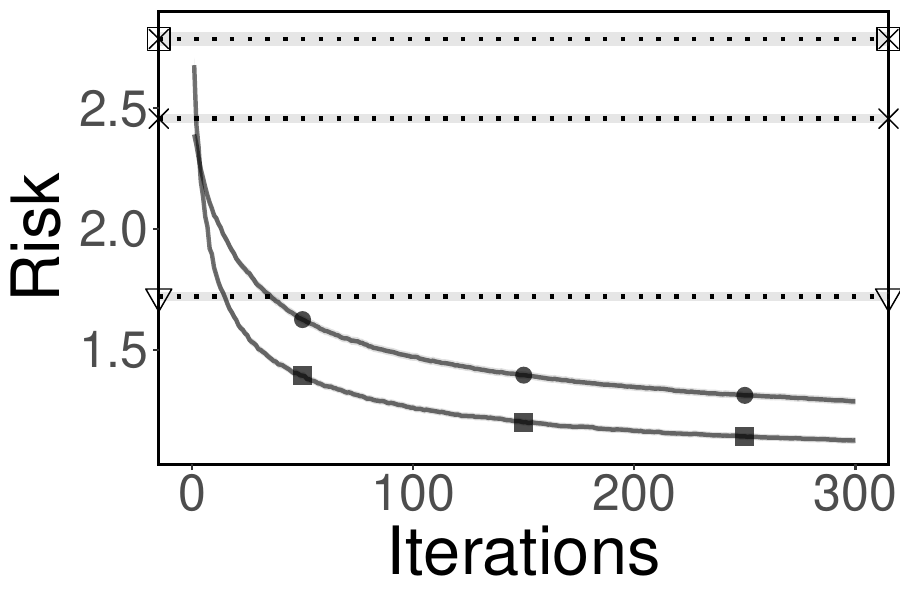}
\caption{
	The curves of estimated divergence risks for each estimator against the number of iterations for the proposed estimators $\widehat{\mu}^\gamma_i$, when $\mu = \lambda_{10}$ in Poisson regression.
	\blackbigcirc : $\widehat{\mu}^0_1 \left( \widehat{\vec{\theta}}_{\mathrm{M}} \right)$,
	$\blacksquare$: $\widehat{\mu}^0_1 \left( \widehat{\vec{\theta}}_{\mathrm{L}} \right)$,
	$\times$: $\widehat{\mu}_{\mathrm{P}} \left( \widehat{\vec{\theta}}_{\mathrm{M}} \right)$,
	$\boxtimes $: $\widehat{\mu}_{\mathrm{P}} \left( \widehat{\vec{\theta}}_{\mathrm{L}} \right)$ and
	$\bigtriangledown$: $\widehat{\mu}_{\mathrm{X}}$.
} \label{fig:poi_d10c}
\end{figure}

Table \ref{tab:accu_berpoi_high} summarizes the resulting minima of divergence risks in a similar manner to that in Table \ref{tab:accu_nor_high}.
Table \ref{tab:accu_berpoi_high} includes the results for $\widehat{\mu}^0_1$ derived from specific three $h_{\mathrm{min}}$, just as in Table \ref{tab:accu_berpoi} of Section \ref{sec:simu_low_berpoi}.
In addition to Table \ref{tab:accu_berpoi_high}, Figures \ref{fig:ber_d10c} and \ref{fig:poi_d10c} present resulting risk curves of $\widehat{\mu}^0_1$ in much the same manner as in Figure \ref{fig:nor_d10c}, when $ \mu = p_{10}$ and $\lambda_{10}$, respectively.
These Figures \ref{fig:ber_d10c} and \ref{fig:poi_d10c} contain the curves for $\widehat{\mu}^0_1$ only derived from one $h_{\mathrm{min}}$ chosen just as in Figures \ref{fig:ber_d3j} and \ref{fig:poi_d3j} of Section \ref{sec:simu_low_berpoi}.
Moreover, Figures \ref{fig:ber_d10_thetas} and \ref{fig:poi_d10_thetas} depict the resulting $\widehat{\vec{\theta}}$ in exactly the same way as in Figure \ref{fig:nor_d10_thetas},
and Table \ref{tab:words_berpoi} lists the resulting $\widehat{\vec{\theta}}$ and $\mathcal{C}_s$ similarly to Table \ref{tab:words_nor},
where the result of $\widehat{\mu}^0_1$ is presented only for the same $h_{\mathrm{min}}$ as in Figures \ref{fig:ber_d10c} and \ref{fig:poi_d10c}.
\begin{table}[tbp]
\small
\setlength{\tabcolsep}{3pt}
\begin{tabular}{crrrrrrrrrr}
\toprule
	$\mu$ &
	$\widehat{\mu}^0_1 \left( \widehat{\vec{\theta}}_{\mathrm{M}} \right)$ &
	$\widehat{\mu}^0_1 \left( \widehat{\vec{\theta}}_{\mathrm{M}} \right)$ &
	$\widehat{\mu}^0_1 \left( \widehat{\vec{\theta}}_{\mathrm{M}} \right)$ &
	$\widehat{\mu}_{\mathrm{P}} \left( \widehat{\vec{\theta}}_{\mathrm{M}} \right)$&
	$\widehat{\mu}^0_1 \left( \widehat{\vec{\theta}}_{\mathrm{L}} \right)$ &
	$\widehat{\mu}^0_1 \left( \widehat{\vec{\theta}}_{\mathrm{L}} \right)$ &
	$\widehat{\mu}^0_1 \left( \widehat{\vec{\theta}}_{\mathrm{L}} \right)$ &
	$\widehat{\mu}_{\mathrm{P}} \left( \widehat{\vec{\theta}}_{\mathrm{L}} \right)$&
	$\widehat{\mu}_{\mathrm{X}}$\\
\midrule
\midrule
	$p_{10}$ & $67[.7]$ & $67[.7]$ & $67[.7]$ & $70[.8]$ & $\uline{46.7[.5]}$ & $\dashuline{47.5[.5]}$ & $\dotuline{48.0[.4]}$ & $80[.6]$ & $50[.4]$ \\
	& [.5, 298] & [.6, 298] & [.7, 296] & & [.2, 290] & [.3, 284] & [.4, 285] & & [1, $10^2$, 4333] \\
\midrule
	$\lambda_{10}$ & $\dashuline{1286 [}15]$ & $1635 [15]$ & $1845 [15]$ & $2457 [18]$ & $\uline{1124 [}14]$ & $\dotuline{1464 [}13]$ & $1718 [14]$ & $2786 [29]$ & $1721 [18]$ \\
	& [.1, 299] & [.2, 299] & [.3, 300] & & [.1, 299] & [.2, 299] & [.3, 300] & & [1, $10^2$, 15000] \\
\midrule
	$p_{300}$ & & & & & $83[.5]$ & $\dotuline{82.3[.5]}$ & $\dashuline{81.7[.5]}$ & $91[.5]$ & $\uline{80[.3]}$ \\
	& & & & & [.5, 280] & [.6, 292] & [.7, 249] & & [1, $10^3$, 13701] \\
\midrule
	$\lambda_{100}$ & & & & & $\uline{2769[29]}$ & $\dashuline{2948[26]}$ & $\dotuline{3095[25]}$ & $3784[31]$ & $3234[40]$ \\
	& & & & & [.1, 269] & [.2, 279] & [.3, 294] & & [1, $10^3$, 6445] \\
\bottomrule
\end{tabular}
\normalsize
\caption{The minima of estimated divergence risks $\left( \times 10^3 \right)$ for higher-dimensional Bernoulli and Poisson regression.} \label{tab:accu_berpoi_high}
\end{table}
\begin{table}[tbp]
\centering
\setlength{\tabcolsep}{3pt}
\begin{tabular}{ccrrrrrr}
\hline
	$\mu$& & $s = 0$ & $s = 1$ & $s = 2$ & $s = 3$ & $s = 4$ & $s = 5, 6, \ldots, d$ \\
\midrule
\midrule
	$p_{10}$&$\widehat{\vec{\theta}}_{\mathrm{M}}$ & 1.574[.009] & $-.01$[.01] & 1.05[.01] & 1.10[.01] & 1.03[.01] & \multirow{2}{*}{Figure \ref{fig:ber_d10_thetas}} \\
	&$\widehat{\vec{\theta}}_{\mathrm{L}}$ & 1.220[.006] & $-.001$[.001] & .130[.007] & .130[.007] & .123[.007] & \\
\cmidrule{2-8}
	& $\mathcal{C}_s \left( \widehat{\vec{\theta}}_{\mathrm{M}} \right)$ & & 72.2[2.7] & 30.2[1.4] & 42.1[1.7] & 37.8[1.7] & 18.0[1.3] \\
	& $\mathcal{C}_s \left( \widehat{\vec{\theta}}_{\mathrm{L}} \right)$ & & 19.5[0.9] & 58.5[1.3] & 81.0[1.4] & 72.6[1.4] & 12.1[0.6] \\
\midrule
	$\lambda_{10}$&$\widehat{\vec{\theta}}_{\mathrm{M}}$ & 1.558[.005] & .004[.008] & .998[.008] & 1.238[.008] & .950[.008] & \multirow{2}{*}{Figure \ref{fig:poi_d10_thetas}} \\
	&$\widehat{\vec{\theta}}_{\mathrm{L}}$ & 1.905[.006] & .002[.004] & .597[.008] & .834[.009] & .536[.008] & \\
\cmidrule{2-8}
	& $\mathcal{C}_s \left( \widehat{\vec{\theta}}_{\mathrm{M}} \right)$ & & 65.2[0.9] & 12.8[0.4] & 30.2[0.7] & 111.9[1.2] & 12.3[0.4] \\
	& $\mathcal{C}_s \left( \widehat{\vec{\theta}}_{\mathrm{L}} \right)$ & & 58.0[0.8] & 35.6[0.7] & 38.5[0.6] & 108.7[1.0] & 10.2[0.4] \\
\midrule
	$p_{300}$& $\widehat{\vec{\theta}}_{\mathrm{L}}$ & 1.210[.006] &$ -.0002[.0004]$ & .035[.004] & .040[.004] & .035[.004] & .0010[0.0007] \\
\cmidrule{2-8}
	& $\mathcal{C}_s \left( \widehat{\vec{\theta}}_{\mathrm{L}} \right)$ & & .6[0.1] & 23.3[1.0] & 34.2[1.2] & 28.3[1.2] & 1.2[0.2] \\
\midrule
	$\lambda_{100}$& $\widehat{\vec{\theta}}_{\mathrm{L}}$ & 1.982[0.006] & .00121[0.002] & .393[0.009] & .590[0.010] & .328[0.008] & $-.003[0.001]$ \\
\cmidrule{2-8}
	& $\mathcal{C}_s \left( \widehat{\vec{\theta}}_{\mathrm{L}} \right)$ & & 13.1[0.5] & 27.4[0.6] & 21.2[0.5] & 62.3[0.8] & 2.3[0.2] \\
\hline
\hline
\end{tabular}
\caption{The results for $\vec{\widehat{\theta}}^\ast_\mathrm{M}$, $\vec{\widehat{\theta}}^\ast_\mathrm{L}$, $\mathcal{C}_s \left( \widehat{\vec{\theta}}_{\mathrm{M}} \right)$ and $\mathcal{C}_s \left( \widehat{\vec{\theta}}_{\mathrm{L}} \right)$ in Bernoulli and Poisson regression.} \label{tab:words_berpoi}
\end{table}
\setlist[enumerate]{label*=\textbf{\arabic*}., listparindent=\parindent}
\begin{enumerate}[label=\textbf{6}.\textbf{\alph*}., leftmargin=*, align = left, resume=desc]
\item $d = 10$:
	\begin{enumerate}
	\item
		Estimation accuracy.
		As seen in Table \ref{tab:accu_berpoi_high}, when both $\mu = p_{10}, \lambda_{10}$, $\widehat{\mu}^0_1 \left( \widehat{\vec{\theta}}_{\mathrm{L}} \right)$ has the smallest risk value among the compared estimators, just as it does for normal regression in Section \ref{sec:simu_high_nor}.
		Moreover, the behavior of each $\widehat{\vec{\theta}}_{\mathrm{M}}$, $\widehat{\vec{\theta}}_{\mathrm{L}}$, $\mathcal{C}_s \left( \widehat{\vec{\theta}}_{\mathrm{M}} \right)$ and $\mathcal{C}_s \left( \widehat{\vec{\theta}}_{\mathrm{L}} \right)$ here is also similar to the results in Section \ref{sec:simu_high_nor}:
		in fact, Figures \ref{fig:ber_d10_thetas} ($\mu = p_{10}$) and \ref{fig:poi_d10_thetas} ($\mu = \lambda_{10}$) show that $\widehat{\mu}_{\mathrm{P}} \left( \widehat{\vec{\theta}}_{\mathrm{L}} \right)$ captures the underlying dependence of $\mu$ on $x_5, x_6, \ldots, x_{10}$ better than $\widehat{\mu}_{\mathrm{P}} \left( \widehat{\vec{\theta}}_{\mathrm{M}} \right)$;
		and as observed in column 8 of Table \ref{tab:words_berpoi},
				$\max_{s = 5, 6, \ldots, 10} \mathcal{C}_s \left( \widehat{\vec{\theta}}_{\mathrm{L}} \right)
		<
			\min_{s = 5, 6, \ldots, d} \mathcal{C}_s \left( \widehat{\vec{\theta}}_{\mathrm{M}} \right)$ when $\mu = p_{10}, \lambda_{10}$;
		as recognized from column 4--7 of Table \ref{tab:words_berpoi},
		$\sum_{s = 1}^{4} \mathcal{C}_s \left( \widehat{\vec{\theta}}_{\mathrm{L}} \right) \thickapprox 231.6 > 182.3 \thickapprox \sum_{s = 1}^{4} \mathcal{C}_s \left( \widehat{\vec{\theta}}_{\mathrm{M}} \right)$ when $\mu = p_{10}$
		and $\sum_{s = 1}^{4} \mathcal{C}_s \left( \widehat{\vec{\theta}}_{\mathrm{L}} \right) \thickapprox 240.8 > 220.1 \thickapprox \sum_{s = 1}^{4} \mathcal{C}_s \left( \widehat{\vec{\theta}}_{\mathrm{M}} \right)$ when $\mu = \lambda_{10}$.

		On the other hand, focusing on the case $\mu = p_{10}$, we can see from Table \ref{tab:accu_berpoi_high} that $\widehat{\mu}^0_1 \left( \widehat{\vec{\theta}}_{\mathrm{L}} \right)$ from the dictionary with $\mathbb{H}_{1, 0.1}$ yields a larger risk value than that with $\mathbb{H}_{1, 0.2}$ of larger bandwidths, which is not analogous to the normal regression setting in Section \ref{sec:simu_high_nor}.
		It is thought that $\widehat{\mu}^0_1 \left( \widehat{\vec{\theta}}_{\mathrm{L}} \right)$ in this case behaves similarly to the description \ref{desc:hmin} in Section \ref{sec:simu_low_berpoi}.
	\item Correcting variable selection in $\widehat{\mu}_{\mathrm{P}} \left( \widehat{\vec{\theta}}_{\mathrm{L}} \right)$.
		As in the description \ref{desc:selection} for normal regression, it is expected also in the Bernoulli and Poisson regression settings here that actual importance of the variables overlooked in the phase of $\widehat{\mu}_{\mathrm{P}} \left( \widehat{\vec{\theta}}_{\mathrm{L}} \right)$ can be evaluated, by considering how many words for such variables appear in the adjustment term of $\widehat{\mu}^0_1 \left( \widehat{\vec{\theta}}_{\mathrm{L}} \right)$:
		in fact, as observed from Figures \ref{fig:ber_d10_thetas} and \ref{fig:poi_d10_thetas}, $\widehat{\vec{\theta}}^\ast_{\mathrm{L} s} \thickapprox 0$ when $s = 1$ and when $s = 5, 6, \ldots, 10$, while $\max_{s = 5, 6, \ldots, 10} \mathcal{C}_s \left( \widehat{\vec{\theta}}_{\mathrm{L}} \right) < \mathcal{C}_1 \left( \widehat{\vec{\theta}}_{\mathrm{L}} \right)$ in Table \ref{tab:words_berpoi}.

		On the other hand, compared with the Poisson regression setting, the difference between the resulting values of $\mathcal{C}_1 \left( \widehat{\vec{\theta}}_{\mathrm{L}} \right)$ and $\max_{s = 5, 6, \ldots, 10} \mathcal{C}_s \left( \widehat{\vec{\theta}}_{\mathrm{L}} \right)$ seems less distinct in the Bernoulli regression setting.
	\end{enumerate}
\item \label{desc:berpoi_high_ber_bad} $n / d = 0.5$.
	First we focus on the case $\mu = \lambda_{100}$.
	Row 8 of Table \ref{tab:accu_berpoi_high} shows that the smallest risk is given by $\widehat{\mu}^0_1 \left( \widehat{\vec{\theta}}_{\mathrm{L}} \right)$.
	Moreover, as seen from Table \ref{tab:words_berpoi}, $\widehat{\vec{\theta}}^\ast_{\mathrm{L} s} \thickapprox 0$ when $s = 1$ and when $s = 5, 6, \ldots, 100$,
	while $\max_{s = 5, 6, \ldots, 100} \mathcal{C}_s \left( \widehat{\vec{\theta}}_{\mathrm{L}} \right) < \mathcal{C}_1 \left( \widehat{\vec{\theta}}_{\mathrm{L}} \right)$.
	Therefore, just as in the description \ref{desc:simu_HDLSS} for normal regression, even in the high-dimension low-sample-size Poisson regression setting here, $\widehat{\mu}^0_1 \left( \widehat{\vec{\theta}}_{\mathrm{L}} \right)$ estimates $\mu$ better than the other estimators compared, and also corrects variable selection in $\widehat{\mu}_{\mathrm{P}} \left( \widehat{\vec{\theta}}_{\mathrm{L}} \right)$.

	On the other hand, as shown in the result for $\mu = p_{300}$ in Table \ref{tab:accu_berpoi_high}, the risk of $\widehat{\mu}^0_1 \left( \widehat{\vec{\theta}}_{\mathrm{L}} \right)$ is larger than that of $\widehat{\mu}_{\mathrm{X}}$, which is not found in the other cases simulated above.
	In addition, as observed in Table \ref{tab:words_berpoi} that $\widehat{\vec{\theta}}_{\mathrm{L} s} \thickapprox 0$ when $s = 1$ and when $s = 5, 6, \ldots, 300$, but $\mathcal{C}_1 \left( \widehat{\vec{\theta}}_{\mathrm{L}} \right) < \max_{s = 5, 6, \ldots, 300} \mathcal{C}_s \left( \widehat{\vec{\theta}}_{\mathrm{L}} \right)$:
	indeed, the coefficient of the variable $x_1$ on which the target $\mu$ actually and essentially depends is estimated as $\widehat{\vec{\theta}}_{\mathrm{L} 1} = -0.0002 \pm 0.0004$, suggesting that $x_1$ is not important for variable selection.
	However, the inequality $\mathcal{C}_1 \left( \widehat{\vec{\theta}}_{\mathrm{L}} \right) < \max_{s = 5, 6, \ldots, 300} \mathcal{C}_s \left( \widehat{\vec{\theta}}_{\mathrm{L}} \right)$ indicates that the adjustment term in $\widehat{\mu}^0_1 \left( \widehat{\vec{\theta}}_{\mathrm{L}} \right)$ includes fewer words for $x_1$.
	Hence, the adjustment does not seem to work well in this Bernoulli regression setting.
\end{enumerate}
\setlist[enumerate]{label*=\textbf{\arabic*}., listparindent=\parindent}
\begin{enumerate}[label=\textbf{6}.\textbf{\alph*}., leftmargin=*, align = left, resume=desc]
\item Figure \ref{fig:ber_d10c}:
	\begin{enumerate}
	\item
		The risk for $\widehat{\mu}^0_1 \left( \widehat{\vec{\theta}}_{\mathrm{M}} \right)$(\blackbigcirc) initially takes larger values than $\widehat{\mu}_\mathrm{P} \left( \widehat{\vec{\theta}}_{\mathrm{M}} \right)$ ($\times$) for sufficiently small $I$, and then decreases with $I$.
		Similar behavior is observed between $\widehat{\mu}^0_1 \left( \widehat{\vec{\theta}}_{\mathrm{L}} \right)$ ($\blacksquare$) and $\widehat{\mu}_\mathrm{P} \left( \widehat{\vec{\theta}}_{\mathrm{L}} \right)$ ($\boxtimes$).
	\item
		Analogous results to the description \ref{desc:nor_d10_LvsM} for normal regression are found even in the Bernoulli regression setting here:
		the risk for $\widehat{\mu}_{\mathrm{P}} \left( \widehat{\vec{\theta}}_{\mathrm{L}} \right)$ ($\boxtimes$) is larger than that for $\widehat{\mu}_{\mathrm{P}} \left( \widehat{\vec{\theta}}_{\mathrm{M}} \right)$ ($\times$),
		while $\widehat{\mu}^0_1 \left( \widehat{\vec{\theta}}_{\mathrm{L}} \right)$ has smaller risk values than $\widehat{\mu}^0_1 \left( \widehat{\vec{\theta}}_{\mathrm{M}} \right)$ for almost all $I$.
	\end{enumerate}
\item Figure \ref{fig:poi_d10c}:
	Results much the same as those described in \ref{desc:nor_d10} are also seen in the Poisson regression setting here:
	both risk curves for $\widehat{\mu}^0_1 \left( \widehat{\vec{\theta}}_{\mathrm{M}} \right)$ and $\widehat{\mu}^0_1 \left( \widehat{\vec{\theta}}_{\mathrm{L}} \right)$ decrease for almost all $I$.
	In addition, $\widehat{\mu}^0_1 \left( \widehat{\vec{\theta}}_{\mathrm{L}} \right)$ has smaller risk values than $\widehat{\mu}^0_1 \left( \widehat{\vec{\theta}}_{\mathrm{M}} \right)$ for almost all $I$,
	even though $\widehat{\mu}_{\mathrm{P}} \left( \widehat{\vec{\theta}}_{\mathrm{L}} \right)$ initially underperforms $\widehat{\mu}_{\mathrm{P}} \left( \widehat{\vec{\theta}}_{\mathrm{M}} \right)$.
\end{enumerate}

\section{Real data applications} \label{sec:real}
In this section, we apply the proposed estimator $\widehat{\mu}$ in Section \ref{sec:esti} to several data sets to evaluate its practical performance,
focusing specifically on predictive performance.
Five-fold cross-validation (CV) is conducted to compare several estimators in both normal and Bernoulli regression settings.

More specifically, we randomly divide the data set $\mathbb{X} = \left\{ \left( Y_1, \vec{X}_1 \right), \ldots, \left( Y_n, \vec{X}_n \right) \right\}$ into $K = 5$ sets $S_1^{[t]}, S_2^{[t]}, \ldots, S_K^{[t]}$ of almost equal size, which is repeated $T = 20$ times for $t = 1, 2, \ldots, T$.
Then we construct an estimator $\Bar{\mu}_t^{(-k)}$ from the observations outside $S_k^{[t]}$ for $k = 1, 2, \ldots, K$,
and calculate the following CV error in each regression setting:
\begin{align}
		\mathrm{CV} \left[ \Bar{\mu} \right]
=
		\frac{1}{2}
	\cdot
		\frac{1}{T}
		\sum_{t = 1}^{T}
		\frac{1}{K}
		\sum_{k = 1}^{K}
		\frac{1}{\left| S_k^{[t]} \right|}
		\sum_{\left( Y, \vec{X} \right) \in S_k^{[t]}}
		\left\{ Y - \Bar{\mu}_t^{(-k)} \left( \vec{X} \right) \right\}^2 \label{eq:CV_nor}
\end{align}
in normal regression;
\begin{align}
		\mathrm{CV} \left[ \Bar{\mu} \right]
=
	\frac{1}{T}
	\sum_{t = 1}^{T}
	\frac{1}{K}
	\sum_{k = 1}^{K}
	\frac{1}{\left| S_k^{[t]} \right|}
	\sum_{\left( Y, \vec{X} \right) \in S_k^{[t]}}
	\left[ Y - \mathbb{I} \left\{ \Bar{\mu}_t^{(-k)} \left( \vec{X} \right) \geq 0.5 \right\} \right]^2 \label{eq:CV_ber}
\end{align}
in Bernoulli regression, representing the prediction error in assigning the label 1 to an input $\vec{X}$ if $\Bar{\mu}_t^{(-k)} \left( \vec{X} \right) \geq 0.5$ and 0 otherwise.
The settings and notation for the estimators in this section are as follows:
\begin{itemize}
\item Parametric estimator $\widehat{\mu}_{\mathrm{P}}$.
	The parametric model $u^{-1} \left( m \left( \cdot, \vec{\theta} \right) \right)$ here is exactly the same as that introduced just before Section \ref{subsec:lower_dim}, where the hyperplane is used.
	As the estimator $\widehat{\vec{\theta}}$, we use the MLE if $n > d$ and the LASSO estimator otherwise.
	The LASSO estimator is implemented just as in Section \ref{sec:simu_high}, by using the package $\texttt{glmnet}$ in $\texttt{R}$.
	Under this setting, we compute $\mathrm{CV} \left[ \widehat{\mu}_{\mathrm{P}} \right]$ in \eqref{eq:CV_nor} and \eqref{eq:CV_ber}.
\item Proposed semiparametric estimator $\widehat{\mu}^\gamma_i$.
	We apply the same settings (for the link function $u$, adjustment type $\gamma$, dictionaries $\mathcal{D}_i$ and sequence $\pi_k$) and notation as in Section \ref{sec:simul}.
	For each $I = 1, 2, \ldots, 300$, we compute $\mathrm{CV} \left[ \widehat{\mu}^\gamma_i \left( I \right) \right]$ in normal regression,
	and $\mathrm{CV} \left[ \widehat{\mu}^\gamma_i \left( h_{\mathrm{min}}, I \right) \right]$ in Bernoulli regression.
\item GBRT $\widehat{\mu}_{\mathrm{X}}$ via XGBoost.
	Its implementation settings are identical to those in Section \ref{sec:simul}.
	Using the same notation as in Section \ref{sec:simul}, we compute
	$ \mathrm{CV} \left[ \widehat{\mu}_{\mathrm{X}} \left( m, I \right) \right]$ for $m \in \left\{ 1, 2, \ldots, 6 \right\} = \mathcal{M}_{\mathrm{X}}$ and $\left\{ 1, 2, \ldots, 20000 \right\} = \mathbb{I}_{\mathrm{X}}$ in normal regression
	and
	$\mathrm{CV} \left[ \widehat{\mu}_{\mathrm{X}} \left( m, \lambda, I \right) \right]$ for $m \in \mathcal{M}_{\mathrm{X}}$, $\lambda \in \left\{ 0, 1, 10, 100, 1000 \right\} = \Lambda_{\mathrm{X}}$ and the searched $I \in \mathbb{I}_{\mathrm{X}}$ in Bernoulli regression.
\end{itemize}

\subsection{Normal regression}
\subsubsection{Low dimension} \label{sec:real_nor_low}
The first data set concerns compressive strength of concrete, studied in \citet*{Y1998}, and is available at https://archive.ics.uci.edu/dataset/165/concrete+compressive+strength.
This data set contains information on eight variables for $n = 1030$ concrete samples, such as their ingredients:
cement ($\mathrm{kg/m}^3$),
blast furnace slag ($\mathrm{kg/m}^3$),
fly ash ($\mathrm{kg/m}^3$),
water ($\mathrm{kg/m}^3$),
superplasticizer ($\mathrm{kg/m}^3$),
coarse aggregate ($\mathrm{kg/m}^3$),
fine aggregate ($\mathrm{kg/m}^3$),
age of testing (days).
The data set also records the resulting compressive strength (MPa) for each sample.
As in \citet*{Y1998}, we predict the compressive strength, using the above $d = 8$ variables as inputs.

\begin{figure}[htb]
\centering
\includegraphics[width = 0.8\textwidth]{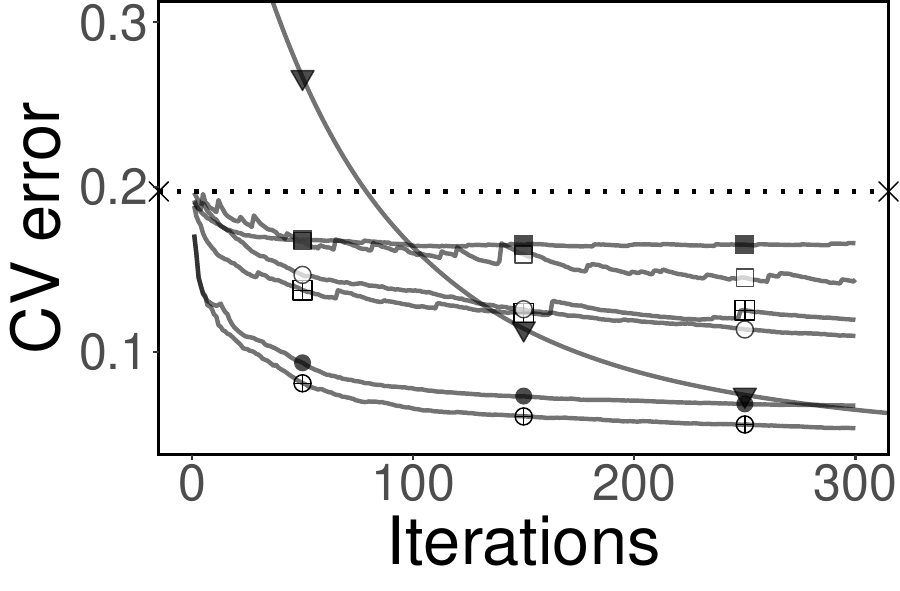}
\caption{
	The curves for prediction errors of each estimator for the concrete data set.
	$\bigcirc$: $\widehat{\mu}^0_0$,
	\blackbigcirc : $\widehat{\mu}^0_1$,
	$\oplus$: $\widehat{\mu}^0_2$,
	$\square$: $\widehat{\mu}^1_0$,
	$\blacksquare$: $\widehat{\mu}^1_1$,
	$\boxplus$: $\widehat{\mu}^1_2$,
	$\times$: $\widehat{\mu}_\mathrm{P}$ and
	\raisebox{-1.1pt}{$\blackbigtriangledown$}: $\widehat{\mu}_{\mathrm{X}}$.
} \label{fig:conc}
\end{figure}

We use the proposed estimators $\widehat{\mu}^\gamma_i$ with $i = 0, 1, 2$ and $\gamma = 0, 1$.
The resulting curves of the prediction error \eqref{eq:CV_nor} are presented in Figure \ref{fig:conc}.
In this figure, the curve for $\widehat{\mu}^0_0$ (respectively $\widehat{\mu}^0_1$, $\widehat{\mu}^0_2$) is drawn with the points ``$\bigcirc$'' (respectively ``\blackbigcirc,'' ``$\oplus$'').
In contrast, the curve for $\widehat{\mu}^1_0$ (respectively $\widehat{\mu}^1_1$, $\widehat{\mu}^1_2$) is depicted with the points ``$\square$'' (respectively ``$\blacksquare$,'' ``$\boxplus $'').
In addition, the curve with \raisebox{-1.1pt}{$\blackbigtriangledown$} is the result of $\widehat{\mu}_{\mathrm{X}} \left( m^\ast, I \right) $ for $1 \leq I \leq 300$, where $m^\ast$ achieves $\min_{m \in \mathcal{M}_{X}, I \in \mathbb{I}_{\mathrm{X}}} \mathrm{CV} \left[ \widehat{\mu}_{\mathrm{X}} \left( m, I \right) \right]$.
The horizontal dotted line with the points ``$\times$'' represents the result for $\widehat{\mu}_\mathrm{P}$.
\setlist[enumerate]{label*=\textbf{\arabic*}., listparindent=\parindent}
\begin{enumerate}[label=\textbf{7}.\textbf{\alph*}., leftmargin=*, align = left, series=desc]
\item
	Observations from Figure \ref{fig:conc} are as follows:
	\begin{enumerate}
	\item
		The resulting CV errors of $\widehat{\mu}^\gamma_i$ with any $\gamma, i$ are smaller than that of $\widehat{\mu}_\mathrm{P}$ ($\times$) for almost any $I$,
		indicating that the proposed estimators adjust $\widehat{\mu}_\mathrm{P}$ well.
	\item
		Among the proposed estimators and $\widehat{\mu}_{\mathrm{X}}$ (\raisebox{-1.1pt}{$\blackbigtriangledown$}), $\widehat{\mu}^0_2$ ($\oplus$) has the smallest CV error for any $I$.
	\end{enumerate}
\item
	We describe results of applying $\widehat{\mu}^0_2$ to the concrete data set.
	The algorithm of the adjustment selects the following words:
	\begin{enumerate}
	\item
		Univariate words are selected 59 times out of 300 total iterations.
		Among them, words corresponding to the variable ``Age'' are chosen 55 times.
	\item
		Bivariate words are selected 241 times, of which the two-dimensional variable including ``Age'' is the most frequently selected and accounts for 193 times.
	\item
		Hence, the proposed estimator $\widehat{\mu}^0_2$ adjusts the parametric estimator $\widehat{\mu}_\mathrm{P}$ by incorporating many words corresponding to the variable ``Age''.
		This suggests that words related to the variable ``Age'' partly contribute to predicting the variable ``compressive strength''.
	\end{enumerate}
\item \label{desc:est_pred_nor}
	Note that $\widehat{\mu}^0_2$ is constructed by optimizing the Bregman divergence, rather than in terms of prediction.
	However, the above descriptions confirm that the resulting $\widehat{\mu}^0_2$ improves the prediction accuracy for $\widehat{\mu}_\mathrm{P}$.
\end{enumerate}

\subsubsection{Higher dimension} \label{sec:real_nor_high}
The second data set is a riboflavin data set from the package \texttt{hdi} in \texttt{R}.
It measures the riboflavin (vitamin $\mathrm{B}_2$) production rate of a bacterium called ``Bacillus subtilis'' across $n = 71$ strains, along with the expression levels of 4088 genes.
More details can be found in \citet*{BKM2014}.
The objective is to predict the production rate from a much larger number of $d = 4088$ variables than the sample size $n$, and to select important genes for this prediction.

\begin{figure}[htb]
\centering
\includegraphics[width = 0.8\textwidth]{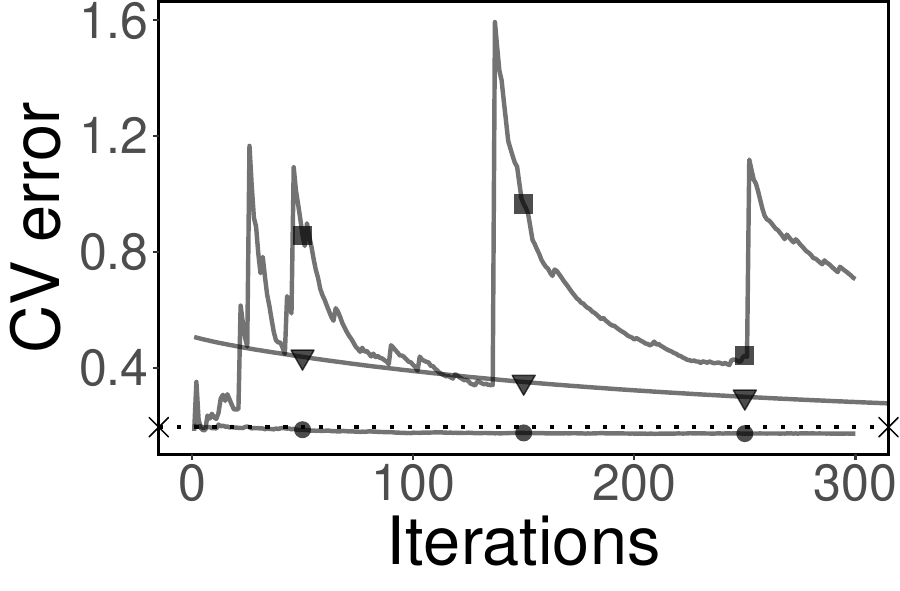}
	\caption{
	The curves for prediction errors of each estimator for the riboflavin data set.
	\blackbigcirc : $\widehat{\mu}^0_1$,
	$\blacksquare$: $\widehat{\mu}^1_1$,
	$\times$: $\widehat{\mu}_\mathrm{P}$ and
	\raisebox{-1.1pt}{$\blackbigtriangledown$}: $\widehat{\mu}_{\mathrm{X}}$.
} \label{fig:ribo}
\end{figure}

We use $\widehat{\mu}^0_1$ and $\widehat{\mu}^1_1$.
The results of the prediction error \eqref{eq:CV_nor} are illustrated in Figure \ref{fig:ribo}, in exactly the same manner as in Figure \ref{fig:conc}.
To confirm the result of variable selection during the nonparametric adjustment, we examine the quantity $\mathcal{C}_s$ for $s = 1, 2, \ldots, d$, which was defined in Section \ref{sec:simu_high}.
This $\mathcal{C}_s$ counts the frequency with which words corresponding to the $s$-th explanatory variable $x_s$ are selected in the construction.
\setlist[enumerate]{label*=\textbf{\arabic*}., listparindent=\parindent}
\begin{enumerate}[label=\textbf{7}.\textbf{\alph*}., leftmargin=*, align = left, resume=desc]
\item Findings from Figure \ref{fig:ribo} are as follows:
	\begin{enumerate}
	\item \label{desc:nor_check}
		The resulting CV error of $\widehat{\mu}^1_1$ ($\blacksquare$) increases drastically at $I = 137, 252$ and fails to improve $\widehat{\mu}_{\mathrm{P}}$ for any $I$.

		Focusing on $I = 137$, we examine this drastic increase.
		We now use the notation introduced in \eqref{eq:CV_nor}.
		We write the proposed estimator $\widehat{\mu}^1_1$ as $\widehat{\mu}_t^{(-k)}$ for a repetition $t$ among the 20 CV runs and for an index $k$ determining the subsets $S_k^{[t]}$, $k = 1, 2, \ldots, 5$.
		In a certain repetition $t_0$ and index $k_0$ of the 5-fold CV, the word selected at $I=137$ takes an extreme value for an input $\vec{X}_\ast$ in the test data $S_{k_0}^{[t_0]}$.
		Because this word is included in the adjustment term,
		$\widehat{\mu}_{t_0}^{(-{k_0})} \left( \vec{X}_\ast \right)$ takes a value far from the test data $Y_\ast$ associated with $\vec{X}_\ast$,
		which causes the drastic increase in the CV error in Figure \ref{fig:ER_ribo}.

		We also check for any problem in constructing $\widehat{\mu}_{t_0}^{(-{k_0})}$ via the algorithm in Section \ref{sec:esti}.
		Figure \ref{fig:ER_ribo} presents the empirical risk of $\widehat{\mu}_{t_0}^{(-{k_0})}$ against each iteration $I$.
		\begin{figure}[htb]
		\centering
		\includegraphics[width = 0.8\textwidth]{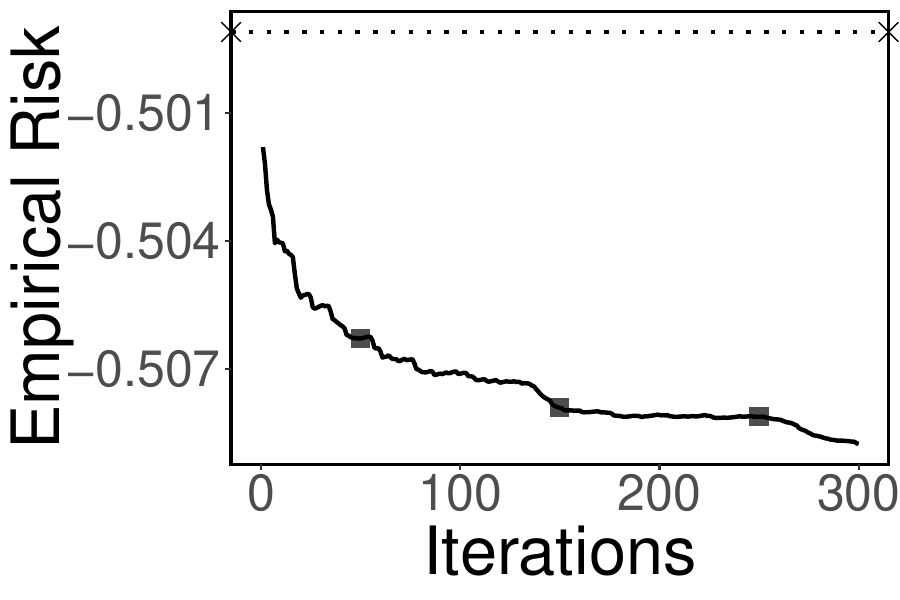}
		\caption{The curves for the empirical risks of $\widehat{\mu}^1_1$ ($\blacksquare$) and $\widehat{\mu}_\mathrm{P}$ ($\times$) for the riboflavin data set.} \label{fig:ER_ribo}
		\end{figure}
		As shown in Figure \ref{fig:ER_ribo}, the empirical risk of $\widehat{\mu}_{t_0}^{(-{k_0})}$ is smaller than that of $\widehat{\mu}_{\mathrm{P}}$ and tends to decrease with $I$.
		That is, $\widehat{\mu}_{t_0}^{(-{k_0})}$ does not seem to exhibit problematic behavior in its construction.
	\item
		For this data set, $\widehat{\mu}_{\mathrm{P}}$ ($\times$) achieves a smaller CV error than $\widehat{\mu}_{\mathrm{X}}$ (\raisebox{-1.1pt}{$\blackbigtriangledown$}) for any $I$.
	\item \label{desc:CV_nor_high}
		This $\widehat{\mu}_{\mathrm{P}}$ is adjusted by the proposed estimator $\widehat{\mu}^0_1$ (\blackbigcirc), which yields the smallest CV error among the compared estimators after $I \thickapprox 50$.
	\end{enumerate}
\item
	We apply $\widehat{\mu}^0_1$ to the riboflavin data set.
	We focus on variable selection, by investigating how many elements are contained in following sets of the explanatory variables $x_s, s = 1, 2, \ldots, 4088$:
	\begin{enumerate}
	\item \label{desc:nor_not_sel}
		First, we see that $\left| \left\{ x_s \middle| \widehat{\theta}_s = 0 \mathrm{~and~} \mathcal{C}_s = 0 \right\} \right| = 3901$, where the set on the left-hand side is the intersection of the following two sets: the variables not selected by the LASSO estimator in $\widehat{\mu}_{\mathrm{P}}$,
		and the variables not selected in the adjustment.
		In other words, about 95\% of the total 4088 variables are selected at neither the parametric nor the nonparametric stage.
	\item \label{desc:num_nor_high}
		The total number of variables selected in the entire procedure of $\widehat{\mu}^0_1$ is 187, which, compared with results in Section 5.1.1 of \citet*{BKM2014}, is larger than that in the FWER-controlling procedure (53).
		On the other hand, this 187 is smaller than that in the FDR-controlling procedure (375) reported therein.
		These 187 variables might be moderate to yield the prediction performance in the description \ref{desc:CV_nor_high}.
	\item \label{desc:nor_para_not_sel_but}
		Other results are that $\left| \left\{ x_s \middle| \widehat{\theta}_s \neq 0 \right\} \right| = 28$,
		whereas
		$\left| \left\{ x_s \middle| \widehat{\theta}_s = 0 \mathrm{~and~} \mathcal{C}_s \neq 0 \right\} \right| = 159$.
		That is, while the parametric regression selects only 28 variables, about 5 times more ones are newly selected in the nonparametric adjustment.
		\begin{enumerate}
		\item \label{desc:sel_nor_BKM}
			The above set $\left\{ x_s \middle| \widehat{\theta}_s \neq 0 \right\}$ includes the variables $x_{624}$, $x_{2564}$ and $x_{4003}$, which correspond to the genes labeled `LYSC\_at', `YOAB\_at' and `YXLD\_at', respectively.
			These three variables were all the selected ones by methods in Sections 3.2.3 and 3.3.1 of \citet*{BKM2014}.
		\item \label{desc:nor_sel_in_adj}
			In this application, the variables with the top four values of $\mathcal{C}_s$ are
			$x_{1479}$ (labeled `YCDI\_at', $\mathcal{C}_{1479} = 16$),
			$x_{285}$ (`DEGS\_at', $\mathcal{C}_{285} = 11$),
			$x_{2020}$ (`YHDK\_at', $\mathcal{C}_{2020} = 9$) and
			$x_{24}$ (`ADDA\_at', $\mathcal{C}_{24} = 8$).
			In addition, their coefficients are all estimated as $\widehat{\theta}_s = 0$ in $\widehat{\mu}_\mathrm{P}$,
			so that these four  belong to $\left\{ x_s \middle| \widehat{\theta}_s = 0 \mathrm{~and~} \mathcal{C}_s \neq 0 \right\}$ of the 159 variables.
		\item \label{desc:nor_80}
			On the other hand, compared with $\mathcal{C}_{1479} = 16$, more than 80\% of the 159 variables in $\left\{ x_s \middle| \widehat{\theta}_s = 0 \mathrm{~and~} \mathcal{C}_s \neq 0 \right\}$ have relatively small $\mathcal{C}_s$: $\mathcal{C}_s = 1$ or $\mathcal{C}_s = 2$.
		\end{enumerate}
	\item \label{desc:var_nor}
		According to the description \ref{desc:nor_not_sel}, some variables are not selected at any stage of $\widehat{\mu}^0_1$, suggesting that they are not important for this application.
		As described in \ref{desc:nor_para_not_sel_but}, there are also variables that are not selected by $\widehat{\mu}_{\mathrm{P}}$ but are selected relatively often in the adjustment, such as $x_{1479}$ (gene `YCDI\_at').
		This result indicates that some variables not selected by $\widehat{\mu}_{\mathrm{P}}$ may also be important, as in the simulation results \ref{desc:selection} and \ref{desc:simu_HDLSS} in Section \ref{sec:simu_high_nor}.
	\end{enumerate}
\item \label{desc:last_nor_high}
	These are results of applying the estimators to this high-dimension low-sample-size data set.
	The proposed estimator $\widehat{\mu}^0_1$ adjusts $\widehat{\mu}_{\mathrm{P}}$.
	This adjustment seems to yield its better performance in predicting the riboflavin production rate than the other estimators (the description \ref{desc:CV_nor_high}).
	At the same time, it can be claimed that $\widehat{\mu}^0_1$ selects some genes potentially important for the prediction, among the huge number of genes that $\widehat{\mu}_{\mathrm{P}}$ does not select.
\end{enumerate}

\subsection{Bernoulli regression}
\subsubsection{Low dimension} \label{sec:real_ber_low}
To confirm the predictive performance of $\widehat{\mu}$ in Bernoulli regression, we use a data set on banknote authentication from https://archive.ics.uci.edu/dataset/267/banknote+authentication.
It includes $n = 1372$ observations on five variables: four features extracted from images of banknote-like specimens, and a binary variable indicating whether each specimen is genuine (0) or forged (1).
The objective is to predict this binary indicator using these features as $d = 4$ explanatory variables.

\begin{figure}[htb]
\centering
\includegraphics[width = \textwidth]{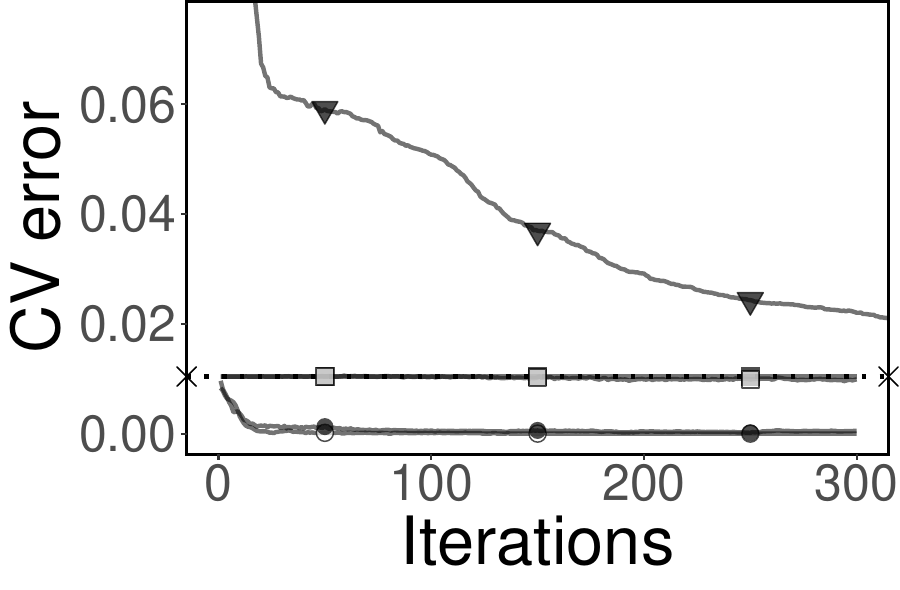}
\caption{
	The curves for prediction errors of each estimator for the banknote authentication data set.
	$\bigcirc$: $\widehat{\mu}^0_0$,
	\blackbigcirc : $\widehat{\mu}^0_1$,
	$\oplus$: $\widehat{\mu}^0_2$,
	$\square$: $\widehat{\mu}^1_0$,
	$\blacksquare$: $\widehat{\mu}^1_1$,
	$\boxplus$: $\widehat{\mu}^1_2$,
	$\times$: $\widehat{\mu}_\mathrm{P}$ and
	\raisebox{-1.1pt}{$\blackbigtriangledown$}: $\widehat{\mu}_{\mathrm{X}}$.
} \label{fig:bank}
\end{figure}

We use the proposed estimators $\widehat{\mu}^\gamma_i$ with $i = 0, 1$ and $\gamma = 0, 1$ and $h_{\mathrm{min}} = 0.1, 0.3, 0.5, 0.7$.
Resulting curves of the prediction error \eqref{eq:CV_ber} are depicted in Figure \ref{fig:bank} in much the same manner as in Figure \ref{fig:conc}.
Among $\left\{ \widehat{\mu}^\gamma_i \left( 0.1, I \right), \widehat{\mu}^\gamma_i \left( 0.3, I \right), \widehat{\mu}^\gamma_i \left( 0.5, I \right), \widehat{\mu}^\gamma_i \left( 0.7, I \right) \right\}$, we draw only one curve of $\widehat{\mu}^\gamma_i \left( h_{\mathrm{min}}^\ast \left( \gamma, i \right), I \right)$ for each $\gamma$, $i$ with the following $h_{\mathrm{min}}^\ast \left( \gamma, i \right)$:
first, we compare the CV results among
\[
		\left\{
			\mathrm{CV} \left[ \widehat{\mu}^\gamma_i \left( 0.1, I \right) \right],
			\mathrm{CV} \left[ \widehat{\mu}^\gamma_i \left( 0.3, I \right) \right],
			\mathrm{CV} \left[ \widehat{\mu}^\gamma_i \left( 0.5, I \right) \right],
			\mathrm{CV} \left[ \widehat{\mu}^\gamma_i \left( 0.7, I \right) \right]
		\right\}
\]
for each $I$.
Then we define $h_{\mathrm{min}}^\ast \left( \gamma, i \right)$ as the largest $h_{\mathrm{min}}$ that most frequently minimizes $\mathrm{CV} \left[ \widehat{\mu}^\gamma_i \left( h_{\mathrm{min}}, I \right) \right]$ through $1 \leq I \leq 300$.

In addition, when applying an estimator $\Bar{\mu}$ to the data set, we check the following numbers of mislabeled cases:
\begin{align}
		\mathrm{F}_p \left[ \Bar{\mu} \right]
&=
		\sum_{i = 1}^{n} \left[ Y_i - \mathbb{I} \left\{ \Bar{\mu} \left( \vec{X}_i \right) \geq p \right\} \right]^2, \label{eq:ber_F} \\
		\mathrm{FP}_p \left[ \Bar{\mu} \right]
&=
		\sum_{i \in \left\{ j \middle| Y_j = 0 \right\}} \left[ Y_i - \mathbb{I} \left\{ \Bar{\mu} \left( \vec{X}_i \right) \geq p \right\} \right]^2, \label{eq:ber_FP} \\
		\mathrm{FN}_p \left[ \Bar{\mu} \right]
&=
		\mathrm{F}_p \left[ \Bar{\mu} \right]
	-
		\mathrm{FP}_p \left[ \Bar{\mu} \right], \label{eq:ber_FN}
\end{align}
for thresholds $p = 0.1, 0.3, 0.5, 0.7, 0.9$.
\setlist[enumerate]{label*=\textbf{\arabic*}., listparindent=\parindent}
\begin{enumerate}[label=\textbf{7}.\textbf{\alph*}., leftmargin=*, align = left, resume=desc]
\item
	Recognitions from Figure \ref{fig:bank} are as follows:
	\begin{enumerate}
	\item
		The semiparametric estimators $\widehat{\mu}^1_0$ ($\square$) and $\widehat{\mu}^1_1$ ($\blacksquare$) with $\gamma = 1$ have almost the same CV error as $\widehat{\mu}_\mathrm{P}$ for any $I$.
	\item
		The resulting CV errors of $\widehat{\mu}^0_0$ ($\bigcirc$) and $\widehat{\mu}^0_1$ (\blackbigcirc) with $\gamma = 0$ are smaller than those of the other estimators for any $I$.
		In particular, $\widehat{\mu}^0_0$ yields slightly smaller CV errors than $\widehat{\mu}^0_1$ for almost all $I$.
	\end{enumerate}
\item
	We now describe results of applying $\widehat{\mu}^0_0$ to the banknote authentication data set.
	The results for mislabeling \eqref{eq:ber_F}--\eqref{eq:ber_FN} of $\widehat{\mu}^0_0$ and $\widehat{\mu}_{\mathrm{P}}$ are summarized in Table \ref{tab:real_ber_low}.
	\begin{table}[tbp]
	\centering
	\small
		\begin{tabular}{cccccc}
			\toprule
				& $p = 0.1$ & $p = 0.3$ & $p = 0.5$ & $p = 0.7$ & $p = 0.9$ \\
			\midrule
			\midrule
			$\widehat{\mu}_{\mathrm{P}}$ & 17 & 13 & 11 & 13 & 17 \\
			& (16, 1) & (10, 3) & (5, 6) & (5, 6) & (0, 17)\\
			\midrule
			$\widehat{\mu}^0_0$ & 0 & 0 & 0 & 0 & 0\\
			& (0, 0) & (0, 0) & (0, 0) & (0, 0) & (0, 0)\\
			\bottomrule
		\end{tabular}
	\caption{
		The results of $\mathrm{F}_p \left[ \widehat{\mu}_{\mathrm{P}} \right]$ and $\mathrm{F}_p \left[ \widehat{\mu}^0_0 \right]$ for the banknote authentication data set.
		For each threshold $p$, rows 2 and 4 present $\mathrm{F}_p \left[ \widehat{\mu}_{\mathrm{P}} \right]$ and $\mathrm{F}_p \left[ \widehat{\mu}^0_0 \right]$, respectively.
		Below them, $\mathrm{FP}_p$ and $\mathrm{FN}_p$ for each estimator are given in parentheses as $\left( \mathrm{FP}_p, \mathrm{FN}_p \right)$.
	}
	\label{tab:real_ber_low}
	\end{table}
	\begin{enumerate}
	\item
		In each column of Table \ref{tab:real_ber_low}, $\mathrm{F}_p \left[ \widehat{\mu}_{\mathrm{P}} \right] > \mathrm{F}_p \left[ \widehat{\mu}^0_0 \right]$ holds for any searched $p$.
		Hence, $\widehat{\mu}^0_0$ fits the data better than $\widehat{\mu}_{\mathrm{P}}$.
		In particular, as observed in row 2, no input is mislabeled by $\widehat{\mu}^0_0$ for any $p$, suggesting that $\widehat{\mu}^0_0$ fits the data well.
	\item
		In banknote authentication, if a forged banknote is incorrectly classified as genuine, it may be allowed to enter circulation.
		Thus it is important for estimators to avoid such false positives.
		As shown in Table \ref{tab:real_ber_low}, the false positives resulting from $\widehat{\mu}_\mathrm{P}$ are correctly adjusted by $\widehat{\mu}^0_0$.
		This may highlight the practical value of using the semiparametric estimator alongside the parametric one.
	\end{enumerate}
\item \label{desc:est_pred_ber}
	The proposed estimator $\widehat{\mu}^0_0$ is constructed by minimizing Bregman divergence, rather than in terms of prediction or labeling.
	Nevertheless, as indicated by the above descriptions, the resulting $\widehat{\mu}^0_0$ improves the performance of $\widehat{\mu}_\mathrm{P}$ well.
\end{enumerate}

\subsubsection{Higher dimension} \label{sec:real_ber_high}
The final data set is a colon cancer data set from the package \texttt{rda} in \texttt{R}.
This data set comprises 40 colon tumor samples (labeled 1) and 22 normal colon tissue samples (labeled 0), with 2000 variables of microarray expression for each sample.
We predict whether samples are normal or not, by using the much larger number of $d = 2000$ variables than $n = 62$, and simultaneously select important genes for this prediction.

\begin{figure}[htb]
\centering
\includegraphics[width = 0.8\textwidth]{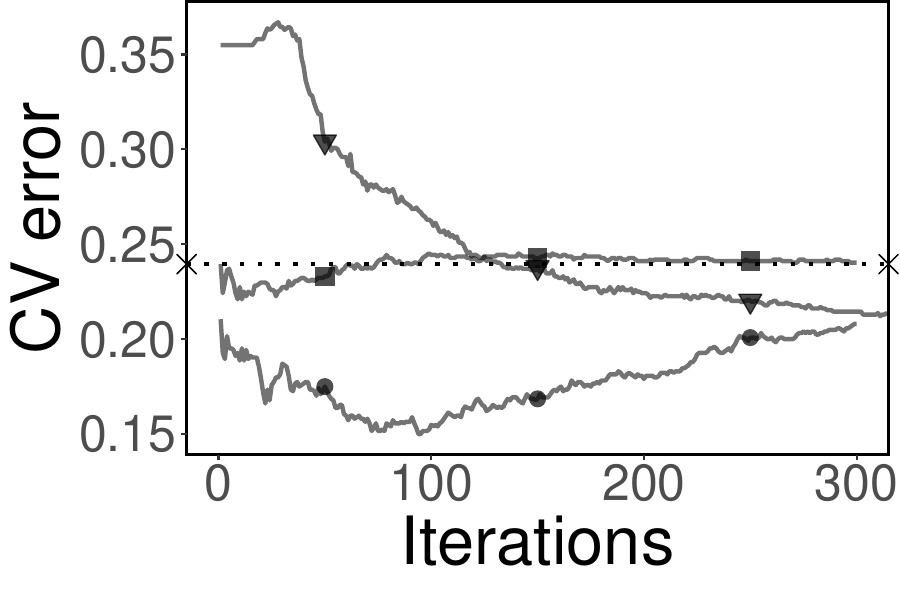}
\caption{
	The curves for prediction errors of each estimator for the colon cancer data set.
	\blackbigcirc : $\widehat{\mu}^0_1$,
	$\blacksquare$: $\widehat{\mu}^1_1$,
	$\times$: $\widehat{\mu}_\mathrm{P}$ and
	\raisebox{-1.1pt}{$\blackbigtriangledown$}: $\widehat{\mu}_{\mathrm{X}}$.
} \label{fig:colon}
\end{figure}

We use the proposed estimators $\widehat{\mu}^0_1$ and $\widehat{\mu}^1_1$ and $h_{\mathrm{min}} = 0.1, 0.2, \ldots, 0.7$.
Resulting curves of the prediction error \eqref{eq:CV_ber} are shown in Figure \ref{fig:colon}, in much the same manner as in Figure \ref{fig:bank}.
\setlist[enumerate]{label*=\textbf{\arabic*}., listparindent=\parindent}
\begin{enumerate}[label=\textbf{7}.\textbf{\alph*}., leftmargin=*, align = left, resume=desc]
\item
	Observations from Figure \ref{fig:colon} are as follows:
	\begin{enumerate}
	\item
		The resulting CV error for $\widehat{\mu}^1_1$ is smaller than that for $\widehat{\mu}_{\mathrm{P}}$ before $I \thickapprox 100$, 
		while these two CV errors are close to each other after $I \thickapprox 100$.
	\item \label{desc:CV_ber_high}
		The semiparametric estimator $\widehat{\mu}^0_1$ has the smallest CV error among those compared for any $I$.
	\end{enumerate}
\item \label{desc:mislabel_ber}
	We apply $\widehat{\mu}^0_1$ to the colon cancer data set.
	The results for mislabeling \eqref{eq:ber_F}--\eqref{eq:ber_FN} of $\widehat{\mu}^0_1$ and $\widehat{\mu}_\mathrm{P}$ are listed in Table \ref{tab:real_ber_high}.
	\begin{table}[tbp]
	\centering
	\small
		\begin{tabular}{cccccc}
			\toprule
				& $p = 0.1$ & $p = 0.3$ & $p = 0.5$ & $p = 0.7$ & $p = 0.9$ \\
			\midrule
			\midrule
			$\mathrm{F}_p \left[ \widehat{\mu}_{\mathrm{P}} \right]$ & 22 & 22 & 8 & 9 & 40\\
			& (2, 20) & (2, 20) & (0, 8) & (0, 9) & (0, 40) \\
			\midrule
			$\mathrm{F}_p \left[ \widehat{\mu}^0_1 \right]$ & 0 & 0 & 0 & 0 & 6\\
			& (0, 0) & (0, 0) & (0, 0) & (0, 0) & (0, 6)\\
			\bottomrule
		\end{tabular}
	\caption{
		The results of $\mathrm{F}_p \left[ \widehat{\mu}_{\mathrm{P}} \right]$ and $\mathrm{F}_p \left[ \widehat{\mu}^0_1 \right]$ for the colon cancer data set.
		For each threshold $p$, rows 2 and 4 present $\mathrm{F}_p \left[ \widehat{\mu}_{\mathrm{P}} \right]$ and $\mathrm{F}_p \left[ \widehat{\mu}^0_1 \right]$, respectively.
		Below them, $\mathrm{FP}_p$ and $\mathrm{FN}_p$ for each estimator are given in parentheses as $\left( \mathrm{FP}_p, \mathrm{FN}_p \right)$.
	}
	\label{tab:real_ber_high}
	\end{table}
	It is observed from each column of Table \ref{tab:real_ber_high} that $\widehat{\mu}^0_1$ fits the data better than $\widehat{\mu}_{\mathrm{P}}$ for any threshold $p$.
	\item
	Next, we focus on variable selection of $\widehat{\mu}^0_1$, by examining the following counts of explanatory variables $x_s, s = 1, 2, \ldots, 2000$ as in Section \ref{sec:real_nor_high}:
	\begin{enumerate}
	\item \label{desc:ber_not_sel}
		We observe $\left| \left\{ x_s \middle| \widehat{\theta}_s = 0 \mathrm{~and~} \mathcal{C}_s = 0 \right\} \right| = 1905$,
		indicating that approximately 95\% of the total 2000 variables are selected in neither the parametric nor the nonparametric phase.
	\item \label{desc:num_ber_high}
		The number of variables selected in the entire procedure of $\widehat{\mu}^0_1$ is 95,
		which is larger than that by PAM (19) in Section 5 of \citet*{GHT2007}
		and, on the other hand, is smaller than that by SCRDA (143) reported therein.
		These 95 variables might be moderate to yield the prediction performance as in the description \ref{desc:CV_ber_high}.
	\item \label{desc:ber_para_not_sel_but}
		Other results are that $\left| \left\{ x_s \middle| \widehat{\theta}_s \neq 0 \right\} \right| = 4$,
		whereas
		$\left| \left\{ x_s \middle| \widehat{\theta}_s = 0 \mathrm{~and~} \mathcal{C}_s \neq 0 \right\} \right| = 91$.
		While the parametric regression selects only 4 variables, the nonparametric adjustment newly selects about twenty times as many.

		\begin{enumerate}
		\item \label{desc:ber_sel_in_adj}
			The top four values of $\mathcal{C}_s$ are achieved by the following variables:
				$x_{897}$ ($\mathcal{C}_{897} = 36$),
				$x_{1635}$ ($\mathcal{C}_{1635} = 19$),
				$x_{765}$ ($\mathcal{C}_{765} = 16$) and
				$x_{513}$ ($\mathcal{C}_{513} = 8$),
			all of whose coefficients are estimated as $\widehat{\theta}_s = 0$ in $\widehat{\mu}_\mathrm{P}$.
			Thus, these four are included in $\left\{ x_s \middle| \widehat{\theta}_s = 0 \mathrm{~and~} \mathcal{C}_s \neq 0 \right\}$ of the 91 variables.
			\item \label{desc:ber_80}
			On the other hand, compared with $\mathcal{C}_{897} = 36$, over 80\% out of the 91 variables in $\left\{ x_s \middle| \widehat{\theta}_s = 0 \mathrm{~and~} \mathcal{C}_s \neq 0 \right\}$ have relatively small $\mathcal{C}_s$: $1 \leq \mathcal{C}_s \leq 4$.
		\end{enumerate}
	\item
		According to the description \ref{desc:ber_not_sel}, some variables are completely ignored by $\widehat{\mu}^0_1$, which suggests that they may not be important for this application.
		On the other hand, as described in \ref{desc:ber_para_not_sel_but}, some variables overlooked by $\widehat{\mu}_{\mathrm{P}}$ are selected relatively often during the adjustment.
		This result indicates that some variables not selected by $\widehat{\mu}_{\mathrm{P}}$ might nevertheless play an important role.
	\end{enumerate}
\item \label{desc:last_ber_high}
	In the simulations for high-dimensional Bernoulli regression of Section \ref{sec:simu_high_berpoi} (specifically, the description \ref{desc:berpoi_high_ber_bad}), the resulting $\widehat{\mu}^0_1$ had lower estimation accuracy than the compared estimator.
	In addition, the simulated $\mathcal{C}_s$ did not seem to differ clearly among the variables $\left\{ x_s \middle| \widehat{\theta}_s = 0 \mathrm{~and~} \mathcal{C}_s \neq 0 \right\}$.
	On the other hand, we obtain some positive results for $\widehat{\mu}^0_1$ in the present application:
	actually,
	it can be said that $\widehat{\mu}^0_1$ predicts the labels for colon samples more effectively than the other estimators (\ref{desc:CV_ber_high} and \ref{desc:mislabel_ber}).
	It can also be claimed that differences in $\mathcal{C}_s$, or differences in importance, occur among the variables $\left\{ x_s \middle| \widehat{\theta}_s = 0 \mathrm{~and~} \mathcal{C}_s \neq 0 \right\}$ (\ref{desc:nor_sel_in_adj} and \ref{desc:ber_80}).
\end{enumerate}

\section{Concluding remarks} \label{sec:concl}
We conclude this work with a summary and remarks.
\begin{itemize}
\item
We have examined semiparametric regression estimators that combine a parametric crude guess with its nonparametric adjustment, via the stagewise empirical risk minimization algorithm.
They are mainly determined by the following triplet $\left( m \left( \cdot, \widehat{\vec{\theta}} \right), U, \mathcal{D} \right)$ (Section \ref{sec:esti}): 
\begin{itemize}
	\item
	A parametric regression estimator $m \left( \cdot, \widehat{\vec{\theta}} \right)$ with a model function $m \left( \cdot, \vec{\theta} \right)$ and an estimator $\widehat{\vec{\theta}}$ for the finite-dimensional parameter $\vec{\theta}$.
	This was used as the parametric crude guess, in which the model was allowed to be misspecified.
	\item
	A strictly convex function $U$ associated with the Bregman divergence, and a dictionary $\mathcal{D}$.
	These were used in the stagewise empirical risk minimization algorithm to adjust the initial $m \left( \cdot, \widehat{\vec{\theta}} \right)$ nonparametrically.
	Different choices of $U$ and $\mathcal{D}$ yielded various forms of the proposed estimators.
\end{itemize}
\item
In the theoretical study, we derived a non-asymptotic error bound for the proposed estimators (Theorem \ref{thm:mu} in Section \ref{sec:theo}).
According to Theorem \ref{thm:mu}, the divergence risk for the proposed estimators can be bounded above only by the sum of a bias term and asymptotically vanishing terms.
\begin{itemize}
	\item
	Although previous studies by \citet*{K2007}, \citet*{NE2013}, \citet*{NN2024} and \citet*{YN2019} developed similar risk bounds, 
	they did not consider data-dependent dictionaries in their theoretical analyses.
	The theoretical evaluation of the proposed estimators with such dictionaries is a contribution of this study.
\end{itemize}
\item
In addition to the theoretical study, we investigated practical behavior of the proposed estimators through the numerical study.
\begin{itemize}
	\item
	Simulations revealed that several of the proposed estimators offered competitive or, in some cases, better estimation performance compared with other estimators (Section \ref{sec:simul}).
	Similar superiority in predictive performance was also observed in real data applications (Section \ref{sec:real}).
	\item 
	Moreover, the numerical study suggested that the proposed estimators can be used to correct variable selection in parametric regression in several settings (Sections \ref{sec:simu_high}, \ref{sec:real_nor_high} and \ref{sec:real_ber_high}).
\end{itemize}
\item
Although we have demonstrated the applicability of the proposed estimators, there is still room for further study. 
For example:
\begin{itemize}
	\item 
	We can utilize proposed estimators with parametric models other than the hyperplane and with dictionaries other than those proposed in this work.
	Investigating their behavior would be of significant interest.
	\item
	It would also be interesting and important to provide theoretical support for variable selection by the proposed estimators, which requires describing the probability distributions of the related statistics.
\end{itemize}
\end{itemize}

\begin{appendices}
\section*{Appendix: Proofs} \label{sec:appe_proofs}
\addcontentsline{toc}{section}{Appendix: Proofs}
\stepcounter{section}
\renewcommand{\theequation}{A.\arabic{equation}}
\setcounter{equation}{0}
\renewcommand{\thesubsection}{A.\arabic{subsection}}
\setcounter{subsection}{0}

\subsection{Proof of Lemma \ref{lem:ineq1}}
The proof of Lemma \ref{lem:ineq1} is divided into the following two parts.
\begin{Lem} \label{lem:ineq1-1}
For any $\widetilde{\Lambda} \in \mathcal{W}$, it follows that
\begin{align}
		&\widehat{L} \left( \mu_{k + 1} \left( \vec{\theta}_1 \right) \right)
	-
		\widehat{L} \left( \mu \left( \cdot, \widetilde{\Lambda}, \vec{\theta}_2 \right) \right) \nonumber \\
&\leq
		\left( 1 - \pi_{k + 1} \right)
		\left\{
			\widehat{L} \left( \mu_k \left( \vec{\theta}_1 \right) \right)
		-
			\widehat{L} \left( \mu \left( \cdot, \widetilde{\Lambda}, \vec{\theta}_2 \right) \right)
		\right\} \nonumber \\
	&\hspace{0.5cm}+
		\pi_{k + 1}^2
		\left\{
			A_1 \left( \vec{\theta}_1, \vec{\theta}_2 \right)
		+
			\left\| \vec{\theta}_1 - \vec{\theta}_2 \right\|_2
			A_3 \left( \vec{\theta}_1, \vec{\theta}_2 \right)
		\right\}
	+
		\pi_{k + 1}
		\left\| \vec{\theta}_1 - \vec{\theta}_2 \right\|_2
		A_2 \left( \vec{\theta}_1, \vec{\theta}_2 \right)
	+
		\pi_{k + 1} \varepsilon \label{eq:L_k_k+1}
\end{align}
for $k \geq 0$, where $A_1$, $A_2$, $A_3$ are in Lemma \ref{lem:ineq1}.
\end{Lem}
\begin{Lem} \label{lem:ineq1-2}
If the following inequality
\begin{align}
		&\widehat{L} \left( \mu_{\ell + 1} \left( \vec{\theta}_1 \right) \right)
	-
		\widehat{L} \left( \mu \left( \cdot, \widetilde{\Lambda}, \vec{\theta}_2 \right) \right) \nonumber\\
&\hspace{0.5cm}\leq
		\left( 1 - \pi_{\ell + 1} \right)
		\left\{
			\widehat{L} \left( \mu_\ell \left( \vec{\theta}_1 \right) \right)
		-
			\widehat{L} \left( \mu \left( \cdot, \widetilde{\Lambda}, \vec{\theta}_2 \right) \right)
		\right\}
	+
		\pi_{\ell + 1}^2 a
	+
		\pi_{\ell + 1} b, \quad
\ell \geq 0 \label{eq:L_recur}
\end{align}
holds for some $\widetilde{\Lambda} \in \mathcal{W}$, $a \geq 0$ and $b \geq 0$,
then
\begin{align}
		\widehat{L} \left( \mu_{k} \left( \vec{\theta}_1 \right) \right)
\leq
		\widehat{L} \left( \mu \left( \cdot, \widetilde{\Lambda}, \vec{\theta}_2 \right) \right)
	+
		\frac{\tau^2}{k + \tau} a
	+
		b, \quad
k \in \mathbb{N}. \label{eq:ind}
\end{align}
\end{Lem}

\subsubsection*{Proof of Lemma \ref{lem:ineq1-1}}
We pick $\widetilde{\Lambda} = \left[ \widetilde{\lambda}_{B} \right]_{B \in \mathcal{D}} \in \mathcal{W}$.
From the definitions of $\mu_{k+1}$, $\xi_{k+1}$, selected $I$ words $B_1, \ldots, B_I$ through algorithm in Section \ref{sec:esti} and the operator $C$ and $T$, it follows that
\begin{align}
		&\widehat{L} \left( \mu_{k + 1} \left( \vec{\theta}_1 \right) \right)
	-
		\widehat{L} \left( \mu \left( \cdot, \widetilde{\Lambda}, \vec{\theta}_2 \right) \right) \nonumber\\
&=
		\widehat{L} \left( u^{-1} \left( T \left( \vec{\theta}_1, \xi_{k + 1} \right) \right) \right)
	-
		\widehat{L} \left( \mu \left( \cdot, \widetilde{\Lambda}, \vec{\theta}_2 \right) \right) \nonumber\\
&=
		\widehat{L} \left( u^{-1} \left( T \left( \vec{\theta}_1, C \left( \pi_{k + 1} \middle| B_{k + 1}, \xi_k \right) \right) \right) \right)
	-
		\widehat{L} \left( \mu \left( \cdot, \widetilde{\Lambda}, \vec{\theta}_2 \right) \right) \nonumber\\
&\leq
		\left\{
			\inf_{\widetilde{B} \in \mathcal{D}}
			\widehat{L} \left( u^{-1} \left(
				T \left( \vec{\theta}_1, C \left( \pi_{k + 1} \middle| \widetilde{B}, \xi_k \right) \right)
				\right) \right)
		+
			\pi_{k + 1} \varepsilon
		\right\}
		-
			\widehat{L} \left( \mu \left( \cdot, \widetilde{\Lambda}, \vec{\theta}_2 \right) \right) \nonumber\\
&=
		\sum_{B \in \mathcal{D}}
		\widetilde{\lambda}_{B}
		\left[
			\inf_{\widetilde{B} \in \mathcal{D}}
			\widehat{L} \left( u^{-1} \left(
				T \left( \vec{\theta}_1, C \left( \pi_{k + 1} \middle| \widetilde{B}, \xi_k \right) \right)
				\right) \right)
		-
			\widehat{L} \left( \mu \left( \cdot, \widetilde{\Lambda}, \vec{\theta}_2 \right) \right)
		\right]
	+
		\pi_{k + 1} \varepsilon \nonumber\\
&\leq
		\sum_{B \in \mathcal{D}}
		\widetilde{\lambda}_{B}
		\left[
			\widehat{L} \left( u^{-1} \left(
				T \left( \vec{\theta}_1, C \left( \pi_{k + 1} \middle| B, \xi_k \right) \right)
				\right) \right)
		-
			\widehat{L} \left( \mu \left( \cdot, \widetilde{\Lambda}, \vec{\theta}_2 \right) \right)
		\right]
	+
		\pi_{k + 1} \varepsilon \nonumber\\
&=
		\sum_{B \in \mathcal{D}}
		\widetilde{\lambda}_{B}
		\left\{
			\widehat{L}
			\left(
				u^{-1}
				\left(
					C \left( \pi_{k + 1} \middle| T \left( \vec{\theta}_1, B \right), T \left( \vec{\theta}_1, \xi_k \right) \right)
				\right)
			\right)
		-
			\widehat{L} \left( \mu \left( \cdot, \widetilde{\Lambda}, \vec{\theta}_2 \right) \right)
		\right\}
	+
		\pi_{k + 1}. \varepsilon \label{eq:est_L_1}
\end{align}
Then by definition of $\widehat{L}$, we see that
\begin{align}
		&\widehat{L} \left( u^{-1} \left(
			C \left(
				\pi_{k + 1}
			\middle|
				T \left( \vec{\theta}_1, B \right),
				T \left( \vec{\theta}_1, \xi_k \right)
			\right)
		\right) \right)
	-
		\widehat{L} \left( \mu \left( \cdot, \widetilde{\Lambda}, \vec{\theta}_2 \right) \right) \nonumber\\
&=
		\frac{1}{n}
		\sum_{i = 1}^{n}
		\biggl\{
			U^\ast \left(
				C \left( \pi_{k + 1} \middle|
				T \left( \vec{\theta}_1, B \right) \left( \vec{X}_i \right),
				T \left( \vec{\theta}_1, \xi_k \right) \left( \vec{X}_i \right)
			\right) \right) \nonumber \\
		&\hspace{2cm}-
			Y_i
			C \left(
				\pi_{k + 1} \middle|
				T \left( \vec{\theta}_1, B \right) \left( \vec{X}_i \right),
				T \left( \vec{\theta}_1, \xi_k \right) \left( \vec{X}_i \right) \right)
		\biggr\}
	-
		\widehat{L} \left( \mu \left( \cdot, \widetilde{\Lambda}, \vec{\theta}_2 \right) \right) \nonumber\\
&=
		\frac{1}{n}
		\sum_{i = 1}^{n}
		\biggl[
			U^\ast \left(
				C \left( \pi_{k + 1} \middle|
				T \left( \vec{\theta}_1, B \right) \left( \vec{X}_i \right),
				T \left( \vec{\theta}_1, \xi_k \right) \left( \vec{X}_i \right)
			\right) \right) \nonumber\\
		&\hspace{2cm}-
			Y_i
			\left\{
				\pi_{k + 1} T \left( \vec{\theta}_1, B \right) \left( \vec{X}_i \right)
			+
				\left( 1 - \pi_{k + 1} \right)
				T \left( \vec{\theta}_1, \xi_k \right) \left( \vec{X}_i \right)
			\right\}
		\biggr]
	-
		\widehat{L} \left( \mu \left( \cdot, \widetilde{\Lambda}, \vec{\theta}_2 \right) \right) \nonumber\\
&=
		\frac{1}{n}
		\sum_{i = 1}^{n}
		\biggl[
			\left( 1 - \pi_{k + 1} \right)
			\left\{
				U^\ast \left( T \left( \vec{\theta}_1, \xi_k \right) \left( \vec{X}_i \right) \right)
			-
				Y_i T \left( \vec{\theta}_1, \xi_k \right) \left( \vec{X}_i \right)
			\right\} \nonumber\\
		&\hspace{1.5cm}+
			U^\ast \left( C \left( \pi_{k + 1} \middle| T \left( \vec{\theta}_1, B \right) \left( \vec{X}_i \right), T \left( \vec{\theta}_1, \xi_k \right) \left( \vec{X}_i \right) \right) \right)
		-
			\left( 1 - \pi_{k + 1} \right)
			U^\ast \left( T \left( \vec{\theta}_1, \xi_k \right) \left( \vec{X}_i \right) \right) \nonumber \\
		&\hspace{1.5cm}-
			Y_i \pi_{k + 1} T \left( \vec{\theta}_1, B \right) \left( \vec{X}_i \right)
		\biggr]
	-
		\widehat{L} \left( \mu \left( \cdot, \widetilde{\Lambda}, \vec{\theta}_2 \right) \right) \nonumber\\
&=
		\left( 1 - \pi_{k + 1} \right)
		\left\{
			\widehat{L} \left( \mu_k \left( \vec{\theta}_1 \right) \right)
		-
			\widehat{L} \left( \mu \left( \cdot, \widetilde{\Lambda}, \vec{\theta}_2 \right) \right)
		\right\} \nonumber\\
	&\hspace{0.5cm}+
		\frac{1}{n}
		\sum_{i = 1}^{n}
		\biggl[
			U^\ast \left(
				C \left( \pi_{k + 1}
				\middle|
				T \left( \vec{\theta}_1, B \right) \left( \vec{X}_i \right),
				T \left( \vec{\theta}_1, \xi_k \right) \left( \vec{X}_i \right) \right)
			\right)
		-
			\left( 1 - \pi_{k + 1} \right)
			U^\ast \left( T \left( \vec{\theta}_1, \xi_k \right) \left( \vec{X}_i \right) \right) \nonumber\\
	&\hspace{2cm}-
			Y_i
			\pi_{k + 1}
			T \left( \vec{\theta}_1, B \right) \left( \vec{X}_i \right)
		\biggr]
		-
		\pi_{k + 1}
		\widehat{L} \left( \mu \left( \cdot, \widetilde{\Lambda}, \vec{\theta}_2 \right) \right) \nonumber\\
&=
		\left( 1 - \pi_{k + 1} \right)
		\left\{
			\widehat{L} \left( \mu_k \left( \vec{\theta}_1 \right) \right)
		-
			\widehat{L} \left( \mu \left( \cdot, \widetilde{\Lambda}, \vec{\theta}_2 \right) \right)
		\right\} \nonumber\\
	&\hspace{0.5cm}+
		\frac{1}{n}
		\sum_{i = 1}^{n}
		\biggl[
			U^\ast \left(
				C \left( \pi_{k + 1}
				\middle|
				T \left( \vec{\theta}_1, B \right) \left( \vec{X}_i \right),
				T \left( \vec{\theta}_1, \xi_k \right) \left( \vec{X}_i \right) \right)
			\right)
		-
			\left( 1 - \pi_{k + 1} \right)
			U^\ast \left( T \left( \vec{\theta}_1, \xi_k \right) \left( \vec{X}_i \right) \right) \nonumber\\
		&\hspace{1.7cm}-
			Y_i
			\pi_{k + 1}
			T \left( \vec{\theta}_1, B \right) \left( \vec{X}_i \right)
		\biggr]
	-
		\pi_{k + 1}
		\left\{
			U^\ast \left( u \left( \mu \left( \vec{X}_i, \widetilde{\Lambda}, \vec{\theta}_2 \right) \right) \right)
		-
			Y_i
			u \left( \mu \left( \vec{X}_i, \widetilde{\Lambda}, \vec{\theta}_2 \right) \right)
		\right\} \nonumber\\
&=
		\left( 1 - \pi_{k + 1} \right)
		\left\{
			\widehat{L} \left( \mu_k \left( \vec{\theta}_1 \right) \right)
		-
			\widehat{L} \left( \mu \left( \cdot, \widetilde{\Lambda}, \vec{\theta}_2 \right) \right)
		\right\} \nonumber \\
	&\hspace{0.5cm}+
		\frac{1}{n}
		\sum_{i = 1}^{n}
		\Biggl[
			U^\ast \left( C \left( \pi_{k + 1} \middle| T \left( \vec{\theta}_1, B \right) \left( \vec{X}_i \right), T \left( \vec{\theta}_1, \xi_k \right) \left( \vec{X}_i \right) \right) \right) \nonumber\\
		&\hspace{2cm}-
			C \left( \pi_{k + 1}
			\middle|
			U^\ast \left( u \left( \mu \left( \vec{X}_i, \widetilde{\Lambda}, \vec{\theta}_2 \right) \right) \right),
			U^\ast \left( T \left( \vec{\theta}_1, \xi_k \right) \left( \vec{X}_i \right) \right)
			\right) \nonumber\\
		&\hspace{2cm}-
			\pi_{k + 1}
			Y_i
			\left\{
				T \left( \vec{\theta}_1, B \right) \left( \vec{X}_i \right)
			-
				u \left( \mu \left( \vec{X}_i, \widetilde{\Lambda}, \vec{\theta}_2 \right) \right)
			\right\}
		\Biggr]. \label{eq:est_L^coco}
\end{align}
We shall further bound \eqref{eq:est_L^coco} from above.
The convexity of $U^\ast$ implies the following inequalities
$C \left( \pi_{k + 1} \middle| U^\ast \left( a \right), U^\ast \left( b \right) \right) \geq U^\ast \left( C \left( \pi_{k + 1} \middle| a, b \right) \right)$
and
$U^\ast \left( b \right) - U^\ast \left( a \right) \leq u^{-1} \left( b \right) \left( b - a \right)$ for $a, b \in \mathbb{R}$, so that
\begin{align}
		&U^\ast \left(
		C \left( \pi_{k + 1} \middle|
			T \left( \vec{\theta}_1, B \right) \left( \vec{X}_i \right),
			T \left( \vec{\theta}_1, \xi_k \right) \left( \vec{X}_i \right)
		\right) \right) \nonumber\\
	&\hspace{0.5cm}-
		C \left( \pi_{k + 1} \middle|
			U^\ast \left( u \left( \mu \left( \vec{X}_i, \widetilde{\Lambda}, \vec{\theta}_2 \right) \right) \right),
			U^\ast \left( T \left( \vec{\theta}_1, \xi_k \right) \left( \vec{X}_i \right) \right)
		\right) \nonumber \\
&\leq
		U^\ast \left( C \left( \pi_{k + 1} \middle| T \left( \vec{\theta}_1, B \right) \left( \vec{X}_i \right), T \left( \vec{\theta}_1, \xi_k \right) \left( \vec{X}_i \right) \right) \right) \nonumber\\
	&\hspace{0.5cm}-
		U^\ast \left(
			C \left( \pi_{k + 1}
			\middle|
				u \left( \mu \left( \vec{X}_i, \widetilde{\Lambda}, \vec{\theta}_2 \right) \right),
				T \left( \vec{\theta}_1, \xi_k \right) \left( \vec{X}_i \right)
			\right)
		\right) \nonumber \\
&\leq
		u^{-1} \left(
				C \left( \pi_{k + 1} \middle| T \left( \vec{\theta}_1, B \right) \left( \vec{X}_i \right), T \left( \vec{\theta}_1, \xi_k \right) \left( \vec{X}_i \right) \right)
			\right) \nonumber \\
	&\hspace{0.5cm}\times
			\pi_{k + 1}
		\left\{
			T \left( \vec{\theta}_1, B \right) \left( \vec{X}_i \right)
		-
			u \left( \mu \left( \vec{X}_i, \widetilde{\Lambda}, \vec{\theta}_2 \right) \right)
		\right\}. \label{eq:est_L_U*-coco}
\end{align}
In addition, by the mean value theorem, there exists $a_k \left( \vec{X}_i, \vec{\theta}_1, B \right) \in \left( 0,1 \right)$ for $i = 1, \ldots, n$ such that
\begin{align}
		&u^{-1} \left(
			C \left( \pi_{k + 1} \middle|
			T \left( \vec{\theta}_1, B \right) \left( \vec{X}_i \right),
			T \left( \vec{\theta}_1, \xi_k \right) \left( \vec{X}_i \right) \right)
		\right) \nonumber\\
&=
		u^{-1} \left( T \left( \vec{\theta}_1, \xi_k \right) \left( \vec{X}_i \right) \right)
	+
		\left( U^\ast \right)''
		\biggl( b_k \left( \vec{X}_i, \vec{\theta}_1, B \right)
		\biggr) \nonumber\\
	&\hspace{0.5cm}\times
		\left\{
			C \left( \pi_{k + 1} \middle|
			T \left( \vec{\theta}_1, B \right) \left( \vec{X}_i \right),
			T \left( \vec{\theta}_1, \xi_k \right) \left( \vec{X}_i \right) \right)
		-
			T \left( \vec{\theta}_1, \xi_k \right) \left( \vec{X}_i \right)
		\right\} \nonumber\\
&=
		u^{-1} \left( T \left( \vec{\theta}_1, \xi_k \right) \left( \vec{X}_i \right) \right)
	+
		\pi_{k + 1}
		\left( U^\ast \right)''
		\left( b_k \left( \vec{X}_i, \vec{\theta}_1, B \right) \right)
		\left\{
			T \left( \vec{\theta}_1, B \right) \left( \vec{X}_i \right)
		-
			T \left( \vec{\theta}_1, \xi_k \right) \left( \vec{X}_i \right)
		\right\}, \label{eq:u_mean}
\end{align}
where we have used a simplified notation
\begin{align*}
		b_k \left( \vec{X}_i, \vec{\theta}_1, B \right)
&=
		T \left( \vec{\theta}_1, \xi_k \right) \left( \vec{X}_i \right)
	+
		a_k \left( \vec{X}_i, \vec{\theta}_1, B \right)
		\pi_{k + 1}
		\left\{
			T \left( \vec{\theta}_1, B \right) \left( \vec{X}_i \right)
		-
			T \left( \vec{\theta}_1, \xi_k \right) \left( \vec{X}_i \right)
		\right\} \nonumber\\
&=
		T \left( \vec{\theta}_1,
			C \left( \pi_{k + 1} a_k \left( \vec{X}_i, \vec{\theta}_1, B \right) \middle| B, \xi_k \right)
		\right).
\end{align*}
Moreover, the common term in \eqref{eq:est_L^coco} and \eqref{eq:est_L_U*-coco} can be expressed as
\begin{align}
		&T \left( \vec{\theta}_1, B \right) \left( \vec{X}_i \right)
	-
		u \left( \mu \left( \vec{X}_i, \widetilde{\Lambda}, \vec{\theta}_2 \right) \right) \nonumber \\
&=
		T \left( \vec{\theta}_1, B \right) \left( \vec{X}_i \right)
	-
		T \left( \vec{\theta}_2,
			\sum_{\widetilde{B} \in \mathcal{D}} \widetilde{\lambda}_{\widetilde{B}} T \left( \vec{\theta}_2, \widetilde{B} \right)
		\right) \left( \vec{X}_i \right) \nonumber \\
&=
		\left\{
			m \left( \vec{X}_i, \vec{\theta}_1 \right)
		+
			m \left( \vec{X}_i, \vec{\theta}_1 \right)^\gamma
			B \left( \vec{X}_i \right)
		\right\}
	-
		\left\{
			m \left( \vec{X}_i, \vec{\theta}_2 \right)
		+
			m \left( \vec{X}_i, \vec{\theta}_2 \right)^\gamma
				\sum_{\widetilde{B} \in \mathcal{D}} \widetilde{\lambda}_{\widetilde{B}} T \left( \vec{\theta}_2, \widetilde{B} \right) \left( \vec{X}_i \right)
		\right\} \nonumber \\
	&\hspace*{0.5cm}+
		m \left( \vec{X}_i, \vec{\theta}_2 \right)^\gamma
		B \left( \vec{X}_i \right)
	-
		m \left( \vec{X}_i, \vec{\theta}_2 \right)^\gamma
		B \left( \vec{X}_i \right) \nonumber \\
&=
		m \left( \vec{X}_i, \vec{\theta}_1 \right)
	-
		m \left( \vec{X}_i, \vec{\theta}_2 \right)
	+
		\left\{
			m \left( \vec{X}_i, \vec{\theta}_1 \right)^\gamma
		-
			m \left( \vec{X}_i, \vec{\theta}_2 \right)^\gamma
		\right\}
	\cdot
		B \left( \vec{X}_i \right) \nonumber \\
	&\hspace{0.5cm}+
		m \left( \vec{X}_i, \vec{\theta}_2 \right)^\gamma
		\left\{
			B \left( \vec{X}_i \right)
		-
			\sum_{\widetilde{B} \in \mathcal{D}} \widetilde{\lambda}_{\widetilde{B}}
			\widetilde{B} \left( \vec{X}_i \right)
		\right\}. \label{eq:T_diff}
\end{align}
Those \eqref{eq:est_L_1}, \eqref{eq:est_L^coco}, \eqref{eq:est_L_U*-coco}, \eqref{eq:u_mean}, \eqref{eq:T_diff} and the equality
\[
		\sum_{B \in \mathcal{D}}
		\widetilde{\lambda}_{B}
		\left\{
			B \left( \vec{X}_i \right)
		-
			\sum_{\widetilde{B} \in \mathcal{D}}
			\widetilde{\lambda}_{\widetilde{B}}
			\widetilde{B} \left( \vec{X}_i \right)
		\right\}
=
		0
\]
imply that
\begin{align*}
		&\widehat{L} \left( \mu_{k + 1} \left( \vec{\theta}_1 \right) \right)
	-
		\widehat{L} \left( \mu \left( \cdot, \widetilde{\Lambda}, \vec{\theta}_2 \right) \right) \\
&\leq
		\left( 1 - \pi_{k + 1} \right)
		\left\{
			\widehat{L} \left( \mu_k \left( \vec{\theta}_1 \right) \right)
		-
			\widehat{L} \left( \mu \left( \cdot, \widetilde{\Lambda}, \vec{\theta}_2 \right) \right)
		\right\}
	+
		\pi_{k + 1} \varepsilon\\
	&\hspace*{0.5cm}+
		\sum_{B \in \mathcal{D}}
		\widetilde{\lambda}_{B}
		\frac{1}{n}
		\sum_{i = 1}^{n}
		\biggl(
			\pi_{k + 1}
			\biggl[
				u^{-1} \left( T \left( \vec{\theta}_1, \xi_k \right) \left( \vec{X}_i \right) \right) \\
			&\hspace*{3.8cm}+
				\pi_{k + 1}
				\left( U^\ast \right)'' \left( b_k \left( \vec{X}_i, \vec{\theta}_1, B \right) \right)
			\cdot
				\left\{
					T \left( \vec{\theta}_1, B \right) \left( \vec{X}_i \right)
				-
					T \left( \vec{\theta}_1, \xi_k \right) \left( \vec{X}_i \right)
				\right\}
			\biggr] \nonumber \\
		&\hspace{3.5cm}\times
			\left\{
				T \left( \vec{\theta}_1, B \right) \left( \vec{X}_i \right)
			-
				u \left( \mu \left( \vec{X}_i, \widetilde{\Lambda}, \vec{\theta}_2 \right) \right)
			\right\} \\
		&\hspace{3.5cm}-
			\pi_{k + 1}
			Y_i
			\left\{
				T \left( \vec{\theta}_1, B \right) \left( \vec{X}_i \right)
			-
				u \left( \mu \left( \vec{X}_i, \widetilde{\Lambda}, \vec{\theta}_2 \right) \right)
			\right\}
		\biggr) \\
&=
		\left( 1 - \pi_{k + 1} \right)
		\left\{
			\widehat{L} \left( \mu_k \left( \vec{\theta}_1 \right) \right)
		-
			\widehat{L} \left( \mu \left( \cdot, \widetilde{\Lambda}, \vec{\theta}_2 \right) \right)
		\right\}
	+
		\pi_{k + 1} \varepsilon \\
	&\hspace*{0.5cm}+
		\sum_{B \in \mathcal{D}}
		\widetilde{\lambda}_{B}
		\frac{1}{n}
		\sum_{i = 1}^{n}
		\pi_{k + 1}
		\biggl[
			u^{-1} \left( T \left( \vec{\theta}_1, \xi_k \right) \left( \vec{X}_i \right) \right)
		-
			Y_i \\
		&\hspace*{3.8cm}+
			\pi_{k + 1}
			\left( U^\ast \right)'' \left( b_k \left( \vec{X}_i, \vec{\theta}_1, B \right) \right)
		\cdot
			\left\{
				T \left( \vec{\theta}_1, B \right) \left( \vec{X}_i \right)
			-
				T \left( \vec{\theta}_1, \xi_k \right) \left( \vec{X}_i \right)
			\right\}
		\biggr] \nonumber \\
	&\hspace{3.5cm}\times
		\left\{
			T \left( \vec{\theta}_1, B \right) \left( \vec{X}_i \right)
		-
			u \left( \mu \left( \vec{X}_i, \widetilde{\Lambda}, \vec{\theta}_2 \right) \right)
		\right\} \\
&=
		\left( 1 - \pi_{k + 1} \right)
		\left\{
			\widehat{L} \left( \mu_k \left( \vec{\theta}_1 \right) \right)
		-
			\widehat{L} \left( \mu \left( \cdot, \widetilde{\Lambda}, \vec{\theta}_2 \right) \right)
		\right\}
	+
		\pi_{k + 1} \varepsilon \\
	&\hspace*{0.5cm}+
		\pi_{k + 1}
		\sum_{B \in \mathcal{D}}
		\widetilde{\lambda}_{B}
		\frac{1}{n}
		\sum_{i = 1}^{n}
		\left\{
			u^{-1} \left( T \left( \vec{\theta}_1, \xi_k \right) \left( \vec{X}_i \right) \right)
		-
			Y_i
		\right\}
		\left\{
			T \left( \vec{\theta}_1, B \right) \left( \vec{X}_i \right)
		-
			u \left( \mu \left( \vec{X}_i, \widetilde{\Lambda}, \vec{\theta}_2 \right) \right)
		\right\} \\
	&\hspace*{0.5cm}+
		\pi_{k + 1}^2
		\sum_{B \in \mathcal{D}}
		\widetilde{\lambda}_{B}
		\frac{1}{n}
		\sum_{i = 1}^{n}
			\left( U^\ast \right)'' \left( b_k \left( \vec{X}_i, \vec{\theta}_1, B \right) \right)
		\cdot
			\left\{
				T \left( \vec{\theta}_1, B \right) \left( \vec{X}_i \right)
			-
				T \left( \vec{\theta}_1, \xi_k \right) \left( \vec{X}_i \right)
			\right\} \nonumber \\
		&\hspace{3.5cm}\times
			\left\{
				T \left( \vec{\theta}_1, B \right) \left( \vec{X}_i \right)
			-
				u \left( \mu \left( \vec{X}_i, \widetilde{\Lambda}, \vec{\theta}_2 \right) \right)
			\right\} \\
&=
		\left( 1 - \pi_{k + 1} \right)
		\left\{
			\widehat{L} \left( \mu_k \left( \vec{\theta}_1 \right) \right)
		-
			\widehat{L} \left( \mu \left( \cdot, \widetilde{\Lambda}, \vec{\theta}_2 \right) \right)
		\right\}
	+
		\pi_{k + 1} \varepsilon \\
	&\hspace{0.4cm}+
		\pi_{k + 1}
		\frac{1}{n}
		\sum_{i = 1}^{n}
		\sum_{B \in \mathcal{D}}
			\widetilde{\lambda}_{B}
			\left\{
				u^{-1} \left( T \left( \vec{\theta}_1, \xi_k \right) \left( \vec{X}_i \right) \right)
			-
				Y_i
			\right\} \\
		&\hspace*{3cm}\times
			\left[
				m \left( \vec{X}_i, \vec{\theta}_1 \right)
			-
				m \left( \vec{X}_i, \vec{\theta}_2 \right)
			+
				\left\{
					m \left( \vec{X}_i, \vec{\theta}_1 \right)^\gamma
				-
					m \left( \vec{X}_i, \vec{\theta}_2 \right)^\gamma
				\right\}
			\cdot
				B \left( \vec{X}_i \right)
			\right] \\
	&\hspace{0.4cm}+
		\pi_{k + 1}^2
		\sum_{B \in \mathcal{D}}
		\widetilde{\lambda}_{B}
		\frac{1}{n}
		\sum_{i = 1}^{n}
			\left( U^\ast \right)'' \left( b_k \left( \vec{X}_i, \vec{\theta}_1, B \right) \right)
			m \left( \vec{X}_i, \vec{\theta}_1 \right)^\gamma
			\left\{
				B \left( \vec{X}_i \right)
			-
				\sum_{j = 1}^k
				\lambda_{j, k}
				B_j \left( \vec{X}_i \right)
			\right\} \\
	&\hspace{1cm}\times
		\Biggl[
			m \left( \vec{X}_i, \vec{\theta}_1 \right)
		-
			m \left( \vec{X}_i, \vec{\theta}_2 \right)
		+
			\left\{
				m \left( \vec{X}_i, \vec{\theta}_1 \right)^\gamma
			-
				m \left( \vec{X}_i, \vec{\theta}_2 \right)^\gamma
			\right\}
		\cdot
			B \left( \vec{X}_i \right) \\
		&\hspace{2cm}+
			m \left( \vec{X}_i, \vec{\theta}_2 \right)^\gamma
			\left\{
				B \left( \vec{X}_i \right)
			-
				\sum_{\widetilde{B} \in \mathcal{D}}
				\widetilde{\lambda}_{\widetilde{B}}
				\widetilde{B} \left( \vec{X}_i \right)
			\right\}
		\Biggr] \\
&=
		\left( 1 - \pi_{k + 1} \right)
		\left\{
			\widehat{L} \left( \mu_k \left( \vec{\theta}_1 \right) \right)
		-
			\widehat{L} \left( \mu \left( \cdot, \widetilde{\Lambda}, \vec{\theta}_2 \right) \right)
		\right\}
	+
		\pi_{k + 1} \varepsilon \\
	&\hspace{0.4cm}+
		\pi_{k + 1}
		\frac{1}{n}
		\sum_{i = 1}^{n}
			\left\{
				u^{-1} \left( T \left( \vec{\theta}_1, \xi_k \right) \left( \vec{X}_i \right) \right)
			-
				Y_i
			\right\} \\
		&\hspace*{2.4cm}\times
			\left[
				m \left( \vec{X}_i, \vec{\theta}_1 \right)
			-
				m \left( \vec{X}_i, \vec{\theta}_2 \right)
			+
				\left\{
					m \left( \vec{X}_i, \vec{\theta}_1 \right)^\gamma
				-
					m \left( \vec{X}_i, \vec{\theta}_2 \right)^\gamma
				\right\}
			\cdot
				\sum_{B \in \mathcal{D}}
				\widetilde{\lambda}_{B}
				B \left( \vec{X}_i \right)
			\right] \\
	&\hspace{0.4cm}+
		\pi_{k + 1}^2
		\sum_{B \in \mathcal{D}}
		\widetilde{\lambda}_{B}
		\frac{1}{n}
		\sum_{i = 1}^{n}
			\left( U^\ast \right)'' \left( b_k \left( \vec{X}_i, \vec{\theta}_1, B \right) \right)
			m \left( \vec{X}_i, \vec{\theta}_1 \right)^\gamma
			\left\{
				B \left( \vec{X}_i \right)
			-
				\sum_{j = 1}^k
				\lambda_{j, k}
				B_j \left( \vec{X}_i \right)
			\right\} \\
		&\hspace{1cm}\times
			\Biggl[
				m \left( \vec{X}_i, \vec{\theta}_1 \right)
			-
				m \left( \vec{X}_i, \vec{\theta}_2 \right)
			+
				\left\{
					m \left( \vec{X}_i, \vec{\theta}_1 \right)^\gamma
				-
					m \left( \vec{X}_i, \vec{\theta}_2 \right)^\gamma
				\right\}
			\cdot
				B \left( \vec{X}_i \right) \\
			&\hspace{2cm}+
				m \left( \vec{X}_i, \vec{\theta}_2 \right)^\gamma
				\left\{
					B \left( \vec{X}_i \right)
				-
					\sum_{\widetilde{B} \in \mathcal{D}}
					\widetilde{\lambda}_{\widetilde{B}}
					\widetilde{B} \left( \vec{X}_i \right)
				\right\}
			\Biggr] \\
&\leq
		\left( 1 - \pi_{k + 1} \right)
		\left\{
			\widehat{L} \left( \mu_k \left( \vec{\theta}_1 \right) \right)
		-
			\widehat{L} \left( \mu \left( \cdot, \widetilde{\Lambda}, \vec{\theta}_2 \right) \right)
		\right\}
	+
		\pi_{k + 1} \varepsilon \\
	&\hspace{0.5cm}+
		\pi_{k + 1}
		\frac{1}{n}
		\sum_{i = 1}^{n}
		\sum_{B \in \mathcal{D}}
			\widetilde{\lambda}_{B}
			\left|
				Y_i
			-
				u^{-1} \left( T \left( \vec{\theta}_1, \xi_k \right) \left( \vec{X}_i \right) \right)
			\right| \\
		&\hspace{3.5cm}\times
			\left|
				m \left( \vec{X}_i, \vec{\theta}_1 \right)
			-
				m \left( \vec{X}_i, \vec{\theta}_2 \right)
			+
				\left\{
					m \left( \vec{X}_i, \vec{\theta}_1 \right)^\gamma
				-
					m \left( \vec{X}_i, \vec{\theta}_2 \right)^\gamma
				\right\}
			\cdot
				B \left( \vec{X}_i \right)
			\right| \\
	&\hspace{0.4cm}+
		\pi_{k + 1}^2
		\sum_{B \in \mathcal{D}}
		\widetilde{\lambda}_{B}
		\sum_{\widetilde{B} \in \mathcal{D}}
		\widetilde{\lambda}_{\widetilde{B}}
		\sum_{j = 1}^k
		\lambda_{j, k}
		\frac{1}{n}
		\sum_{i = 1}^{n}
			\left( U^\ast \right)'' \left( b_k \left( \vec{X}_i, \vec{\theta}_1, B \right) \right) \\
		&\hspace{6.5cm}\times
			\left|
				m \left( \vec{X}_i, \vec{\theta}_1 \right)
			\right|^\gamma
			\left|
				B \left( \vec{X}_i \right)
			-
				B_j \left( \vec{X}_i \right)
			\right| \\
		&\hspace{1cm}\times
			\biggl\{
				\left|
					m \left( \vec{X}_i, \vec{\theta}_1 \right)
				-
					m \left( \vec{X}_i, \vec{\theta}_2 \right)
				+
					\left\{
						m \left( \vec{X}_i, \vec{\theta}_1 \right)^\gamma
					-
						m \left( \vec{X}_i, \vec{\theta}_2 \right)^\gamma
					\right\}
				\cdot
					B \left( \vec{X}_i \right)
				\right| \\
			&\hspace{2cm}+
				\left|
					m \left( \vec{X}_i, \vec{\theta}_2 \right)^\gamma
					\left\{
						B \left( \vec{X}_i \right)
					-
						\widetilde{B} \left( \vec{X}_i \right)
					\right\}
				\right|
			\biggr\} \\
&\leq
		\left( 1 - \pi_{k + 1} \right)
		\left\{
			\widehat{L} \left( \mu_k \left( \vec{\theta}_1 \right) \right)
		-
			\widehat{L} \left( \mu \left( \cdot, \widetilde{\Lambda}, \vec{\theta}_2 \right) \right)
		\right\}
	+
		\pi_{k + 1} \varepsilon \\
	&\hspace{0.5cm}+
		\pi_{k + 1}
		\left\| \vec{\theta}_1 - \vec{\theta}_2 \right\|_2
		\sup_{B \in \mathcal{D}}
		\sup_{\vec{x} \in \mathbb{D}}
		\sup_{\beta \in [0,1]}
			\left\|
				\left.
					\left\{
						1
					+
						\gamma
						m \left( \vec{x}, \vec{\theta} \right)^{\gamma - 1}
						B \left( \vec{x} \right)
					\right\}
					\frac{\partial}{\partial \vec{\theta}}
					m \left( \vec{x}, \vec{\theta} \right)
				\right|_{
						\vec{\theta}
				=
						\vec{\theta}_2 + \beta \left( \vec{\theta}_1 - \vec{\theta}_2 \right)
				}
			\right\|_2 \\
		&\hspace{1cm}\times
			\frac{1}{n}
			\sum_{i = 1}^{n}
			\left|
				Y_i
			-
				u^{-1} \left( T \left( \vec{\theta}_1, \xi_k \right) \left( \vec{X}_i \right) \right)
			\right| \\
	&\hspace{0.5cm}+
		\pi_{k + 1}^2
		\left\| \vec{\theta}_1 - \vec{\theta}_2 \right\|_2
		\sup_{B \in \mathcal{D}}
		\sup_{\vec{x} \in \mathbb{D}}
		\sup_{\beta \in [0,1]}
		\left\|
			\left.
				\left\{
					1
				+
					\gamma
					m \left( \vec{x}, \vec{\theta} \right)^{\gamma - 1}
					B \left( \vec{x} \right)
				\right\}
				\frac{\partial}{\partial \vec{\theta}}
				m \left( \vec{x}, \vec{\theta} \right)
			\right|_{
					\vec{\theta}
			=
					\vec{\theta}_2 + \beta \left( \vec{\theta}_1 - \vec{\theta}_2 \right)
			}
			\right\|_2 \\
	&\hspace{1cm}\times
		\sum_{B \in \mathcal{D}}
		\widetilde{\lambda}_{B}
		\sum_{j = 1}^k
		\lambda_{j, k}
		\frac{1}{n}
		\sum_{i = 1}^{n}
		\left|
			m \left( \vec{x}, \vec{\theta}_1 \right)
		\right|^\gamma
		\left( U^\ast \right)'' \left( b_k \left( \vec{X}_i, \vec{\theta}_1, B \right) \right)
		\left|
			B \left( \vec{X}_i \right)
		-
			B_j \left( \vec{X}_i \right)
		\right| \\
	&\hspace{0.5cm}+
		\pi_{k + 1}^2
		\sum_{B \in \mathcal{D}}
		\widetilde{\lambda}_{B}
		\sum_{\widetilde{B} \in \mathcal{D}}
		\widetilde{\lambda}_{\widetilde{B}}
		\sum_{j = 1}^k
		\lambda_{j, k}
		\frac{1}{n}
		\sum_{i = 1}^{n}
			\left|
				m \left( \vec{X}_i, \vec{\theta}_1 \right)
				m \left( \vec{X}_i, \vec{\theta}_2 \right)
			\right|^\gamma \\
		&\hspace{2cm}\times
			\left( U^\ast \right)'' \left( b_k \left( \vec{X}_i, \vec{\theta}_1, B \right) \right)
			\left|
				B \left( \vec{X}_i \right)
			-
				B_j \left( \vec{X}_i \right)
			\right|
			\left|
					B \left( \vec{X}_i \right)
				-
					\widetilde{B} \left( \vec{X}_i \right)
			\right| \\
&\leq
		\left( 1 - \pi_{k + 1} \right)
		\left\{
			\widehat{L} \left( \mu_k \left( \vec{\theta}_1 \right) \right)
		-
			\widehat{L} \left( \mu \left( \cdot, \widetilde{\Lambda}, \vec{\theta}_2 \right) \right)
		\right\}
	+
		\pi_{k + 1} \varepsilon \\
	&\hspace{0.5cm}+
		\pi_{k + 1}
		\left\| \vec{\theta}_1 - \vec{\theta}_2 \right\|_2
		\sup_{B \in \mathcal{D}}
		\sup_{\vec{x} \in \mathbb{D}}
		\sup_{\beta \in [0,1]}
			\left\|
				\left.
				\left\{
					1
				+
					\gamma
					m \left( \vec{x}, \vec{\theta} \right)^{\gamma - 1}
					B \left( \vec{x} \right)
				\right\}
				\frac{\partial}{\partial \vec{\theta}}
				m \left( \vec{x}, \vec{\theta} \right)
				\right|_{
						\vec{\theta}
					=
						\vec{\theta}_2 + \beta \left( \vec{\theta}_1 - \vec{\theta}_2 \right)
				}
			\right\|_2 \\
		&\hspace{1cm}\times
			\frac{1}{n}
			\sum_{i = 1}^{n}
			\max
			\left\{
			\left|
				Y_i
			-
				u^{-1} \left( T \left( \vec{\theta}_1, \sup_{B \in \mathcal{D}} B \right) \left( \vec{X}_i \right) \right)
			\right|,
			\left|
				Y_i
			-
				u^{-1} \left( T \left( \vec{\theta}_1, \inf_{B \in \mathcal{D}} B \right) \left( \vec{X}_i \right) \right)
			\right|
			\right\} \\
	&\hspace{0.5cm}+
		2
		\pi_{k + 1}^2
		\sup_{B \in \mathcal{D}} \sup_{\vec{x} \in \mathbb{D}} \left| B \left( \vec{x} \right) \right|
		\left\| \vec{\theta}_1 - \vec{\theta}_2 \right\|_2 \\
	&\hspace{1cm}\times
		\sup_{\vec{x} \in \mathbb{D}}
		\left|
			m \left( \vec{x}, \vec{\theta}_1 \right)
		\right|^\gamma
		\sum_{B \in \mathcal{D}}
		\widetilde{\lambda}_{B}
		\sup_{\beta \in [0,1]}
		\left( U^\ast \right)''
		\left(
			C \left( \beta
			\middle|
				T \left( \vec{\theta}_1, B \right),
				T \left( \vec{\theta}_1, \xi_k \right)
			\right) \left( \vec{x} \right)
		\right)\\
	&\hspace{1cm}\times
		\sup_{B \in \mathcal{D}}
		\sup_{\vec{x} \in \mathbb{D}}
		\sup_{\beta \in [0,1]}
		\left\|
			\left.
			\left\{
				1
			+
				\gamma
				m \left( \vec{x}, \vec{\theta} \right)^{\gamma - 1}
				B \left( \vec{x} \right)
			\right\}
			\frac{\partial}{\partial \vec{\theta}}
			m \left( \vec{x}, \vec{\theta} \right)
			\right|_{
					\vec{\theta}
			=
					\vec{\theta}_2 + \beta \left( \vec{\theta}_1 - \vec{\theta}_2 \right)
			}
		\right\|_2 \\
	&\hspace{0.5cm}+
		4 \pi_{k + 1}^2
		\sup_{B \in \mathcal{D}}
		\sup_{\vec{x} \in \mathbb{D}}
		\left| B \left( \vec{x} \right) \right|^2
		\sup_{\vec{x} \in \mathbb{D}}
			\left|
				m \left( \vec{x}, \vec{\theta}_1 \right)
				m \left( \vec{x}, \vec{\theta}_2 \right)
			\right|^\gamma \\
		&\hspace{2cm}\times
			\sum_{B \in \mathcal{D}}
			\widetilde{\lambda}_{B}
			\sup_{\beta \in [0,1]}
			\left( U^\ast \right)''
			\left(
				C \left( \beta
				\middle|
					T \left( \vec{\theta}_1, B \right),
					T \left( \vec{\theta}_1, \xi_k \right)
				\right) \left( \vec{x} \right)
			\right) \\
&\leq
		\left( 1 - \pi_{k + 1} \right)
		\left\{
			\widehat{L} \left( \mu_k \left( \vec{\theta}_1 \right) \right)
		-
			\widehat{L} \left( \mu \left( \cdot, \widetilde{\Lambda}, \vec{\theta}_2 \right) \right)
		\right\}
	+
		\pi_{k + 1} \varepsilon \\
	&\hspace{0.5cm}+
		\pi_{k + 1}^2
		\Biggl[
			4 \sup_{B \in \mathcal{D}}
			\sup_{\vec{x} \in \mathbb{D}}
			\left| B \left( \vec{x} \right) \right|^2
			\sup_{\vec{x} \in \mathbb{D}}
			\left|
				m \left( \vec{x}, \vec{\theta}_1 \right)
				m \left( \vec{x}, \vec{\theta}_2 \right)
			\right|^\gamma
			\sup_{\Lambda \in \mathcal{W}}
			\sup_{\vec{x} \in \mathbb{D}}
			\left( U^\ast \right)''
			\left( u \left(
				\mu \left( \vec{x}, \Lambda, \vec{\theta}_1 \right)
			\right) \right)^2 \\
		&\hspace{1.5cm}+
			2
			\sup_{B \in \mathcal{D}}
			\sup_{\vec{x} \in \mathbb{D}}
			\left| B \left( \vec{x} \right) \right|
			\left\| \vec{\theta}_1 - \vec{\theta}_2 \right\|_2
			\sup_{\vec{x} \in \mathbb{D}}
			\left|
				m \left( \vec{x}, \vec{\theta}_1 \right)
			\right|^\gamma
			\sup_{\Lambda \in \mathcal{W}}
			\sup_{\vec{x} \in \mathbb{D}}
				\left( U^\ast \right)''
				\left( u \left(
					\mu \left( \vec{x}, \Lambda, \vec{\theta}_1 \right)
				\right) \right)^2\\
			&\hspace{2.5cm}\times
				\sup_{B \in \mathcal{D}}
				\sup_{\vec{x} \in \mathbb{D}}
				\sup_{\beta \in [0,1]}
				\left\|
					\left.
					\left\{
						1
					+
						\gamma
						m \left( \vec{x}, \vec{\theta} \right)^{\gamma - 1}
						B \left( \vec{x} \right)
					\right\}
					\frac{\partial}{\partial \vec{\theta}}
					m \left( \vec{x}, \vec{\theta} \right)
					\right|_{
							\vec{\theta}
					=
							\vec{\theta}_2 + \beta \left( \vec{\theta}_1 - \vec{\theta}_2 \right)}
				\right\|_2
		\Biggr]\\
	&\hspace{0.5cm}+
		\pi_{k + 1}
		\left\| \vec{\theta}_1 - \vec{\theta}_2 \right\|_2
		\sup_{B \in \mathcal{D}}
		\sup_{\vec{x} \in \mathbb{D}}
		\sup_{\beta \in [0,1]}
			\left\|
				\left.
				\left\{
					1
				+
					\gamma
					m \left( \vec{x}, \vec{\theta} \right)^{\gamma - 1}
					B \left( \vec{x} \right)
				\right\}
				\frac{\partial}{\partial \vec{\theta}}
				m \left( \vec{x}, \vec{\theta} \right)
				\right|_{
						\vec{\theta}
				=
						\vec{\theta}_2 + \beta \left( \vec{\theta}_1 - \vec{\theta}_2 \right)}
			\right\|_2 \\
		&\hspace{1cm}\times
			\frac{1}{n}
			\sum_{i = 1}^{n}
			\sup_{B \in \mathcal{D}}
			\left|
				Y_i
			-
				u^{-1} \left( T \left( \vec{\theta}_1, B \right) \left( \vec{X}_i \right) \right)
			\right| \\
&=
		\left( 1 - \pi_{k + 1} \right)
		\left\{
			\widehat{L} \left( \mu_k \left( \vec{\theta}_1 \right) \right)
		-
			\widehat{L} \left( \mu \left( \cdot, \widetilde{\Lambda}, \vec{\theta}_2 \right) \right)
		\right\} \\
	&\hspace{0.5cm}+
		\pi_{k + 1}^2
		\left\{
			A_1 \left( \vec{\theta}_1, \vec{\theta}_2 \right)
		+
			\left\| \vec{\theta}_1 - \vec{\theta}_2 \right\|_2
			A_3 \left( \vec{\theta}_1, \vec{\theta}_2 \right)
		\right\}
	+
		\pi_{k + 1}
		\left\| \vec{\theta}_1 - \vec{\theta}_2 \right\|_2
		A_2 \left( \vec{\theta}_1, \vec{\theta}_2 \right)
	+
		\pi_{k + 1} \varepsilon. \tag*{$\Box$}
\end{align*}

\subsubsection*{Proof of Lemma \ref{lem:ineq1-2}}
When $k = 1$,
the inequalities
$\tau^2 / \left( 1 + \tau \right) = \tau - 1 + 1 / \left( 1 + \tau \right) \geq 1$
and
\eqref{eq:L_recur} with $\ell = 0$ imply that
\begin{align}
		\widehat{L} \left( \mu_{1} \left( \vec{\theta}_1 \right) \right)
	-
		\widehat{L} \left( \mu \left( \cdot, \widetilde{\Lambda}, \vec{\theta}_2 \right) \right)
\leq
		a + b
\leq
		\frac{\tau^2}{1 + \tau} a + b.
\end{align}
If \eqref{eq:ind} holds with arbitrary $k \in \mathbb{N}$, then
\begin{align*}
		&\widehat{L} \left( \mu_{k + 1} \left( \vec{\theta}_1 \right) \right)
	-
		\widehat{L} \left( \mu \left( \cdot, \widetilde{\Lambda}, \vec{\theta}_2 \right) \right) \\
&\leq
		\left( 1 - \pi_{k + 1} \right)
		\left\{
			\widehat{L} \left( \mu_k \left( \vec{\theta}_1 \right) \right)
		-
			\widehat{L} \left( \mu \left( \cdot, \widetilde{\Lambda}, \vec{\theta}_2 \right) \right)
		\right\}
	+
		\pi_{k + 1}^2 a
	+
		\pi_{k + 1} b \\
&\leq
		\left( 1 - \pi_{k + 1} \right)
		\left\{
			\frac{\tau^2}{k + \tau} a
		+
			b
		\right\}
	+
		\pi_{k + 1}^2 a
	+
		\pi_{k + 1} b \\
&=
		\pi_{k + 1}^2 a
	+
		\left( 1 - \pi_{k + 1} \right)
		\frac{\tau^2}{k + \tau} a
	+
		b \\
&=
		\left\{
			\frac{\tau^2}{\left( k + \tau \right)^2}
		+
			\frac{k}{k + \tau}
			\frac{\tau^2}{k + \tau}
		\right\}
		a
	+
		b \\
&=
		\frac{k + 1}{\left( k + \tau \right)^2}
		\tau^2
		a
	+
		b \\
&=
		\frac{1}{k + 1 + \tau}
		\cdot
		\frac{\left( k + \tau \right)^2 - \left( k + \tau \right)^2 + \left( k + 1 \right) \left( k + 1 + \tau \right)}{\left( k + \tau \right)^2}
		\tau^2
		a
	+
		b \\
&=
		\frac{1}{k + 1 + \tau}
		\cdot
		\left\{
			1
		-
			\frac{\tau^2 - 3 + \left( k - 1 \right) \left( \tau - 2 \right) }{\left( k + \tau \right)^2}
		\right\}
		\tau^2
		a
	+
		b \\
&\leq
		\frac{\tau^2}{k + 1 + \tau} a
	+
		b,
\end{align*}
so that induction gives \eqref{eq:ind}. \qed

\subsubsection*{Proof of Lemma \ref{lem:ineq1}}
In the inequality \eqref{eq:L_k_k+1} in Lemma \ref{lem:ineq1-2}, we put
		$a
=
		A_1 \left( \vec{\theta}_1, \vec{\theta}_2 \right)
	+
		\left\| \vec{\theta}_1 - \vec{\theta}_2 \right\|_2
		A_3 \left( \vec{\theta}_1, \vec{\theta}_2 \right)$
and
		$b
=
		\left\| \vec{\theta}_1 - \vec{\theta}_2 \right\|_2
		A_2 \left( \vec{\theta}_1, \vec{\theta}_2 \right)
	+
	\varepsilon$
based on \eqref{eq:L_k_k+1}.
Then, for any $\widetilde{\Lambda} \in \mathcal{W}$,
\begin{align}
		\widehat{L} \left( \mu \left( \cdot, \Lambda, \vec{\theta}_1 \right) \right)
&\leq
		\widehat{L} \left( \mu \left( \cdot, \widetilde{\Lambda}, \vec{\theta}_2 \right) \right)
	+
		\frac{\tau^2}{I + \tau}
		\Biggl\{
			A_1 \left( \vec{\theta}_1, \vec{\theta}_2 \right)
		+
			\left\| \vec{\theta}_1 - \vec{\theta}_2 \right\|_2
			A_3 \left( \vec{\theta}_1, \vec{\theta}_2 \right)
		\Biggr\} \nonumber \\
	&\hspace{0.5cm}+
		\left\| \vec{\theta}_1 - \vec{\theta}_2 \right\|_2
		A_2 \left( \vec{\theta}_1, \vec{\theta}_2 \right)
	+
		\varepsilon \nonumber \\
&=
		\mathsf{L} \left( \widetilde{\Lambda} \right) \label{eq:any_lam}
\end{align}
This inequality \eqref{eq:any_lam} menas that $\widehat{L} \left( \mu \left( \cdot, \Lambda, \vec{\theta}_1 \right) \right)$ is a lower bound of the set
		$\left\{
				\mathsf{L} \left( \widetilde{\Lambda} \right)
		\middle|
				\widetilde{\Lambda} \in \mathcal{W}
		\right\}
\subset \mathbb{R}$.
Therefore, Lemma \ref{lem:ineq1} holds. \qed

\subsection{Proof of Lemma \ref{lem:ineq2}}
Pick $\widetilde{\Lambda} \in \mathcal{W}$.
The assumption \eqref{eq:assu_lem2} implies that
\begin{align}
		\widehat{L} \left( \mu \left( \cdot, \Bar{\Lambda} \left( \mathbb{X} \right), \vec{\theta}_1 \left( \mathbb{X} \right) \right) \right)
\leq
		\widehat{L} \left( \mu \left( \cdot, \widetilde{\Lambda},\vec{\theta}_2 \left( \mathbb{X} \right) \right) \right)
	+
		\delta \left( \mathbb{X} \right). \label{eq:ineqq1}
\end{align}
From this inequality \eqref{eq:ineqq1} and the fact
\begin{align*}
		&D_U \left( \mu, \mu \left( \cdot, \Bar{\Lambda} \left( \mathbb{X} \right), \vec{\theta}_1 \left( \mathbb{X} \right) \right) \right)
	-
		\widehat{L} \left( \mu \left( \cdot, \Bar{\Lambda} \left( \mathbb{X} \right), \vec{\theta}_1 \left( \mathbb{X} \right) \right) \right) \nonumber \\
&=
		\int_{\mathbb{R} \times \mathbb{R}^{d}}
			\left\{
				\ell \left( y, \vec{x} \middle| \mu \left( \cdot, \Bar{\Lambda} \left( \mathbb{X} \right), \vec{\theta}_1 \left( \mathbb{X} \right) \right) \right)
			-
				\ell \left( y, \vec{x} \middle| \mu \right)
			\right\}
		f \left( y, \vec{x} \right) d y d \vec{x} \nonumber \\
	&\hspace{0.5cm}-
		\frac{1}{n}
		\sum_{i = 1}^{n}
		\ell \left( Y_i, \vec{X}_i \middle| \mu \left( \cdot, \Bar{\Lambda} \left( \mathbb{X} \right), \vec{\theta}_1 \left( \mathbb{X} \right) \right) \right) \nonumber \\
&=
	-
		\mathcal{E}_n \left( \mu \left( \cdot, \Bar{\Lambda} \left( \mathbb{X} \right), \vec{\theta}_1 \left( \mathbb{X} \right) \right) \right)
	-
		\int_{\mathbb{R} \times \mathbb{R}^{d}}
			\ell \left( y, \vec{x} \middle| \mu \right)
		f \left( y, \vec{x} \right) d y d \vec{x},
\end{align*}
it follows that
\begin{align}
		&D_U \left( \mu, \mu \left( \cdot, \Bar{\Lambda} \left( \mathbb{X} \right), \vec{\theta}_1 \left( \mathbb{X} \right) \right) \right) \nonumber \\
&=
		D_U \left( \mu, \mu \left( \cdot, \widetilde{\Lambda}, \vec{\theta}_2 \left( \mathbb{X} \right) \right) \right)
	-
		\left\{
			D_U \left( \mu, \mu \left( \cdot, \widetilde{\Lambda}, \vec{\theta}_2 \left( \mathbb{X} \right) \right) \right)
		-
			\widehat{L} \left( \mu \left( \cdot, \Bar{\Lambda} \left( \mathbb{X} \right), \vec{\theta}_1 \left( \mathbb{X} \right) \right) \right)
		\right\} \nonumber \\
	&\hspace{0.5cm}+
		D_U \left( \mu, \mu \left( \cdot, \Bar{\Lambda} \left( \mathbb{X} \right), \vec{\theta}_1 \left( \mathbb{X} \right) \right) \right)
	-
		\widehat{L} \left( \mu \left( \cdot, \Bar{\Lambda} \left( \mathbb{X} \right), \vec{\theta}_1 \left( \mathbb{X} \right) \right) \right) \nonumber \\
&\leq
		D_U \left( \mu, \mu \left( \cdot, \widetilde{\Lambda}, \vec{\theta}_2 \left( \mathbb{X} \right) \right) \right)
	-
		\left\{
			D_U \left( \mu, \mu \left( \cdot, \widetilde{\Lambda}, \vec{\theta}_2 \left( \mathbb{X} \right) \right) \right)
		-
			\widehat{L} \left( \mu \left( \cdot, \widetilde{\Lambda} \left( \mathbb{X} \right), \vec{\theta}_2 \left( \mathbb{X} \right) \right) \right)
		\right\} \nonumber \\
	&\hspace{0.5cm}+
		D_U \left( \mu, \mu \left( \cdot, \Bar{\Lambda} \left( \mathbb{X} \right), \vec{\theta}_1 \left( \mathbb{X} \right) \right) \right)
	-
		\widehat{L} \left( \mu \left( \cdot, \Bar{\Lambda} \left( \mathbb{X} \right), \vec{\theta}_1 \left( \mathbb{X} \right) \right) \right)
	+ \delta \left( \mathbb{X} \right)\nonumber \\
&=
		D_U \left( \mu, \mu \left( \cdot, \widetilde{\Lambda}, \vec{\theta}_2 \left( \mathbb{X} \right) \right) \right)
	+
		\mathcal{E}_n \left( \mu \left( \cdot, \widetilde{\Lambda} \left( \mathbb{X} \right), \vec{\theta}_2 \left( \mathbb{X} \right) \right) \right)
	-
		\mathcal{E}_n \left( \mu \left( \cdot, \Bar{\Lambda} \left( \mathbb{X} \right), \vec{\theta}_1 \left( \mathbb{X} \right) \right) \right)
	+
		\delta \left( \mathbb{X} \right) \nonumber \\
&\leq
		D_U \left( \mu, \mu \left( \cdot, \widetilde{\Lambda}, \vec{\theta}_2 \left( \mathbb{X} \right) \right) \right)
	+
		\sup_{\Lambda \in \mathcal{W}}
		\mathcal{E}_n \left( \mu \left( \cdot, \Lambda, \vec{\theta}_2 \left( \mathbb{X} \right) \right) \right)
	+
		\sup_{\Lambda \in \mathcal{W}}
		\left[
		- \mathcal{E}_n \left( \mu \left( \cdot, \Lambda, \vec{\theta}_1 \left( \mathbb{X} \right) \right) \right)
		\right] \nonumber \\
	&\hspace{0.5cm}+
		\delta \left( \mathbb{X} \right) \nonumber \\
&=
		\mathsf{D} \left( \widetilde{\Lambda} \right). \label{eq:ineqq3}
\end{align}
This inequality \eqref{eq:ineqq3} shows that $D_U \left( \mu, \mu \left( \cdot, \Bar{\Lambda} \left( \mathbb{X} \right), \vec{\theta}_1 \left( \mathbb{X} \right) \right) \right)$ is a lower bound of the following set
$\left\{
		\mathsf{D} \left( \widetilde{\Lambda} \right)
\middle|
		\widetilde{\Lambda} \in \mathcal{W}
\right\}$.
Therefore,
\begin{align*}
		D_U \left( \mu, \mu \left( \cdot, \Bar{\Lambda} \left( \mathbb{X} \right), \vec{\theta}_1 \left( \mathbb{X} \right) \right) \right)
&\leq
		\inf_{\Lambda \in \mathcal{W}}
		D_U \left( \mu, \mu \left( \cdot, \Lambda, \vec{\theta}_2 \left( \mathbb{X} \right) \right) \right) +
		\sup_{\Lambda \in \mathcal{W}}
		\mathcal{E}_n \left( \mu \left( \cdot, \Lambda, \vec{\theta}_2 \left( \mathbb{X} \right) \right) \right) \nonumber \\
	&\hspace{0.5cm}+
		\sup_{\Lambda \in \mathcal{W}}
		\left[ - \mathcal{E}_n \left( \mu \left( \cdot, \Lambda, \vec{\theta}_1 \left( \mathbb{X} \right) \right) \right) \right]
	+
		\delta \left( \mathbb{X} \right).
\end{align*}
Lemma \ref{lem:ineq2} is obtained by taking expectation. \qed

\subsection{Lemmas for Theorem \ref{thm:mu} and their proofs} \label{sec:lemmas}
In this subsection, we arrange and prove lemmas used in the next subsection, when investigating the second term of the bound in \eqref{eq:mu_hat}.
Let $\sigma_1, \ldots, \sigma_n$ be Rademacher random variables, satisfying $\mathbb{P} \left( \sigma_i = 1 \right) = \mathbb{P} \left( \sigma_i = -1 \right) = 1/2$, independent of the observations $\mathbb{X}$.
We also denote the expectations taken with respect to $\vec{\sigma} = \left[ \sigma_1 \cdots \sigma_n \right]$ and the observations $\mathbb{X}$ by $\mathbb{E}_{\vec{\sigma}}$ and $\mathbb{E}_{\mathbb{X}}$ respectively.
The following lemma will be used to bound the term in \eqref{eq:mu_hat} with what is called a Rademacher complexity.
\begin{Lem}{(symmetrization)} \label{lem:sym}
For a function class $\mathcal{F}$ and both $\alpha = 0, 1$,
\begin{align*}
		&\mathbb{E}_{\mathbb{X}}
		\left[
		\sup_{\varphi \in \mathcal{F}}
		\left[
			\left( -1 \right)^\alpha
			\left\{
				\frac{1}{n}
				\sum_{i = 1}^{n}
				\varphi \left( Y_i, \vec{X}_i \right)
			-
				\int_{\mathbb{R} \times \mathbb{R}^{d}}
				\varphi \left( y, \vec{x} \right)
				f \left( y, \vec{x} \right) d y d \vec{x}
			\right\}
		\right]
		\right] \\
&\leq
		2
		\mathbb{E}_{\mathbb{X}}
		\left[
			\mathbb{E}_{\vec{\sigma}}
			\left[
				\sup_{\varphi \in \mathcal{F}}
				\frac{1}{n}
				\sum_{i = 1}^{n}
				\sigma_i
				\cdot
				\varphi \left( Y_i, \vec{X}_i \right)
			\right]
		\right].
\end{align*}
\end{Lem}
\begin{proof}
Let $\mathbb{X}' = \left\{ \left( Y'_1, \vec{X}'_1 \right), \ldots, \left( Y'_n, \vec{X}'_n \right) \right\}$ be independent copies of $\mathbb{X}$.
Then
\begin{align}
		&\mathbb{E}_{\mathbb{X}}
		\left[
		\sup_{\varphi \in \mathcal{F}}
		\left[
			\left( -1 \right)^\alpha
			\left\{
				\frac{1}{n}
				\sum_{i = 1}^{n}
				\varphi \left( Y_i, \vec{X}_i \right)
			-
				\int_{\mathbb{R} \times \mathbb{R}^{d}}
				\varphi \left( y, \vec{x} \right)
				f \left( y, \vec{x} \right) d y d \vec{x}
			\right\}
		\right]
		\right] \nonumber \\
&=
		\mathbb{E}_{\mathbb{X}}
		\left[
		\sup_{\varphi \in \mathcal{F}}
		\left[
			\left( -1 \right)^\alpha
			\left\{
				\frac{1}{n}
				\sum_{i = 1}^{n}
				\varphi \left( Y_i, \vec{X}_i \right)
			-
				\mathbb{E}_{\mathbb{X}'}
				\left[
						\frac{1}{n}
						\sum_{i = 1}^{n}
						\varphi \left( Y'_i, \vec{X}'_i \right)
				\right]
				\right\}
		\right]
		\right] \nonumber \\
&\leq
		\mathbb{E}_{\mathbb{X}, \mathbb{X}'}
		\left[
		\sup_{\varphi \in \mathcal{F}}
		\left[
			\left( -1 \right)^\alpha
				\frac{1}{n}
				\sum_{i = 1}^{n}
				\left\{
					\varphi \left( Y_i, \vec{X}_i \right)
				-
					\varphi \left( Y'_i, \vec{X}'_i \right)
				\right\}
		\right]
		\right]. \label{eq:symm1}
\end{align}
For the term \eqref{eq:symm1}, we shall show that the following equality
\begin{align}
		&\mathbb{E}_{\mathbb{X}, \mathbb{X}'}
		\left[
		\sup_{\varphi \in \mathcal{F}}
		\left[
			\left( -1 \right)^\alpha
				\frac{1}{n}
				\sum_{i = 1}^{n}
				\left\{
					\varphi \left( Y_i, \vec{X}_i \right)
				-
					\varphi \left( Y'_i, \vec{X}'_i \right)
				\right\}
		\right]
		\right] \nonumber \\
&=
		\mathbb{E}_{\mathbb{X}, \mathbb{X}'}
		\left[
			\mathbb{E}_{\vec{\sigma}}
		\left[
			\sup_{\varphi \in \mathcal{F}}
			\frac{1}{n}
			\sum_{i = 1}^n
				\sigma_i
				\cdot
				\left\{
					\varphi \left( Y_i, \vec{X}_i \right)
				-
					\varphi \left( Y'_i, \vec{X}'_i \right)
				\right\}
		\right]
		\right] \label{eq:symm2}
\end{align}
holds:
we define
\begin{align*}
		\mathsf{E}^{[1]}_j
&=
		\mathbb{E}_{\mathbb{X}, \mathbb{X}'}
		\Biggl[
		\mathbb{E}_{\left[ \sigma_1 ~ \cdots ~ \sigma_{j - 1} \right] }
		\Biggl[
			\sup_{\varphi \in \mathcal{F}}
			\frac{1}{n}
			\Biggl[
				\sum_{i = 1}^{j - 1}
				\sigma_i
				\left\{
					\varphi \left( Y_i, \vec{X}_i \right)
				-
					\varphi \left( Y'_i, \vec{X}'_i \right)
				\right\} \\
			&\hspace{5cm}+
				\left( -1 \right)^\alpha
				\sum_{i = j}^{n}
				\left\{
					\varphi \left( Y_i, \vec{X}_i \right)
				-
					\varphi \left( Y'_i, \vec{X}'_i \right)
				\right\}
			\Biggr]
		\Biggr]
		\Biggr] \\
&=
		\mathbb{E}_{\mathbb{X}, \mathbb{X}'}
		\left[
		\mathbb{E}_{\left[ \sigma_1 ~ \cdots ~ \sigma_{j - 1} \right] }
		\left[
			\sup_{\varphi \in \mathcal{F}}
			\frac{1}{n}
			\left[
				\Phi^{(j)}
			+
				\left( -1 \right)^\alpha
				\left\{
					\varphi \left( Y_j, \vec{X}_j \right)
				-
					\varphi \left( Y'_j, \vec{X}'_j \right)
				\right\}
			\right]
		\right]
		\right], \\
		\Phi^{(j)}
&=
		\sum_{i = 1}^{j - 1}
		\sigma_i
		\left\{
			\varphi \left( Y_i, \vec{X}_i \right)
		-
			\varphi \left( Y'_i, \vec{X}'_i \right)
		\right\}
	+
		\left( -1 \right)^\alpha
		\sum_{i = j + 1}^{n}
		\left\{
			\varphi \left( Y_i, \vec{X}_i \right)
		-
			\varphi \left( Y'_i, \vec{X}'_i \right)
		\right\}
\end{align*}
for $j = 1, 2, \ldots, n + 1$, where we regard $\mathsf{E}^{[1]}_1$ as the left-hand side of \eqref{eq:symm2}, and $\mathsf{E}^{[1]}_{n + 1}$ coincides with right-hand side of \eqref{eq:symm2}.
Then we can prove the sequence $\left( \mathsf{E}^{[1]}_j \right)_{j = 1}^{n + 1}$ to be constant for $j = 1, 2, \ldots, n + 1$, since the equality
\begin{align*}
		&\mathbb{E}_{\mathbb{X}, \mathbb{X}', \left[ \sigma_1 ~ \cdots ~ \sigma_{j - 1} \right]}
		\left[
			\sup_{\varphi \in \mathcal{F}}
			\left[
					\Phi^{(j)}
				+
					\left( -1 \right)^\alpha
					\left\{
						\varphi \left( Y_j, \vec{X}_j \right)
					-
						\varphi \left( Y'_j, \vec{X}'_j \right)
					\right\}
			\right]
		\right] \\
&=
		\mathbb{E}_{\mathbb{X}, \mathbb{X}', \left[ \sigma_1 ~ \cdots ~ \sigma_{j - 1} \right]}
		\left[
			\sup_{\varphi \in \mathcal{F}}
			\left[
				\Phi^{(j)}
			+
				\left( -1 \right)^\alpha
				\left\{
					\varphi \left( Y'_j, \vec{X}'_j \right)
				-
					\varphi \left( Y_j, \vec{X}_j \right)
				\right\}
			\right]
		\right]
\end{align*}
holds for $j = 1, 2, \ldots, n + 1$
by the transformation from $\left[ Y_j ~ \vec{X}_j ~ Y'_j ~ \vec{X}'_j \right]$ to $\left[ Y'_j ~ \vec{X}'_j ~ Y_j ~ \vec{X}_j \right]$, and hence
\begin{align*}
		\mathsf{E}^{[1]}_j
&=
		\mathbb{E}_{\mathbb{X}, \mathbb{X}', \left[ \sigma_1 ~ \cdots ~ \sigma_{j - 1} \right]}
		\left[
			\sup_{\varphi \in \mathcal{F}}
			\frac{1}{n}
			\left[
				\Phi^{(j)}
			+
				\left( -1 \right)^\alpha
				\left\{
					\varphi \left( Y_j, \vec{X}_j \right)
				-
					\varphi \left( Y'_j, \vec{X}'_j \right)
				\right\}
			\right]
		\right] \\
&=
	\mathbb{E}_{\mathbb{X}, \mathbb{X}', \left[ \sigma_1 ~ \cdots ~ \sigma_{j - 1} \right]}
	\Biggl[
		\frac{1}{2}
		\sup_{\varphi \in \mathcal{F}}
		\frac{1}{n}
		\left[
				\Phi^{(j)}
			+
				\left( -1 \right)^\alpha
				\left\{
					\varphi \left( Y_j, \vec{X}_j \right)
				-
					\varphi \left( Y'_j, \vec{X}'_j \right)
				\right\}
		\right] \\
	&\hspace{3cm}+
		\frac{1}{2}
		\sup_{\varphi \in \mathcal{F}}
		\frac{1}{n}
		\left[
				\Phi^{(j)}
			+
				\left( -1 \right)^\alpha
				\left\{
					\varphi \left( Y_j, \vec{X}_j \right)
				-
					\varphi \left( Y'_j, \vec{X}'_j \right)
				\right\}
		\right]
	\Biggr] \\
&=
	\mathbb{E}_{\mathbb{X}, \mathbb{X}', \left[ \sigma_1 ~ \cdots ~ \sigma_{j - 1} \right]}
	\left[
		\frac{1}{2}
		\sup_{\varphi \in \mathcal{F}}
		\frac{1}{n}
		\left[
				\Phi^{(j)}
			+
				\left( -1 \right)^\alpha
				\left\{
					\varphi \left( Y_j, \vec{X}_j \right)
				-
					\varphi \left( Y'_j, \vec{X}'_j \right)
				\right\}
		\right]
	\right] \\
	&\hspace{2cm}+
		\mathbb{E}_{\mathbb{X}, \mathbb{X}', \left[ \sigma_1 ~ \cdots ~ \sigma_{j - 1} \right]}
		\left[
			\frac{1}{2}
			\sup_{\varphi \in \mathcal{F}}
			\frac{1}{n}
			\left[
					\Phi^{(j)}
				+
					\left( -1 \right)^\alpha
					\left\{
						\varphi \left( Y_j, \vec{X}_j \right)
					-
						\varphi \left( Y'_j, \vec{X}'_j \right)
					\right\}
			\right]
		\right] \\
&=
		\mathbb{E}_{\mathbb{X}, \mathbb{X}', \left[ \sigma_1 ~ \cdots ~ \sigma_{j - 1} \right]}
		\left[
			\frac{1}{2}
			\sup_{\varphi \in \mathcal{F}}
			\frac{1}{n}
			\left[
					\Phi^{(j)}
				+
					\left( -1 \right)^\alpha
					\left\{
						\varphi \left( Y_j, \vec{X}_j \right)
					-
						\varphi \left( Y'_j, \vec{X}'_j \right)
					\right\}
			\right]
		\right] \\
	&\hspace{2cm}+
		\mathbb{E}_{\mathbb{X}, \mathbb{X}', \left[ \sigma_1 ~ \cdots ~ \sigma_{j - 1} \right]}
		\left[
			\frac{1}{2}
			\sup_{\varphi \in \mathcal{F}}
			\frac{1}{n}
			\left[
					\Phi^{(j)}
				+
					\left( -1 \right)^\alpha
					\left\{
						\varphi \left( Y'_j, \vec{X}'_j \right)
					-
						\varphi \left( Y_j, \vec{X}_j \right)
					\right\}
			\right]
	\right] \\
&=
	\mathbb{E}_{\mathbb{X}, \mathbb{X}', \left[ \sigma_1 ~ \cdots ~ \sigma_{j - 1} \right]}
	\Biggl[
		\frac{1}{2}
		\sup_{\varphi \in \mathcal{F}}
		\frac{1}{n}
		\left[
				\Phi^{(j)}
			+
				\left( -1 \right)^\alpha
				\left\{
					\varphi \left( Y_j, \vec{X}_j \right)
				-
					\varphi \left( Y'_j, \vec{X}'_j \right)
				\right\}
		\right] \\
	&\hspace{2cm}+
		\frac{1}{2}
		\sup_{\varphi \in \mathcal{F}}
		\frac{1}{n}
		\left[
				\Phi^{(j)}
			+
				\left( -1 \right)^{\alpha + 1}
				\left\{
					\varphi \left( Y_j, \vec{X}_j \right)
				-
					\varphi \left( Y'_j, \vec{X}'_j \right)
				\right\}
		\right]
	\Biggr] \\
&=
		\mathbb{E}_{\mathbb{X}, \mathbb{X}', \left[ \sigma_1 ~ \cdots ~ \sigma_{j - 1} \right]}
		\Biggl[
			\mathbb{P} \left( \sigma_j = \left( -1 \right)^\alpha \right)
			\sup_{\varphi \in \mathcal{F}}
			\frac{1}{n}
			\left[
					\Phi^{(j)}
				+
					\left( -1 \right)^\alpha
					\left\{
						\varphi \left( Y_j, \vec{X}_j \right)
					-
						\varphi \left( Y'_j, \vec{X}'_j \right)
					\right\}
			\right] \\
		&\hspace{2cm}+
			\mathbb{P} \left( \sigma_j = \left( -1 \right)^{\alpha + 1} \right)
			\sup_{\varphi \in \mathcal{F}}
			\frac{1}{n}
			\left[
					\Phi^{(j)}
				+
					\left( -1 \right)^{\alpha + 1}
					\left\{
						\varphi \left( Y_j, \vec{X}_j \right)
					-
						\varphi \left( Y'_j, \vec{X}'_j \right)
					\right\}
			\right]
		\Biggr] \\
&=
		\mathbb{E}_{\mathbb{X}, \mathbb{X}', \left[ \sigma_1 ~ \cdots ~ \sigma_{j - 1} \right]}
		\left[
			\mathbb{E}_{\sigma_j}
			\left[ \sup_{\varphi \in \mathcal{F}}
				\frac{1}{n}
				\left[
					\Phi^{(j)}
				+
					\sigma_j
					\left\{
						\varphi \left( Y_j, \vec{X}_j \right)
					-
						\varphi \left( Y'_j, \vec{X}'_j \right)
					\right\}
				\right]
			\right]
		\right] \\
&=
		\mathbb{E}_{\mathbb{X}, \mathbb{X}'}
		\Biggl[
		\mathbb{E}_{\left[ \sigma_1 ~ \cdots ~ \sigma_j \right] }
		\Biggl[
			\sup_{\varphi \in \mathcal{F}}
			\frac{1}{n}
			\Biggl[
				\sum_{i = 1}^{j}
				\sigma_i
				\left\{
					\varphi \left( Y_i, \vec{X}_i \right)
				-
					\varphi \left( Y'_i, \vec{X}'_i \right)
				\right\} \\
			&\hspace{5cm}+
				\left( -1 \right)^\alpha
				\sum_{i = j + 1}^{n}
				\left\{
					\varphi \left( Y_i, \vec{X}_i \right)
				-
					\varphi \left( Y'_i, \vec{X}'_i \right)
				\right\}
			\Biggr]
		\Biggr]
		\Biggr] \\
&=
		\mathbb{E}_{\mathbb{X}, \mathbb{X}'}
		\left[
		\mathbb{E}_{\left[ \sigma_1 ~ \cdots ~ \sigma_{j} \right] }
		\left[
			\sup_{\varphi \in \mathcal{F}}
			\frac{1}{n}
			\left[
				\Phi^{(j + 1)}
			+
				\left( -1 \right)^\alpha
				\left\{
					\varphi \left( Y_{j + 1}, \vec{X}_{j + 1} \right)
				-
					\varphi \left( Y'_{j + 1}, \vec{X}'_{j + 1} \right)
				\right\}
			\right]
		\right]
		\right] \\
&=
		\mathsf{E}^{[1]}_{j + 1},
\end{align*}
so that the equality \eqref{eq:symm2} is shown.
From (i) the inequality \eqref{eq:symm1},
(ii) the equality \eqref{eq:symm2} and
(iii) the equality
		$\mathbb{E}_{\vec{\sigma}} \left[ g \left( \vec{\sigma} \right) \right]
=
		\mathbb{E}_{\vec{\sigma}}
		\left[ g \left( - \vec{\sigma} \right) \right]$
for a function $g$ via the transformation from $\vec{\sigma}$ to $- \vec{\sigma}$, it follows that
\begin{align*}
		&\mathbb{E}_{\mathbb{X}}
		\left[
		\sup_{\varphi \in \mathcal{F}}
		\left[
			\left( -1 \right)^\alpha
			\left\{
				\frac{1}{n}
				\sum_{i = 1}^{n}
				\varphi \left( Y_i, \vec{X}_i \right)
			-
				\int_{\mathbb{R} \times \mathbb{R}^{d}}
				\varphi \left( y, \vec{x} \right)
				f \left( y, \vec{x} \right) d y d \vec{x}
			\right\}
		\right]
		\right] \\
&\leq
		\mathbb{E}_{\mathbb{X}, \mathbb{X}'}
		\left[
		\sup_{\varphi \in \mathcal{F}}
		\left[
			\left( -1 \right)^\alpha
			\left\{
				\frac{1}{n}
				\sum_{i = 1}^{n}
				\left\{
					\varphi \left( Y_i, \vec{X}_i \right)
				-
					\varphi \left( Y'_i, \vec{X}'_i \right)
				\right\}
				\right\}
		\right]
		\right] \\
&=
		\mathbb{E}_{\mathbb{X}, \mathbb{X}'}
		\left[
		\mathbb{E}_{\vec{\sigma}}
		\left[
		\sup_{\varphi \in \mathcal{F}}
		\left[
			\frac{1}{n}
			\sum_{i = 1}^{n}
			\sigma_i
		\cdot
			\left\{
				\varphi \left( Y_i, \vec{X}_i \right)
			-
				\varphi \left( Y'_i, \vec{X}'_i \right)
			\right\}
		\right]
		\right]
		\right] \\
&\leq
		\mathbb{E}_{\mathbb{X}, \mathbb{X}'}
		\left[
		\mathbb{E}_{\vec{\sigma}}
		\left[
		\sup_{\varphi \in \mathcal{F}}
		\left[
			\frac{1}{n}
			\sum_{i = 1}^{n}
			\sigma_i
		\cdot
			\varphi \left( Y_i, \vec{X}_i \right)
		\right]
		\right]
		\right]
	+
		\mathbb{E}_{\mathbb{X}, \mathbb{X}'}
		\left[
		\mathbb{E}_{\vec{\sigma}}
		\left[
		\sup_{\varphi \in \mathcal{F}}
		\left[
			\frac{1}{n}
			\sum_{i = 1}^{n}
			\left( - \sigma_i \right)
		\cdot
			\varphi \left( Y'_i, \vec{X}'_i \right)
		\right]
		\right]
		\right] \\
&=
		2
		\mathbb{E}_{\mathbb{X}}
		\left[
		\mathbb{E}_{\vec{\sigma}}
		\left[
		\sup_{\varphi \in \mathcal{F}}
		\left[
			\frac{1}{n}
			\sum_{i = 1}^{n}
			\sigma_i
		\cdot
			\varphi \left( Y_i, \vec{X}_i \right)
		\right]
		\right]
		\right]. \tag*{$\Box$}
\end{align*}
\end{proof}
When addressing function classes in the supremum for Rademacher complexity,
we will utilize the next three lemmas \ref{lem:lip}--\ref{lem:oper}:
\begin{Lem} \label{lem:lip}
For $i = 1, \ldots, n$, let $\ell_{\mathbb{X} i}: \mathbb{R} \to \mathbb{R}$ be functions depending on $\mathbb{X}$ and satisfying the following condition: there exists $L_{\mathbb{X} i} > 0$ such that
		$\left|
			\ell_{\mathbb{X} i} \left( z_1 \right) - \ell_{\mathbb{X} i} \left( z_2 \right)
		\right|
\leq
		L_{\mathbb{X} i} \left| z_1 - z_2 \right|$
for any $z_1, z_2 \in \mathbb{R}$.
For such $\ell_{\mathbb{X} i}$ and a function class $\mathcal{G}$, the following inequality of random variables holds:
\begin{align}
		\mathbb{E}_{\vec{\sigma}}
		\left[
			\sup_{g \in \mathcal{G}}
			\frac{1}{n}
			\sum_{i = 1}^{n}
			\sigma_i
		\cdot
			\ell_{\mathbb{X} i} \left( g \left( \vec{X}_i \right) \right)
		\right]
\leq
		\mathbb{E}_{\vec{\sigma}}
		\left[
			\sup_{g \in \mathcal{G}}
			\frac{1}{n}
			\sum_{i = 1}^{n}
			\sigma_i
		\cdot
			L_{\mathbb{X} i}
		\cdot
			g \left( \vec{X}_i \right)
		\right]. \label{eq:seq_E}
\end{align}
\end{Lem}
\begin{proof}
Using the notation
\begin{align*}
		&\mathsf{E}^{[2]}_j
=
		\mathbb{E}_{\vec{\sigma}}
		\left[
			\sup_{g \in \mathcal{G}}
			\frac{1}{n}
			\left[
				\sum_{i = 1}^{j - 1}
				\sigma_i
				\cdot
				L_{\mathbb{X} i}
				\cdot
				g \left( \vec{X}_i \right)
		+
				\sum_{i = j}^{n}
				\sigma_i
				\cdot
				\ell_{\mathbb{X} i} \left( g \left( \vec{X}_i \right) \right)
			\right]
		\right] \\
&\hspace{0.5cm}=
		\mathbb{E}_{\vec{\sigma}}
		\left[
			\sup_{g \in \mathcal{G}}
			\frac{1}{n}
			\left[
				\mathfrak{S}^{(j)}_{\mathbb{X}} \left( \vec{\sigma}, g \right)
			+
				\sigma_j
			\cdot
				\ell_{\mathbb{X} j} \left( g \left( \vec{X}_j \right) \right)
				\right]
		\right], \\
		&\mathfrak{S}^{(j)}_{\mathbb{X}} \left( \vec{\sigma}, g \right)
=
		\sum_{i = 1}^{j - 1}
		\sigma_i
		\cdot
		L_{\mathbb{X} i}
		\cdot
		g \left( \vec{X}_i \right)
	+
		\sum_{i = j + 1}^{n}
		\sigma_i
		\cdot
		\ell_{\mathbb{X} i} \left( g \left( \vec{X}_i \right) \right),
\end{align*}
for $j = 1, 2, \ldots, n + 1$,
\eqref{eq:seq_E} is equivalent to $\mathsf{E}^{[2]}_1 \leq \mathsf{E}^{[2]}_{n + 1}$.
We shall show $\mathsf{E}^{[2]}_j \leq \mathsf{E}^{[2]}_{j + 1}$ for $j = 1, 2, \ldots, n + 1$:
let $\mathbb{E}_{\vec{\sigma}^{(j)}}$ be the expectation taken with respect to $\left[ \sigma_1 \cdots \sigma_{j - 1} ~ \sigma_{j + 1} \cdots \sigma_n \right]$.
Then, for any $j = 1, 2, \ldots, n + 1$, from the following equality of sets
\begin{align*}
		&\left\{
			\mathfrak{S}^{(j)}_{\mathbb{X}} \left( \vec{\sigma}, g \right)
		+
			\mathfrak{S}^{(j)}_{\mathbb{X}} \left( \vec{\sigma}, \widetilde{g} \right)
		+
			L_{\mathbb{X} j}
		\cdot
			\left|
				g \left( \vec{X}_j \right)
			-
				\widetilde{g} \left( \vec{X}_i \right)
			\right|
		\middle|
			g \in \mathcal{G},
			\widetilde{g} \in \mathcal{G}
		\right\}\\
&=
		\left\{
				\mathfrak{S}^{(j)}_{\mathbb{X}} \left( \vec{\sigma}, g \right)
			+
				\mathfrak{S}^{(j)}_{\mathbb{X}} \left( \vec{\sigma}, \widetilde{g} \right)
			+
				L_{\mathbb{X} j}
			\cdot
				\left\{
					g \left( \vec{X}_j \right)
				-
					\widetilde{g} \left( \vec{X}_i \right)
				\right\}
		\middle|
			g \in \mathcal{G},
			\widetilde{g} \in \mathcal{G}
		\right\},
\end{align*}
it follows that
\begin{align*}
		\mathsf{E}^{[2]}_j
&=
		\mathbb{E}_{\vec{\sigma}}
		\left[
			\sup_{g \in \mathcal{G}}
			\frac{1}{n}
			\left[
				\mathfrak{S}^{(j)}_{\mathbb{X}} \left( \vec{\sigma}, g \right)
			+
				\sigma_j
			\cdot
				\ell_{\mathbb{X} j} \left( g \left( \vec{X}_j \right) \right)
				\right]
		\right] \\
&=
		\mathbb{E}_{\vec{\sigma}^{(j)}}
		\Biggl[
			\frac{1}{2}
			\sup_{g \in \mathcal{G}}
			\frac{1}{n}
			\left[
				\mathfrak{S}^{(j)}_{\mathbb{X}} \left( \vec{\sigma}, g \right)
			+
				\ell_{\mathbb{X} j} \left( g \left( \vec{X}_j \right) \right)
			\right] \\
		&\hspace{1.5cm}+
			\frac{1}{2}
			\sup_{\widetilde{g} \in \mathcal{G}}
			\frac{1}{n}
			\left[
				\mathfrak{S}^{(j)}_{\mathbb{X}} \left( \vec{\sigma}, \widetilde{g} \right)
			+
				\left( -1 \right)
			\cdot
				\ell_{\mathbb{X} j} \left( \widetilde{g} \left( \vec{X}_i \right) \right)
			\right]
		\Biggr] \\
&=
		\mathbb{E}_{\vec{\sigma}^{(j)}}
		\left[
			\frac{1}{2}
			\sup_{\substack{g \in \mathcal{G}, \\
			\widetilde{g} \in \mathcal{G}}}
			\frac{1}{n}
			\left[
				\mathfrak{S}^{(j)}_{\mathbb{X}} \left( \vec{\sigma}, g \right)
			+
				\mathfrak{S}^{(j)}_{\mathbb{X}} \left( \vec{\sigma}, \widetilde{g} \right)
			+
				\left\{
					\ell_{\mathbb{X} j} \left( g \left( \vec{X}_j \right) \right)
				-
					\ell_{\mathbb{X} j} \left( \widetilde{g} \left( \vec{X}_i \right) \right)
				\right\}
			\right]
		\right] \\
&\leq
		\mathbb{E}_{\vec{\sigma}^{(j)}}
		\left[
			\frac{1}{2}
			\sup_{\substack{g \in \mathcal{G}, \\
			\widetilde{g} \in \mathcal{G}}}
			\frac{1}{n}
			\left[
				\mathfrak{S}^{(j)}_{\mathbb{X}} \left( \vec{\sigma}, g \right)
			+
				\mathfrak{S}^{(j)}_{\mathbb{X}} \left( \vec{\sigma}, \widetilde{g} \right)
			+
				L_{\mathbb{X} j}
			\cdot
				\left|
					g \left( \vec{X}_j \right)
				-
					\widetilde{g} \left( \vec{X}_i \right)
				\right|
			\right]
		\right] \\
&=
		\mathbb{E}_{\vec{\sigma}^{(j)}}
		\left[
			\frac{1}{2}
			\sup_{\substack{g \in \mathcal{G}, \\
			\widetilde{g} \in \mathcal{G}}}
			\frac{1}{n}
			\left[
				\mathfrak{S}^{(j)}_{\mathbb{X}} \left( \vec{\sigma}, g \right)
			+
				\mathfrak{S}^{(j)}_{\mathbb{X}} \left( \vec{\sigma}, \widetilde{g} \right)
			+
				L_{\mathbb{X} j}
			\cdot
				\left\{
					g \left( \vec{X}_j \right)
				-
					\widetilde{g} \left( \vec{X}_i \right)
				\right\}
			\right]
		\right] \\
&=
		\mathbb{E}_{\vec{\sigma}^{(j)}}
		\Biggl[
			\frac{1}{2}
			\sup_{g \in \mathcal{G}}
			\frac{1}{n}
			\left[
				\mathfrak{S}^{(j)}_{\mathbb{X}} \left( \vec{\sigma}, g \right)
			+
				L_{\mathbb{X} j}
			\cdot
				g \left( \vec{X}_j \right)
			\right] \\
		&\hspace{1.5cm}+
			\frac{1}{2}
			\sup_{\widetilde{g} \in \mathcal{G}}
			\frac{1}{n}
			\left[
				\mathfrak{S}^{(j)}_{\mathbb{X}} \left( \vec{\sigma}, \widetilde{g} \right)
			+
				\left( -1 \right)
			\cdot
				L_{\mathbb{X} j}
			\cdot
				\widetilde{g} \left( \vec{X}_i \right)
			\right]
		\Biggr] \\
&=
		\mathbb{E}_{\vec{\sigma}}
		\left[
			\sup_{g \in \mathcal{G}}
			\frac{1}{n}
			\left[
				\mathfrak{S}^{(j)}_{\mathbb{X}} \left( \vec{\sigma}, g \right)
			+
				\sigma_j
			\cdot
				L_{\mathbb{X} j}
			\cdot
				g \left( \vec{X}_j \right)
			\right]
		\right] \\
&=
		\mathbb{E}_{\vec{\sigma}}
		\left[
			\sup_{g \in \mathcal{G}}
			\frac{1}{n}
			\left[
				\sum_{i = 1}^{j}
				\sigma_i
				\cdot
				L_{\mathbb{X} i}
				\cdot
				g \left( \vec{X}_i \right)
			+
				\sum_{i = j+1}^{n}
				\sigma_i
				\cdot
				\ell_{\mathbb{X} i} \left( g \left( \vec{X}_i \right) \right)
			\right]
		\right] \\
&=
		\mathbb{E}_{\vec{\sigma}}
		\left[
			\sup_{g \in \mathcal{G}}
			\frac{1}{n}
			\left[
				\mathfrak{S}^{(j + 1)}_{\mathbb{X}} \left( \vec{\sigma}, g \right)
			+
				\sigma_{j + 1}
			\cdot
				\ell_{\mathbb{X} j + 1} \left( g \left( \vec{X}_{j + 1} \right) \right)
			\right]
		\right] \\
&=
		\mathsf{E}^{[2]}_{j + 1}. \tag*{$\Box$}
\end{align*}
\end{proof}
\begin{Lem} \label{lem:hull}
For real numbers $L_{\mathbb{X} 1}, \ldots L_{\mathbb{X} n}$ depending on $\mathbb{X}$ and for a function class $\mathcal{G}$,
\begin{align*}
		\mathbb{E}_{\vec{\sigma}}
		\left[
			\sup_{g \in \mathrm{co} \left( \mathcal{G} \right) }
			\frac{1}{n}
			\sum_{i = 1}^{n}
			\sigma_i
		\cdot
			L_{\mathbb{X} i}
		\cdot
			g \left( \vec{X}_i \right)
		\right]
=
		\mathbb{E}_{\vec{\sigma}}
		\left[
			\sup_{g \in \mathcal{G}}
			\frac{1}{n}
			\sum_{i = 1}^{n}
				\sigma_i
			\cdot
				L_{\mathbb{X} i}
			\cdot
				g \left( \vec{X}_i \right)
		\right].
\end{align*}
\end{Lem}
\begin{proof}
Since $\mathrm{co} \left( \mathcal{G} \right) \supset \mathcal{G}$, the following inequality holds:
\begin{align*}
		\sup_{g \in \mathrm{co} \left( \mathcal{G} \right) }
		\frac{1}{n}
		\sum_{i = 1}^{n}
			\sigma_i
		\cdot
			L_{\mathbb{X} i}
		\cdot
			g \left( \vec{X}_i \right)
\geq
		\sup_{g \in \mathcal{G}}
		\frac{1}{n}
		\sum_{i = 1}^{n}
			\sigma_i
		\cdot
			L_{\mathbb{X} i}
		\cdot
			g \left( \vec{X}_i \right).
\end{align*}
Moreover, for any $\Bar{g} = \sum_{\widetilde{g} \in \mathcal{G}} \lambda_{\widetilde{g}} \cdot \widetilde{g} \in \mathrm{co} \left( \mathcal{G} \right)$,
\begin{align*}
		\sum_{i = 1}^{n}
			\sigma_i
		\cdot
			L_{\mathbb{X} i}
		\cdot
			\Bar{g} \left( \vec{X}_i \right)
&=
		\sum_{i = 1}^{n}
			\sigma_i
		\cdot
			L_{\mathbb{X} i}
		\cdot
			\left\{
				\sum_{\widetilde{g} \in \mathcal{G}}
				\lambda_{\widetilde{g}}
			\cdot
				\widetilde{g} \left( \vec{X}_i \right)
			\right\} \\
&=
		\sum_{\widetilde{g} \in \mathcal{G}}
		\lambda_{\widetilde{g}}
		\sum_{i = 1}^{n}
			\sigma_i
		\cdot
			L_{\mathbb{X} i}
		\cdot
			\widetilde{g} \left( \vec{X}_i \right) \\
&\leq
		\sup_{g \in \mathcal{G}}
		\sum_{i = 1}^{n}
			\sigma_i
		\cdot
			L_{\mathbb{X} i}
		\cdot
			g \left( \vec{X}_i \right),
\end{align*}
so that the other inequality also holds:
\begin{align*}
		\sup_{g \in \mathrm{co} \left( \mathcal{G} \right) }
		\frac{1}{n}
		\sum_{i = 1}^{n}
			\sigma_i
		\cdot
			L_{\mathbb{X} i}
		\cdot
			g \left( \vec{X}_i \right)
\leq
		\sup_{g \in \mathcal{G}}
		\frac{1}{n}
		\sum_{i = 1}^{n}
			\sigma_i
		\cdot
			L_{\mathbb{X} i}
		\cdot
			g \left( \vec{X}_i \right) \tag*{$\Box$}
\end{align*}
\end{proof}
\begin{Lem} \label{lem:oper}
For some function classes $\mathcal{A}$ and $\mathcal{B}$,
let $\mathcal{A} + \mathcal{B} = \left\{ g_1 + g_2 \middle| g_1 \in \mathcal{A}, g_2 \in \mathcal{B} \right\}$
and
$\mathcal{A} \cdot \mathcal{B} = \left\{ g_1 \cdot g_2 \middle| g_1 \in \mathcal{A}, g_2 \in \mathcal{B} \right\}$.
Then
\begin{align*}
		\sup_{g \in \mathcal{A} + \mathcal{B}}
		\frac{1}{n}
		\sum_{i = 1}^{n}
			\sigma_i
		\cdot
			L_{\mathbb{X} i}
		\cdot
			g \left( \vec{X}_i \right)
=
		\sup_{g \in \mathcal{A}}
		\frac{1}{n}
		\sum_{i = 1}^{n}
			\sigma_i
		\cdot
			L_{\mathbb{X} i}
		\cdot
			g \left( \vec{X}_i \right)
	+
		\sup_{g \in \mathcal{B}}
		\frac{1}{n}
		\sum_{i = 1}^{n}
			\sigma_i
		\cdot
			L_{\mathbb{X} i}
		\cdot
			g \left( \vec{X}_i \right).
	\end{align*}
Moreover, for
		$\sup_{g \in \mathcal{A}}
		\sup_{\vec{x} \in \mathbb{R}^d}
		\left| g \left( \vec{x} \right) \right|
=
		C_{\mathcal{A}}$
and
		$\sup_{g \in \mathcal{B}}
		\sup_{\vec{x} \in \mathbb{R}^d}
		\left| g \left( \vec{x} \right) \right|
=
		C_{\mathcal{B}}$,
\begin{align*}
		&\mathbb{E}_{\vec{\sigma}}
		\left[
			\sup_{g \in \mathcal{A} \cdot \mathcal{B}}
			\frac{1}{n}
			\sum_{i = 1}^{n}
				\sigma_i
			\cdot
				L_{\mathbb{X} i}
			\cdot
				g \left( \vec{X}_i \right)
		\right] \\
&\leq
		\left(
			C_{\mathcal{A}}
		+
			C_{\mathcal{B}}
		\right)
		\left\{
			\mathbb{E}_{\vec{\sigma}}
			\left[
				\sup_{g \in \mathcal{A}}
				\frac{1}{n}
				\sum_{i = 1}^{n}
					\sigma_i
				\cdot
					L_{\mathbb{X} i}
				\cdot
					g \left( \vec{X}_i \right)
			\right]
		+
			\mathbb{E}_{\vec{\sigma}}
			\left[
				\sup_{g \in \mathcal{B}}
				\frac{1}{n}
				\sum_{i = 1}^{n}
					\sigma_i
				\cdot
					L_{\mathbb{X} i}
				\cdot
					g \left( \vec{X}_i \right)
			\right]
		\right\}.
\end{align*}
\end{Lem}
\begin{proof}
The definition of $\mathcal{A} + \mathcal{B}$ implies that
\begin{align*}
		\sup_{g \in \mathcal{A} + \mathcal{B}}
		\frac{1}{n}
		\sum_{i = 1}^{n}
			\sigma_i
		\cdot
			L_{\mathbb{X} i}
		\cdot
			g \left( \vec{X}_i \right)
&=
		\sup_{g_1 \in \mathcal{A}, g_2 \in \mathcal{B}}
		\frac{1}{n}
		\sum_{i = 1}^{n}
		\sigma_i
		\cdot
			L_{\mathbb{X} i}
		\cdot
			\left\{
				g_1 \left( \vec{X}_i \right)
			+
				g_2 \left( \vec{X}_i \right)
			\right\} \\
&=
		\sup_{g_1 \in \mathcal{A}}
		\frac{1}{n}
		\sum_{i = 1}^{n}
			\sigma_i
		\cdot
			L_{\mathbb{X} i}
		\cdot
			g_1 \left( \vec{X}_i \right)
	+
		\sup_{g_2 \in \mathcal{B}}
		\frac{1}{n}
		\sum_{i = 1}^{n}
			\sigma_i
		\cdot
			L_{\mathbb{X} i}
		\cdot
			g_2 \left( \vec{X}_i \right) \\
&=
		\sup_{g \in \mathcal{A}}
		\frac{1}{n}
		\sum_{i = 1}^{n}
			\sigma_i
		\cdot
			L_{\mathbb{X} i}
		\cdot
			g \left( \vec{X}_i \right)
	+
		\sup_{g \in \mathcal{B}}
		\frac{1}{n}
		\sum_{i = 1}^{n}
			\sigma_i
		\cdot
			L_{\mathbb{X} i}
		\cdot
			g \left( \vec{X}_i \right).
\end{align*}
By applying Lemma \ref{lem:lip} with
$\ell_{\mathbb{X} i} \left( z \right) = s \left( z \right) = z^2$ for $z \in \left[ - \left( C_{\mathcal{A}} + C_{\mathcal{B}} \right), C_{\mathcal{A}} + C_{\mathcal{B}} \right]$, which satisfies
		$\left|s \left( z_1 \right) - s \left( z_2 \right) \right|
=
		\left|z_1^2 - z_2^2 \right|
=
		\left| z_1 + z_2 \right| \left| z_1 - z_2 \right|
\leq
		2 \left( C_{\mathcal{A}} + C_{\mathcal{B}} \right)
		\left| z_1 - z_2 \right|$
for any
$z_1, z_2 \in \left[ - \left( C_{\mathcal{A}} + C_{\mathcal{B}} \right), C_{\mathcal{A}} + C_{\mathcal{B}} \right]$,
we see that
\begin{align*}
		&\mathbb{E}_{\vec{\sigma}}
		\left[
		\sup_{g \in \mathcal{A} \cdot \mathcal{B}}
		\frac{1}{n}
		\sum_{i = 1}^{n}
			\sigma_i
		\cdot
			L_{\mathbb{X} i}
		\cdot
			g \left( \vec{X}_i \right) \right] \\
&=
		\mathbb{E}_{\vec{\sigma}}
		\left[
			\sup_{g_1 \in \mathcal{A}, g_2 \in \mathcal{B}}
			\frac{1}{n}
			\sum_{i = 1}^{n}
				\sigma_i
			\cdot
				L_{\mathbb{X} i}
			\cdot
				g_1 \left( \vec{X}_i \right)
				g_2 \left( \vec{X}_i \right)
		\right] \\
&=
		\mathbb{E}_{\vec{\sigma}}
		\left[
			\sup_{g_1 \in \mathcal{A}, g_2 \in \mathcal{B}}
			\frac{1}{n}
			\sum_{i = 1}^{n}
				\sigma_i
			\cdot
				L_{\mathbb{X} i}
			\cdot
				\frac{1}{4}
				\left[
					\left\{
						g_1 \left( \vec{X}_i \right)
					+
						g_2 \left( \vec{X}_i \right)
					\right\}^2
				-
					\left\{
						g_1 \left( \vec{X}_i \right)
					-
						g_2 \left( \vec{X}_i \right)
					\right\}^2
				\right]
		\right] \\
&\leq
		\frac{1}{4}
		\mathbb{E}_{\vec{\sigma}}
		\left[
			\sup_{g_1 \in \mathcal{A}, g_2 \in \mathcal{B}}
			\frac{1}{n}
			\sum_{i = 1}^{n}
				\sigma_i
			\cdot
				L_{\mathbb{X} i}
			\cdot
				\left\{
					g_1 \left( \vec{X}_i \right)
				+
					g_2 \left( \vec{X}_i \right)
				\right\}^2
		\right] \\
	&\hspace{0.5cm}+
		\frac{1}{4}
		\mathbb{E}_{\vec{\sigma}}
		\left[
			\sup_{g_1 \in \mathcal{A}, g_2 \in \mathcal{B}}
			\frac{1}{n}
			\sum_{i = 1}^{n}
				\left( - \sigma_i \right)
			\cdot
				L_{\mathbb{X} i}
			\cdot
				\left\{
					g_1 \left( \vec{X}_i \right)
				-
					g_2 \left( \vec{X}_i \right)
				\right\}^2
		\right] \\
&=
		\frac{1}{4}
		\mathbb{E}_{\vec{\sigma}}
		\left[
			\sup_{g_1 \in \mathcal{A}, g_2 \in \mathcal{B}}
			\frac{1}{n}
			\sum_{i = 1}^{n}
				\sigma_i
			\cdot
				L_{\mathbb{X} i}
			\cdot
				\left\{
					g_1 \left( \vec{X}_i \right)
				+
					g_2 \left( \vec{X}_i \right)
				\right\}^2
		\right] \\
	&\hspace{0.5cm}+
		\frac{1}{4}
		\mathbb{E}_{\vec{\sigma}}
		\left[
			\sup_{g_1 \in \mathcal{A}, g_2 \in \mathcal{B}}
			\frac{1}{n}
			\sum_{i = 1}^{n}
				\sigma_i
			\cdot
				L_{\mathbb{X} i}
					\left\{
						g_1 \left( \vec{X}_i \right)
					-
						g_2 \left( \vec{X}_i \right)
					\right\}^2
		\right] \\
&=
		\frac{1}{4}
		\mathbb{E}_{\vec{\sigma}}
		\left[
			\sup_{g_1 \in \mathcal{A}, g_2 \in \mathcal{B}}
			\frac{1}{n}
			\sum_{i = 1}^{n}
				\sigma_i
			\cdot
				L_{\mathbb{X} i}
			\cdot
					s \left(
						g_1 \left( \vec{X}_i \right)
					+
						g_2 \left( \vec{X}_i \right)
					\right)
		\right] \\
	&\hspace{0.5cm}+
		\frac{1}{4}
		\mathbb{E}_{\vec{\sigma}}
		\left[
			\sup_{g_1 \in \mathcal{A}, g_2 \in \mathcal{B}}
			\frac{1}{n}
			\sum_{i = 1}^{n}
			\sigma_i
			\cdot
				L_{\mathbb{X} i}
			\cdot
				s \left(
					g_1 \left( \vec{X}_i \right)
				-
					g_2 \left( \vec{X}_i \right)
				\right)
		\right] \\
&\leq
		\frac{\left( C_{\mathcal{A}} + C_{\mathcal{B}} \right)}{2}
		\mathbb{E}_{\vec{\sigma}}
		\left[
			\sup_{g_1 \in \mathcal{A}, g_2 \in \mathcal{B}}
			\frac{1}{n}
			\sum_{i = 1}^{n}
			\sigma_i
				\cdot
			L_{\mathbb{X} i}
		\cdot
			\left\{
				g_1 \left( \vec{X}_i \right)
			+
				g_2 \left( \vec{X}_i \right)
			\right\}
		\right] \\
	&\hspace{0.5cm}+
		\frac{\left( C_{\mathcal{A}} + C_{\mathcal{B}} \right)}{2}
		\mathbb{E}_{\vec{\sigma}}
		\left[
			\sup_{g_1 \in \mathcal{A}, g_2 \in \mathcal{B}}
			\frac{1}{n}
			\sum_{i = 1}^{n}
				\sigma_i
			\cdot
				L_{\mathbb{X} i}
			\cdot
				\left\{
					g_1 \left( \vec{X}_i \right)
				-
					g_2 \left( \vec{X}_i \right)
				\right\}
		\right] \\
&=
		\left( C_{\mathcal{A}} + C_{\mathcal{B}} \right)
		\left\{
			\mathbb{E}_{\vec{\sigma}}
			\left[
				\sup_{g \in \mathcal{A}}
				\frac{1}{n}
				\sum_{i = 1}^{n}
					\sigma_i
				\cdot
					L_{\mathbb{X} i}
				\cdot
						g \left( \vec{X}_i \right)
			\right]
		+
			\mathbb{E}_{\vec{\sigma}}
			\left[
				\sup_{g \in \mathcal{B}}
				\frac{1}{n}
				\sum_{i = 1}^{n}
					\sigma_i
				\cdot
					L_{\mathbb{X} i}
				\cdot
					g \left( \vec{X}_i \right)
			\right]
		\right\}. \tag*{$\Box$}
\end{align*}
\end{proof}
Let $N \left( \epsilon, \mathcal{G}, \rho \right)$ be the $\epsilon$-covering number of a class $\mathcal{G}$ with respect to a pseudometric $\rho$, see Definition 5.1 and Section 5.3.3 of \citet*{W2019}, and Definition 2.1.5 of \citet*{VW1996}.
We will finally bound the second term on the right-hand side of \eqref{eq:mu_hat} from above with the covering number, by utilizing the next lemma:
\begin{Lem} \label{lem:dudley}
For
		$\sup_{g \in \mathcal{G}}
		\sup_{\vec{x} \in \mathbb{R}^d}
		g \left( \vec{x} \right)
\leq
		\Delta$
and any $n \in \mathbb{N}$,
\begin{align*}
		\mathbb{E}_{\vec{\sigma}}
		\left[
			\sup_{g \in \mathcal{G}}
			\frac{1}{n}
			\sum_{i = 1}^{n}
				\sigma_i
			\cdot
				L_{\mathbb{X} i}
			\cdot
				g \left( \vec{X}_i \right)
		\right]
\leq
		32
		\frac{\sqrt{\sum_{i = 1}^{n} L_{\mathbb{X} i}}}{n}
		\int_0^{2 \Delta}
		\sqrt{\ln N \left( \epsilon, \mathcal{G}, \| \cdot \|_\infty \right)}
		d \epsilon.
\end{align*}
\end{Lem}
\begin{proof}
We shall first prove that in terms of $\vec{\sigma}$, the following
\[
		\left\{
			\frac{1}{n}
			\sum_{i = 1}^{n}
			\sigma_i
				\cdot
			L_{\mathbb{X} i}
				\cdot
			g \left( \vec{X}_i \right)
		\middle|
			g \in \mathcal{G}
		\right\}
=
		\left\{
			Z_g
		\middle|
			g \in \mathcal{G}
		\right\}
\]
is a 0-mean sub-Gaussian process with respect to a pseudometric
\[
		\rho \left( g, \widetilde{g} \right)
	=
		\sqrt{
		(1 / n)
		\sum_{i = 1}^{n}
			L_{\mathbb{X} i}^2
		\cdot
			\left\{
				g \left( \vec{X}_i \right)
			-
				\widetilde{g} \left( \vec{X}_i \right)
		\right\}^2}.
\]
The 0-mean property is satisfied since
\begin{align*}
		\mathbb{E}_{\vec{\sigma}} \left[ Z_g \right]
&=
		\mathbb{E}_{\vec{\sigma}}
		\left[
			\frac{1}{\sqrt{n}}
			\sum_{i = 1}^{n}
				\sigma_i
			\cdot
				L_{\mathbb{X} i}
			\cdot
				g \left( \vec{X}_i \right)
		\right] \\
&=
		\frac{1}{\sqrt{n}}
		\sum_{i = 1}^{n}
			\mathbb{E}_{\vec{\sigma}} \left[ \sigma_i \right]
			L_{\mathbb{X} i}
		\cdot
			g \left( \vec{X}_i \right) \\
&=
		\frac{1}{\sqrt{n}}
		\sum_{i = 1}^{n}
			0
		\cdot
			L_{\mathbb{X} i}
		\cdot
			g \left( \vec{X}_i \right) \\
&= 0.
\end{align*}
To prove the sub-Gaussianity, we now show that
$\left\{
	Z_g
\middle|
	g \in \mathcal{G}
\right\}$
satisfies Definition 5.16 of \citet*{W2019}:
that is, for any $g, \widetilde{g} \in \mathcal{G}$ and any $\lambda \in \mathbb{R}$,
\[
		\mathbb{E} \left[ \exp \left[ \lambda \left( Z_g - Z_{\widetilde{g}} \right) \right] \right]
\leq
		\mathbb{E} \left[ \exp \left[ \lambda^2 \rho \left( g, \widetilde{g} \right)^2 / 2 \right] \right].
\]
This is shown by using the inequality $\mathbb{E}_{\sigma_i} \left[ e^{\sigma_i x} \right] \leq e^{x^2 / 2}$ in Example 2.3 of \citet*{W2019} as follows:
\begin{align*}
		\mathbb{E}_{\vec{\sigma}}
		\left[ e^{\lambda \cdot \left( Z_g - Z_{\widetilde{g}} \right)} \right]
&=
		\mathbb{E}_{\vec{\sigma}}
		\left[
			\exp \left[
				\lambda
			\cdot
				\frac{1}{\sqrt{n}}
				\sum_{i = 1}^{n}
				\sigma_i
			\cdot
				L_{\mathbb{X} i}
			\cdot
			\left\{ g \left( \vec{X}_i \right) - \widetilde{g} \left( \vec{X}_i \right) \right\}
			\right]
		\right] \\
&=
		\mathbb{E}_{\vec{\sigma}}
		\left[
			\prod_{i = 1}^{n}
			\exp \left[
				\lambda
			\cdot
				\frac{1}{\sqrt{n}}
				\sigma_i
			\cdot
				L_{\mathbb{X} i}
			\cdot
				\left\{ g \left( \vec{X}_i \right) - \widetilde{g} \left( \vec{X}_i \right) \right\}
			\right]
		\right] \\
&=
		\prod_{i = 1}^{n}
		\mathbb{E}_{\sigma_i}
		\left[
			\exp \left[
				\sigma_i
			\cdot
				\lambda
			\cdot
				\frac{1}{\sqrt{n}}
				L_{\mathbb{X} i}
			\cdot
				\left\{ g \left( \vec{X}_i \right) - \widetilde{g} \left( \vec{X}_i \right) \right\}
			\right]
		\right] \\
&\leq
		\prod_{i = 1}^{n}
		\exp \left[
			\frac{\lambda^2}{2}
			\left[
				\frac{1}{\sqrt{n}}
				L_{\mathbb{X} i}
			\cdot
				\left\{ g \left( \vec{X}_i \right) - \widetilde{g} \left( \vec{X}_i \right) \right\}
			\right]^2
		\right] \\
&=
		\exp \left[
			\frac{\lambda^2}{2}
			\cdot
			\frac{1}{n}
			\sum_{i = 1}^{n}
			L_{\mathbb{X} i}^2
			\cdot
			\left\{ g \left( \vec{X}_i \right) - \widetilde{g} \left( \vec{X}_i \right) \right\}^2
		\right] \\
&=
		\exp \left[
			\frac{\lambda^2}{2}
			\rho \left( g, \widetilde{g} \right)^2
		\right].
\end{align*}
Hence, we have finished proving that
$\left\{
	Z_g
\middle|
	g \in \mathcal{G}
\right\}$
is a 0-mean sub-Gaussian process with respect to $\rho$,
and therefore we can apply Theorem 5.22 of \citet*{W2019} to this process.
By noting that
\begin{align*}
		\rho \left( g, \widetilde{g} \right)
&=
		\sqrt{
		\frac{1}{n}
		\sum_{i = 1}^{n}
			L_{\mathbb{X} i}^2
		\cdot
			\left\{ g \left( \vec{X}_i \right) - \widetilde{g} \left( \vec{X}_i \right) \right\}^2
		} \\
&\leq
		\sqrt{
		\frac{1}{n}
		\sum_{i = 1}^{n}
			L_{\mathbb{X} i}^2
		\cdot
			\sup_{\vec{x} \in \mathbb{D}} \left| g \left( \vec{x} \right) - \widetilde{g} \left( \vec{x} \right) \right|^2
		} \\
&\leq
		\sqrt{
			\frac{1}{n}
			\sum_{i = 1}^{n}
			L_{\mathbb{X} i}^2
		}
	\cdot
		\sup_{\vec{x} \in \mathbb{D}}
		\left| g \left( \vec{x} \right) - \widetilde{g} \left( \vec{x} \right) \right| \\
&=
		C \cdot \left\| g - \widetilde{g} \right\|_\infty.
\end{align*}
and that
\[
		N \left( \epsilon, \mathcal{G}, \rho \right) \leq
		N \left( \epsilon, \mathcal{G}, C \cdot \left\| \cdot \right\|_\infty \right)
=
		N \left( \epsilon / C, \mathcal{G}, \left\| \cdot \right\|_\infty \right)
\]
(since for any $\widetilde{g} \in \mathcal{G}$,
	$\left\{ g \in \mathcal{G} \middle| \rho \left( g, \widetilde{g} \right) < \epsilon \right\}
	\supset
	\left\{ g \in \mathcal{G} \middle| C \cdot \left\| g - \widetilde{g} \right\|_\infty < \epsilon \right\}$),
Theorem 5.22 of \citet*{W2019} implies the following evaluation for any $\widetilde{g} \in \mathcal{G}$ and any $\delta > 0$:
\begin{align}
		&\mathbb{E}_{\vec{\sigma}}
		\left[ \sup_{g \in \mathcal{G}} Z_g \right] \nonumber \\
&=
		\mathbb{E}_{\vec{\sigma}}
		\left[ \sup_{g \in \mathcal{G}} \left[ Z_g - Z_{\widetilde{g}} + Z_{\widetilde{g}} \right] \right] \nonumber \\
&=
		\mathbb{E}_{\vec{\sigma}}
		\left[
			\sup_{g \in \mathcal{G}}
			\left[ Z_g - Z_{\widetilde{g}} \right]
		\right]
	+
		\mathbb{E}_{\vec{\sigma}}
		\left[ Z_{\widetilde{g}} \right] \nonumber \\
&\leq
		\mathbb{E}_{\vec{\sigma}}
		\left[
			\sup_{g, \widetilde{g} \in \mathcal{G}}
			\left[ Z_g - Z_{\widetilde{g}} \right]
		\right]
	+
		0 \nonumber \\
&\leq
		2 \mathbb{E}_{\vec{\sigma}}
		\left[
			\sup_{\substack{g, \widetilde{g} \in \mathcal{G}\nonumber \\
			\rho \left( g, \widetilde{g} \right) \leq \delta}}
			\left[ Z_g - Z_{\widetilde{g}} \right]
		\right]
		+
		32 \int_{\delta / 4}^{\sup_{g, \widetilde{g} \in \mathcal{G}} \rho \left( g, \widetilde{g} \right)}
			\sqrt{\ln N \left( \epsilon, \mathcal{G}, \left\| \cdot \right\|_\infty \right)}
		d \epsilon \nonumber \\
&\leq
		2 \mathbb{E}_{\vec{\sigma}}
		\left[
			\sup_{\substack{g, \widetilde{g} \in \mathcal{G}\nonumber \\
			\rho \left( g, \widetilde{g} \right) \leq \delta}}
				\frac{1}{\sqrt{n}}
				\sum_{i = 1}^{n}
				\left| \sigma_i \right|
				\left| L_{\mathbb{X} i} \right|
				\left| g \left( \vec{X}_i \right) - \widetilde{g} \left( \vec{X}_i \right) \right|
		\right] \\
	&\hspace{0.5cm}+
		32 \int_{\delta / 4}^{\sup_{g, \widetilde{g} \in \mathcal{G}}
			\rho \left( g, \widetilde{g} \right)}
			\sqrt{\ln N \left( \frac{\epsilon}{C}, \mathcal{G}, \left\| \cdot \right\|_\infty \right)}
		d \epsilon \nonumber \\
&\leq
		2 \mathbb{E}_{\vec{\sigma}}
		\left[
			\sup_{\substack{g, \widetilde{g} \in \mathcal{G}\nonumber \\
			\rho \left( g, \widetilde{g} \right) \leq \delta}}
				\frac{1}{\sqrt{n}}
				\sqrt{\sum_{i = 1}^{n} 1^2}
			\cdot
				\sqrt{
				\sum_{i = 1}^{n}
				L_{\mathbb{X} i}^2
			\cdot
				\left\{ g \left( \vec{X}_i \right) - \widetilde{g} \left( \vec{X}_i \right) \right\}^2}
		\right] \nonumber \\
	&\hspace{0.5cm}+
		32 C
		\int_{\delta / (4 C)}^{\sup_{g, \widetilde{g} \in \mathcal{G}} \rho \left( g, \widetilde{g} \right) / C}
			\sqrt{\ln N \left( \epsilon, \mathcal{G}, \left\| \cdot \right\|_\infty \right)}
		d \epsilon \nonumber \\
&\leq
		2 \mathbb{E}_{\vec{\sigma}}
		\left[
			\sup_{\substack{g, \widetilde{g} \in \mathcal{G}\nonumber \\
			\rho \left( g, \widetilde{g} \right) \leq \delta}}
				\sqrt{n}
			\cdot
				\rho \left( g, \widetilde{g} \right)
		\right]
	+
		32 C
		\int_{\delta / (4 C)}^{\sup_{g, \widetilde{g} \in \mathcal{G}}
		\left\| g - \widetilde{g} \right\|_\infty
		C / C}
			\sqrt{\ln N \left( \epsilon, \mathcal{G}, \left\| \cdot \right\|_\infty \right)}
		d \epsilon \nonumber \\
&\leq
		2 \delta \sqrt{n}
	+
		32 C
		\int_{\delta / (4 C)}^{2 \sup_{g \in \mathcal{G}} \left\| g \right\|_{\infty}}
			\sqrt{\ln N \left( \epsilon, \mathcal{G}, \left\| \cdot \right\|_\infty \right)}
		d \epsilon. \label{eq:last}
\end{align}
By multiplying $1 / \sqrt{n}$ and $\delta \searrow 0$ in \eqref{eq:last}, we obtain the inequality in Lemma \ref{lem:dudley}. \qed
\end{proof}
Let $\mathcal{M} = \left\{ m \left( \cdot, \vec{\theta} \right) \colon \mathbb{D} \to \mathbb{R} \middle| \vec{\theta} \in \Theta_c \right\}$, where the set $\Theta_c \subset \mathbb{R}^p$ for $c > 0$ is already introduced in Section \ref{sec:theo}.
We evaluate the convering number of $\mathcal{M}$ here:
\begin{Lem} \label{lem:cov_M}
For $C_3 = 2
		\sup_{\vec{x} \in \mathbb{R}^d}
		\sup_{\vec{\theta} \in \Theta_c}
		\left\| \left( \partial / \partial \vec{\theta} \right) m \left( \vec{x}, \vec{\theta} \right) \right\|_2$
and
$C_4 = \left( 3c \cdot C_3 \right)^p$,
\begin{align*}
		N \left( \epsilon, \mathcal{M}, \left\| \cdot \right\|_\infty \right)
&\leq
		\mathbb{I}
		\left(
			\epsilon
		<
			c \cdot C_3 \right)
	\cdot
		C_4
	\cdot
		\left(\frac{1}{\epsilon} \right)^p
	+
		\mathbb{I} \left( \epsilon \geq c \cdot C_3 \right).
\end{align*}
\end{Lem}
\begin{proof}
From (i) Lemma 9.18 of \citet*{K2008},
(ii) Theorem 2.7.11 of \citet*{VW1996} (or Theorem 9.23 of \citet*{K2008}) with $\mathcal{F} = \mathcal{M}$, $T = \Theta_c$, $d (s, t) = \left\| s - t \right\|_2$ and $F \left( \vec{x} \right) = \sup_{\vec{x} \in \mathbb{R}^d} \sup_{\vec{\theta} \in \Theta_c} \left\| \left( \partial / \partial \vec{\theta} \right) m \left( \vec{x}, \vec{\theta} \right) \right\|_2$,
(iii) Lemma 6.27 (and the statement in Lemma 6.26 for the notation) of \citet*{MRT2018} with $\mathfrak{X} = \Theta_c$ and $R = c$,
and
(iv) a trivaial fact $N \left( \epsilon, \Theta_c, \left\| \cdot \right\|_\infty \right) = 1$ for $\epsilon \geq c$,
it follows that
\begin{align*}
		&N \left( \epsilon, \mathcal{M}, \left\| \cdot \right\|_\infty \right) \\
&\leq
		N \left( \frac{\epsilon}{2 \sup_{\vec{x} \in \mathbb{R}^d} \sup_{\vec{\theta} \in \Theta_c} \left\| \left( \partial / \partial \vec{\theta} \right) m \left( \vec{x}, \vec{\theta} \right) \right\|_2}, \Theta_c, \left\| \cdot \right\|_2 \right) \\
&\leq
		\mathbb{I} \left( \epsilon < c \cdot 2 \sup_{\vec{x} \in \mathbb{R}^d} \sup_{\vec{\theta} \in \Theta_c} \left\| \left( \partial / \partial \vec{\theta} \right) m \left( \vec{x}, \vec{\theta} \right) \right\|_2 \right)
		\left( 3c \frac{2 \sup_{\vec{x} \in \mathbb{R}^d} \sup_{\vec{\theta} \in \Theta_c} \left\| \left( \partial / \partial \vec{\theta} \right) m \left( \vec{x}, \vec{\theta} \right) \right\|_2}{\epsilon} \right)^p \\
	&\hspace{0.5cm}+
		\mathbb{I} \left( \epsilon \geq c \cdot 2 \sup_{\vec{x} \in \mathbb{R}^d} \sup_{\vec{\theta} \in \Theta_c} \left\| \left( \partial / \partial \vec{\theta} \right) m \left( \vec{x}, \vec{\theta} \right) \right\|_2 \right). \tag*{$\Box$}
\end{align*}
\end{proof}

\subsection{Proof of Theorem \ref{thm:mu}}
Putting
$\vec{\theta}_1 \left( \mathbb{X} \right) = \vec{\theta}_2 \left( \mathbb{X} \right) = \widehat{\vec{\theta}}$
in Lemma \ref{lem:ineq1}, it is shown that
\begin{align*}
		\widehat{L} \left( \widehat{\mu} \right)
&\leq
		\inf_{\Lambda \in \mathcal{W}}
		\widehat{L} \left( \mu \left( \cdot, \Lambda, \widehat{\vec{\theta}} \right) \right)
	+
		\frac{\tau^2}{I + \tau} A_1 \left( \widehat{\vec{\theta}}, \widehat{\vec{\theta}} \right)
	+
		\varepsilon.
\end{align*}
This inequality allows us to apply Lemma \ref{lem:ineq2} with
$\vec{\theta}_1 \left( \mathbb{X} \right) = \vec{\theta}_2 \left( \mathbb{X} \right) = \widehat{\vec{\theta}}$,
$\mu \left( \cdot, \Bar{\Lambda} \left( \mathbb{X} \right), \vec{\theta}_1 \left( \mathbb{X} \right) \right) = \widehat{\mu}$
and
$\delta \left( \mathbb{X} \right)
=
		\left\{ \tau^2 / \left( I + \tau \right) \right\} A_1 \left( \widehat{\vec{\theta}}, \widehat{\vec{\theta}} \right)
	+
		\varepsilon$.
Then
\begin{align*}
		&\mathbb{E} \left[ D_U \left( \mu, \widehat{\mu} \right) \right] \\
&\leq
		\mathbb{E}
		\left[
			\inf_{g \in u^{-1} \left( \mathrm{co} \left( T \left( \widehat{\vec{\theta}}, \mathcal{D} \right) \right) \right) }
			D_U \left( \mu, g \right)
		\right]
	+
		\mathbb{E}
		\left[
			\sup_{\Lambda \in \mathcal{W}}
			\mathcal{E}_n \left( \mu \left( \cdot, \Lambda, \widehat{\vec{\theta}} \right) \right)
		\right] \\
	&\hspace{0.5cm} +
		\mathbb{E}
		\left[
			\sup_{\Lambda \in \mathcal{W}}
			\left[ - \mathcal{E}_n \left( \mu \left( \cdot, \Lambda, \widehat{\vec{\theta}} \right) \right) \right]
		\right]
	+
		\mathbb{E}
		\left[
			\frac{\tau^2}{I + \tau} A_1 \left( \widehat{\vec{\theta}}, \widehat{\vec{\theta}} \right)
		+
			\varepsilon
		\right] \\
&=
		\mathbb{E}
		\left[
			\inf_{g \in u^{-1} \left( \mathrm{co} \left( T \left( \widehat{\vec{\theta}}, \mathcal{D} \right) \right) \right) }
			D_U \left( \mu, g \right)
		\right]
	+
		\sum_{\alpha = 0}^1
		\mathbb{E}
		\left[
			\sup_{g \in u^{-1} \left( \mathrm{co} \left( T \left( \widehat{\vec{\theta}}, \mathcal{D} \right) \right) \right)}
			\left[
				\left( - 1 \right)^\alpha
				\mathcal{E}_n \left( g \right)
			\right]
		\right]
	+
		\frac{\tau^2}{I + \tau}
		C_1 \left( n \right) \\
	&\hspace{0.5cm}+
		\varepsilon.
\end{align*}
which shows the inequlity \eqref{eq:mu_hat}. 

From now on, we investigate the rate of convergence for the term
\[
		\mathbb{E}
		\left[
			\sup_{g \in u^{-1} \left( \mathrm{co} \left( T \left( \widehat{\vec{\theta}}, \mathcal{D} \right) \right) \right)}
			\left[
				\left( - 1 \right)^\alpha
				\mathcal{E}_n \left( g \right)
			\right]
		\right]
\]
with $\alpha = 0$ and 1.
By denoting
\begin{align*}
		&u^{-1} \left( \mathrm{co} \left( T \left( \Theta_c, \mathcal{B} \right) \right) \right)
=
		\bigcup_{\vec{\theta} \in \Theta_c}
		u^{-1} \left( \mathrm{co} \left( T \left( \vec{\theta}, \mathcal{B} \right) \right) \right), \\
		&\ell \left( \cdot, \cdot
			\middle|
				u^{-1} \left( \mathrm{co} \left( T \left( \Theta_c, \mathcal{B} \right) \right) \right)
			\right)
=
		\left\{
			\ell \left( \cdot, \cdot
			\middle|
				g
			\right) \colon
			\mathbb{R}^d \times \mathbb{R} \to \mathbb{R}
		\middle|
			g \in u^{-1} \left( \mathrm{co} \left( T \left( \Theta_c, \mathcal{B} \right) \right) \right)
		\right\}, \\
		&F \left( y \right)
=
		\sup_{z \in \left[ - \Delta_1, \Delta_1 \right]}
		\left| U^\ast \left( z \right) \right|
	+
		\left| y \right| \Delta_1,
\end{align*}
and using Lemma \ref{lem:sym}, it follows for both $\alpha = 0, 1$ that
\begin{align}
		&\mathbb{E}
		\left[
			\sup_{g \in u^{-1} \left( \mathrm{co} \left( T \left( \widehat{\vec{\theta}}, \mathcal{D} \right) \right) \right)}
			\left[
				\left( - 1 \right)^\alpha
				\mathcal{E}_n \left( g \right)
			\right]
		\right] \nonumber \\
&\leq
		\mathbb{E}
		\left[
			\sup_{g \in u^{-1} \left( \mathrm{co} \left( T \left( \widehat{\vec{\theta}}, \mathcal{B} \right) \right) \right)}
			\left[
				\left( - 1 \right)^\alpha
				\mathcal{E}_n \left( g \right)
			\right]
		\right] \nonumber \\
&=
		\mathbb{E}
		\Biggl[
			\left[
				\mathbb{I} \left\{ \left\| \widehat{\vec{\theta}} - \vec{\theta}^\ast \right\|_2 \leq c \right\}
			+
				\mathbb{I} \left\{ \left\| \widehat{\vec{\theta}} - \vec{\theta}^\ast \right\|_2 > c \right\}
			\right]
			\sup_{g \in u^{-1} \left( \mathrm{co} \left( T \left( \widehat{\vec{\theta}}, \mathcal{B} \right) \right) \right)}
			\left[
				\left( - 1 \right)^\alpha
				\mathcal{E}_n \left( g \right)
			\right]
		\Biggr] \nonumber \\
&\leq
		\mathbb{E}
		\Biggl[
			\mathbb{I} \left\{ \left\| \widehat{\vec{\theta}} - \vec{\theta}^\ast \right\|_2 \leq c \right\}
			\sup_{g \in u^{-1} \left( \mathrm{co} \left( T \left( \widehat{\vec{\theta}}, \mathcal{B} \right) \right) \right)}
			\left[
				\left( - 1 \right)^\alpha
				\mathcal{E}_n \left( g \right)
			\right]
		\Biggr] \nonumber \\
		&\hspace{0.5cm}+
			\mathbb{E}
			\Biggl[
				\mathbb{I} \left\{ \left\| \widehat{\vec{\theta}} - \vec{\theta}^\ast \right\|_2 > c \right\}
				\sup_{g \in u^{-1} \left( \mathrm{co} \left( T \left( \widehat{\vec{\theta}}, \mathcal{B} \right) \right) \right)}
				\left[
					\left| \mathcal{E}_n \left( g \right) \right|
				\right]
			\Biggr] \nonumber \\
&\leq
		\mathbb{E}
		\left[
			\sup_{\varphi \in \ell \left( \cdot, \cdot
			\middle|
				u^{-1} \left( \mathrm{co} \left( T \left( \Theta_c, \mathcal{B} \right) \right) \right)
			\right)}
			\left\{
			\frac{1}{n}
			\sum_{i = 1}^{n}
			\varphi \left( Y_i, \vec{X}_i \right)
		-
			\int_{\mathbb{R} \times \mathbb{R}^{d}}
				\varphi \left( y, \vec{x} \right)
			f \left( y, \vec{x} \right) d y d \vec{x}
			\right\}
		\right] \nonumber \\
	&\hspace{0.5cm}+
		\mathbb{P} \left(
			\left\| \widehat{\vec{\theta}} - \vec{\theta}^\ast \right\|_2 > c
		\right)^{1 / 2}
		\mathbb{E}
		\Biggl[
			\sup_{g \in u^{-1} \left( \mathrm{co} \left( T \left( \widehat{\vec{\theta}}, \mathcal{B} \right) \right) \right)}
			\left[
				\left| \mathcal{E}_n \left( g \right) \right|^2
			\right]
		\Biggr]^{1 / 2} \nonumber \\
&\leq
		2
		\mathbb{E}
		\left[
			\mathbb{E}_{\vec{\sigma}}
			\left[
				\sup_{\varphi \in \ell \left( \cdot, \cdot
				\middle|
					u^{-1} \left( \mathrm{co} \left( T \left( \Theta_c, \mathcal{B} \right) \right) \right)
				\right)}
				\frac{1}{n}
				\sum_{i = 1}^{n}
					\sigma_i
				\cdot
					\varphi \left( Y_i, \vec{X}_i \right)
			\right]
		\right] \nonumber \\
		&\hspace{0.5cm}+
		\frac{1}{c}
		\mathbb{E} \left[ \left\| \widehat{\vec{\theta}} - \vec{\theta}^\ast \right\|_2^2 \right]^{1 / 2}
		\mathbb{E}
		\left[
			\left\{
				\frac{1}{n}
				\sum_{i = 1}^{n}
				F \left( Y_i \right)
			+
				\int_{\mathbb{R} \times \mathbb{R}^{d}}
					F \left( y \right)
				f \left( y, \vec{x} \right) d y d \vec{x}
			\right\}^2
		\right]^{1 / 2}\nonumber \\
&=
		2
		\mathbb{E}
		\left[
			\mathbb{E}_{\vec{\sigma}}
			\left[
				\sup_{g \in \mathrm{co} \left( T \left( \Theta_c, \mathcal{B} \right) \right) }
				\frac{1}{n}
				\sum_{i = 1}^{n}
					\sigma_i
				\cdot
					\left\{
						U^\ast \left( g \left( \vec{X}_i \right) \right)
					-
						Y_i
					\cdot
						g \left( \vec{X}_i \right)
					\right\}
			\right]
		\right] \nonumber \\
	&\hspace{0.5cm}+
		\frac{1}{c}
	\cdot
		O \left( \frac{1}{\sqrt{n}} \right)
	\cdot
		\mathbb{E}
		\left[
			\left|
				2 \sup_{z \in \left[ - \Delta_1, \Delta_1 \right]}
				\left| U^\ast \left( z \right) \right|
			+
				\Delta_1
				\left(
					\frac{1}{n}
					\sum_{i = 1}^{n}
					\left| Y_i \right|
				+
					\int_{\mathbb{R} \times \mathbb{R}^{d}}
						\left| y \right|
					f \left( y, \vec{x} \right) d y d \vec{x}
				\right)
			\right|^2
		\right]^{1 / 2} \nonumber \\
&=
		\mathsf{E}
	+
		O \left( \frac{1}{\sqrt{n}} \right). \label{eq:E+O}
\end{align}
To evaluate the above $\mathsf{E}$, we will apply Lemma \ref{lem:lip} with
$\ell_{\mathbb{X} i} \left( z \right) = U^\ast \left( z \right) - Y_i \cdot z$, $i = 1, \ldots, n$.
Denoting
		$\Delta'_2
=
		\sup_{\vec{\theta} \in \Theta_c}
		\sup_{\vec{x} \in \mathbb{R}^d}
		\left| m \left( \vec{x}, \vec{\theta} \right) \right|
	+
		\Delta_2
		\sup_{\vec{\theta} \in \Theta_c}
		\sup_{\vec{x} \in \mathbb{R}^d}
		\left| m \left( \vec{x}, \vec{\theta} \right) \right|^\gamma$,
the functions $\ell_{\mathbb{X} i} \left( z \right)$, $i = 1, \ldots, n$ with $z \in \left[ - \Delta'_2, \Delta'_2 \right]$ can be evaluated as follows:
from the mean value theorem, there exists $c_{z_1, z_2} \in \left( 0, 1 \right)$ such that
\begin{align*}
		\left|
			\ell_{\mathbb{X} i} \left( z_1 \right)
		-
			\ell_{\mathbb{X} i} \left( z_2 \right)
		\right|
&=
		\left| u^{-1} \left( z_1 + c_{z_1, z_2} \left( z_1 - z_2 \right) \right) - Y_i \right|
	\cdot
		\left| z_1 - z_2 \right| \\
&\leq
		\sup_{z \in \left[ - \Delta'_2, \Delta'_2 \right]} \left| u^{-1} \left( z \right) - Y_i \right|
	\cdot
		\left| z_1 - z_2 \right|.
\end{align*}
Thus, putting
\begin{align*}
		C_1
&=
		1
	+
		\mathbb{I} \left( \gamma = 1 \right)
		\left\{
		\sup_{\vec{\theta} \in \Theta_c}
		\sup_{\vec{x} \in \mathbb{R}^d}
		\left| m \left( \vec{x}, \vec{\theta} \right) \right|^\gamma + \Delta_2 \right\}, \\
		C_2
&=
		\mathbb{I} \left( \gamma = 0 \right)
	+
		\mathbb{I} \left( \gamma = 1 \right)
		\left\{
		\sup_{\vec{\theta} \in \Theta_c}
		\sup_{\vec{x} \in \mathbb{R}^d}
		\left| m \left( \vec{x}, \vec{\theta} \right) \right|^\gamma + \Delta_2 \right\}
\end{align*}
and
$\mathcal{M}^\gamma = \left\{ m \left( \cdot, \vec{\theta} \right)^\gamma \colon \mathbb{D} \to \mathbb{R} \middle| \vec{\theta} \in \Theta_c \right\}$,
where $\gamma = 1$ leads to $\mathcal{M}^\gamma = \mathcal{M}$, already defined in previous subsection.
Lemma \ref{lem:lip} and Lemma \ref{lem:cov_M} imply that
\begin{align}
		\mathsf{E}
&\leq
		\mathbb{E}
		\left[
			\mathbb{E}_{\vec{\sigma}}
			\left[
				\sup_{g \in \mathrm{co} \left( T \left( \Theta_c, \mathcal{B} \right) \right) }
				\frac{1}{n}
				\sum_{i = 1}^{n}
				\sigma_i
				\sup_{z \in \left[ - \Delta'_2, \Delta'_2 \right]}
				\left| u^{-1} \left( z \right) - Y_i \right|
					g \left( \vec{X}_i \right)
			\right]
		\right] \nonumber \\
&=
		\mathbb{E}
		\left[
			\mathbb{E}_{\vec{\sigma}}
			\left[
				\sup_{g \in T \left( \Theta_c, \mathcal{B} \right) }
				\frac{1}{n}
				\sum_{i = 1}^{n}
				\sigma_i
				\sup_{z \in \left[ - \Delta'_2, \Delta'_2 \right]}
				\left| u^{-1} \left( z \right) - Y_i \right|
					g \left( \vec{X}_i \right)
			\right]
		\right] \nonumber \\
&\leq
		\mathbb{E}
		\left[
			\mathbb{E}_{\vec{\sigma}}
			\left[
				\sup_{g \in \mathcal{M} + \mathcal{M}^\gamma \cdot \mathcal{B}}
				\frac{1}{n}
				\sum_{i = 1}^{n}
				\sigma_i
				\sup_{z \in \left[ - \Delta'_2, \Delta'_2 \right]}
				\left| u^{-1} \left( z \right) - Y_i \right|
					g \left( \vec{X}_i \right)
			\right]
		\right] \nonumber \\
&=
		\mathbb{E}
		\left[
			\mathbb{E}_{\vec{\sigma}}
			\left[
				\sup_{g \in \mathcal{M} }
				\frac{1}{n}
				\sum_{i = 1}^{n}
				\sigma_i
				\sup_{z \in \left[ - \Delta'_2, \Delta'_2 \right]}
				\left| u^{-1} \left( z \right) - Y_i \right|
					g \left( \vec{X}_i \right)
			\right]
		\right] \nonumber \\
	&\hspace{1cm}+
		\mathbb{E}
		\left[
			\mathbb{E}_{\vec{\sigma}}
			\left[
				\sup_{g \in \mathcal{M}^\gamma \cdot \mathcal{B}}
				\frac{1}{n}
				\sum_{i = 1}^{n}
				\sigma_i
				\sup_{z \in \left[ - \Delta'_2, \Delta'_2 \right]}
				\left| u^{-1} \left( z \right) - Y_i \right|
					g \left( \vec{X}_i \right)
			\right]
		\right] \nonumber \\
&=
		\mathbb{E}
		\left[
			\mathbb{E}_{\vec{\sigma}}
			\left[
				\sup_{g \in \mathcal{M} }
				\frac{1}{n}
				\sum_{i = 1}^{n}
				\sigma_i
				\sup_{z \in \left[ - \Delta'_2, \Delta'_2 \right]}
				\left| u^{-1} \left( z \right) - Y_i \right|
					g \left( \vec{X}_i \right)
			\right]
		\right] \nonumber \\
	&\hspace{1cm}+
		\mathbb{I} \left( \gamma = 0 \right)
		\mathbb{E}
		\left[
			\mathbb{E}_{\vec{\sigma}}
			\left[
				\sup_{g \in \mathcal{B}}
				\frac{1}{n}
				\sum_{i = 1}^{n}
				\sigma_i
				\sup_{z \in \left[ - \Delta'_2, \Delta'_2 \right]}
				\left| u^{-1} \left( z \right) - Y_i \right|
					g \left( \vec{X}_i \right)
			\right]
		\right] \nonumber \\
	&\hspace{1cm}+
		\mathbb{I} \left( \gamma = 1 \right)
		\mathbb{E}
		\left[
			\mathbb{E}_{\vec{\sigma}}
			\left[
				\sup_{g \in \mathcal{M} \cdot \mathcal{B}}
				\frac{1}{n}
				\sum_{i = 1}^{n}
				\sigma_i
				\sup_{z \in \left[ - \Delta'_2, \Delta'_2 \right]}
				\left| u^{-1} \left( z \right) - Y_i \right|
					g \left( \vec{X}_i \right)
			\right]
		\right] \nonumber \\
&\leq
		\mathbb{E}
		\left[
			\mathbb{E}_{\vec{\sigma}}
			\left[
				\sup_{g \in \mathcal{M} }
				\frac{1}{n}
				\sum_{i = 1}^{n}
				\sigma_i
				\sup_{z \in \left[ - \Delta'_2, \Delta'_2 \right]}
				\left| u^{-1} \left( z \right) - Y_i \right|
					g \left( \vec{X}_i \right)
			\right]
		\right] \nonumber \\
	&\hspace{1cm}+
		\mathbb{I} \left( \gamma = 0 \right)
		\mathbb{E}
		\left[
			\mathbb{E}_{\vec{\sigma}}
			\left[
				\sup_{g \in \mathcal{B}}
				\frac{1}{n}
				\sum_{i = 1}^{n}
				\sigma_i
				\sup_{z \in \left[ - \Delta'_2, \Delta'_2 \right]}
				\left| u^{-1} \left( z \right) - Y_i \right|
					g \left( \vec{X}_i \right)
			\right]
		\right] \nonumber \\
	&\hspace{1cm}+
		\mathbb{I} \left( \gamma = 1 \right)
		\left\{
		\sup_{\vec{\theta} \in \Theta_c}
		\sup_{\vec{x} \in \mathbb{R}^d}
		\left| m \left( \vec{x}, \vec{\theta} \right) \right|^\gamma + \Delta_2 \right\} \nonumber \\
	&\hspace{2cm}\times
		\Biggl\{
			\mathbb{E}
			\left[
				\mathbb{E}_{\vec{\sigma}}
				\left[
					\sup_{g \in \mathcal{M} }
					\frac{1}{n}
					\sum_{i = 1}^{n}
					\sigma_i
					\sup_{z \in \left[ - \Delta'_2, \Delta'_2 \right]}
					\left| u^{-1} \left( z \right) - Y_i \right|
						g \left( \vec{X}_i \right)
				\right]
			\right] \nonumber \\
		&\hspace{2.5cm}+
			\mathbb{E}
			\left[
				\mathbb{E}_{\vec{\sigma}}
				\left[
					\sup_{g \in \mathcal{B}}
					\frac{1}{n}
					\sum_{i = 1}^{n}
					\sigma_i
					\sup_{z \in \left[ - \Delta'_2, \Delta'_2 \right]}
					\left| u^{-1} \left( z \right) - Y_i \right|
						g \left( \vec{X}_i \right)
				\right]
			\right]
		\Biggr\} \nonumber \\
&=
		\left[
			1
		+
			\mathbb{I} \left( \gamma = 1 \right)
			\left\{
			\sup_{\vec{\theta} \in \Theta_c}
			\sup_{\vec{x} \in \mathbb{R}^d}
			\left| m \left( \vec{x}, \vec{\theta} \right) \right|^\gamma
		+
			\Delta_2 \right\}
		\right] \nonumber \\
	&\hspace{2cm}\times
		\mathbb{E}
		\left[
			\mathbb{E}_{\vec{\sigma}}
			\left[
				\sup_{g \in \mathcal{M} }
				\frac{1}{n}
				\sum_{i = 1}^{n}
				\sigma_i
				\sup_{z \in \left[ - \Delta'_2, \Delta'_2 \right]}
				\left| u^{-1} \left( z \right) - Y_i \right|
					g \left( \vec{X}_i \right)
			\right]
		\right] \nonumber \\
	&\hspace{1cm}+
		\left[
			\mathbb{I} \left( \gamma = 0 \right)
		+
			\mathbb{I} \left( \gamma = 1 \right)
			\left\{
				\sup_{\vec{\theta} \in \Theta_c}
				\sup_{\vec{x} \in \mathbb{R}^d}
				\left| m \left( \vec{x}, \vec{\theta} \right) \right|^\gamma + \Delta_2
			\right\}
		\right] \nonumber \\
	&\hspace{2cm}\times
		\mathbb{E}
		\left[
			\mathbb{E}_{\vec{\sigma}}
			\left[
				\sup_{g \in \mathcal{B}}
				\frac{1}{n}
				\sum_{i = 1}^{n}
				\sigma_i
				\sup_{z \in \left[ - \Delta'_2, \Delta'_2 \right]}
				\left| u^{-1} \left( z \right) - Y_i \right|
					g \left( \vec{X}_i \right)
			\right]
		\right] \nonumber \\
&=
		C_1
		\mathbb{E}
		\left[
			\mathbb{E}_{\vec{\sigma}}
			\left[
				\sup_{g \in \mathcal{M} }
				\frac{1}{n}
				\sum_{i = 1}^{n}
				\sigma_i
				\sup_{z \in \left[ - \Delta'_2, \Delta'_2 \right]}
				\left| u^{-1} \left( z \right) - Y_i \right|
					g \left( \vec{X}_i \right)
			\right]
		\right] \nonumber \\
	&\hspace{1cm}+
		C_2
		\mathbb{E}
		\left[
			\mathbb{E}_{\vec{\sigma}}
			\left[
				\sup_{g \in \mathcal{B}}
				\frac{1}{n}
				\sum_{i = 1}^{n}
				\sigma_i
				\sup_{z \in \left[ - \Delta'_2, \Delta'_2 \right]}
				\left| u^{-1} \left( z \right) - Y_i \right|
					g \left( \vec{X}_i \right)
			\right]
		\right]. \label{eq:sfE1}
\end{align}
We will apply Lemma \ref{lem:dudley} to \eqref{eq:sfE1}.
From triangle inequality, $\sqrt{a + b} \leq \sqrt{a} + \sqrt{b}$ ($a, b \geq 0$) and Jensen's inequality,
it holds that
\begin{align*}
		\mathbb{E} \left[ \sqrt{\sum_{i = 1}^{n} L_{\mathbb{X} i}} \right]
&=
		\mathbb{E} \left[ \sqrt{\sum_{i = 1}^{n} \sup_{z \in \left[ - \Delta'_2, \Delta'_2 \right]}
				\left| u^{-1} \left( z \right) - Y_i \right|} \right] \\
&\leq
		\sqrt{n}
		\sup_{z \in \left[ - \Delta'_2, \Delta'_2 \right]} \left| u^{-1} \left( z \right) \right|^{1/2}
	+
		\mathbb{E} \left[ \sqrt{\sum_{i = 1}^{n} \left| Y_i \right|} \right] \\
&\leq
		\sqrt{n}
		\sup_{z \in \left[ - \Delta'_2, \Delta'_2 \right]} \left| u^{-1} \left( z \right) \right|^{1/2}
	+
		\mathbb{E} \left[ \sum_{i = 1}^{n} \left| Y_i \right| \right]^{1/2} \\
&=
		\sqrt{n}
		\left\{
			\sup_{z \in \left[ - \Delta'_2, \Delta'_2 \right]} \left| u^{-1} \left( z \right) \right|^{1/2}
		+
			\mathbb{E} \left[ \left| Y_i \right| \right]^{1/2}
		\right\} \\
&=
		\left\langle \mathrm{\text{Constant for $n$}} \right\rangle
	\cdot
		\sqrt{n}.
\end{align*}
Noting that, by the change of variable $t = \ln \left( 1 / \epsilon \right) + C_4$, we have
\begin{align*}
		\int_0^{\min \{ c \cdot C_3, 2 \Delta_2 \}}
		\sqrt{\ln \frac{1}{\epsilon} + C_4} d \epsilon
&=
		\int_\infty^{\ln \left[ 1 / \min \{ c \cdot C_3, 2 \Delta_2 \} \right] + C_4} \sqrt{t} \cdot (-e^{C_4} \cdot e^{-t}) dt \\
&=
		e^{C_4} \int_{\ln \left[ 1 / \min \{ c \cdot C_3, 2 \Delta_2 \} \right] + {C_4}}^{\infty} t^{1/2} \cdot e^{-t} dt \\
&\leq
		e^{C_4} \cdot \Gamma \left( \frac{3}{2} \right) < \infty,
\end{align*}
Lemma \ref{lem:dudley} and Lemma \ref{lem:cov_M} imply that
\begin{align}
		\mathsf{E}
&\leq
		C_1
		\mathbb{E}
		\left[
			\mathbb{E}_{\vec{\sigma}}
			\left[
				\sup_{g \in \mathcal{M} }
				\frac{1}{n}
				\sum_{i = 1}^{n}
				\sigma_i
				\sup_{z \in \left[ - \Delta'_2, \Delta'_2 \right]}
				\left| u^{-1} \left( z \right) - Y_i \right|
					g \left( \vec{X}_i \right)
			\right]
		\right] \nonumber\\
	&\hspace{1cm}+
		C_2
		\mathbb{E}
		\left[
			\mathbb{E}_{\vec{\sigma}}
			\left[
				\sup_{g \in \mathcal{B}}
				\frac{1}{n}
				\sum_{i = 1}^{n}
				\sigma_i
				\sup_{z \in \left[ - \Delta'_2, \Delta'_2 \right]}
				\left| u^{-1} \left( z \right) - Y_i \right|
					g \left( \vec{X}_i \right)
			\right]
		\right] \nonumber\\
&\leq
		C_1
		\mathbb{E}
		\left[
			32
		\frac{\sqrt{\sum_{i = 1}^{n}
				\sup_{z \in \left[ - \Delta'_2, \Delta'_2 \right]}
				\left| u^{-1} \left( z \right) - Y_i \right| }}
		{n}
		\int_0^{2 \Delta_2}
			\sqrt{\ln N \left( \epsilon, \mathcal{M}, \rho \right)}
		d \epsilon
		\right] \nonumber\\
	&\hspace{1cm}+
		C_2
		\mathbb{E}
		\left[
			32
		\frac{\sqrt{\sum_{i = 1}^{n}
				\sup_{z \in \left[ - \Delta'_2, \Delta'_2 \right]}
				\left| u^{-1} \left( z \right) - Y_i \right| }}
		{n}
		\int_0^{2 \Delta_2}
			\sqrt{\ln N \left( \epsilon, \mathcal{B}, \rho \right)}
		d \epsilon
		\right] \nonumber\\
&\leq
		\frac{\left\langle \mathrm{\text{Constant for $n$}} \right\rangle}{\sqrt{n}}
		\max \left\{C_1, C_2 \right\}
	\cdot
		32 \nonumber\\
	&\hspace{0.5cm}\times
		\left\{
			\mathbb{E}
			\left[
			\int_0^{2 \Delta_2}
				\sqrt{\ln N \left( \epsilon, \mathcal{M}, \| \cdot \|_\infty \right)}
			d \epsilon
			\right]
		+
			\mathbb{E}
			\left[
			\int_0^{2 \Delta_2}
				\sqrt{\ln N \left( \epsilon, \mathcal{B}, \| \cdot \|_\infty \right)}
			d \epsilon
			\right]
		\right\}\nonumber\\
&\leq
		\frac{\left\langle \mathrm{\text{Constant for $n$}} \right\rangle}{\sqrt{n}}
		\Biggl\{
			\int_0^{2 \Delta_2}
				\sqrt{\ln
				\left[
					\mathbb{I}
					\left(
						\epsilon
					<
						c \cdot C_3 \right)
					\cdot
					C_4
					\cdot
					\left( \frac{1}{\epsilon} \right)^p
				+
					\mathbb{I} \left( \epsilon \geq c \cdot C_3 \right)
					\right]}
			d \epsilon \nonumber\\
		&\hspace{4cm}+
			\int_0^{2 \Delta_2}
				\sqrt{\ln N \left( \epsilon, \mathcal{B}, \| \cdot \|_\infty \right)}
			d \epsilon
		\Biggr\}\nonumber\\
&=
		\frac{\left\langle \mathrm{\text{Constant for $n$}} \right\rangle}{\sqrt{n}}
		\Biggl[
			\left\{
				\int_0^{\min \left\{ c \cdot C_3, 2 \Delta_2 \right\}}
					\sqrt{p \ln \frac{1}{\epsilon}
					+
					\ln C_4}
				d \epsilon
			+
				\int_{c \cdot C_3}^{\max \left\{ c \cdot C_3, 2 \Delta_2 \right\}}
					1
				d \epsilon
			\right\} \nonumber\\
		&\hspace{1.5cm}+
			\left\{
				\int_0^{\min \left\{ 1, 2 \Delta_2 \right\}}
					\sqrt{\ln N \left( \epsilon, \mathcal{B}, \| \cdot \|_\infty \right)}
				d \epsilon
			+
				\int_1^{\max \left\{ 1, 2 \Delta_2 \right\}}
					sqrt{\ln N \left( \epsilon, \mathcal{B}, \| \cdot \|_\infty \right)}
				d \epsilon
			\right\}
		\Biggr] \nonumber\\
&=
		\frac{\left\langle \mathrm{\text{Constant for $n$}} \right\rangle}{\sqrt{n}}
		\Biggl[
			\left\{
				e^{C_4} \cdot \Gamma \left( \frac{3}{2} \right)
			+
				\max \left\{ c \cdot C_3, 2 \Delta_2 \right\}
			-
				1
			\right\} \nonumber\\
		&\hspace{1cm}+
			\left\{
				\int_0^{\min \left\{ 1, 2 \Delta_2 \right\}}
				\sqrt{\ln N \left( \epsilon, \mathcal{B}, \| \cdot \|_\infty \right)}
				d \epsilon
			+
				\sqrt{\ln N \left( 1, \mathcal{B}, \| \cdot \|_\infty \right)}
				\left\{
					\max \left\{ 1, 2 \Delta_2 \right\}
				-
					1
				\right\}
			\right\}
		\Biggr] \nonumber\\
&\leq
		\frac{1}{\sqrt{n}} \cdot \left\langle \mathrm{\text{Constant for $n$}} \right\rangle. \label{eq:sfE2}
\end{align}
The above \eqref{eq:E+O}, \eqref{eq:sfE1} and \eqref{eq:sfE2} finally yield the conclusion. \qed

\subsection{Proof of Theorem \ref{thm:mu0}}
Putting
$\vec{\theta}_1 \left( \mathbb{X} \right) = \vec{\theta}_2 \left( \mathbb{X} \right) = \widehat{\vec{\theta}}$,
$\delta \left( \mathbb{X} \right) = \varepsilon$
and
$\mu \left( \cdot, \Bar{\Lambda} \left( \mathbb{X} \right), \vec{\theta}_1 \left( \mathbb{X} \right) \right) = \mu^0$
in Lemma \ref{lem:ineq2},
\begin{align*}
		&\mathbb{E} \left[ D_U \left( \mu, \widehat{\mu} \right) \right] \\
&\leq
		\mathbb{E}
		\left[
			\inf_{g \in u^{-1} \left( \mathrm{co} \left( T \left( \widehat{\vec{\theta}}, \mathcal{D} \right) \right) \right) }
			D_U \left( \mu, g \right)
		\right]\nonumber \\
	&\hspace{0.5cm}+
		\mathbb{E}
		\left[ \sup_{\Lambda \in \mathcal{W}}
			\mathcal{E}_n \left( \mu \left( \cdot, \Lambda, \widehat{\vec{\theta}} \left( \mathbb{X} \right) \right) \right)
		\right]
	+
		\mathbb{E}
		\left[
			\sup_{\Lambda \in \mathcal{W}}
			\left[ - \mathcal{E}_n \left( \mu \left( \cdot, \Lambda, \widehat{\vec{\theta}} \left( \mathbb{X} \right) \right)	\right) \right]
		\right]
	+
		\mathbb{E} \left[ \varepsilon \right] \\
&=
		\mathbb{E}
		\left[
			\inf_{g \in u^{-1} \left( \mathrm{co} \left( T \left( \widehat{\vec{\theta}}, \mathcal{D} \right) \right) \right) }
			D_U \left( \mu, g \right)
		\right]
	+
		\sum_{\alpha = 0}^1
		\mathbb{E}
		\left[
			\sup_{g \in u^{-1} \left( \mathrm{co} \left( T \left( \widehat{\vec{\theta}}, \mathcal{D} \right) \right) \right)}
			\left[
				\left( - 1 \right)^\alpha
				\mathcal{E}_n \left( g \right)
			\right]
		\right]
	+
		\varepsilon,
\end{align*}
which shows the inequlity \eqref{eq:mu_hat}. \qed

\subsection{Proof of Proposition \ref{prop:D012}} \label{sec:proof_theo_dict}
Defining $\mathcal{P}_0 = \mathbb{R} \times \mathbb{R}^d \times \mathbb{H}_0$, we put
\begin{align*}
		B_0 \left( \cdot \middle| \vec{\theta}, \mathcal{Z}, \mathcal{P}_0 \right)
&=
		B_0 \left( \cdot \middle| \vec{\theta}, \mathcal{Z}, \left( v, \vec{w} \right), H \right) \\
&=
		\frac{
			K_H \left( \vec{w} - \cdot \right)
			m \left( \vec{w}, \vec{\theta} \right)^\gamma
			\left\{ v - u^{-1}\left( m \left( \vec{w}, \vec{\theta} \right) \right) \right\}
		}{
			(1 / n) \sum_{j = 1}^{n}
			K_H \left( \vec{w}_j - \cdot \right)
			m \left( \vec{w}_j, \vec{\theta} \right)^{2 \gamma}
			\left( U^\ast \right)'' \left( m \left( \vec{w}_j, \vec{\theta} \right) \right)
		}, \\
& \vec{\theta} \in \Theta, \left( \left( v, \vec{w} \right), H \right) \in \mathcal{P}_0,
\end{align*}
which has the same form as the word introduced below Lemma \ref{lem:ineq2} in Section \ref{sec:theo}.
Then we set $\mathcal{P}^\ast_0 = \left\{ \left( \left( y_i, \vec{x}_i \right), H_s \right) \in \mathbb{R} \times \mathbb{R}^d \times \mathbb{H}_0 \middle| i = 1, \ldots n, ~ s = 1, \ldots, S_0 \right\}$ for some $S_0 \in \mathbb{N}$
and
\begin{align}
		&B_0 \left( \cdot \middle| \widehat{\vec{\theta}} \left( \mathcal{X} \right), \mathcal{X}, \left( y_i, \vec{x}_i \right), H_s \right) \nonumber \\
&=
		\frac{
			K_{H_s} \left( \vec{x}_i - \cdot \right)
			m \left( \vec{x}_i, \widehat{\vec{\theta}} \left( \mathcal{X} \right) \right)^\gamma
			\left\{ y_i - u^{-1}\left( m \left( \vec{x}_i, \widehat{\vec{\theta}} \left( \mathcal{X} \right) \right) \right) \right\}
		}{
			(1 / n) \sum_{j = 1}^{n}
			K_{H_s} \left( \vec{x}_j - \cdot \right)
			m \left( \vec{x}_j, \widehat{\vec{\theta}} \left( \mathcal{X} \right) \right)^{2 \gamma}
			\left( U^\ast \right)'' \left( m \left( \vec{x}_j, \widehat{\vec{\theta}} \left( \mathcal{X} \right) \right) \right)
		}, \label{eq:B0} \\
&\left( \left( y_i, \vec{x}_i \right), H_s \right) \in \mathcal{P}^\ast_0, \nonumber
\end{align}
which coincides with a word in $\mathcal{D}_0$.

Words for $\mathcal{D}_1$ and $\mathcal{D}_2$ can be written in a similar way:
setting $\mathcal{P}_1 = \mathbb{R} \times \mathbb{R} \times \mathbb{H}_1$,
\begin{align*}
		&B_1 \left( \cdot \middle| \vec{\theta}, \mathcal{Z}, \left( v, w \right), H \right)
=
		\frac{
			K_H \left( w - \cdot \right)
			m_k \left( w, \vec{\theta} \right)^\gamma
			\left\{ v - u^{-1} \left( m_k \left( w, \vec{\theta} \right) \right) \right\}
		}{
			(1 / n) \sum_{j = 1}^{n}
			K_H \left( w_{jk} - \cdot \right)
			m_k \left( w_{jk}, \vec{\theta} \right)^{2 \gamma}
			\left( U^\ast \right)'' \left( m_k \left( w_{jk}, \vec{\theta} \right) \right)
		}, \\
& \vec{\theta} \in \Theta, \left( \left( v, w \right), H \right) \in \mathcal{P}_2, 1 \leq k \leq d,
\end{align*}
and
		$\mathcal{P}^\ast_1
=
		\left\{
				\left( \left( y_i, x_{ik} \right), H_s \right) \in \mathbb{R} \times \mathbb{R} \times \mathbb{H}_1
		\middle|
			i = 1, \ldots n, ~ 1 \leq k \leq d, ~ s = 1, \ldots, S_1
		\right\}$
for $S_1 \in \mathbb{N}$,
\begin{align}
		&B_1 \left( \cdot \middle| \widehat{\vec{\theta}} \left( \mathcal{X} \right), \mathcal{X}, \left( y_i, x_{ik} \right), H_s \right) \nonumber \\
&=
		\frac{
			K_{H_s} \left( x_{ik} - \cdot \right)
			m_k \left( x_{ik}, \widehat{\vec{\theta}} \left( \mathcal{X} \right) \right)^\gamma
			\left\{ y_i - u^{-1} \left( m_k \left( x_{ik}, \widehat{\vec{\theta}} \left( \mathcal{X} \right) \right) \right) \right\}
		}{
			(1 / n) \sum_{j = 1}^{n}
			K_{H_s} \left( x_{jk} - \cdot \right)
			m_k \left( x_{jk}, \widehat{\vec{\theta}} \left( \mathcal{X} \right) \right)^{2 \gamma}
			\left( U^\ast \right)'' \left( m_k \left( x_{jk}, \widehat{\vec{\theta}} \left( \mathcal{X} \right) \right) \right)
		}, \label{eq:B1} \\
&\left( \left( y_i, x_{ik} \right), H_s \right) \in \mathcal{P}^\ast_1; \nonumber
\end{align}
setting $\mathcal{P}_2 = \mathbb{R} \times \mathbb{R}^2 \times \mathbb{H}_2$,
\begin{align*}
		&B_2 \left( \cdot \middle| \vec{\theta}, \mathcal{Z}, \left( v, \vec{w} \right), H \right) \\
&=
		\frac{
			K_H \left( \vec{w} - \cdot \right)
			m_{k \ell} \left( \vec{w}, \vec{\theta} \right)^\gamma
			\left\{ v - u^{-1} \left( m_{k \ell} \left( \vec{w}, \vec{\theta} \right) \right) \right\}
		}{
			(1 / n) \sum_{j = 1}^{n}
			K_H \left( \vec{w}_{j k \ell} - \cdot \right)
			m_{k \ell} \left( \vec{w}_{j k \ell}, \vec{\theta} \right)^{2 \gamma}
			\left( U^\ast \right)'' \left( m_{k \ell} \left( \vec{w}_{j k \ell}, \vec{\theta} \right) \right)
		}, \\
&\vec{\theta} \in \Theta, \left( \left( v, \vec{w} \right), H \right) \in \mathcal{P}_2, 1 \leq k < \ell \leq d,
\end{align*}
and
		$\mathcal{P}^\ast_2
=
		\left\{
			\left( \left( y_i, \vec{x}_{i k \ell} \right), H_s \right) \in \mathbb{R} \times \mathbb{R}^2 \times \mathbb{H}_2
		\middle|
			i = 1, \ldots n, ~ 1 \leq k < \ell \leq d, ~ s = 1, \ldots, S_2
		\right\}$
for $S_2 \in \mathbb{N}$,
\begin{align}
		&B_2 \left( \cdot \middle| \widehat{\vec{\theta}} \left( \mathcal{X} \right), \mathcal{X}, \left( y_i, \vec{x}_{i k \ell} \right), H_s \right) \nonumber \\
&=
		\frac{
			K_{H_s} \left( \vec{x}_{i k \ell} - \cdot \right)
			m_{k \ell} \left( \vec{x}_{i k \ell}, \widehat{\vec{\theta}} \left( \mathcal{X} \right) \right)^\gamma
			\left\{ y - u^{-1} \left( m_{k \ell} \left( \vec{x}_{i k \ell}, \widehat{\vec{\theta}} \left( \mathcal{X} \right) \right) \right) \right\}
		}{
			(1 / n) \sum_{j = 1}^{n}
			K_{H_s} \left( \vec{x}_{j k \ell} - \cdot \right)
			m_{k \ell} \left( \vec{x}_{jk \ell}, \widehat{\vec{\theta}} \left( \mathcal{X} \right) \right)^{2 \gamma}
			\left( U^\ast \right)'' \left( m_{k \ell} \left( \vec{x}_{jk \ell}, \widehat{\vec{\theta}} \left( \mathcal{X} \right) \right) \right)
		}, \label{eq:B2} \\
&\left( \left( y_i, \vec{x}_{i k \ell} \right), H_s \right) \in \mathcal{P}^\ast_2. \nonumber
\end{align}

Now we fix an element in $\mathcal{P}^\ast_0$, $\mathcal{P}^\ast_1$, $\mathcal{P}^\ast_2$, and briefly denote the above
\[
		B_0 \left( \cdot \middle| \widehat{\vec{\theta}} \left( \mathcal{X} \right), \mathcal{X}, \left( y_i, \vec{x}_i \right), H_s \right),
		B_1 \left( \cdot \middle| \widehat{\vec{\theta}} \left( \mathcal{X} \right), \mathcal{X}, \left( y_i, x_{ik} \right), H_s \right),
		B_2 \left( \cdot \middle| \widehat{\vec{\theta}} \left( \mathcal{X} \right), \mathcal{X}, \left( y_i, \vec{x}_{i k \ell} \right), H_s \right)\]
on $\mathbb{D}$ by $B_0 \left( \vec{x} \right), B_1 \left( \vec{x} \right), B_2 \left( \vec{x} \right)$ for $\vec{x} \in \mathbb{D}$.
From \eqref{eq:B0}, \eqref{eq:B1} and \eqref{eq:B2}, the functions
$B_0 $,
$B_1 $,
$B_2 $
belong to the class $C^\infty \left( \mathbb{D} \right)$ of smooth functions on $\mathbb{D}$, since they depend on the input in $\mathbb{D}$ only through the smooth function $K$.

We shall show the words $B_0 $, $B_1 $, $B_2 $ belong to a function class $C_M^\alpha \left( \mathbb{D} \right)$, which appears in Section 2.7.1 of \citet*{VW1996}.
Since $B_0, B_1, B_2 \in C^\infty \left( \mathbb{D} \right)$, and continuous functions on compact sets are bounded,
we can see
\[
		\sup_{\vec{x} \in \mathbb{D}}
		D^{\vec{k}}
		B_i \left( \vec{x} \right) < \infty, i = 0,1,2
\]
for any multi-index $\vec{k} = \left[ k_1 \cdots k_d \right]$, where $ D^{\vec{k}} = \partial^{|\vec{k}|} / \left( \partial \vec{x}^{k_1} \cdots \partial \vec{x}^{k_d} \right)$.
Moreover, the mean value theorem implies that there exists $t \in \left( 0,1 \right)$ such that
\begin{align*}
		\sup_{\vec{x}, \vec{y} \in \mathbb{D}}
		\frac{
		\left|
			D^{\vec{k}}
			B_i \left( \vec{x} \right)
		-
			D^{\vec{k}}
			B_i \left( \vec{y} \right)
		\right|}
		{\left\| \vec{x} - \vec{y} \right\|_2}
&=
		\sup_{\vec{x}, \vec{y} \in \mathbb{D}}
		\frac{
		\left|
			\left. (\partial / \partial \vec{z}^\top)
				D^{\vec{k}}
				B_i \left( \vec{z} \right) \right|_{\vec{z} = t \vec{x} + \left( 1 - t \right) \vec{y}}
		\left( \vec{x} - \vec{y} \right)
		\right|}
		{\left\| \vec{x} - \vec{y} \right\|_2} \\
&\leq
		\sup_{\vec{x}, \vec{y} \in \mathbb{D}}
		\sup_{t \in \left( 0,1 \right)}
			\frac{
				\left\|
					\left. (\partial / \partial \vec{z})
						D^{\vec{k}}
						B_i \left( \vec{z} \right) \right|_{\vec{z} = t \vec{x} + \left( 1 - t \right) \vec{y}}
				\right\|_2
				\left\| \vec{x} - \vec{y} \right\|_2
			}{
			\left\| \vec{x} - \vec{y} \right\|_2} \\
&\leq
		\sup_{\vec{x} \in \mathrm{co} \left( \mathbb{D} \right)}
		\left\|
			\frac{\partial}{\partial \vec{x}}
			D^{\vec{k}}
			B_i \left( \vec{x} \right)
		\right\|_2 \\
&<\infty.
\end{align*}
Thus by putting
\begin{align*}
		M
&=
		\max_{|\vec{k}| \leq \alpha - 1}
		\sup_{\vec{x} \in \mathbb{D}}
		D^{\vec{k}}
		B_i \left( \vec{x} \right)
	+
		\max_{|\vec{k}| = \alpha - 1}
		\sup_{\vec{x} \in \mathbb{D}}
		\sup_{\vec{x} \in \mathrm{co} \left( \mathbb{D} \right)}
		\left\|
			\frac{\partial}{\partial \vec{x}}
			D^{\vec{k}}
			B_i \left( \vec{x} \right)
		\right\|_2 \\
&<
	\infty,
\end{align*}
it holds that
$\mathcal{D}_0 \left( \Delta_1, \Delta_2, c \right), \mathcal{D}_1 \left( \Delta_1, \Delta_2, c \right), \mathcal{D}_2 \left( \Delta_1, \Delta_2, c \right) \subset C_M^\alpha \left( \mathbb{D} \right)$ for $\alpha > d / 2$.
We now evaluate the integral of the covering number of $C_M^\alpha \left( \mathbb{D} \right)$.
Theorem 2.7.1 of \citet*{VW1996} implies that there exists a constant $C$ such that
\begin{align*}
		\int_0^1
			\sqrt{\ln \mathcal{N} \left( \varepsilon, C_M^\alpha \left( \mathbb{D} \right), \| \cdot \|_\infty \right)}
		d \varepsilon
&=
		\int_0^1
			\sqrt{\ln \mathcal{N} \left( \varepsilon, M^{-1} \cdot C_1^\alpha \left( \mathbb{D} \right), \| \cdot \|_\infty \right)}
		d \varepsilon \\
&=
		\int_0^1
			\sqrt{\ln \mathcal{N} \left( \varepsilon M, C_1^\alpha \left( \mathbb{D} \right), \| \cdot \|_\infty \right)}
		d \varepsilon \\
&\leq
		\int_0^1
			\sqrt{\left[ C \left( \varepsilon M \right)^{- d / \alpha} \right]}
		d \varepsilon \\
&=
		\sqrt{C} M^{-d /(2 \alpha)} \left[ \frac{\varepsilon^{- d / (2 \alpha)}}{1- d /(2 \alpha)} \right]_0^1 \\
&< \infty. \tag*{$\Box$}
\end{align*}

\subsection{Proof of Lemma \ref{lem:theta}}
Note that
\begin{align*}
		\argmin_{\vec{\theta} \in \mathbb{R}^{d+1}} D_U \left( \mu, u^{-1} \left( \left[ 1 ~ \cdot^\top \right] \vec{\theta} \right) \right)
&=
		\argmin_{\vec{\theta}}
			\int_{\mathbb{R}^d}
				d_U \left( \mu \left( \vec{x} \right), u^{-1} \left( m \left( \vec{x}, \vec{\theta} \right) \right) \right)
			q \left( \vec{x} \right) d \vec{x} \\
&=
		\argmin_{\vec{\theta}}
		\int_{\left[ -1, 1 \right]^d}
			\left\{
				U^\ast \left(
					\left[ 1 ~ \vec{x}^\top \right]
					\vec{\theta}
				\right)
			-
				\mu \left( \vec{x} \right)
			\cdot
				\left[ 1 ~ \vec{x}^\top \right]
				\vec{\theta}
			\right\}
		d \vec{x} \\
&=
			\argmin_{\vec{\theta}}
			L \left( \vec{\theta} \right)
\end{align*}
is unique, since the function $\vec{\theta} \mapsto L \left( \vec{\theta} \right)$ is strictly convex.
For $\vec{\theta}^\ast$, we set
\[
		\vec{\theta}'
=
		\left[ \theta^\ast_0 ~ - \theta^\ast_1 ~ \theta^\ast_2 ~ \theta^\ast_3 ~ \theta^\ast_4 ~ - \theta^\ast_5 ~ - \theta^\ast_6 ~ \cdots ~ - \theta^\ast_d \right]^\top.
\]
Then from the fact
\[
		\mu \left( \left[ - x_1 ~ x_2 ~ x_3 ~ x_4 ~ - x_5 ~ - x_6 ~ \cdots ~ - x_d \right]^\top \right) = \mu \left( \vec{x} \right)
\]
and by the definition of $\eta \left( \vec{x} \right)$, it follows that
\begin{align*}
		&L \left( \vec{\theta}' \right) \\
&=
		\int_{\left[ -1, 1 \right]^d}
			\left\{
				U^\ast \left(
					\left[ 1 ~ \vec{x}^\top \right]
					\vec{\theta}'
				\right)
			-
				\mu \left( \vec{x} \right)
			\cdot
				\left[ 1 ~ \vec{x}^\top \right]
				\vec{\theta}'
			\right\}
		d \vec{x} \\
&=
		\int_{\left[ -1, 1 \right]^d}
			\bigl\{
				U^\ast \left(
					\left[ - x_1 ~ x_2 ~ x_3 ~ x_4 ~ - x_5 ~ - x_6 ~ \cdots ~ - x_d \right]
					\vec{\theta}'
				\right) \\
			&\hspace{0.3cm}-
				\mu \left( \left[ - x_1 ~ x_2 ~ x_3 ~ x_4 ~ - x_5 ~ - x_6 ~ \cdots ~ - x_d \right] \right)
			\cdot
				\left[1 ~ - x_1 ~ x_2 ~ x_3 ~ x_4 ~ - x_5 ~ - x_6 ~ \cdots ~ - x_d \right]
				\vec{\theta}'
			\bigr\}
		1 d \vec{x} \\
&=
		\int_{\left[ -1, 1 \right]^d}
			\left\{
				U^\ast \left(
					\left[ 1 ~ \vec{x}^\top \right]
					\vec{\theta}^\ast
				\right)
			-
				\mu \left( \vec{x} \right)
			\cdot
				\left[ 1 ~ \vec{x}^\top \right]
				\vec{\theta}^\ast
			\right\}
		d \vec{x} \\
&=
		L \left( \vec{\theta}^\ast \right).
\end{align*}
The uniqueness of $\vec{\theta}^\ast$ implies $\vec{\theta}' = \vec{\theta}^\ast$, and then $\theta^\ast_1 = \theta^\ast_5 = \cdots = \theta^\ast_d =0$.

If $u = \mathrm{id}$, we can directly calculate $\vec{\theta}^\ast$ as follows:
\begin{align*}
		\vec{\theta}^\ast
&=
		\argmin_{\vec{\theta}}
		\int_{\mathbb{R}^d}
			\left\{
			\mu \left( \vec{x} \right)
			-
			\vec{\theta}^\top
			\begin{bmatrix}
				1 \\
				\vec{x}
			\end{bmatrix}
			\right\}^2
		q \left( \vec{x} \right) d \vec{x} \\
&=
		\argmin_{\vec{\theta}}
		\int_{\left[ -1, 1 \right]^d}
			\left\{
			\mu \left( \vec{x} \right)
			-
			\vec{\theta}^\top
			\begin{bmatrix}
				1 \\
				\vec{x}
			\end{bmatrix}
			\right\}^2
		d \vec{x} \\
&=
		\argmin_{\vec{\theta}}
		\left[
			2 \vec{\theta}^\top
			\int_{\left[ -1, 1 \right]^d}
				\mu \left( \vec{x} \right)
				\begin{bmatrix}
					1 \\
					\vec{x}
				\end{bmatrix}
			d \vec{x}
		+
			\vec{\theta}^\top
			\int_{\left[ -1, 1 \right]^d}
				\begin{bmatrix}
					1 \\
					\vec{x}
				\end{bmatrix}
				\left[ 1 ~ \vec{x}^\top \right]
			d \vec{x}
			\vec{\theta}
		\right] \\
&=
		\sol_{\vec{\theta}}
		\left[
			2\int_{\left[ -1, 1 \right]^d}
				\mu \left( \vec{x} \right)
				\begin{bmatrix}
					1 \\
					\vec{x}
				\end{bmatrix}
			d \vec{x}
		+
			2
			\int_{\left[ -1, 1 \right]^d}
				\begin{bmatrix}
					1 \\
					\vec{x}
				\end{bmatrix}
				\left[ 1 ~ \vec{x}^\top \right]
			d \vec{x}
			\vec{\theta} = 0
		\right] \\
&=
		\sol_{\vec{\theta}}
		\left[
			\begin{bmatrix}
				2^{d-1}
				\cdot
				\left\{ 2 / 3 + 0 + \left( e - e^{-1} \right) + 0 \right\} \\
				2^{d-1}
				\begin{bmatrix}
					0+0+0+0 \\
					0+2 / 3+0+0 \\
					0+0+2 / e+0 \\
					0+0+0+2 / \pi \\
				\end{bmatrix}\\
				\vec{0}_{d-4}
			\end{bmatrix}
		+
			\begin{bmatrix}
				2^d & \vec{0}_d^\top \\
				\vec{0}_d & \left( 2^d / 3 \right) I_d
			\end{bmatrix}
			\vec{\theta}
		= 0
		\right] \\
&=
		\begin{bmatrix}
			1 / 3 + \sinh (1) \\
			0 \\
			1 \\
			3 / e \\
			3 / \pi\\
		\vec{0}_{d-4}
		\end{bmatrix}.
\end{align*}
\end{appendices}

\addcontentsline{toc}{section}{Acknowledgments}
\section*{Acknowledgments}
The research of the second author is supported in part by MEXT, Japan KAKENHI Grant Number 23K28043.

\bibliographystyle{apalike}

\end{document}